\newcommand{\Up}{\Upsilon}
\newcommand{\la}{\lambda}
\newcommand{\si}{\sigma}
\newcommand{\ve}{\varepsilon}
\newcommand{\B}{\bar}
\newcommand{\ti}{\tilde}
\newcommand{\ph}{\phantom}
\newcommand{\ul}{\underline}

\newcommand{\beq}{\begin{equation}}
\newcommand{\eeq}{\end{equation}}
\newcommand{\ba}{\begin{array}}
\newcommand{\ea}{\end{array}}
\newcommand{\bt}{\begin{tabular}}
\newcommand{\et}{\end{tabular}}
\newcommand{\bea}{\begin{eqnarray}}
\newcommand{\eea}{\end{eqnarray}}
\newcommand{\bean}{\begin{eqnarray*}}
\newcommand{\eean}{\end{eqnarray*}}
\newcommand{\eref}[1]{(\ref{#1})}
\newcommand{\tref}[1]{Table~\ref{#1}}

\newcommand{\nn}{\nonumber}
\newcommand{\comment}[1]{}

\newcommand{\overbar}[1]{\overline{#1}}

\newtheorem{theorem}{\sf THEOREM}

\newcommand{\rk}{\mathop{{\rm rk}}}
\newcommand{\ind}{\mathop{{\rm ind}}}
\newcommand{\ch}{\mathop{{\rm ch}}}
\newcommand{\td}{\mathop{{\rm Td}}}
\newcommand{\coker}{\mathop{{\rm coker}}}
\newcommand{\im}{\mathop{{\rm im}}}
\newcommand{\IC}{\mathbb{C}}
\newcommand{\IP}{\mathbb{P}}
\newcommand{\IZ}{\mathbb{Z}}
\newcommand{\cO}{{\cal O}}
\newcommand{\cL}{{\cal L}}
\newcommand{\cN}{{\cal N}}
\newcommand{\cV}{{\cal V}}
\newcommand{\cB}{{\cal B}}
\newcommand{\cC}{{\cal C}}
\newcommand{\cK}{{\cal K}}
\newcommand{\cW}{{\cal W}}

\newcommand{\II}{\mathbb{I}}

\newcommand{\cA}{{\cal A}}

\newcommand{\tconf}[1]{{\tiny \left[\begin{matrix}#1\end{matrix}\right]}}

\newcommand{\sseq}[3]{0 \to #1 \to #2 \to #3 \to 0}

\def\cjn1{{\cA, \cC^*\otimes \wedge^j \cN^*}}
\def\bjn1{{\cA, \cB^*\otimes \wedge^j \cN^*}}
\def\vjn1{{\cA, \cV^*\otimes \wedge^j \cN^*}}
\def\cjn2{{\cA, \cC\otimes \wedge^j \cN^*}}
\def\bjn2{{\cA, \cB\otimes \wedge^j \cN^*}}
\def\vjn2{{\cA, \cV\otimes \wedge^j \cN^*}}
\def\sumi{\sum\limits_{i=1}^{r_B}}
\def\sumj{\sum\limits_{j=1}^{r_C}}

\def\cjn{{\cA, S^2 \cC^* \otimes \wedge^j \cN^*}}
\def\cbjn{{\cA, \cC^* \otimes \cB^*\otimes \wedge^j \cN^*}}
\def\bjn{{\cA, S^2 \cB^* \otimes \wedge^j \cN^*}}

\def\cv2s{\wedge^2 {\cal V}^*}

\documentclass[11pt]{ociamthesis}

\usepackage[sumlimits]{amsmath}
\usepackage{amsfonts,amssymb,latexsym,mathrsfs,enumerate,bbm,bbold,epsfig,setspace}
\usepackage{graphicx,cancel}

\title{Heterotic and M-theory Compactifications for String Phenomenology}
\author{Lara Briana Anderson}
\college{Magdalen College}
\degree{Doctor of Philosophy}
\degreedate{Trinity 2008}

\begin{document}

\baselineskip=21pt plus1pt

\setcounter{secnumdepth}{3}
\setcounter{tocdepth}{3}

\maketitle

\begin{table}[c]
{\emph{``Although the universe is under no obligation to make sense, students in pursuit of the PhD are."}} \\
\begin{flushright}
-Robert P. Kirshner
\end{flushright}
\end{table}
\begin{abstractseparate}

In this thesis, we explore two approaches to string phenomenology. In the first half of the work, we investigate compactifications of M-theory on spaces with co-dimension four, orbifold singularities. We construct M-theory on $\mathbb{C}^2/\mathbb{Z}_N$ by coupling $11$-dimensional supergravity to a seven-dimensional Yang-Mills theory located on the orbifold fixed-plane. The resulting action is supersymmetric to leading non-trivial order in the $11$-dimensional Newton constant. We thereby reduce M-theory on a $G_2$ orbifold with $\mathbb{C}^2/\mathbb{Z}_N$ singularities, explicitly incorporating the additional gauge fields at the singularities. We derive the K\"ahler potential, gauge-kinetic function and superpotential for the resulting $N=1$ four-dimensional theory. Blowing-up of the orbifold is described by a Higgs effect induced by continuation along D-flat directions. Using this interpretation, we show our results are consistent with the corresponding ones obtained for smooth $G_2$ spaces. Finally, we consider switching on flux and Wilson lines on singular loci of the $G_2$ space, and discuss the relation to $N=4$ SYM theory.

In the second half, we develop an algorithmic framework for $E_8\times E_8$ heterotic compactifications with monad bundles, including new and efficient techniques for proving stability and calculating particle spectra. We begin by considering ÔcyclicÕ Calabi-Yau manifolds where we classify positive monad bundles, prove stability, and compute the complete particle spectrum for all bundles. Next, we generalize the construction to bundles on complete intersection Calabi-Yau manifolds.  We show that the class of positive monad bundles, subject to the heterotic anomaly condition, is finite ($\sim 7000$ models). We compute the particle spectrum for these models and develop new techniques for computing the cohomology of line bundles on CICYs. There are no anti-generations of particles and the spectrum is manifestly moduli-dependent. We further investigate the slope-stability of positive monad bundles and develop a new method for proving stability of $SU(n)$ vector bundles on CICYs.
\end{abstractseparate}
\begin{acknowledgements}
I would like to acknowledge the excellent support and guidance of my supervisors Professor Philip Candelas and Professor Andr\'e Lukas. Your help, patience and insight over the past years has been invaluable. I am very lucky to have had the opportunity to work with both of you and I sincerely grateful for all that you have done for me. 

\bigskip
I would also like to thank my collaborators Adam Barrett and Yang-Hui He. Special appreciation goes to Yang, who has provided endless help and enthusiasm in the course of our work together. I would also like to acknowledge the significant contributions of Babiker Hassanain, Tanya Elliott, and Mario Serna of Office 6.2, who have taught me so much, preserved my sanity, and explored physics with me.

\bigskip
For interesting and helpful conversations I am grateful to James Gray, Bal\'{a}zs Szendr\"{o}i, Lionel Mason, Dominic Joyce, Adrian Langer, Xenia de la Ossa, Graham Ross, Vincent Bouchard, and Eran Palti.

\bigskip
I gratefully acknowledge the Rhodes Trust, the US National Science Foundation, and Magdalen College, Oxford for granting me scholarships and making my studies possible.

\end{acknowledgements}
\begin{dedication}

I dedicate this thesis to my family in gratitude for their constant support and kindness (and for being their extraordinary selves) and to Dr. Wheeler, for showing me what was possible.
\end{dedication}
\begin{romanpages}
\tableofcontents
\listoffigures
\listoftables
\end{romanpages}

\chapter{Introduction}

\section{The Problem of String Phenomenology}

As a physical model, string theory is still only in the beginning of its development. While it has shown remarkable success as a natural extension of many of our mathematical descriptions of nature, it has yet to prove that it can make unique, falsifiable predictions. If we are ever to decide whether string theory is just beautiful mathematics without a basis in the physical world, or a new model of fundamental physics, it is necessary to develop the mathematical toolkit of ``string phenomenology" - the aspect of string theory that attempts to bridge the gap between abstract mathematics and measurable physics.
String phenomenology and its associated mathematics are the specific area of this thesis. The questions that it addresses, such as the nature of the underlying geometry of string and M-theory and the mathematical structure of physically realistic models, are particularly interesting in the current era of development in theoretical physics. As our mathematical control of string theory grows richer, and we enter a new era of experimental exploration - with upcoming experiments producing unprecedented data in particle physics and cosmology - it is critical to attempt to answer the question, \emph{can string theory give a description of the real world?}

There are three main approaches to string phenomenology. The earliest arises from compactifications of the $10$-dimensional heterotic string. The second approach is to arrive at four-dimensional physics through the dynamics of $D$-brane solutions in type IIA (and IIB) string theories. The third, and least developed approach comes from M-theory, the still not fully understood $11$-dimensional theory believed to contain all other string theories as limits in its moduli space. In this thesis, we shall investigate aspects of two of these approaches: Heterotic model building and compactifications of M-theory. 

Before we begin, however, we shall mention a few features that all these approaches have in common. From string/M-theory compactifications, we hope to obtain realistic four-dimensional physics. Ultimately, we would hope that a `realistic' model would agree with all available observational data ranging from particle physics to cosmology. However, as a starting point, we shall focus on several broadly defined `physical' features and require that our constructions exhibit these. One of the most important properties is that the models that we derive should include the symmetries and particle spectra of the Standard Model of particle physics. Ideally, the model should also agree with its detailed structure as well, including fermion masses, Yukawa couplings, discrete symmetries, etc. 

Another attractive feature in a string model is a solution to the Hierarchy problem (the observed large hierarchy between the scale of electroweak symmetry breaking and the reduced Planck scale). Current thought suggests that one of the best solutions to this problem is $N=1$ supersymmetry in four-dimensions. Such $N=1$ supersymmetry allows for the unification of gauge coupling at a realistic scale and consistent supersymmetric extensions of the Standard Model (i.e. the minimally supersymmetric Standard Model (MSSM)). To this end, in all the models considered in this work, we shall consider it a goal to produce realistic particle spectra with $N=1$ supersymmetry in four-dimensions. Other important and outstanding problems in string phenomenology such as the cosmological constant problem, supersymmetry breaking mechanisms, and the stabilization of moduli are also critical to realistic models but we will not address them in this work\footnote{For an overview of current issues in string phenomenology see \cite{quevedo-1996,Quevedo:1997uy,mohapatra-1999,munoz-2003,delamacorra-1995,Kachru:1997pc} and for related topics \cite{Hassanain:2007js,Elliott:arXiv0704.0188}.}.

\section{An Overview}
A deeper understanding of M-theory and its associated geometry would open up a new and fascinating area of investigation for string phenomenology.  Beginning with an $11$-dimensional formulation of M-theory, there are several available routes to arrive at four-dimensional physics. The first is to reduce the dimension by one (compactifying on a circle or an interval), and proceed with Calabi-Yau three-fold compactification in either type II or Heterotic string theory. The second choice is to reduce the dimension to four directly by compactifying on a seven-dimensional space. Selecting for $N=1$ symmetry in four-dimensions, the natural choice for such a seven-fold is a manifold with $G_2$ holonomy. While compactification on a smooth $G_2$ space produces unrealistic 4-dimensional physics (abelian multiplets and uncharged chiral matter) \cite{Papadopoulos:1995da,Comp1}, it has been shown that singular $G_2$ spaces could produce much more interesting results \cite{Horava:1996ma,Acharya:1998pm,Acharya:2002kv}. As we shall review in Chapter \ref{G2Intro}, near the neighborhood of certain singularities, enhanced gauge symmetries and matter are possible. Specifically, it was discovered that non-abelian multiplets are present in the neighborhood of co-dimension four singularities and chiral matter can exist near co-dimension seven (conical) singularities. Using arguments from string duality, it is believed that at points of intersection of these four and seven singularities there should be chiral multiplets, charged under the appropriate non-abelian symmetries. In such a region, it is possible to imagine physically realistic compactifications of M-theory. Near such singular points in the space, one can ask how is the low-energy effective action for M-theory modified? That is, what is M-theory near these singularities? How does the low energy theory differ from $11$-dimensional supergravity? In the first half of this thesis, we explore these questions and study M-theory compactifications on singular $G_2$ spaces.

It is well known that M-theory compactifications on singular spaces can yield interesting four-dimensional physics. The compactification of M-theory on singular $G_2$ spaces is one of several known examples of enhanced symmetries and matter content in the neighborhood of a singularity.  In pioneering work \cite{Horava:1996ma}, Ho\v{r}ava and Witten explicitly derived correction terms to the low energy M-theory lagrangian due to the effects of another type of singularity, $S^{1}/\mathbb{Z}_{2}$. By  formulating $11$-dimensional M-theory on the space $\mathbb{R}^{1,9} \times S^{1}/\mathbb{Z}_2$ (a $10$-dimensional manifold with boundary), they were able to construct correction terms to the lagrangian of $11$-dimensional supergravity in the neighborhood of the fixed points of a $S^{1}/\mathbb{Z}_2$ orbifold. This work formed the basis for important developments, including heterotic M-theory and ``braneworld" scenarios \cite{Horava:1996ma,Lukas:1998tt}.

Taking inspiration from the Ho\v{r}ava-Witten constructions, it is natural to ask - what is the structure of M-theory near the singularities present in $G_2$ compactifications? Although the symmetries and matter content can be predicted from string dualities, can the explicit low-energy theory be found as in Ho\v{r}ava-Witten theory? Clearly, from the viewpoint of string phenomenology, such a lagrangian must be constructed if we ever hope to study the four-dimensional physics arising from these models. To answer this question, in the spirit of the Ho\v{r}ava-Witten construction, in Chapters \ref{c:G2Ch1} and \ref{c:G2Ch2} we develop the low-energy M-theory action on a space with co-dimension four ($\mathbb{C}^{2}/\mathbb{Z}_{N}$ orbifold) singularities. These ADE-type singularities are an example of the co-dimension four singularities expected to produce enhanced non-Abelian symmetries in $G_2$ compactifications. While we do not yet include the co-dimension seven singularities necessary for chiral matter, the explicit coupling of the enhanced $SU(N)$ fields to the effective action of M-theory is an important first step towards our goal of realistic four-dimensional physics.
	
In Chapter \ref{c:G2Ch1}, we construct $11$-dimensional supergravity on the orbifold $\mathbb{C}^{2}/\mathbb{Z}_{N} \times \mathbb{R}^{1,6}$ coupled to a seven-dimensional $SU(N)$ super-Yang-Mills theory (SYM) located on the orbifold fixed plane $\{0\}\times\mathbb{R}^{1,6}$. We require that the full coupled theory should be locally supersymmetric at the seven-dimensional orbifold fixed planes. This constraint, together with the specific structure of seven-dimensional supergravity, allows us to determine the unique supersymmetric coupling of the $SU(N)$ multiplets to leading order in a perturbative expansion. 

The construction of this theory is accomplished by ``up-lifting" information from the known action of seven-dimensional Einstein Yang-Mills (EYM) theory and identifying $11$- and $7$- dimensional degrees of freedom appropriately. Concretely, we first constrain the field content of $11$-dimensional supergravity to be consistent with the $\mathbb{Z}_N$ orbifolding symmetry (we shall denote this lagrangian by $\mathcal{L}_{11}$). When constrained to the orbifold fixed plane, the resulting theory is a seven-dimensional $\mathcal{N}=1$ supergravity theory coupled to a single $U(1)$ vector multiplet for $N >2$ or three $U(1)$ multiplets for $N=2$. This theory must couple supersymmetrically to the enhanced field content near the orbifold singularity. The additional states on the orbifold fixed plane should form a seven-dimensional vector multiplet with gauge group $SU(N)$. We explicitly couple these fields to the truncated seven-dimensional 'bulk' theory to obtain a seven-dimensional EYM supergravity, $\mathcal{L}_{SU(N)}$, with gauge group $U(1)\times SU(N)$ for $N>2$ or $U(1)^{3} \times SU(N)$ for $N=2$. Our construction of the theory explicitly distinguishes the bulk degrees of freedom (which can be identified with components of $11$-dimensional fields) and the degrees of freedom in the $SU(N)$ vector multiplets. Furthermore, we prove in general that the action of the full theory
\beq\label{starting_action}
S_{11-7} = \int_{\mathcal{M}_{11}} d^{11}x \{ \mathcal{L}_{11}+ \delta^{(4)}(y^{A})(\mathcal{L}_{SU(N)} - \kappa^{8/9} \mathcal{L}_{11}|_{y=0}) \}
\eeq
is supersymmetric to leading non-trivial order in an expansion in $\kappa$, the $11$-dimensional Newton constant. 

The results of this Chapter hold for M-theory near a $\mathbb{C}^{2}/\mathbb{Z}_{N}$ singularity and could be applied to any geometry containing orbifold singularities of ADE type (for example, M-theory on certain singular limits of $K3$). However, our primary motivation for this construction was its application to the $G_2$ compactifications described above. Thus, this result forms the starting point for ``phenomenological"  $G_2$ compactifications of M-theory near co-dimension $4$ singularities.

Having constructed the effective action for M-theory near these co-dimension 4 singularities, in Chapter \ref{c:G2Ch2} we proceed with this program and study the compactification of M-theory (via the action \eref{starting_action}) on $G_2$ spaces with singularities that produce enhanced $SU(N)$ gauge fields. The relevant $G_2$ spaces were first described by Joyce \cite{joyce1,joyce2} and are constructed by dividing a seven-torus $\mathcal{T}^{7}$ by a discrete symmetry group $\Gamma$, such that the resulting singularities are $A$-type and co-dimension $4$. In this construction the singular locus of the co-dimension $4$ singularity within the $G_2$ space is a three torus. Using these spaces, we shall compactify the $11$-dimensional theory down to four dimensions and compute the K\"ahler potential and superpotential of the $N=1$ theory. 

We analyze a number of features of this theory further. We begin with an explicit comparison of our results for M-theory on singular $G_2$ spaces with those for compactification on the associated smooth $G_2$ spaces obtained by blowing-up the singularities. It turns out that this blowing-up can be described by a Higgs effect induced by continuation along D-flat directions in the four-dimensional effective theory. Further, we investigate the effects of introducing backgrounds involving Wilson lines and flux and present an explicit Gukov-type formula for the superpotential. Finally, we consider one of the $SU(N)$ gauge sectors of the theory with gravity switched off. In this limit, these subsectors have the field content of $N=4$ supersymmetry. Viewing the theory in this way, we demonstrate that the famous S-duality symmetry of $N=4$ super-Yang-Mills theory appears in our model as T-duality on the singular $\mathcal{T}^3$ locus. We speculate about a possible extension of this S-duality to the full supergravity theory.

\vspace{10 mm}
In the second half of this thesis, we explore a different route to four-dimensional physics. The $E_{8}\times E_{8}$ Heterotic string theory provided one of the earliest approaches to string phenomenology and still remains one of the most viable stringy approaches to realistic particle physics \cite{Gross:1985fr, gsw}. However, a string model that exactly describes not only the spectra of the Standard Model, but detailed properties such as Yukawa couplings, mu-terms and discrete symmetries, remain elusive. 

One of the main obstacles to this goal is the inherent mathematical difficulty of heterotic string constructions. A heterotic compactification requires a Calabi-Yau $3$-fold, $X$, and two vector bundles $V$ and $\tilde{V}$ (with structure groups in $E_{8}$) on $X$. In most cases, the construction of vector bundles is highly involved and in order to find the four-dimensional effective theory, the topological data of the bundle must be determined in detail. For a general vector bundle, this is often a difficult task. For example, the property of stability of the vector bundle (essential if the model is to be $N=1$ supersymmetric) is notoriously difficult to prove in algebraic geometry. Further, even after such involved calculations have been accomplished, a single model (manifold $+$ bundles) is highly likely to be unphysical when compared with the detailed structure of the standard model.

In this thesis, we take a new approach to finding realistic heterotic vacua. Rather than attempting to fine-tune the construction of a single string model to match the Standard Model particle spectra, it is possible to take an algorithmic approach to (heterotic) string model building. In Chapters \ref{MonCh1}, \ref{c:posCICY}, \ref{c:line_bundle_coho} and \ref{StabCh}, we outline an algorithmic and systematic search for phenomenologically correct vacua. Using techniques well-known to mathematics, such as the monad construction of vector bundles and new methods in computational algebraic geometry, we have begun a new scan of heterotic vacua on a large scale. With a combination of analytic and computer methods, we have generated a database of over $4000$ complete intersection Calabi-Yau manifolds (CICYs) and thousands of vector bundles with broadly desirable physical characteristics Ð such as three generations of matter, particle spectra close to the minimal supersymmetric Standard Model (MSSM), and consistent supersymmetric vacua. With these constructions in place, we hope eventually to be able to scan through literally hundreds of billions of candidate models in the vast landscape of string vacua.

Before describing the construction of our heterotic models, it is useful to outline several of the guiding principles of our program. In order to build a large data set of physically realistic heterotic models, it is necessary to select a class of Calabi-Yau $3$-folds and a method of constructing vector bundles over them. In selecting which bundles and manifolds to consider, we are guided by two main motivations. 
\begin{enumerate}
\item We seek a construction of vector bundles which allows us to systematically build a very large data set of heterotic models (ultimately on the order of hundreds of thousands or millions of models) which can be scanned for physical suitability. 
\item The classes of bundles and manifolds must be well adapted to such scans. That is, the constructions must be well suited to explicit computation of the topological quantities which determine the physical constraints and particle spectra described above (i.e. we must be able to check for bundle stability, compute the particle spectra, and explicitly introduce discrete symmetries and Wilson lines for symmetry breaking). Further, since we are interested in immense data sets, the types of constructions we investigate should be well adapted to computer implementation and scans. 
\end{enumerate}
To this end, we have selected the monad construction of vector bundles and one of the most explicit constructions of Calabi-Yau manifolds - the Complete Intersection Calabi-Yau spaces (or CICYs). These spaces are defined as complete intersection hypersurfaces in products of projective spaces. 

Beginning in Chapter \ref{HetIntro}, we will review the basic framework for a compactification of the heterotic string on a Calabi-Yau $3$-fold. In particular, we shall provide an overview of the essential elements of a heterotic model  and the constraints placed on the Calabi-Yau $3$-fold $X$ and two holomorphic vector bundles $V$ and $\tilde{V}$, on $X$. We shall outline how the structure, symmetries and topological data of the geometry determine the effective four-dimensional theory. Finally, we will describe the constraints placed on the geometry by imposing such `physical' requirements as $3$-generations of particles, and the existence of $N=1$ supersymmetry in four-dimensions. Further introductory information regarding the monad construction of vector bundles and the Calabi-Yau manifolds used in this work is provided in Appendix \ref{mon-appen}.

In Chapter \ref{MonCh1} we introduce the elements of our `algorithmic approach' to heterotic model building by investigating bundles over the simplest known class of Calabi-Yau manifolds. These are the so-called `cyclic' Calabi-Yau manifolds, defined as complete intersection hypersurfaces in a single Projective space. These include the famous 'Quintic' Calabi-Yau manifold defined by a quintic hypersurface in $\mathbb{P}^4$. There are five such cyclic manifolds, characterized by the property that $h^{1,1}(TX)=1$, and equipped with a single K\"ahler form, $J$. 

Over these spaces, we define the monad construction of vector bundles. The monads used in this work are short exact sequences of the form
\bea\label{a-mon}
\nn && 0 \to V \to B \stackrel{f}{\longrightarrow} C \to 0 \ , \quad
\mbox{with} \\
B &:=& \bigoplus_{i=1}^{r_B} \cO_X(b_i) \ , \quad
C := \bigoplus_{j=1}^{r_C} \cO_X(c_j) \ .
\eea
The exactness of \eref{a-mon} defines a new vector bundle $V$ from two `component' bundles. Given two vector bundles $B$ and $C$ and a map, $f$ between them, $V$ is defined as
\beq
V=\ker(f) \ .
\eeq
By using some of the simplest vector bundles - direct sums of line bundles ($\cO_X({b}_i),\cO_X({c}_j)$), we are able to generate new and far more complex bundles, $V$. We shall constrain this monad construction to yield rank $3,4$ and $5$ $SU(n)$ bundles, suitable for heterotic compactification - that is anomaly free, stable bundles which can yield three-generations of particles. After imposing the physical constraints on these models as outlined in Section \ref{s:het}, we show that there are a finite class of physically relevant monad bundles \eref{a-mon} on the cyclic Calabi-Yau manifolds. In particular, we find $37$ examples over these $5$ manifolds. In a significant result, we prove stability of these bundles using a criterion due to Hoppe \cite{hoppe}. We proceed to compute the particle spectrum of all these models, including gauge singlets. In all cases these models contain only generations, with no anti-generations of particles.

While the results of Chapter \ref{MonCh1} are a good starting point for our program of heterotic model building, ultimately we are interested in much larger data sets. To this end, in Chapter \ref{c:posCICY} we extend the monad construction to define bundles over generic Complete Intersection Calabi-Yau (CICY) manifolds. There are $7890$ known manifolds defined as complete intersection hypersurfaces in products of projective space. Of these, $4515$ are so called ``favorable" CICYs, which possess a simple K\"ahler structure: the second cohomology of $X$ descends directly from that on the ambient projective space. That is, for a favorable CICY defined in an ambient space $\cA = \mathbb{P}^{n_1}\times \ldots \mathbb{P}^{n_m}$, we have $h^{1,1}(TX)=m$ and the number of K\"ahler forms $J$ on $X$ is the same as on $\cA$. On these favorable CICYs, we consider monad bundles of the form  \eref{a-mon} where $B$ and $C$ are composed entirely of 'positive' line bundles\footnote{positive in the sense of the Kodaira vanishing theorem. See Section \ref{s:line} and \cite{AG1,AG2}.} of the form $\cO_X(e^{r}_{j})$ with $e^{r}_{j} >0$. For such positive monads,  we once again prove that the number of physically relevant bundles is finite. We show that of the over $4000$ manifolds at our disposal, only $36$ admit positive monad models satisfying the anomaly cancellation condition. Over these $36$ manifolds we find $7118$ bundles. For these models we can compute the spectra using general techniques. As in the cyclic case, we find no anti-generations. In general, the Higgs particle content depends on the location in bundle moduli space. While we do not yet prove stability of these bundles, we show that for all bundles $H^0(X,V)=H^0(X,V^*)=0$, which is a non-trivial check of stability for $SU(n)$ bundles.

In the final two Chapters, we introduce several new mathematical results which are essential to the development of our program of heterotic model building. In Chapter \ref{c:line_bundle_coho}, we outline a new algorithm for computing the complete cohomology of line bundles over CICYs. As is clear from \eref{a-mon}, this is a necessary and powerful calculational tool if we are to compute the cohomology of a monad bundle $V$. To accomplish this, we introduce the techniques of Koszul and Leray spectral sequences which allow us to use information about the cohomology of objects on the ambient space $\cA$ to compute cohomologies on the variety $X$. We also make use of a powerful computational variant of the Bott-Borel-Weil theorem and develop the frame-work to explicitly construct maps between bundle cohomologies on $\cA$. We develop algorithmic techniques for computing the kernels and images of such maps.

Finally, in Chapter \ref{StabCh}, we address the issue of stability for non-cyclic CICYs. We present a proof of the stability of positive monad bundles defined as complete intersection hypersurfaces in a product of two projective spaces, $\mathbb{P}^{n_1} \times \mathbb{P}^{n_2}$. In the mathematics literature, most proofs of stability attempt to show that a given bundle is stable over its entire K\"ahler cone. However, this is a stronger constraint than necessary from a physics point of view.  For realistic four-dimensional physics within a heterotic model, we need only prove that the bundles are stable \emph{somewhere} in the K\"ahler cone. Using this insight we provide a generalization of Hoppe's criterion and key observations about the possible sub-sheaves which might de-stabilize $V$. Using several simple cohomological criteria, we are able to show that the bundles we consider are stable in an open (and explicitly determined) region of the K\"ahler cone.

The work presented in this thesis is drawn from five research papers. Chapters $3$ and $4$ are based on the following two papers
\begin{itemize}
\item L.B. Anderson, A.B. Barrett, and A. Lukas ``M-theory on the Orbifold $\mathbb{C}^{2}/\mathbb{Z}_N$", Phys. Rev. D{\bf 73} (2006) 0106011, [arXiv:hep-th/0602055]. \cite{Anderson:2006pb}
\item L.B. Anderson, A.B. Barrett, A. Lukas and M.Yamaguchi, ``Four-dimensional Effective M-theory on a Singular $G_2$ Manifold", Phys. Rev. D {\bf 74} (2006) 086008, [arXiv:hep-th/0606285].  \cite{Anderson:2006mv}
\end{itemize}
Chapters $6$ and $7$ are based on
\begin{itemize}
\item L.B. Anderson, Y.H. He, and A.Lukas ``Heterotic Compactification, An Algorithmic Approach", JHEP {\bf 07} (2007) 049, [arXiv:hep-th/0702210]. \cite{Anderson:2007nc}
\item L.B. Anderson, Y.H. He, and A. Lukas ``Monad Bundles in Heterotic String Compactifications", JHEP {\bf 07} (2008) 104, [arXiv:0805.2875]. \cite{positives}
\end{itemize}
Finally Chapters $8$, and $9$ are based on an article in preparation:
\begin{itemize}
\item L.B. Anderson, Y.H. He and A. Lukas ``Vector Bundle Stability In Heterotic Monad Models", to appear. \cite{stability}
\end{itemize}
In addition to these papers, this thesis includes Chapter \ref{G2Intro}, a review of M-theory on manifolds of $G_2$ holonomy, and Chapter \ref{HetIntro} which provides an overview of compactifications of heterotic string theory on Calabi-Yau $3$-manifolds.

\chapter{M-theory on manifolds of $G_2$ holonomy}\label{G2Intro}

\section{Introduction}
Over the past decades, it was discovered that the five formulations of $10$-dimensional string theory are in fact all contained within the parameter space of a larger, $11$-dimensional theory, known as M-theory \cite{Witten:1995ex}. While its fundamental degrees of freedom are still not completely understood, M-theory has emerged as an important tool in our understanding of string theory and its relationship to the observable world. Like string theory, M-theory incorporates both general relativity and quantum field theory, and thus has the potential to provide insight into the fundamental forces of nature. However, due to its $11$-dimensional formulation, it is clear that we must compactify the theory in order to produce a physically relevant model. That is, we must decompose the $11$-dimensional space as a product $M_{10}=M_{4} \times X_{7}$, where $M_4$ is macroscopic (Minkowski) four-dimensional space and $X_7$ is a compact seven-dimensional manifold. Furthermore, in order for our M-theory compactification to have a chance at describing realistic particle physics in four-dimensions, we are interested in choosing $X$ so that $N=1$ supersymmetry is preserved in $4$-dimensions.\footnote{Of course, since we do not observe superpartners to known particles at the energy levels explored by present day experimental particle physics, we hope to eventually break this symmetry by some mechanism.}

There are two main approaches to producing vacua with four macroscopic dimensions with $N=1$ supersymmetry from M-theory. The first of these is to compactify M-theory on a space $S^{1}/\mathbb{Z}_{2}\times X$ where $X$ is a Calabi-Yau manifold \cite{Horava:1996ma}. This approach relates M-theory to the strongly coupled $E_{8}\times E_{8}$ heterotic string. This approach has generated many interesting applications \cite{Horava:1996vs}, \cite{Lukas:1998tt} and we will discuss one aspect of it briefly in Section \ref{s:hor_witt_review}. The second approach, and the main topic for the following two chapters, is to take $X$ to be a seven-dimensional manifold with $G_2$ holonomy. 

In this section we will provide a brief overview of $G_2$ holonomy spaces and compactifications of M-theory. We will not attempt a comprehensive review, but rather attempt to provide some of the key concepts used in later chapters. For a more detailed treatment, we recommend to the reader the review articles \cite{Acharya:2004qe} and \cite{Metzger:2003ud} and the book by Joyce \cite{Joyce}, as well as recent literature \cite{Berglund:2002hw, Gutowski:2002dk, Ferretti:2001qz, House:2004hv, deCarlos:2004ci, Acharya:2000gb, Acharya:2000ps,Witten:2001bf, Chong:2002yh, He:2002fp, Bilal:2003pz}. In addition, the following contain useful reviews of M-theory compactifications and properties of $G_2$ manifolds \cite{Lukas:2003dn, duff-2002, Atiyah:2001qf, adam}. To begin then, we first examine properties necessary in an M-theory compactification to guarantee $N=1$ supersymmetric solutions.

The low-energy effective limit of M-theory is given by $11$-dimensional supergravity \cite{Cremmer:1978km}. The field content is simply a supergravity multiplet consisting of the vielbein $\tilde{e}_{M}^{\underline{N}}$ (and associated metric $\tilde{g}_{MN}$), an antisymmetric three-form field $C$ (with field strength $G=dC$) and a gravitino $\Psi_{M}$ (a spin-$3/2$ Majorana fermion). Here the indices run $M,N=0,1 \ldots 10$ and underlined indices refer to flat space. The action \cite{Cremmer:1978km} is given by\footnote{In this thesis we shall not compute any four Fermi terms (and associated cubic fermion terms in the supersymmetry transformations) and so we shall neglect them throughout.}
\bea\label{m-theory}
S_{11} & = & \frac{1}{\kappa_{11}^2}\int_{M_{11}}d^{11}x\sqrt{-\tilde{g}}\bigg(\frac{1}{2}\tilde{R}-\frac{1}{2}\bar{\Psi}\Gamma^{NMP}\nabla_{N}\Psi_{P} \\
& &-\frac{1}{96}G_{MNPQ}G^{MNPQ}-\frac{1}{192}\bigg(\bar{\Psi}\Gamma^{MNPQRS}\Psi_{S}+12\bar{\Psi}^N\Gamma^{PQ}\Psi^{R}\bigg)G_{NPQR}\bigg) \\
& &-\frac{1}{12\kappa^2}\int_{M_{11}} C\wedge G\wedge G + \ldots
\eea
where we have neglected four fermion terms. Here $\kappa_{11}$ is the $11$-dimensional Yang-Mills coupling, $\Gamma^{M_{1}\ldots M_{n}}$ are anti-symmetrized products of $11$-dimensional gamma matrices, and $\nabla_{M}=\partial_{M}+\frac{1}{4}{\omega_{M}}^{\underline{PQ}}\Gamma_{\underline{PQ}}$ is the spinor covariant derivative, defined in terms of the spin connection, $\omega$.

The supersymmetry variations of the fields, parameterized by a $32$ real component Majorana spinor $\eta$, are
\bea\label{m-theory-var}
\delta\tilde{e}_{M}^{\underline{N}} & = & \bar{\eta}\Gamma^{\underline{N}}\Psi_M \\
\delta C_{MNP} & =  & -3 \bar{\eta}\Gamma_{[MN}\Psi_{P]} \\
\delta\Psi_{M} & = &2\nabla_{M}\eta + \frac{1}{144}(\Gamma_{M}^{NPQR}-8\delta^{N}_{M}\Gamma^{PQR})\eta G_{NPQR}
\eea

A vacuum solution of the theory consists of a choice of background space, $M_{11}$, together with a set of field configurations (for $\tilde{g}$, $C$, and $\Psi_M$) which satisfy the equations of motion of $11$-dimensional supergravity. Setting the gravitino $\Psi_{M}=0$, the equations of motion are
\bea\label{m-theory-EOM}
d\ast G & = & -\frac{1}{2}G\wedge G \\
R_{MN} & = & \frac{1}{12}(G_{MPQR}G_{N}^{PQR}-\frac{1}{12}\tilde{g}_{MN}G_{PQRS}G^{PQRS}).
\eea
In addition, a Bianchi identity holds on the four-form field strength $G$ of the three-form $C$
\beq
dG=0
\eeq
We will investigate a vacuum solution on the space $M^{4}\times X$ with metric 
\beq
ds^{2}=\eta_{\mu\nu}dx^{\mu}dx^{\nu}+g_{AB}(x^{C})dx^{A}dx^{B}
\eeq
where $\eta_{\mu\nu}=\text{diag}(-1,1,1,1)$ is the Minkowski metric on $M^4$ and $A,B=4,5 \ldots 10$. We consider vacuum solutions with vanishing gravitino and three-form, $\Psi_{M}=0$, $C=0$. With this choice of vanishing $G$-flux, the equations of motion imply that $R_{MN}=0$, thus the internal metric $g_{AB}$ must be Ricci-flat.

Next, we observe that in order for a vacuum to exhibit $N=1$ supersymmetry, it is necessary that the supersymmetry transformations preserve the vacuum structure, that is, $\delta\psi=0$, for all fermionic fields $\psi$. For $11$-dimensional supergravity, this is just the gravitino $\Psi_{M}$. Thus for supersymmetry, we require
\beq\label{gravitino_var}
0=\delta\Psi_{M} = 2\nabla_{M}\eta + \frac{1}{144}(\Gamma_{M}^{NPQR}-8\delta^{N}_{M}\Gamma^{PQR})\eta G_{NPQR}
\eeq
Taking a vacuum solution with trivial $G$-flux, this reduces to the condition
\beq
\nabla_{M}\eta=0
\eeq
Then, considering our compactification, we can decompose the $\bf{32}$ component $SO(1,10)$ spinor into a product representation $\bf{4}\otimes\bf{8}$ of $SO(1,3) \times SO(7)$:
\beq
\eta(x^{\mu},x^{A})=\epsilon(x^{\mu})\otimes\psi(x^{A})
\eeq
where $\epsilon$ is a spinor on $M^4$ and $\psi$ is a spinor on $X$.

Therefore, the condition $\nabla_{M}\eta=0$ leads to a condition on the internal space, namely that $X$ must possess a single, covariantly constant spinor, $\psi$, such that $\nabla_{A}\psi=0$. Such a condition can be translated into a statement regarding the holonomy of the internal space \cite{berger}. It is possible to compute the number of covariantly constant spinors $\psi$ by decomposing the $\bf{8}$ spinor representation of $SO(7)$ into irreducible representations of the holonomy group and counting the number of singlets obtained \cite{freund}.

According to Berger's classification of the holonomy of compact, irreducible Riemannian manifolds (see \cite{berger}), there is only one holonomy group in seven-dimensions that comes equipped with the single covariantly constant spinor necessary for $N=1$ supersymmetry: the exceptional Lie group, $G_2$. (For $G_2$, the spinor decomposition is $\bf{8} \rightarrow \bf{7}\oplus \bf{1}$). For the spinor satisfying $\nabla_{A}\psi=0$, we have the integrability condition that $0=[\nabla_{A},\nabla_{B}]\psi$ which implies that $R_{AB}=0$. That is, the $G_2$ holonomy space is Ricci-flat.

Thus, in our search for realistic compactifications of M-theory, we are lead naturally to the notion of seven-dimensional $G_2$ spaces. We will define what we mean by $G_2$ holonomy manifolds in the following section.

\section{Manifolds of $G_2$ holonomy}
The exceptional Lie group $G_2$ is a compact, simply connected and $14$-dimensional. It is the automorphism group of the Octonians and can be simply characterized as the subgroup of $GL(7, \mathbb{R})$ which preserves a canonical three-form $\phi_0$ defined by
\bea\label{flat_three_form}
\phi_0 = dx^{1}\wedge dx^{2}\wedge dx^{7}+dx^{1}\wedge dx^{3}\wedge dx^{6}+dx^{1}\wedge dx^{4}\wedge dx^{5}+dx^{2}\wedge dx^{3} \wedge dx^{5} \\
+dx^{4}\wedge dx^{2}\wedge dx^{6}+dx^{3}\wedge dx^{4}\wedge dx^{7}+dx^{5}\wedge dx^{6}\wedge dx^{7}
\eea
on $\mathbb{R}^7$ with coordinates $X^A$, where $A,B=1, \ldots 7$. If $X$ is a seven-dimensional oriented manifold, a general $G_2$ invariant three-form on $X$ can be defined as a smooth three-form $\phi$ which is locally isomorphic to the flat $3$-form, $\phi_0$ in \eref{flat_three_form}. Such a three-form can be said be `induced' from the covariantly constant spinor characteristic of $G_2$ holonomy via 
\beq
\phi_{ABC}=i\bar{\psi}\gamma_{ABC}\psi~.
\eeq
The isomorphism between $\phi$ and $\phi_{0}$ induces a metric $g$ on $X$ which is referred to as the metric associated with $\phi$ and can be explicitly computed from it. Defining
\beq\label{metric1}
\gamma_{AB}=\frac{1}{144}\phi_{ACD}\phi_{BEF}\phi_{GHI}\epsilon^{CDEFGHI}
\eeq
with the ``pure-number" Levi-Civita tensor $\epsilon$, the associated metric $g$ is
\bea\label{metric2}
g_{AB}=\text{det}(\gamma)^{-1/9}\gamma_{AB}, && \sqrt{\text{det}(g)}=\text{det}(\gamma)^{1/9}.
\eea
We shall refer to a pair $(\phi,g)$ as a $G_2$-\emph{structure} on $X$. Note that a number of useful properties of $\phi$ can be derived from its flat counterpart. For instance, $\phi_{ABC}\phi^{ABC}=42$ where the indices have been raised with the metric $g$. The volume of the manifold can be measure by $\phi$ as
\beq
\text{vol}(X)=\int_{X}\sqrt{\text{det}(g)}~d^{7}x=\frac{1}{7}\int_{X}\phi\wedge\Theta(\phi).
\eeq
where $\Theta$ is the Hodge dual of $\phi$, $\Theta(\phi)=\ast\phi$, taken with respect to the metric $g$ associated with $\phi$. Because the metric depends cubicly on $\phi$, by inspection of \eref{metric1} and \eref{metric2}, it is clear that $\Theta$ is a highly non-linear map on $\phi$.

A $G_2$ structure is said to have vanishing torsion if $\phi$ is covariantly constant with respect to the Levi-Civita connection, $\nabla$, induced by the associated metric $g$. The following statements are equivalent
\begin{enumerate}
\item $(\phi,g)$ is torsion-free
\item $\nabla\phi=0$ on $X$, where $\nabla$ is the Levi-Civita connection of $g$
\item $d\phi=d\ast\phi=0$ on $X$
\item $Hol(g) \subseteq G_2$ and $\phi$ is the induced three-from.
\end{enumerate}
Thus, if the $G_2$ structure is torsion-free, the holonomy of $X$ is a sub-group of $G_2$. If in addition, the first fundamental group $\pi_{1}(X)$ is finite, the holonomy is precisely $G_2$. It is easy to show that if $Hol(g) \subseteq G_2$, then $g$ is Ricci-flat. We shall define a $G_2$-\emph{manifold} as a triple, $(X,\phi,g)$, where $X$ is seven-dimensional and $(\phi,g)$ is a torsion-free $G_2$-structure on $X$.

From the perspective of string/M-theory compactifications, $G_2$ manifolds are difficult, because unlike Calabi-Yau compactifications of $10$-dimensional theories, there are no existence theorems for $G_2$ holonomy metrics. For Calabi-Yau manifolds, Yau's theorem \cite{calabi},\cite{yau} guarantees the existence of a Ricci-flat metric on spaces of $SU(3)$ holonomy. Unfortunately, for $G_2$ spaces, any construction must be explicit.

There are have been several systematic attempts to construct $G_2$ manifolds. The first of these is Joyce's construction which generates $G_2$ spaces from orbifolded tori. (We will focus on these in subsequent chapters). In the Joyce construction \cite{joyce1},\cite{joyce2},\cite{Joyce}, $G_2$ holonomy manifolds are created by considering spaces of the form $T^{7}/\Gamma$ where $\Gamma$ is a discrete group. The quotienting by $\Gamma$ in general produces singularities (orbifold fixed points) in the space, which must be repaired in such a way as to give a smooth $G_2$ holonomy manifold. Roughly speaking,  this is accomplished by blowing-up the singularity, i.e. cutting out a patch around the singularity and replacing it with a smooth cycle of the same symmetry. The local structure of this space (i.e. symmetry and $2$-cycles) will be described in these constructions by asympotically locally Euclidean (ALE) spaces (in our examples, Eguchi-Hanson and Gibbons-Hawking spaces). We will explore these geometries in more detail in the next sections. For now, we note that the moduli space of such $G_2$ manifolds will consist of the radii of the torus and from the radii and orientation of cycles associated with the blow-ups.

Joyce's constructions produce a large number of compact $G_2$ manifolds (for an alternative compact construction, see \cite{kovalev}). Most other constructions of $G_2$ spaces produce non-compact manifolds. These include some of the earliest $G_2$ holonomy spaces discovered \cite{bryant},\cite{gibbons_page_pope}. Other important classes of constructions include Hitchin's homogenous quotient spaces \cite{hitchin_g2}  and spaces related to $G_2$ holonomy such as so-called ``Weak" $G_2$ manifolds (see e.g. \cite{Bilal:2001an}, \cite{Bilal:2003bf}).

\subsection{An example $G_2$ manifold}\label{s:g_2example}
To illustrate some of the ideas given above and to set the stage for the calculations of chapters \ref{c:G2Ch1} and \ref{c:G2Ch2}, here we shall give an example of a simple $G_2$ space, $T^{7}/\mathbb{Z}_{2}\times\mathbb{Z}_{2}\times\mathbb{Z}_{2}$ one of the earliest of Joyce's examples \cite{Joyce}.

We begin with the seven-torus, $T^7$ which comes equipped with a torsion-free $G_2$ structure defined by
\begin{eqnarray}
\phi & = & R^1R^2R^3\mathrm{d}x^1\wedge\mathrm{d}x^2\wedge\mathrm{d}x^3+
R^1R^4R^5\mathrm{d}x^1\wedge\mathrm{d}x^4\wedge\mathrm{d}x^5-
R^1R^6R^7\mathrm{d}x^1\wedge\mathrm{d}x^6\wedge\mathrm{d}x^7 \nonumber \\
 & &+R^2R^4R^6\mathrm{d}x^2\wedge\mathrm{d}x^4\wedge\mathrm{d}x^6 +
R^2R^5R^7\mathrm{d}x^2\wedge\mathrm{d}x^5\wedge\mathrm{d}x^7+
R^3R^4R^7\mathrm{d}x^3\wedge\mathrm{d}x^4\wedge\mathrm{d}x^7 \nonumber \\
 & & -R^3R^5R^6\mathrm{d}x^3\wedge\mathrm{d}x^5\wedge\mathrm{d}x^6.
\end{eqnarray}
where $R^{A}$ are the seven radii of the torus. The holonomy of $T^7$ is trivial. To define an orbifold, we will divide $T^7$ by the discrete symmetry $\Gamma=\mathbb{Z}_{2}\times\mathbb{Z}_{2}\times\mathbb{Z}_{2} \subset G_2$ which will render the first fundamental group finite. We define the action of the group $\Gamma$, with generators $\alpha$, $\beta$, $\gamma$, on the torus as
\bea
\alpha: (x^1,\ldots ,x^7) & \rightarrow & (x^{1},x^{2},x^{3},-x^{4},-x^{5},-x^{6},-x^{7}) \\
\beta: (x^1,\ldots ,x^7) & \rightarrow & (x^{1},-x^{2},-x^{3},\frac{1}{2}-x^{4},-x^{5},x^{6},x^{7}) \\
\gamma: (x^1,\ldots ,x^7) & \rightarrow & (-x^{1},-x^{2},x^{3},x^{4},\frac{1}{2}-x^{5},\frac{1}{2}-x^{6},x^{7})
\eea
The orbifold singularities develop at the fixed points of the above transformations. In this example, there are $12$ disjoint singularities, each of co-dimension four. An example of a singular locus is $(x_1,x_2,x_3,0,0,0,0)$ which is unaffected by the generator $\alpha$. Near each singular point, the fixed locus is simply a three-torus, $T^3$. That is, near the singularity, the space takes the form $T^{3}\times \mathbb{C}^2/\mathbb{Z}_2$.

Repairing a singularity of the form $T^{3}\times \mathbb{C}^2/\mathbb{Z}_2$ involves replacing this space with $T^3 \times \mathbb{EH}$, where $\mathbb{EH}$ is an Eguchi-Hanson space \cite{Eguchi:1978xp}, \cite{Eguchi:1980jx}. The homology of $\mathbb{EH}$ consists of a single $2$-cycle centered at the origin. The space is asymptotically locally Euclidean, that is, it approaches the flat space $\mathbb{C}^2/\mathbb{Z}_2$ at infinity. We shall return to such spaces in the following sections when we discuss M-theory on singular $G_2$ spaces.

\section{M-theory compactified on a smooth $G_2$-manifold}\label{smooth_g2}
In the previous section, we argued that $G_2$ manifolds are a natural choice for M-theory compactifications, and indeed the explicit compactification of low energy effective M-theory on a smooth $G_2$ manifold was carried out in \cite{Papadopoulos:1995da},\cite{Comp1}. Unfortunately, the resulting $N=1$ theory in four-dimensions is disappointing as a physical model. The four-dimensional effective theory obtained is supergravity coupled to $b^2(X)$ Abelian vector multiplets and $b^3(X)$ massless neutral chiral multiplets (here $b^{k}(X)$ represents the $k^{th}$ Betti number). This is clearly not a suitable model for real-world particle physics. However,  as we shall see in the next section, the situation turns out to be much more promising when $X$ is allowed to develop singularities. We will 
look at the more complicated construction of M-theory on singular $G_2$ spaces in the following pages, but first, we will take a look at those elements of the theory that can be derived in general for $G_2$ manifolds. 

To begin, we briefly review the Kaluza-Klein compactification on a generic $G_2$ space. As mentioned above, on a $G_2$ manifold $X$, there is an isomorphism between torsion-free $G_2$ structures and Ricci-flat metrics \cite{Lukas:2003dn}. Thus, Ricci-flat deformations of the metric can be described by the torsion-free deformations of the $G_2$ structure and hence, by the third cohomology $H^3(X,\mathbb{R})$. Consequently, the number of independent metric moduli is given by the third Betti number $b^3(X)$. To define these moduli explicitly, we
introduce to $X$ an integral basis $\{C^I\}$ of three-cycles, and a dual basis $\{\Phi_J\}$ of harmonic three forms satisfying
\begin{equation} \label{dual}
\int_{C^I}\Phi_J=\delta^I_J,
\end{equation}
where $I,J,\ldots=1,\ldots,b^3(X)$.
 We can then expand the torsion-free $G_2$ structure $\varphi$ as
\begin{equation}
\varphi=\sum_{I}a^I\Phi_I.
\end{equation}
Then, by equation \eqref{dual}, the $a^I$ can be computed in terms of certain underlying geometrical parameters by performing the period integrals
\begin{equation}
a^I=\int_{C^I}\varphi.
\end{equation}
Let us also introduce an integral basis $\{D^P\}$ of two-cycles, where $P,Q,\ldots=1,\ldots b^2(X)$, and a dual basis $\{\omega_P\}$ of two-forms satisfying
\begin{equation}
\int_{D^P}\omega_Q=\delta^P_Q\,.
\end{equation}
Then, the three-form field $C$ of 11-dimensional supergravity can be expanded in terms of the basis $\{\Phi_I\}$ and $\{\omega_P\}$ as\footnote{Note that $b^1(X)=0$ for $G_2$ manifolds and that we are neglecting the purely external part of $C$ which corresponds to flux vacua which shall not be considered for now.}
\begin{equation} \label{Cexp}
C=\nu^I\Phi_I+A^P\omega_P.
\end{equation}

The expansion coefficients $\nu^I$ represent $b^3(X)$ axionic fields in the four-dimensional effective theory, while the Abelian gauge fields $A^P$, with field strengths $F^P$, are part of $b^2(X)$ Abelian vector multiplets. The $\nu^I$ pair up with the metric moduli $a^P$ to form the bosonic parts of $b^3(X)$ four-dimensional chiral superfields
\begin{equation}
T^I=a^I+i\nu^I.
\end{equation}

With this in hand, we are ready to construct the effective $N=1$ supergravity theory in four-dimensions. We first note that since the isometry group of a compact $G_2$ manifold is trivial, we will not obtain any four-dimensional gauge fields from a Kaluza-Klein expansion of the metric \cite{Metzger:2003ud}. Furthermore, for a non-flux background, the superpotential and D-term potential vanish (for the flux case see section \ref{Ch2_flux} and \cite{Beasley:2002db}, \cite{Acharya:2004qe}), so the four-dimensional theory is entirely determined by the K\"ahler potential and gauge-kinetic function. Using general properties of $G_2$ manifolds, it was shown in Ref. \cite{Beasley:2002db} that the K\"ahler metric descends from the K\"ahler potential
\begin{equation} \label{Kapp}
K=-\frac{3}{\kappa_4^2}\ln \left( \frac{V}{v_7} \right)\, ,
\end{equation}
where $V$ is the volume of $X$ as measured by the dynamical Ricci flat metric $g$, and $v_7$ is a reference volume, thus
\begin{equation}
V=\int_{X}\mathrm{d}^7x\sqrt{ \det g}\, , \hspace{0.5cm} v_7=\int_{X}\mathrm{d}^7x\sqrt{\det g_0}\, .
\end{equation}
The four-dimensional Newton constant $\kappa_4$ is related to the 11-dimensional version by
\begin{equation}
\kappa_{11}^2=\kappa_4^2v_7\, .
\end{equation}
We note here that \eref{Kapp} demonstrates that the K\"ahler potential is a function of the metric moduli, $a^I$ only, and does not depend on the axions, $\nu^I$. Written in terms of superfields, this is the statement that $K$ depends on the real parts, $T^{I}+\bar{T}^I$ only.

With this in hand, we now examine the gauge-kinetic function. Reduction of the Chern-Simons term of 11-dimensional supergravity by inserting the gauge field part of \eqref{Cexp} leads to the four-dimensional term \cite{Beasley:2002db}
\begin{equation}
\int_{\mathcal{M}^4}c_{IPQ}\nu^{I}F^P\wedge F^Q\,,
\end{equation}
where the coefficients $c_{IPQ}$ are given by
\begin{equation}
c_{IPQ}\sim \int_{X} \Phi_I \wedge \omega_P \wedge \omega_Q.
\end{equation}
This implies that the gauge-kinetic function $f_{PQ}$, which couples $F^P$ and $F^Q$, takes the form
\begin{equation} \label{gkfapp}
f_{PQ} \sim \sum_IT^Ic_{IPQ}.
\end{equation}

Examining the content of this four-dimensional theory, it is clear that this is not suitable for realistic particle physics. The gauge symmetry is purely Abelian and the fermionic superpartners of the Abelian gauge fields are neutral. In addition, the effective action contains no terms which couple the chiral multiplets to the Abelian gauge fields, so the chiral multiplets are uncharged as well. This leaves us with a decidedly uninteresting theory.\footnote{This was, of course, to be expected and is a general feature of compactification of $11$-dimensional supergravity compactified on smooth seven-dimensional manifolds \cite{Papadopoulos:1995da},\cite{Comp1}} However, as we shall see in the next section, far more suitable theories are possible if we consider a more general $G_2$ space containing singularities.

\section{M-theory on singular $G_2$-spaces}

Having outlined the unsatisfying four-dimensional effective theory obtained by compactifying M-theory on a smooth $G_2$ manifold, we are led naturally to the question: can we find something better for singular spaces? The answer is positive and much interesting work has been done over the past two decades on this topic. (See e.g. \cite{Acharya:1998pm},\cite{Beasley:2002db},\cite{Atiyah:2001qf},\cite{Witten:2001uq},\cite{Acharya:2001gy},\cite{Friedmann:2002ty}).

In the following sections we will outline how string dualities lead to the understanding that non-Abelian gauge symmetries and charged chiral matter could arise near two particular types of singularities: with non-Abelian symmetries originating from co-dimension four, orbifold-type singularities and charged chiral matter from co-dimension seven, conical singularities. Further, we will take a momentary step back from our $G_2$ compatifications and review the construction of M-theory on another type of singular space, $S^{1}/{\mathbb{Z}}_2$, the famous Horava-Witten theory. These results are central to the first half of this thesis and in the following Chapters, we shall explicitly construct the coupling of some of these additional symmetries and states to the effective action of M-theory. To accomplish this, we will use the Ho\v{r}ava-Witten construction as an inspiration for how to study M-theory on singular $G_2$ spaces.

\subsection{Arguments from String dualities}
There are two string dualities that are of substantial use in studying M-theory on $G_2$ spaces. These involve dualities relating $M$-theory to the type $IIA$ and heterotic strings, respectively. We shall not use the type $IIA$ duality in this work and only mention it briefly here. In the first of these dualities, $M$-theory compactified on an $S^1$ is dual to type $IIA$ string theory. This is a particularly fruitful correspondence when we include the dynamics of $D6$-branes in the type $IIA$ theory. With enhanced gauge symmetry arising from the interaction (and intersections) of stacks of branes, we expect enhanced symmetries on the $M$-theory side.  That is, for any $D6$-brane state in the $IIA$ theory, we can uplift it to some neighborhood of ADE singularities in M-theory, with chiral fermions arising at special points on the branes \cite{Behrndt:2002xm},\cite{Acharya:2004qe}, \cite{Cvetic:2001nr}. We refer the reader to \cite{Acharya:2004qe} for more detailed treatments of this correspondence. As for the second of these dualities, there are several ways in which the heterotic correspondence will be of use to us. We will use one form of it to look at singularities and enhanced symmetries in the following paragraphs.

In 1986 \cite{Duff:1986cx} it was observed that $11$-dimensional supergravity on $\mathbb{R}^7 \times K3$ and the $10$-dimensional heterotic strong on $\mathbb{R}^7 \times T^3$ not only have the same supersymmetry, but also the same moduli spaces of vacua, namely
\beq
\mathcal{M}=\frac{SO(19,3)}{SO(19)\times SO(3)}
\eeq
This coincidence was not explained fully \cite{Witten:1995ex}, \cite{Hull:1994ys} until the development of M-theory and string dualities. By U-duality, $11$-dimensional M-theory compactified on a four-dimensional $K3$ manifold is believed to be dual to the heterotic string compactified on a $3$-torus, $T^3$ \cite{Acharya:2004qe}. We will not go into the detailed derivation of this, but merely state that a similar Kaluza-Klein reduction to that in the last section can be done on each side of the duality, leading to identical symmetries and fields (see \cite{Acharya:2004qe,Metzger:2003ud,Hull:1994ys,Witten:1995ex} and Chapter \ref{HetIntro} for a discussion of the $E_8 \times E_8$ heterotic theory).

The duality arguments described above are based at generic points in moduli space. But what happens at special points? At special points in the moduli space of the heterotic string on $T^3$, additional massless modes and non-Abelian gauge symmetries can arise \cite{Acharya:2004qe}. They are generated from the Wilson lines through a process similar to the Higgs mechanism (see \cite{Acharya:2004qe} for a full discussion). If M-theory on $K3$ is really equivalent to the heterotic string on $T^3$, there must be points in the moduli space of $K3$ where enhanced symmetries arise as well. Indeed, just such regions have long been known to exist. These points in moduli space exactly correspond to regions where the $K3$ develops orbifold type singularities. We will not go into the details of this argument here, but rather, we will outline below the basic structure of ADE orbifold-type singularities and how they can give rise to enhanced symmetries.

\subsection{Co-dimension $4$ Singularities}\label{co-dim4}
To begin our review of orbifold singularities, we first recall that $K3$ is a compact four-dimensional manifold with $SU(2)$ holonomy\footnote{Up to diffeomorphisms this is the unique simply connected manifold of this type.}. Preservation of this holonomy means that an orbifold singularity in $K3$ must have the local form $\mathbb{C}^{2}/\Gamma$ where $\Gamma$ is a finite subgroup of $SU(2)$. The finite subgroups of $SU(2)$ are fully determined by the ADE classification which may be described in terms of simply laced, semi-simple Lie algebras: $A_n$, $D_k$, $E_6$, $E_7$, and $E_8$. There is a one-to-one correspondence \cite{mckay},\cite{rossman}, \cite{Ono:1988kh} between the set of possible groups $\Gamma$ and the Dynkin diagrams of the Lie algebras:  The first two of these are infinite correspondences, $su(n+1) \sim A_n$ and $so(2k) \sim D_k$ and the three exceptional subgroups correspond to the three exceptional Lie algebras of $E$-type. We shall denote the subgroups $\Gamma$ as $\Gamma_{A_n}$, $\Gamma_{D_k}$, $\Gamma_{E_i}$. Each can be described explicitly and the following isomorphisms hold:

\[ \begin{array}{ccccc}
\Gamma_{A_n} & = & \mathbb{Z}_{n+1} & \sim & \mbox{cyclic group of order $n$} \\
\Gamma_{D_k} & = & \mathbb{D}_{k-2} & \sim & \mbox{ binary dihedral group of order $4k-8$} \\
\Gamma_{E_6} & = & \mathbb{T} & \sim & \mbox{binary tetrahedral group of order $24$} \\
\Gamma_{E_7} & = & \mathbb{O} & \sim & \mbox{binary octohedral group of order $48$} \\
\Gamma_{E_8} & = & \mathbb{I} & \sim & \mbox{binary icosahedral group of order $120$} 
\end{array} \]
As an example, if one chooses the standard action of $SU(2)$ on the coordinates of $\mathbb{C}^2$, then $\Gamma_{A_{n-1}}$ is isomorphic to $Z_{n}$ and be represented by
\beq
\left(
\begin{array}{ccc}
e^{2i\pi/n} & 0 \\
0 & e^{-2i\pi/n}
\end{array}
\right)
\eeq

Non-Abelian gauge symmetries arise in the neighborhood of ADE singularities according to Lie algebras described above. That is, near a $\mathbb{C}^2/\mathbb{Z}_n$ singularity, the enhanced symmetry is expected to be $SU(n+1)$

As an example of how new states arise in the neighborhood of orbifold-type singularities, we can consider $A_1 \sim \mathbb{Z}_2$ which produces an ADE singularity of the form described in section \ref{s:g_2example}. To examine the behavior of the singularity, we must replace it with a blow-up, that is, an Eguchi-Hanson space \cite{Eguchi:1978xp},\cite{Eguchi:1980jx} with a single $2$-cycle, $\gamma$, at the origin (and associated harmonic form, $\omega$).  The $2$-cycle is parameterized by three scalars \cite{Metzger:2003ud}. The singularity is obtained in the limit in which the radius of the two-cycle shrinks to zero. Examining the spectrum of M-theory in this localized neighborhood, we expect enhanced symmetry in the zero-volume limit. First, we observe that the $2$-cycle itself generates a $U(1)$ gauge field, $A$, which descends from the three form as
\beq
C=A\wedge\omega
\eeq
This vector field combines with the three scalar moduli of the cycle, $\gamma$, to form the bosonic part of a seven-dimensional vector multiplet. These are the only immediately apparent new massless states. However, additional massive states can arise from $M2$-branes wrapping $\gamma$. Recalling the standard coupling of a wrapped $M2$-brane to the $C$-field, the $M2$-brane volume can be expressed as $\gamma \times s$ where $s$ is a path in $M^4 \times \mathbb{R}^3$. From a seven-dimensional perspective, these wrapped states appear as particles (which are massive so long as the two-cycle is finite). An $M2$-brane has a mass which is essentially its tension times its world-volume and couples to the $C$-field as
\beq 
T\int_{V} C = \pm \int_{s} A \int_{\gamma} T\omega = \pm \int_{s} TA
\eeq
where $T$ is the tension.\footnote{The membrane will have minimal energy if it wraps the minimal two-cycle available in a given space.} As a result, there are two membrane states generated which are charged under the $U(1)$ gauge field $A$, corresponding to opposite orientations of the membrane with respect to the cycle. These states are of opposite $U(1)$ charge. In the singular limit, as the $2$-cycle collapses, they become massless and combine with the $U(1)$ gauge field to form an enhanced multiplet of an $SU(2)$ super Yang-Mills theory. For the higher $A_n$ series orbifolds, the process is similar to that described above, except that instead of an Eguchi-Hanson space, we replace our singular geometry ($\mathbb{C}^2/{Z_{n+1}}$) with a Gibbons-Hawking space \cite{Gibbons:1979zt},\cite{Hitchin:1900zr}, complete with a chain of $n$ $2$-cycles at the origin. Each of these can be wrapped by membrane states as described above, leading to enhanced $SU(n+1)$ symmetry. 

We began this discussion of orbifold singularities by considering $ADE$ singularities in $K3$ spaces. However, the effects described in the preceding sections are all local. That is, these results can be generalized from $M$-theory on a $K3$ manifold to $M$-theory on any space containing $ADE$ singularities. In the following chapters, we will investigate $M$-theory near these singular regions in general and look in particular at such singularities as they arise in compact $G_2$ manifolds (specifically those in the Joyce constructions described in section \ref{s:g_2example}).

\subsection{Co-dimension $7$ singularities}

To produce realistic physics from a M-theory compactification on a $G_2$ space, we must consider one more type of singularity. It can be shown that in addition to the non-Abelian symmetries generated by the orbifold singularities of the previous section, co-dimension seven singularities must also be present in the $G_2$ space in order to have charged chiral matter. By a co-dimension seven singularity we refer to a space which locally has a conical metric. Such conical singularities arise when the metric  on $X$ locally degenerates to the form:
\beq\label{conical}
ds^2 = dr^2 +r^{2}d\Omega^2
\eeq
where $d\Omega^2$ is the metric on a compact six-manifold, $Y$. The radial variable, $r \in (0,\infty)$ produces an isolated singularity in $X$ at $r=0$. While singularities such as \eref{conical} are among the simplest isolated singularities one could study, they are still difficult to construct in $G_2$ manifolds. The first such metrics appeared in \cite{bryant} and \cite{gibbons_page_pope} and have been further developed in recent years (see e.g. \cite{Acharya:2001gy,Witten:2001uq}). 
In fact, all known examples of $G_2$ manifolds with conical singularities are non-compact. It is an open problem to find compact manifolds of $G_2$ holonomy that contain both co-dimension four $ADE$ and co-dimension seven conical singularities, though the duality of M-theory with type $IIA$ theory tells us that such spaces  should exist. 

The detailed derivation of the enhanced matter content arising in M-theory near a conical singularity is somewhat lengthy and can actually be argued in three separate ways. The first of these is by duality to the type $IIA$ string \cite{Acharya:2004qe}, the second, through duality with the heterotic theory \cite{Acharya:2004qe,Metzger:2003ud} and finally through arguments based on anomalies \cite{Witten:2001uq}. In the interest of space, we shall outline only the last argument here, following the development in \cite{Witten:2001uq}. For a more detailed discussion of charged fermions and conical singularities see \cite{Acharya:2004qe}.

The basic approach will be to define $X$ as a cone on $Y$, that is a space with a conical singularity of the form \eref{conical}. We shall consider M-theory over $M_{4} \times X$ and look at the variation of terms in the M-theory effective action under its gauge symmetries. These will be shown to be non-zero if the conical base, $Y$ obeys certain conditions. Since these anomalous variations must be cancelled, we shall find that chiral fermions must exist in the spectrum.

We begin with the anomaly that arises from the bulk $U(1)$ gauge fields, $A^P$, $P=1,\ldots b^2(X)$ discussed in \eref{Cexp} and write the zero-mode expansion for the three-form
\beq
C=\sum_{P}A^{P}\wedge \omega_P
\eeq
in a basis of two-forms $\{\omega_{P}\}$ on $X$. The anomaly arises from the Chern-Simons term of $11$-dimensional supergravity \eref{m-theory}. Under a gauge transformation of $C$, $C \rightarrow C + d\epsilon$, where $\epsilon$ is a two-form and the Chern-Simons term is changed by
\beq
\delta S \sim \int_{M^{4}\times X} d(\epsilon \wedge G \wedge G)
\eeq
With the conical structure of \eref{conical}, we may formally treat $X$ as a manifold with boundary at $r=0$ and hence
\beq
\delta S \sim  \int_{M^{4}\times Y} \epsilon \wedge G \wedge G  .
\eeq
If we now make the Kaluza-Klein ansatz for the $2$-form, $\epsilon = \sum_{P}\epsilon^{P} \omega_{P}$ we obtain
\beq
\delta S \sim \int_{Y} \omega_{P} \wedge \omega_{Q}\wedge \omega_{R}\int_{M^{4}}\epsilon^{P} F^{Q}\wedge F^{R}
\eeq
where $F^{P}= dA^{P}$ are the field strengths of the $U(1)$ gauge fields. 

If the integrals over $Y$ (which are topological from a four-dimensional perspective) are non-zero, we obtain a four-dimensional interaction characteristic of an anomaly in an abelian gauge theory. Thus, for a consistent theory, we expect the spectrum at the singularity to contain chiral fermions which can exactly cancel $\delta S$. This cancellation would be accomplished by the usual chiral anomaly of the form
\beq
q_{p}q_{Q}q_{R}\int_{M^4} \epsilon^{P}F^{Q}\wedge F^{R} 
\eeq
where $q_{P}$ are the $U(1)$ charges of the chiral superfield.

In the previous section we saw that enhanced, non-abelian gauge symmetries arise in the neighborhood of co-dimension four $ADE$ singularities. In such a neighborhood, $X \sim W \times \mathbb{C}^{2}/\gamma$ where $W$ is the three-dimensional locus of the orbifold singularity. If in addition, conical singularities on $X$ are to support chiral fermions charged under the $ADE$ gauge group, then clearly the conical singularities must be points in $W$. This was shown by anomaly arguments in \cite{Witten:2001uq}. At present, no compact $G_2$ spaces have been found explicitly which contains the appropriate intersecting singularities. The development of such geometries is crucial to generating realistic four-dimensional particle spectra from $M$-theory. 

\section{Review of Ho\v{r}ava-Witten Theory}\label{s:hor_witt_review}
Before we proceed further in our development of M-theory on $G_2$ manifolds, we take a temporary detour in our construction to review M-theory on another type of singular geometry, the famous Ho\v{r}ava-Witten theory \cite{Horava:1996ma},\cite{Witten:1996mz},\cite{Horava:1996vs}. With the construction of M-theory on the orbifold
$S^1/\mathbb{Z}_2\times\mathbb{R}^{1,9}$, Ho\v rava and
Witten~\cite{Horava:1996ma} demonstrated for the first time that physically realistic M-theory compactifications are possible on singular
spaces.  In fact, they showed that new states in the form of two
10-dimensional $E_8$ super-Yang-Mills multiplets, located on the two
10-dimensional fixed planes of this orbifold, had to be added to the
theory for consistency and they explicitly constructed the
corresponding supergravity theory by coupling 11-dimensional
supergravity in the bulk to these super-Yang-Mills theories. It soon
became clear that this theory allows for phenomenologically
interesting Calabi-Yau compactifications~\cite{Witten:1996mz}-\cite{Lukas:1998tt}
and, as the strong-coupling limit of the heterotic string, should be
regarded as a promising avenue towards particle phenomenology from
M-theory (see Chapter \ref{HetIntro}). In this construction, Ho\v{r}ava and Witten constructed \emph{explicit correction terms} to $11$-dimensional supergravity in the neighborhood of the orbifold singularities by anomaly considerations and demanding local supersymmetry. Their work will provide the primary motivation and inspiration for our constructions of M-theory on co-dimension four orbifold singularities which will be developed in Chapters \ref{c:G2Ch1} and \ref{c:G2Ch2}.

The effective action was derived in \cite{Horava:1996ma} by finding the unique supersymmetric coupling of the $10$-dimensional $E_8$ vector multiplets living on the boundary to the $11$-dimensional supergravity theory propagating in the bulk. The action obtained is not gauge-invariant at the classical level, but one-loop anomalies insure that the quantum theory is invariant. Interestingly, this cancellation is possible precisely because the gauge group is $E_8$ and additionally the gauge coupling takes a fixed value with respect to the gravitational coupling. Ho\v{r}rava-Witten theory takes the basic form
\beq\label{coupled_HW}
\mathcal{S}_{11-10} = \mathcal{S}_{11} + {\sum_{k=1}}^2 \mathcal{S}_{10}^{(k)}
\eeq
where $\mathcal{S}_{11}$ is the action of $11$-dimensional supergravity, \eref{m-theory}, and $\mathcal{S}_{10}^{k}$, $k=1,2$ are two copies of the $10$-dimensional SYM action, living on the world-boundary (orbifold fixed planes), $M_{10}^{k}$. The theory is constructed as an expansion in $\zeta_{10} = \kappa_{11}/\lambda_{10}$ where $\lambda_{10}$ is the $10$-dimensional gauge coupling and $\kappa_{11}$ is the $11$-dimensional Newton constant. The anomaly analysis fixes $\lambda_{10}$ in terms of $\kappa_{11}$ as
\beq
\lambda_{10}^{2} = 2\pi (4\pi \kappa_{11}^{2})^{2/3}.
\eeq
The bulk action, $11$-dimensional supergravity, appears at zeroth order in this expansion, with the Yang-Mills theories occurring at higher order.

The coordinates on the circle are taken to be $x^{10} \in [-1/2,1/2]$ with the endpoints identified. The equivalence classes in $S^{1}/\mathbb{Z}_{2}$ are the pairs of points with coordinates $x^{10}$ and $-x^{10}$ (i.e. $\mathbb{Z}_2$ acts as $x^{10} \rightarrow -x^{10}$. This map has fixed points at $0$ and $1/2$, and thus the space $M_{10} \times S^{1}/\mathbb{Z}_2$ has two $10$-dimensional fixed planes ($\mathcal{M}^{(k)}_{10}$). By defining the M-theory effective action over the orbifold $S^{1}/\mathbb{Z}_{2}$, for consistency, we must guarantee that the fields are well-defined over this space and the presence of the discrete symmetry will truncate certain fields. The bosonic fields in \eref{m-theory} must be either even or odd under this $\mathbb{Z}_2$ action (i.e. for all bosonic fields, $\varphi$, $\varphi(x^{10}) = \pm \varphi(-x^{10})$). For example, under this action, $C_{\hat{M}\hat{N}\hat{P}}$ is projected out for $\hat{M},\hat{N}=0,1, \ldots 9$, while $C_{\hat{M}\hat{N}10}$ survives.

At the orbifold planes, $11$-dimensional supergravity is enhanced by a $10$-dimensional SYM theory. The leading terms are of order $\zeta^{2}_{10} \sim \kappa^{2/3}_{11}$ relative to those of the bulk supergravity. The additions to the action are given by,
\bea\label{horava_witten_action}
\mathcal{S}^{(k)}_{10} &  =  & \frac{1}{{\lambda_{10}}^{2}} \int_{M^{(k)}_{10}} d^{10}x \sqrt{-g} ( -\frac{1}{4}F^{a}_{\hat{M}\hat{N}}F_{a}^{\hat{M}\hat{N}} -\frac{1}{2}\bar{\lambda}^{a}\Gamma^{\hat{M}}\mathcal{D}_{\hat{M}}\lambda_{a} \\
& & -\frac{1}{8}\bar{\Psi}_{\hat{M}}\Gamma^{\hat{N}\hat{P}}\Gamma^{\hat{M}}F^{a}_{\hat{N}\hat{P}}\lambda_{a}+\frac{1}{48}\bar{\lambda}^{a}\Gamma^{\hat{M}\hat{N}\hat{P}}\lambda_{a}G_{\hat{M}\hat{N}\hat{P}10})
\eea
The added field content consists of the $E_8$ gauge vector, $A^{a}$ and gaugino, $\lambda_a$. The other fields are contributed from the bulk theory in the straightforward way. The coupled action, \eref{coupled_HW} is locally supersymmetric to order $\zeta^{2}_{10} \sim \kappa^{2/3}_{11}$ under the following transformations
\bea\label{HW_variation}
\delta C_{\hat{M}\hat{N}10} & = & -\frac{\kappa_{11}^{2}}{2\lambda_{10}^{2}}\delta(x^{10})A^{a}_{[\hat{M}}\bar{\eta}\Gamma_{\hat{N}]}\lambda_{a} \\
\delta A^{a}_{\hat{M}} & = & \bar{\eta}\Gamma_{\hat{M}}\lambda^{a} \\
\delta \lambda^{1} & = & -\frac{1}{8}\Gamma^{\hat{M}\hat{N}}F^{a}_{\hat{M}\hat{N}}\eta
\eea
where $\delta(x^{10})$ is the Dirac Delta function. Supersymmetry additionally requires a modification of the Bianchi identity: $dG_{\hat{M}\hat{N}\hat{P}\hat{Q}10} = -\frac{3\kappa^{2}_{11}}{48\lambda^{2}_{10}}\delta(x^{10})F^{a}_{[\hat{M}\hat{N}}F_{\hat{P}\hat{Q}]a}$. If supersymmetry is checked at the next order in the perturbation, $\zeta^{4}_{10} \sim \kappa^{4/3}_{11}$, a cancellation occurs of terms that are formally proportional to $\delta(0)$. In \cite{Horava:1996ma}, Ho\v{r}ava and Witten interpret the occurrence of these terms as a problem with the classical treatment of M-theory. In the full quantum theory, one would expect that $E_8$ gauge fields would propagate not \emph{exactly} at the orbifold fixed planes $x^{10}=0,1/2$ but rather in a region of some small but finite thickness (of order $\kappa^{2/9}$). Such a built-in cut off would replace $\delta(0)$ with a finite constant times $\kappa_{11}^{-2/9}$.  We shall note many similar features to those mentioned above in our construction of M-theory on ADE singularities, including the truncation of $11$-dimensional fields to an orbifold fixed plane, an order by order expansion  in a parameter analogous to $\zeta_{10}$, and the appearance of delta-function-squared terms. 
\bigskip

With this review in hand, we now have all the tools we need to explicitly construct M-theory on co-dimension four singularities and to examine the four-dimensional effective theory that arises from a $G_2$ compactification with such singular regions. We shall turn to this in the following chapters.

\chapter{M-theory on the orbifold $\mathbb{C}^{2}/\mathbb{Z}_{N}$}\label{c:G2Ch1}

\section{Introduction}
In the previous Chapter, we saw that while compactification of
$11$-dimensional supergravity on smooth seven-dimensional manifolds \cite{Papadopoulos:1995da,Comp1} does not provide a viable framework for particle
phenomenology, more interesting physics is possible in the case of singular spaces \cite{Horava:1996ma,Acharya:1998pm,Acharya:2002kv}. In particular, phenomenologically
interesting theories can be obtained by M-theory compactification on singular spaces with $G_2$
holonomy. In such compactifications, certain co-dimension four singularities within the
$G_2$ space lead to low-energy non-Abelian gauge fields~\cite{Acharya:1998pm,Acharya:2004qe} while co-dimension seven singularities can lead to matter fields~\cite{Acharya:2001gy,Atiyah:2001qf,Acharya:2004qe}. 

Our focus in the present chapter will be the
non-Abelian gauge fields arising from co-dimension four
singularities. As seen in Section \ref{co-dim4}, the structure of the $G_2$ space close to such a
singularity is of the form $\mathbb{C}^2/\Gamma \times B$, where $\Gamma$ is one
of the discrete ADE subgroups of $\rm{SU}(2)$ and $B$ is a
three-dimensional space. We will, for simplicity, focus on A-type
singularities, that is $\Gamma=\mathbb{Z}_N$, which lead to gauge fields
with gauge group $\rm{SU}(N)$. A large class of singular $G_2$ spaces
containing such singularities has been obtained in
\cite{Barrett:2004tc}, by orbifolding seven tori~\cite{Joyce}. It
was shown that, within this class of examples, the possible values
of $N$ are 2, 3, 4 and 6. However, in this chapter, we keep $N$
general, given that there may be other constructions which lead to
more general $N$. The gauge fields are located at the fixed point
of $\mathbb{C}^2/\mathbb{Z}_N$ (the origin of $\mathbb{C}^2$) times
$B\times\mathbb{R}^{1,3}$, where $\mathbb{R}^{1,3}$ is the
four-dimensional uncompactified space-time, and are, hence,
seven-dimensional in nature. One would, therefore, expect there to
exist a supersymmetric theory which couples M-theory on the orbifold
$\mathbb{C}^2/\mathbb{Z}_N$ to seven-dimensional super-Yang-Mills
theory. It is the main purpose of the present chapter to construct this
theory explicitly.

Although motivated by the prospect of applications to $G_2$
compactifications, we will formulate this problem in a slightly more
general context, seeking to understand the general structure of
low-energy M-theory on orbifolds of ADE type. Concretely, we will
construct 11-dimensional supergravity on the orbifold
$\mathbb{C}^2/\mathbb{Z}_N\times \mathbb{R}^{1,6}$ coupled to
seven-dimensional $\rm{SU}(N)$ super-Yang-Mills theory located on the
orbifold fixed plane $\{{\bf{0}}\}\times{\mathbb{R}}^{1,6}$. For ease of
terminology, we will also refer to this orbifold plane, somewhat
loosely as the ``brane''. This result can then be applied to
compactifications of M-theory on $G_2$ spaces with
$\mathbb{C}^2/\mathbb{Z}_N$ singularities, as well as to other
problems (for example M-theory on certain singular limits of K3). We
stress that this construction is very much in the spirit of the Ho\v
rava-Witten theory~\cite{Horava:1996ma}, which couples 11-dimensional
supergravity on $S^1/\mathbb{Z}_2\times \mathbb{R}^{1,9}$ to
10-dimensional super-Yang-Mills theory.

Let us briefly outline our method to construct this theory which
relies on combining information from the known actions of
11-dimensional~\cite{Cremmer:1978km,gsw} and seven-dimensional
supergravity~\cite{Bergshoeff:1985mr}-\cite{Park:1988id}. Firstly, we constrain the field content of
11-dimensional supergravity (the ``bulk fields'') to be compatible
with the $\mathbb{Z}_N$ orbifolding symmetry. We will call the
Lagrangian for this constrained version of 11-dimensional supergravity
${\cal L}_{11}$. As a second step this action is truncated to seven
dimensions, by requiring all fields to be independent of the
coordinates $y$ of the orbifold $\mathbb{C}^2/\mathbb{Z}_N$ (or,
equivalently, constraining it to the orbifold plane at $y=0$). The
resulting Lagrangian, which we call ${\cal L}_{11}|_{y=0}$, is
invariant under half of the original 32 supersymmetries and represents
a seven-dimensional ${\cal N}=1$ supergravity theory which turns out
to be coupled to a single $U(1)$ vector multiplet for $N>2$ or three
$U(1)$ vector multiplets for $N=2$. As a useful by-product, we obtain
an explicit identification of the (truncated) 11-dimensional bulk
fields with the standard fields of 7-dimensional Einstein-Yang-Mills
(EYM) supergravity. We know that the additional states on the
orbifold fixed plane should form a seven-dimensional vector multiplet
with gauge group ${\rm SU}(N)$.  In a third step, we couple these
additional states to the truncated seven-dimensional bulk theory
${\cal L}_{11}|_{y=0}$ to obtain a seven-dimensional EYM supergravity
${\cal L}_{SU(N)}$ with gauge group $U(1)\times SU(N)$ for $N>2$ or
$U(1)^3\times SU(N)$ for $N=2$. We note that, given a fixed gauge group
the structure of ${\cal L}_{SU(N)}$ is essentially determined by
seven-dimensional supergravity. We further write this theory in a form
which explicitly separates the bulk degrees of freedom (which we have
identified with 11-dimensional fields) from the degrees of freedom in
the $SU(N)$ vector multiplets. Given this preparation we prove in
general that the action
\begin{equation}
S_{11-7}=\int_{\mathcal{M}_{11}} \mathrm{d}^{11}x \left[ \mathcal{L}_{11} + \delta^{(4)}
(y^A)\left( \mathcal{L}_{\mathrm{SU(N)}}-\kappa^{8/9}\mathcal{L}_{11}\right) \right]
\label{S117}
\end{equation}
is supersymmetric to leading non-trivial order in an expansion in
$\kappa$, the 11-dimensional Newton constant. Inserting the various
Lagrangians with the appropriate field identifications into this
expression then provides us with the final result.

The plan of the Chapter is as follows. In Section \ref{Bulk} we remind
the reader of the action of 11-dimensional supergravity. As
mentioned above this is to be our bulk theory. We then go on to
discuss the constraints that arise on the fields from putting this
theory on the orbifold. We also lay out our conventions for rewriting
11-dimensional fields according to a seven plus four split of the
coordinates. In Section \ref{reduction} we examine our bulk Lagrangian
constrained to the orbifold plane and recast it in standard
seven-dimensional form. The proof that the action~\eqref{S117} is
indeed supersymmetric to leading non-trivial order is presented in
Section \ref{construct}. Finally, in Section \ref{results} we present
the explicit result for the coupled 11-/7-dimensional action and the
associated supersymmetry transformations. We end with a discussion of
our results and an outlook on future directions. Three appendices
present technical background material. In Appendix \ref{spinors} we
detail our conventions for spinors in eleven, seven and four
dimensions and describe how we decompose 11-dimensional
spinors. We also give some useful spinor identities. In Appendix
\ref{Pauli} we have collected some useful group-theoretical
information related to the cosets $SO(3,M)/SO(3)\times SO(M)$ of $d=7$
EYM supergravity which will be used in the reduction of the bulk
theory to seven-dimensions. Further, Appendix \ref{bigEYM} is a self-contained
introduction to EYM supergravity in seven dimensions.


\section{Eleven-dimensional supergravity on the orbifold} \label{Bulk}
In this section we begin our discussion of M-theory on
$\mathcal{M}^N_{11}=\mathbb{R}^{1,6}\times\mathbb{C}^2/\mathbb{Z}_N$
by describing the bulk action and the associated bulk supersymmetry
transformations. We recall that fields propagating on orbifolds are
subject to certain constraints on their configurations and proceed by
listing and explaining these. First however we lay out our
conventions, and briefly describe the decomposition of spinors
in a four plus seven split of the coordinates.\\

We take space-time to have mostly positive signature, that is
$(-++\ldots+)$, and use indices $M,N,\ldots=0,1,\ldots,10$ to label
the 11-dimensional coordinates $(x^M)$. It is often convenient to
split these into four coordinates $y^A$, where $A,B,\ldots=7,8,9,10$,
in the directions of the orbifold $\mathbb{C}^2/\mathbb{Z}_N$ and seven
remaining coordinates $x^\mu$, where $\mu,\nu,\ldots=0,1,2,\ldots,6$,
on $\mathbb{R}^{1,6}$.  Frequently, we will also use complex
coordinates $(z^p,\bar{z}^{\bar{p}})$ on $\mathbb{C}^2/\mathbb{Z}_N$,
where $p,q,\ldots=1,2$, and $\bar{p},\bar{q},\ldots=\bar{1},\B{2}$
label holomorphic and anti-holomorphic coordinates, respectively.
Underlined versions of all the above index types denote the associated
tangent space indices.

All 11-dimensional spinors in this chapter are Majorana. Having
split coordinates into four- and seven-dimensional parts it is useful
to decompose 11-dimensional Majorana spinors accordingly as tensor
products of $SO(1,6)$ and $SO(4)$ spinors. To this end, we introduce a
basis of left-handed $SO(4)$ spinors $\{\rho^i\}$ and their
right-handed counterparts $\{\rho^{\bar{\jmath}}\}$ with indices
$i,j,\ldots =1,2$ and $\bar{\imath},\bar{\jmath},\ldots
=\bar{1},\bar{2}$. Up to an overall rescaling, this basis can be
defined by the relations $\gamma^{\underline{A}}\rho^i =
\left(\gamma^{\underline{A}}\right)_{\B{\jmath}}^{\ph{j}i}\rho^{\B{\jmath}}$.
An 11-dimensional spinor $\psi$ can then be written as
\begin{equation}
\psi=\psi_i (x,y)\otimes\rho^i + \psi_{\bar{\jmath}}(x,y)\otimes\rho^{\bar{\jmath}},
 \label{spinor7}
\end{equation}
where the 11-dimensional Majorana condition on $\psi$ forces
$\psi_i(x,y)$ and $\psi_{\bar{\jmath}}(x,y)$ to be $SO(1,6)$
symplectic Majorana spinors. In the following, for any 11-dimensional
Majorana spinor we will denote its associated seven-dimensional 
symplectic Majorana spinors by the same symbol with additional $i$
and $\bar{\imath}$ indices. A full account of spinor conventions used
in this chapter, together with a derivation of the above decomposition
can be found in Appendix \ref{spinors}.

To begin, we remind the reader of the structure of $11$-dimensional
supergravity~\cite{Cremmer:1978km, gsw, Horava:1996ma} reviewed in Chapter \ref{G2Intro}.  The field content consists of the vielbein
${e_M}^{\underline{N}}$ and associated metric
$g_{MN}=\eta_{\underline{M}\underline{N}}e_M^{\phantom{M}\underline{M}}
e_N^{\phantom{N}\underline{N}}$,
the three-form field $C$, with field strength $G=\mathrm{d}C$, and the
gravitino $\Psi_M$. The action is given by \eref{m-theory} and the supersymmetry transformations by \eref{m-theory-var}. In order for the above bulk theory to be consistent on the orbifold
$\mathbb{C}^2/\mathbb{Z}_N\times\mathbb{R}^{1,6}$ we need to constrain
fields in accordance with the $\mathbb{Z}_N$ orbifold action. Let us
now discuss in detail how this works.  

We denote by $R$ the $SO(4)$ matrix of order $N$ that generates the
$\mathbb{Z}_N$ symmetry on our orbifold. This matrix acts on the
11-dimensional coordinates as $(x,y)\rightarrow (x,Ry)$ which
implies the existence of a seven-dimensional fixed plane characterized
by $\{ y=0\}$. For a field $X$ to be well-defined on the orbifold
it must satisfy
\begin{equation} \label{theta}
X(x,Ry)=\Theta(R)X(x,y)
\end{equation}
for some linear operator $\Theta(R)$ that represents the generator of
$\mathbb{Z}_N$. In constructing our theory we have to choose, for each
field, a representation $\Theta$ of $\mathbb{Z}_N$ for which we wish
to impose this constraint. For the theory to be well-defined, these
choices of representations must be such that the action
\eqref{m-theory} is invariant under the $\mathbb{Z}_N$ orbifold
symmetry. Concretely, what we do is choose how each index type
transforms under $\mathbb{Z}_N$. We take $R\equiv (R^A_{\ph{A}B})$ to
be the transformation matrix acting on curved four-dimensional indices
$A,B,\ldots$ while the generator acting on tangent space indices
$\underline{A},\underline{B},\ldots$ is some other $SO(4)$ matrix
$T^{\underline{A}}_{\phantom{A}\underline{B}}$. It turns out that this
matrix must be of order $N$ for the four-dimensional components of the
vielbein to remain non-singular at the orbifold fixed
plane. Seven-dimensional indices $\mu ,\nu ,\ldots$ transform
trivially. Following the correspondence Eq.~\eqref{spinor7},
11-dimensional Majorana spinors $\psi$ are described by two pairs
$\psi_i$ and $\psi_{\bar{\imath}}$ of seven-dimensional symplectic
Majorana spinor.  We should, therefore, specify how the $\mathbb{Z}_N$
symmetry acts on indices of type $i$ and $\bar{\imath}$. Supersymmetry
requires that at least some spinorial degrees of freedom survive at
the orbifold fixed plane. For this to be the case, one of these type
of indices, $i$ say, must transform trivially. Invariance of fermionic
terms in the action \eqref{m-theory} requires that the other indices,
that is those of type $\bar{\imath}$, be acted upon by a $U(2)$
matrix $S_{\bar{\imath}}^{\phantom{i}\bar{\jmath}}$ that satisfies the equation
\begin{equation} \label{STconstraint}
{S_{\bar{\imath}}}^{\bar{k}}\left(\gamma^{\underline{A}}\right)_{\B{k}}^{\ph{k}\jmath}=
{T^{\ul{A}}}_{\underline{B}}
{\left(\gamma^{\underline{B}}\right)_{\B{\imath}}}^{\jmath}.
\end{equation}
Given this basic structure, the constraints satisfied by the fields are as follows
\begin{eqnarray} 
e_\mu^{\ph{\mu}\underline{\nu}}(x,Ry)&=&
e_\mu^{\ph{\mu}\underline{\nu}}(x,y), \label{cond1} \\
e_A^{\ph{A}\underline{\nu}}(x,Ry)&=&
(R^{-1})_A^{\phantom{A}B}e_B^{\ph{B}\underline{\nu}}(x,y),\\
e_\mu^{\ph{\mu}\underline{A}}(x,Ry)&=&T^{\underline{A}}_{\phantom{A}\underline{B}}
e_\mu^{\ph{\mu}\underline{B}}(x,y),\\
e_A^{\ph{A}\underline{B}}(x,Ry)&=&(R^{-1})_A^{\phantom{A}C}T^{\underline{B}}_{\phantom{B}\underline{D}}
e_C^{\ph{C}\underline{D}}(x,y), \label{cond4}
\end{eqnarray}
\begin{eqnarray}
C_{\mu\nu\rho}(x,Ry) & = & C_{\mu\nu\rho}(x,y), \\ \label{Ccond} C_{\mu\nu A}(x,Ry) & = &
(R^{-1})_A^{\phantom{A}B}C_{\mu\nu B}(x,y), \: \:\: \mathrm{etc.} \label{cond6}\\
\Psi_{\mu i}(x,Ry)&=&\Psi_{\mu i}(x,y), \\ \Psi_{\mu
\bar{\imath}}(x,Ry)&=&S_{\bar{\imath}}^{\phantom{i}\bar{\jmath}}\Psi_{\mu
\bar{\jmath}}(x,y),\\ \Psi_{A
i}(x,Ry)&=&(R^{-1})_{A}^{\phantom{A}B}\Psi_{B i}(x,y),\\ \Psi_{A
\bar{\imath}}(x,Ry)&=&(R^{-1})_{A}^{\phantom{A}B}S_{\bar{\imath}}^{\phantom{i}\bar{\jmath}}\Psi_{B
\bar{\jmath}}(x,y). \label{condn} \eea
Furthermore, covariance of the supersymmetry transformation laws with
respect to $\mathbb{Z}_N$ requires
\bea \eta_{i}(x,Ry)&=&\eta_{i}(x,y), \\
\eta_{\bar{\imath}}(x,Ry)&=&S_{\bar{\imath}}^{\phantom{i}\bar{\jmath}}\eta_{\bar{\jmath}}(x,y).
\eea
In complex coordinates $(z^p,\bar{z}^{\bar{p}})$, it is convenient to represent $R$
by the following matrices
\begin{equation} \label{R}
(R^p_{\ph{p}q})=e^{2i\pi /N}\boldsymbol{1}_2, \hspace{0.3cm} (R^{\B{p}}_{\ph{p}\B{q}})=
e^{-2i\pi /N}\boldsymbol{1}_2, \hspace{0.3cm} (R^{\B{p}}_{\ph{p}q})=(R^p_{\ph{p}\B{q}})=0\, . 
\end{equation}
Using this representation, the constraint~\eqref{cond4} implies
\begin{equation}
 {e_p}^{\ul{A}}=e^{-2i\pi /N}{T^{\ul{A}}}_{\ul{B}}\, {e_p}^{\ul{B}}\; .
\end{equation}
Hence, for the vielbein ${e_{\ul{A}}}^{\ul{B}}$ to be non-singular $T$
must have two eigenvalues $e^{2i\pi /N}$. Similarly, the conjugate of the
above equation shows that $T$ should have two eigenvalues $e^{-2i\pi /N}$.
Therefore, in an appropriate basis we can use the following representation
\begin{equation}
 ({T^{\ul{p}}}_{\ul{q}})=e^{2i\pi /N}{\bf 1}_2\; ,\qquad
 ({T^{\ul{\bar p}}}_{\ul{\bar q}})=e^{-2i\pi /N}{\bf 1}_2\; .\label{T}
\end{equation}
Given these representations for $R$ and $T$, the matrix $S$ is uniquely fixed
by Eq.~\eqref{STconstraint} to be
\begin{equation}
 ({S_{\ul{i}}}^{\ul{j}})=e^{2i\pi /N}{\bf 1}_2\; .\label{S}
\end{equation}
We will use the explicit form of $R$, $T$ and $S$ above to analyze the
degrees of freedom when we truncate fields to be $y$ independent.

When the 11-dimensional fields are taken to be independent of the
orbifold $y$ coordinates, the constraints~\eqref{cond1}--\eqref{condn}
turn into projection conditions, which force certain field components
to vanish. As we will see shortly, the surviving field components fit
into seven-dimensional ${\cal N}=1$ supermultiplets, a confirmation
that we have chosen the orbifold $\mathbb{Z}_N$ action on fields
compatible with supersymmetry. More precisely, for the case $N>2$, we
will find a seven-dimensional gravity multiplet and a single
$U(1)$ vector multiplet. Hence, we expect the associated
seven-dimensional ${\cal N}=1$ Einstein-Yang-Mills (EYM) supergravity
to have gauge group $U(1)$.  For $\mathbb{Z}_2$ the situation is
slightly more complicated, since, unlike for $N>2$, some of the field
components which transform bi-linearly under the generators are now
invariant.  This leads to two additional vector multiplets, so that
the associated theory is a seven-dimensional ${\cal N}=1$ EYM
supergravity with gauge group $U(1)^3$.  In the following section, we
will write down this seven-dimensional theory, both for $N=2$ and
$N>2$, and find the explicit identifications of truncated
11-dimensional fields with standard seven-dimensional supergravity
fields.


\section{Truncating the bulk theory to seven dimensions} \label{reduction}
In this section, we describe in detail how the bulk theory is
truncated to seven dimensions. We recall from the introduction that
this constitutes one of the essential steps in the construction of the
theory. As a preparation, we explicitly write down
the components of the 11-dimensional fields that survive on the
orbifold plane and work out how these fit into seven-dimensional
super-multiplets. We then describe, for each orbifold, the
seven-dimensional EYM supergravity with the appropriate field
content. By an explicit reduction of the 11-dimensional theory and
comparison with this seven-dimensional theory, we find a list of
identification between 11- and 7-dimensional fields which is
essential for our subsequent construction.

To discuss the truncated field content, we use the
representations~\eqref{R}, \eqref{T}, \eqref{S} of $R$, $T$ and $S$
and the orbifold conditions \eqref{cond1}-\eqref{condn} for $y$
independent fields. For $N>2$ we find that the
surviving components are given by $g_{\mu\nu}$,
$e_p^{\ph{p}\underline{q}}$, $C_{\mu\nu\rho}$, $C_{\mu p\B{q}}$,
$\Psi_{\mu i}$, $(\Gamma^p\Psi_p)_i$ and
$(\Gamma^{\B{p}}\Psi_{\B{p}})_i$. Meanwhile, the spinor $\eta$ which
parameterizes supersymmetry reduces to $\eta_i$, a single symplectic
Majorana spinor, which corresponds to seven-dimensional ${\cal N}=1$
supersymmetry. Comparing with the structure of seven-dimensional
multiplets (see Appendix~\ref{bigEYM} for a review of
seven-dimensional EYM supergravity), these field components fill out
the seven-dimensional supergravity multiplet and a single $U(1)$
vector multiplet.  For the case of the $\mathbb{Z}_2$ orbifold, a
greater field content survives, corresponding in seven-dimensions to a
gravity multiplet plus three $U(1)$ vector multiplets. The surviving
fields in this case are expressed most succinctly by $g_{\mu\nu}$,
$e_A^{\ph{A}\underline{B}}$, $C_{\mu\nu\rho}$, $C_{\mu AB}$,
$\Psi_{\mu i}$ and $\Psi_{A\B{\imath}}$. The spinor $\eta$ which
parameterizes supersymmetry again reduces to $\eta_i$, a single
symplectic Majorana spinor.

These results imply that the truncated bulk theory is a
seven-dimensional $\mathcal{N}=1$ EYM supergravity with gauge group
$U(1)^n$, where $n=1$ for $N>2$ and $n=3$ for $N=2$. In the following,
we discuss both cases and, wherever possible, use a unified
description in terms of $n$, which can be set to either $1$ or $3$, as
appropriate.  The correspondence between 11-dimensional truncated
fields and seven-dimensional supermultiplets is as follows. The
gravity super-multiplet contains the purely seven-dimensional parts of
the 11-dimensional metric, gravitino and three-form, that is,
$g_{\mu\nu}$, $\Psi_{\mu i}$ and $C_{\mu\nu\rho}$, along with three
vectors from $C_{\mu AB}$, a spinor from $\Psi_{A\B{\imath}}$ and the
scalar $\mathrm{det}(e_A^{\ph{A}\underline{B}})$. The remaining
degrees of freedom, that is, the remaining vector(s) from $C_{\mu
AB}$, the remaining spinor(s) from $\Psi_{A\B{\imath}}$ and the
scalars contained in
$v_A^{\ph{A}\underline{B}}:=\mathrm{det}(e_A^{\ph{A}\underline{B}})^{-1/4}e_A^{\ph{A}\underline{B}}$,
the unit-determinant part of $e_A^{\ph{A}\underline{B}}$, fill out $n$
seven-dimensional vector multiplets. The $n+3$ Abelian gauge fields
transform under the $SO(3,n)$ global symmetry of the $d=7$ EYM
supergravity while the vector multiplet scalars parameterize the coset
$SO(3,n)/SO(3)\times SO(n)$. Let us describe how such coset spaces are
obtained from the vierbein $v_A^{\ph{A}\underline{B}}$, starting with
the generic case $N>2$ with seven-dimensional gauge group $U(1)$, that
is, $n=1$. In this case, the rescaled vierbein
$v_A^{\ph{A}\underline{B}}$ reduces to $v_p^{\ph{p}\underline{q}}$,
which represents a set of $2\times 2$ matrices with determinant one,
identified by $SU(2)$ transformations acting on the tangent space
index. Hence, these matrices form the coset $SL(2,\mathbb{C})/SU(2)$
which is isomorphic to $SO(3,1)/SO(3)$, the correct coset
space for $n=1$. For the special $\mathbb{Z}_2$ case, which implies
$n=3$, the whole of $v_A^{\ph{A}\underline{B}}$ is present and forms
the coset space $SL(4,\mathbb{R})/SO(4)$. This space is
isomorphic to $SO(3,3)/SO(3)^2$ which is indeed the correct
coset space for $n=3$.

We now briefly review seven-dimensional EYM supergravity with
gauge group $U(1)^n$. A more general account of seven-dimensional
supergravity including non-Abelian gauge groups can be found in
Appendix \ref{bigEYM}. The seven-dimensional ${\cal N}=1$ supergravity multiplet
contains the vielbein $\ti{e}_{\mu}^{\ph{\mu}\underline{\nu}}$, the
gravitino $\psi_{\mu i}$, a triplet of vectors $A_{\mu i}^{\ph{\mu
i}j}$ with field strengths $F_i^{\ph{i}j}=\mathrm{d}A_i^{\ph{i}j}$, a
three-form $\ti{C}$ with field strength $\ti{G}=\mathrm{d}\ti{C}$, a
spinor $\chi_i$, and a scalar $\si$. A seven-dimensional
vector multiplet contains a $U(1)$ gauge field $A_\mu$ with field
strength $F=\mathrm{d}A$, a gaugino $\la_i$ and a triplet of scalars
$\phi_i^{\ph{i}j}$. Here, all spinors are symplectic Majorana spinors
and indices $i,j,\ldots=1,2$ transform under the $SU(2)$
R-symmetry. For ease of notation, the three vector fields in the
supergravity multiplet and the $n$ additional Abelian gauge fields from
the vector multiplet are combined into a single $SO(3,n)$ vector
$A_\mu^I$, where $I,J,\ldots =1,\ldots ,n+3$. The coset space
$SO(3,n)/SO(3)\times SO(n)$ is described by a $(3+n)\times (3+n)$ matrix
$\ell_{I}^{\ph{I}\underline{J}}$, which depends on the $3n$ vector
multiplet scalars and satisfies the $SO(3,n)$ orthogonality condition
\begin{equation}
\ell_{I}^{\ph{I}\underline{J}}\ell_{K}^{\ph{K}\underline{L}}\eta_{\underline{J}\underline{L}}=\eta_{IK}
\end{equation}
with
$(\eta_{IJ})=(\eta_{\underline{I}\underline{J}})=\rm{diag}(-1,-1,-1,+1,\ldots,+1)$. Here,
indices $I,J,\ldots=1,\ldots,(n+3)$ transform under $SO(3,n)$. Their
flat counterparts $\underline{I},\underline{J}\ldots$ decompose into a
triplet of $SU(2)$, corresponding to the gravitational directions and
$n$ remaining directions corresponding to the vector multiplets. Thus
we can write $\ell_{I}^{\ph{I}\underline{J}}\to
(\ell_{I}^{\ph{I}u},\ell_I^{\ph{I}\alpha})$, where $u=1,2,3$ and
$\alpha=1,\ldots,n$. The adjoint $SU(2)$ index $u$ can be converted
into a pair of fundamental $SU(2)$ indices by multiplication with the
Pauli matrices, that is,
\begin{equation}
\ell_{I\ph{i}j}^{\ph{I}i}=\frac{1}{\sqrt{2}}\ell_I^{\ph{I}u}(\si_u)^i_{\ph{i}j}.
\end{equation}
The Maurer-Cartan forms $p$ and $q$ of the matrix $\ell$, defined by
\bea 
p_{\mu\alpha \ph{i}j}^{\ph{\mu\alpha }i}&=&\ell^{I}_{\ph{I}\alpha}
\partial_\mu \ell_{I\ph{i}j}^{\ph{\mu}i}, \label{Maurer1}\\
q_{\mu \ph{i}j\ph{k}l}^{\ph{\mu}i\ph{j}k}&=&\ell^{Ii}_{\ph{Ii}j}
 \partial_\mu \ell_{I\ph{k}l}^{\ph{\mu}k}, \\
q_{\mu \phantom{i}j}^{\phantom{\mu}i}&=&\ell^{Ii}_{\ph{Ii}k}
\partial_\mu \ell_{I\ph{k}j}^{\ph{\mu}k}, \label{Maurern}
\eea
will be needed as well.

With everything in place, we can now write down our expression for $\mathcal{L}_7^{(n)}$,
the Lagrangian of seven-dimensional ${\cal N}=1$ EYM supergravity with gauge group
$U(1)^n$ \cite{Park:1988id}. Neglecting four-fermi terms, it is given by
\begin{eqnarray} \label{7dsugra}
\mathcal{L}_7^{(n)}\!\!\! &=&\!\!\!\frac{1}{\kappa^2_7}\sqrt{-\ti{g}}\left\{\frac{1}{2}
R-\frac{1}{2}\bar{\psi}_{\mu
}^{i}\Upsilon ^{\mu \nu \rho }\hat{\mathcal{D}} _{\nu }\psi _{\rho i}-\frac{1}{4}e^{-2\si}
\left( \ell_{I\phantom{i}j}^{\phantom{I}i}\ell_{%
J\phantom{j}i}^{\phantom{J}j}+\ell_{I}^{\ph{I}\alpha}\ell_{J\alpha}\right)
F_{\mu \nu }^{I}F^{J\mu \nu } \right.  \notag \\
&&\hspace{1.5cm}-\frac{1}{96}e^{4\si}\ti{G}_{\mu \nu \rho \sigma }\ti{G}^{\mu \nu \rho \sigma }
-\frac{1}{2}\bar{\chi}^{i}\Upsilon ^{\mu }\hat{\mathcal{D}} _{\mu }\chi _{i}-\frac{5}{%
2}\partial _{\mu }\si \partial ^{\mu }\si +\frac{\sqrt{5}}{2}\left( \bar{\chi}^{i}
\Upsilon ^{\mu \nu }\psi _{\mu i}+\bar{\chi}^{i}\psi _{i}^{\nu }\right)
\partial _{\nu }\si   \notag \\
&&\hspace{1.5cm}-\frac{1}{2}\bar{\lambda}^{\alpha i}\Upsilon ^{\mu }\hat{\mathcal{D}}
_{\mu }\lambda _{\alpha i}-\frac{1}{2}p_{\mu\alpha \phantom{i}j}^{\phantom{\mu\alpha}i}p_{%
\phantom{\mu\alpha j}i}^{\mu\alpha j}-\frac{1}{\sqrt{2}}\left( \bar{\lambda}^{\alpha i}
\Upsilon ^{\mu \nu }\psi _{\mu
j}+\bar{\lambda}^{\alpha i}\psi _{j}^{\nu }\right) p_{\nu\alpha \phantom{j}i}^{%
\phantom{\nu\alpha}j}  \notag \\
&&\hspace{1.5cm}+e^{2\si}\ti{G}_{\mu \nu \rho \sigma }\left[ 
\frac{1}{192}\left( 12\bar{\psi}^{\mu i}\Upsilon ^{\nu \rho }\psi _{i}^{\si}+ \bar{\psi}_{\la}^{i}
\Upsilon ^{\lambda \mu \nu \rho \sigma \tau }\psi _{\tau
i}\right)+\frac{1}{48\sqrt{5}}\left( 4\bar{\chi}^{i}\Upsilon
^{\mu \nu \rho }\psi _{i}^{\si} \right.\right.\nn 
\end{eqnarray}
\begin{eqnarray}
&& \hspace{3.5cm} \left. \left. -\bar{\chi}^{i}\Upsilon ^{\mu \nu \rho \sigma
\tau }\psi _{\tau i}\right) -\frac{1}{320}\bar{\chi}^{i}\Upsilon ^{\mu \nu
\rho \sigma }\chi _{i}+\frac{1}{192}\bar{\lambda}^{\alpha i}\Upsilon ^{\mu \nu \rho\sigma }
\lambda _{\alpha i}\right] \nn \\
&&\hspace{1.5cm} -ie^{-\si}F_{\mu \nu }^{I}\ell_{I%
\phantom{j}i}^{\phantom{I}j}\left[ \frac{1}{4\sqrt{2}}\left( \bar{\psi}%
_{\rho }^{i}\Upsilon ^{\mu \nu \rho \sigma }\psi _{\sigma j}+2\bar{\psi}%
^{\mu i}\psi _{j}^{\nu }\right) +\frac{1}{2\sqrt{10}}\left( \bar{\chi}%
^{i}\Upsilon ^{\mu \nu \rho }\psi _{\rho j}-2\bar{\chi}^{i}\Upsilon ^{\mu
}\psi _{j}^{\nu }\right) \right.  \notag \\
&&\hspace{3.9cm}\left. +\frac{3}{20\sqrt{2}}\bar{\chi}^{i}\Upsilon ^{\mu \nu
}\chi _{j}-\frac{1}{4\sqrt{2}}\bar{\lambda}^{\alpha i}\Upsilon ^{\mu \nu }\lambda
_{\alpha j}\right]  \notag \\
&&\hspace{1.5cm}+e^{-\si }F_{\mu \nu }^{I}\ell_{I\alpha}\left[ 
\frac{1}{4}\left( 2\bar{\lambda}^{\alpha i}\Upsilon ^{\mu }\psi _{i}^{\nu }
-\bar{\lambda}^{\alpha i}\Upsilon ^{\mu \nu \rho }\psi _{\rho i}\right)
 +\frac{1}{2\sqrt{5}}\bar{\lambda}^{\alpha i}\Upsilon ^{\mu \nu }\chi _{i}\right]  \notag \\
&&\hspace{1.5cm}\left. -\frac{1}{96}\epsilon^{\mu\nu\rho\sigma\kappa\lambda\tau}C_{\mu\nu\rho
}F_{\sigma\kappa}^{\tilde{I}}F_{\tilde{I}}{}_{\lambda\tau} \right\}\; .
\end{eqnarray}
In this Lagrangian the covariant derivatives of symplectic Majorana spinors $\epsilon _{i}$
are defined by 
\begin{equation}
\hat{\mathcal{D}} _{\mu }\epsilon _{i}=\partial _{\mu }\epsilon _{i}+\frac{1}{2}q_{\mu
i}^{\phantom{\mu i}j}\epsilon _{j}+\frac{1}{4}\ti{\omega} _{\mu }^{\phantom{\mu}%
\underline{\mu }\underline{\nu }}\Upsilon _{\underline{\mu }\underline{\nu }%
}\epsilon _{i}.
\end{equation}
The associated supersymmetry transformations, parameterized by the spinor $\varepsilon_i$,
are, up to cubic fermion terms, given by
\bea
\delta \sigma &=& \frac{1}{\sqrt{5}}\bar{\chi}^{i}\varepsilon _{i}\; ,  \nn\\
\delta \ti{e} _{\mu }^{%
\underline{\nu}}&=&\B{\varepsilon }^{i}\Upsilon ^{\underline{\nu}}\psi
_{\mu i }\; , \nn\\
\delta \psi _{\mu i}&=&2\hat{\mathcal{D}} _{\mu }\varepsilon _{i}-\frac{e^{2\sigma}}{80}
\left(\Upsilon _{\mu }^{\ph{\mu} \nu \rho \sigma \eta }-\frac{8}{3}\delta _{\mu }^{\nu
}\Upsilon ^{\rho \sigma \eta }\right) \varepsilon _{i}\ti{G}_{\nu \rho \sigma \eta
}+\frac{ie^{-\sigma}}{5\sqrt{2}}\left(
\Upsilon _{\mu }^{\ph{\mu}\nu \rho }-8\delta _{\mu }^{\nu }\Upsilon ^{\rho
}\right) \varepsilon _{j}F_{\nu \rho }^{I}\ell_{I\phantom{i}i}^{\phantom{I}j}\; , \nn\\
\delta \chi _{i}&=& \sqrt{5}\Upsilon ^{\mu }\varepsilon _{i}\partial _{\mu }\si -
\frac{1}{24\sqrt{5}}\Upsilon ^{\mu \upsilon \rho \sigma }\varepsilon
_{i}\ti{G}_{\mu \nu \rho \sigma }e^{2\si}\text{ }-\frac{i}{\sqrt{10}}\Upsilon ^{\mu\nu }
\varepsilon _{j}F_{\mu\nu }^{I}\ell_{I\phantom{i}i}^{\phantom{I}j}e^{-\si}, \nn\\
\delta \ti{C}_{\mu \nu \rho }&=&\left( -3\B{\psi }^{i}_{\left[ \mu \right. }\Upsilon _{\left.
 \nu \rho \right] }\varepsilon _{i}-\frac{2}{\sqrt{5}}\B{\chi }^{i}\Upsilon _{\mu \nu \rho }
\varepsilon_{i}\right) e^{-2\si},  \\
\ell_{I\phantom{i}j}^{\phantom{I}i}\delta A_{\mu }^{I}&=&\left[ i\sqrt{2}\left( 
\B{\psi }_{\mu }^{i}\varepsilon _{j}-\frac{1}{2}%
\delta _{j}^{i}\B{\psi }_{\mu }^{k}\varepsilon _{k}\right) -\frac{2i}{%
\sqrt{10}}\left( \B{\chi }^{i}\Upsilon _{\mu }\varepsilon _{j}-\frac{1%
}{2}{}\delta _{j}^{i}\B{\chi }^{k}\Upsilon _{\mu }\varepsilon
_{k}\right) \right] e^{\si},  \nn\\
\ell_{I}^{\ph{I}\alpha}\delta A_{\mu }^{I}&=&\B{\varepsilon }%
^{i}\Upsilon _{\mu }\lambda _{i}^\alpha e^{\si},  \nn\\
\delta \ell_{I\phantom{i}j}^{\phantom{I}i}&=&-i\sqrt{2}\B{\varepsilon }^{i}
\lambda _{\alpha j}\ell_{I}^{\ph{I}\alpha}+\frac{i}{\sqrt{2}}\B{\varepsilon }^{k}
\lambda _{\alpha k}\ell_{I}^{\ph{I}\alpha}\delta _{j}^{i}\; ,  \nn\\
\delta \ell_{I}^{\ph{I}\alpha}&=&-i\sqrt{2}\B{\varepsilon }^{i}\lambda_{j}^\alpha
\ell_{I\phantom{j}i}^{\phantom{I}j}\; , \nn\\
\delta \lambda _{i}^\alpha &=&-\frac{1}{2}\Upsilon ^{\mu\nu
}\varepsilon _{i}F_{\mu\nu}^{I}\ell_{I}^{\ph{I}\alpha}e^{-\si}+\sqrt{2}i
\Upsilon ^{\mu }\varepsilon _{j}p_{\mu \phantom{\alpha i}j}^{\phantom{\mu}\alpha i}\; . \nn
\eea
We now explain in detail how the truncated  bulk theory corresponds
to the above seven-dimensional EYM supergravity with gauge group $U(1)^n$,
where $n=1$ for the $\mathbb{Z}_N$ orbifold with $N>2$ and $n=3$ for the
special $\mathbb{Z}_2$ case. It is convenient to choose the seven-dimensional
Newton constant $\kappa_7$ as $\kappa_7=\kappa^{5/9}$. The correspondence
between 11- and 7-dimensional Lagrangians can then be written as
\begin{equation} \label{equivalence}
\kappa ^{8/9}\mathcal{L}_{11}\lvert_{y=0}=\mathcal{L}_7^{(n)}\, .
\end{equation}
We have verified by explicit computation that this relation indeed
holds for appropriate identifications of the truncated 11-dimensional fields
with the standard seven-dimensional fields which appear in Eq.~\eqref{7dsugra}.
For the generic $\mathbb{Z}_N$ orbifold
with $N>2$ and $n=1$, they are given by
\begin{eqnarray}
\si &=& \frac{3}{20}\ln\det g_{AB},  \label{id1}\\
\ti{g}_{\mu \nu } &=&e^{\frac{4}{3}\si }g_{\mu \nu }, \label{Weyl}\\
\psi_{\mu i} & = & \Psi_{\mu i}e^{\frac{1}{3}\si}-\frac{1}{5}\Up_\mu
 \left(\Gamma^A\Psi_{A}\right)_i e^{-\frac{1}{3}\si}, \\
\ti{C}_{\mu\nu\rho}&=&C_{\mu\nu\rho}, \\
\chi_i & = & \frac{3}{2\sqrt{5}}\left(\Gamma^A\Psi_{A}\right)_ie^{-\frac{1}{3}\si}, \label{idn}\\
F^I_{\mu\nu}&=&-\frac{i}{2}\mathrm{tr}\left( \si^IG_{\mu\nu}\right), \\
\la_i & =&  \frac{i}{2}\left( \Gamma^p\Psi_p - \Gamma^{\B{p}}\Psi_{\B{p}}\right)_ie^{-\frac{1}{3}\si}, \\
\ell_I^{\ph{I}\underline{J}}&=&\frac{1}{2}\mathrm{tr}\left( \B{\si}_I v \si^J v^{\dagger} \right).
 \label{idn2}
\eea
Furthermore the seven-dimensional supersymmetry spinor $\ve_i$ is related to
its 11-dimensional counterpart $\eta$ by
\begin{equation} \label{spinorrel}
\varepsilon_i=e^{\frac{1}{3}\si}\eta_i.
\end{equation}
In these identities, we have defined the matrices $G_{\mu\nu}\equiv
(G_{\mu\nu p\B{q}})$, $v\equiv
(e^{5\si/6}e^{\B{p}}_{\ph{p}\B{\underline{q}}})$ and made use of the
standard $SO(3,1)$ Pauli matrices $\si^I$, defined in Appendix
\ref{Pauli}.
For the $\mathbb{Z}_2$ orbifold we have $n=3$ and, therefore, two
additional $U(1)$ vector multiplets. Not surprisingly, field identification in
the gravity multiplet sector is unchanged from the generic case and still
given by Eqs.~\eqref{id1}-\eqref{idn}. It
is in the vector multiplet sector, where the additional states
appear, that we have to make a distinction. For the bosonic 
vector multiplet fields we find
\bea
F^I_{\mu\nu}&=&-\frac{1}{4}\mathrm{tr}\left( T^IG_{\mu\nu}\right),
\label{idn3} \\
\ell_I^{\ph{I}\underline{J}}&=&\frac{1}{4}\mathrm{tr}\left( \B{T}_I v
T^J v^{T} \right), \label{idn4} \eea where now $G_{\mu\nu}\equiv
(G_{\mu\nu AB})$ and $v\equiv
(e^{5\si/6}e^{A}_{\ph{A}\underline{B}})$. Here, $T^I$ are the six $SO(4)$
generators, which are explicitly given in Appendix \ref{Pauli}.


\section{General form of the supersymmetric bulk-brane action} \label{construct}
In this section, we present our general method of construction for the
full action, which combines 11-dimensional supergravity with the
seven-dimensional super-Yang-Mills theory on the orbifold plane in a
supersymmetric way. Main players in this construction will be the
constrained 11-dimensional bulk theory ${\cal L}_{11}$, as discussed
in Section~\ref{Bulk}, its truncation to seven dimensions, ${\cal
L}_7^{(n)}$, which has been discussed in the previous section and
corresponds to a $d=7$ EYM supergravity with gauge group $U(1)^n$ and
${\cal L}_{SU(N)}$, a $d=7$ EYM supergravity with gauge group
$U(1)^n\times SU(N)$. The $SU(N)$ gauge group in the latter theory
corresponds, of course, to the additional $SU(N)$ gauge multiplet
which one expects to arise for M-theory on the orbifold
$\mathbb{C}^2/\mathbb{Z}_N$.

Let us briefly discuss the physical origin of these $SU(N)$ gauge
fields on the orbifold fixed plane. It is well-known~\cite{Comp1,Lukas:2003dn} that
the $N-1$ Abelian $U(1)$ gauge fields within $SU(N)$ are already
massless for a smooth blow-up of the orbifold
$\mathbb{C}^2/\mathbb{Z}_N$ by an ALE manifold. More precisely, they
arise as zero modes of the M-theory three-form on the blow-up ALE
manifold. The remaining vector fields arise from membranes wrapping
the two-cycles of the ALE space and become massless only in the
singular orbifold limit, when these two-cycles collapse. For our
purposes, the only relevant fact is that all these $SU(N)$ vector fields are
located on the orbifold fixed plane. This allows us to treat the
Abelian and non-Abelian parts of $SU(N)$ on the same footing,
despite their different physical origin.

We claim that the action for the bulk-brane system is given by
\begin{equation} \label{action1}
S_{11-7}=\int_{\mathcal{M}^N_{11}} \mathrm{d}^{11}x \left[
 \mathcal{L}_{11} + \delta^{(4)}(y^A)\mathcal{L}_{\mathrm{brane}} \right],
\end{equation}
where
\begin{equation} \label{braneformula}
\mathcal{L}_{\mathrm{brane}}=\mathcal{L}_{\mathrm{SU(N)}}-\mathcal{L}_7^{(n)}.
\end{equation}
Here, as before, $\mathcal{L}_{11}$ is the Lagrangian for
11-dimensional supergravity~\eqref{m-theory} with fields
constrained in accordance with the orbifold $\mathbb{Z}_N$ symmetry,
as discussed in Section~\ref{Bulk}. The Lagrangian
$\mathcal{L}_7^{(n)}$ describes a seven-dimensional ${\cal N}=1$ EYM
theory with gauge group $U(1)^n$. Choosing $n=1$ for generic
$\mathbb{Z}_N$ with $N>2$ and $n=3$ for $\mathbb{Z}_2$, this
Lagrangian corresponds to the truncation of the bulk Lagrangian
$\mathcal{L}_{11}$ to seven dimensions, as we have shown in the
previous section. This correspondence implies identifications between
truncated 11-dimensional bulk fields and the fields in
$\mathcal{L}_7^{(n)}$, which have also been worked out explicitly in
the previous section (see Eqs.~\eqref{id1}--\eqref{idn2} for the case
$N>2$ and Eqs.~\eqref{id1}--\eqref{idn} and \eqref{idn3}--\eqref{idn4}
for $N=2$).  These identifications are also considered part of the
definition of the Lagrangian~\eqref{action1}. The new Lagrangian
$\mathcal{L}_{SU(N)}$ is that of seven-dimensional EYM supergravity
with gauge group $U(1)^n\times SU(N)$, where, as usual, $n=1$ for
generic $\mathbb{Z}_N$ with $N>2$ and $n=3$ for $\mathbb{Z}_2$.
This Lagrangian contains the ``old'' states in the gravity multiplet
and the $U(1)^n$ gauge multiplet and the ``new'' states in the
$SU(N)$ gauge multiplet. We will think of the former states as being
identified with the truncated 11-dimensional bulk states by precisely
the same relations we have used for $\mathcal{L}_7^{(n)}$. The idea
of this construction is, of course, that in $\mathcal{L}_{\rm brane}$
the pure supergravity and $U(1)^n$ vector multiplet parts cancel
between ${\mathcal L}_{SU(N)}$  and $\mathcal{L}_7^{(n)}$, so that we
remain with ``pure'' $SU(N)$ theory on the orbifold plane. For this to
work out, we have to choose the seven-dimensional Newton constant
$\kappa_7$ within $\mathcal{L}_{SU(N)}$ to be the same as the one
in $\mathcal{L}_7^{(n)}$, that is
\begin{equation}
\kappa_7=\kappa^{5/9}\, .
\end{equation}
The supersymmetry transformation laws for the action~\eqref{action1}
are schematically given by
\bea \label{susy1}
\delta_{11}&=&\delta_{11}^{\ph{11}11}+\kappa^{8/9}\delta^{(4)}(y^A)\delta_{11}^{\ph{11}\mathrm{brane}}, \\
\delta_7&=&\delta_7^{\ph{7}\mathrm{SU(N)}}, \label{susy2}
\eea
where
\begin{equation} \label{susy3}
\delta_{11}^{\ph{11}\mathrm{brane}}=\delta_{11}^{\ph{11}\mathrm{SU(N)}}-\delta_{11}^{\ph{11}11}.
\end{equation}
Here $\delta_{11}$ and $\delta_7$ denote the supersymmetry transformation of
bulk fields and fields on the orbifold fixed plane, respectively.
A superscript 11 indicates a supersymmetry transformation
law of $\mathcal{L}_{11}$, as given in equations
\eqref{m-theory-var}, and a superscript $SU(N)$ indicates a
supersymmetry transformation law of $\mathcal{L}_{SU(N)}$, as can be
found by substituting the appropriate gauge group into equations
\eqref{7dsusy}. These transformation laws are parameterized by a
single 11-dimensional spinor, with the seven-dimensional spinors
in $\delta_{11}^{\ph{11}\mathrm{SU(N)}}$ and $\delta_{7}^{\ph{7}\mathrm{SU(N)}}$
being simply related to this 11-dimensional spinor by
equation \eqref{spinorrel}. On varying $S_{11-7}$ with respect to
these supersymmetry transformations we find
\bea
\delta S_{11-7}&=&-\int_{\mathcal{M}_{11}} \mathrm{d}^{11}x \delta^{(4)}(y^A)
\left( 1-\kappa^{8/9}\delta^{(4)}(y^A) \right) \delta_{11}^{\ph{11}\mathrm{brane}}
\mathcal{L}_{\mathrm{brane}}. \label{susycalc}
\eea
At first glance, the occurrence of delta-function squared terms is
concerning. However, as in Ho\v rava-Witten theory \cite{Horava:1996ma}, we
can interpret the occurrence of these terms as a symptom of attempting
to treat in classical supergravity what really should be treated in
quantum M-theory. It is presumed that in proper quantum M-theory,
fields on the brane penetrate a finite thickness into the bulk, and that
there would be some kind of built-in cutoff allowing us to replace
$\delta^{(4)}(0)$ by a finite constant times $\kappa^{-8/9}$. If we
could set this constant to one and formally substitute
\begin{equation}
\delta^{(4)}(0)=\kappa^{-8/9}
\end{equation}
then the above integral would vanish.

As in Ref.~\cite{Horava:1996ma}, we can avoid such a regularization if we
work only to lowest non-trivial order in $\kappa$, or, more precisely
to lowest non-trivial order in the parameter
$h=\kappa_7/g_\mathrm{YM}$. Note that $h$ has dimension of inverse
energy. To determine the order in $h$ of various terms in the
Lagrangian we need to fix a convention for the energy dimensions of
the fields. We assign energy dimension 0 to bulk bosonic fields and
energy dimension 1/2 to bulk fermions. This is consistent with the way
we have written down 11-dimensional supergravity \eqref{m-theory}. In
terms of seven-dimensional supermultiplets this tells us to assign
energy dimension 0 to the gravity multiplet and the $U(1)$ vector
multiplet bosons and energy dimension 1/2 to the fermions in these
multiplets. For the $SU(N)$ vector multiplet, that is for the brane
fields, we assign energy dimension 1 to the bosons and 3/2 to the
fermions. With these conventions we can expand
\begin{equation} \label{expansion1}
\mathcal{L}_{SU(N)}=\kappa_7^{-2}\left( \mathcal{L}_{(0)}+h^2\mathcal{L}_{(2)}+
h^4\mathcal{L}_{(4)}+\ldots \right),
\end{equation}
where the $\mathcal{L}_{(m)}$, $m=0,2,4,\ldots$ are independent of $h$.
The first term in this
series is equal to $\mathcal{L}_7^{(n)}$, and therefore the leading
order contribution to $\mathcal{L}_{\mathrm{brane}}$ is precisely the
second term of order $h^2$ in the above series. It turns out that,
up to this order, the action $S_{11-7}$ is supersymmetric
under \eqref{susy1} and \eqref{susy2}. To see this we also expand
the supersymmetry transformation in orders of $h$, that is
\begin{equation}
\delta_{11}^{\ph{11}SU(N)} = \delta_{11}^{(0)}+h^2\delta_{11}^{(2)}+h^4\delta_{11}^{(4)}+\ldots\, .
\end{equation}
Using this expansion and Eq.~\eqref{expansion1} one finds that the
uncanceled variation \eqref{susycalc} is, in fact, of order
$h^4$. This means the action~\eqref{action1} is indeed supersymmetric
up to order $h^2$ and can be used to write down a supersymmetric
theory to this order. This is exactly what we will do in the following
section. We have also checked explicitly that the terms of order $h^4$
in Eq.~\eqref{susycalc} are non-vanishing, so that our method cannot
be extended straightforwardly to higher orders.

A final remark concerns the value of the Yang-Mills gauge coupling
$g_{\rm YM}$.  The above construction does not fix the value of this
coupling and our action is supersymmetric to order $h^2$ for all
values of $g_{\rm YM}$.  However, within M-theory one expects
$g_{\rm YM}$ to be fixed in terms of the 11-dimensional Newton constant
$\kappa$. Indeed, reducing M-theory on a circle to IIA string theory,
the orbifold seven-planes turn into D6 branes whose tension is fixed
in terms of the string tension~\cite{Polchinski}. By a straightforward
matching procedure this fixes the gauge coupling to be~\cite{Friedmann:2002ty}
\begin{equation}\label{couplerel}
g_{\mathrm{YM}}^2=(4\pi)^{4/3}\kappa^{2/3}\; .
\end{equation}


\section{The explicit bulk/brane theory} \label{results}
In this section, we give a detailed description of M-theory on
$\mathcal{M}^N_{11}=\mathbb{R}^{1,6}\times\mathbb{C}^2/\mathbb{Z}_N$,
taking account of the additional states that appear on the brane. We begin
with a reminder of how the bulk fields, truncated to seven dimensions,
are identified with the fields that appear in the seven-dimensional
supergravity Lagrangians from which the theory is constructed.
Then we write down our full Lagrangian, and present the supersymmetry
transformation laws.

As discussed in the previous section, the full Lagrangian is
constructed from three parts, the Lagrangian of 11-dimensional
supergravity $\mathcal{L}_{11}$ with bulk fields constrained by the
orbifold action, $\mathcal{L}_7^{(n)}$, the Lagrangian of
seven-dimensional EYM supergravity with gauge group $U(1)^n$ and
$\mathcal{L}_{SU(N)}$, the Lagrangians for seven-dimensional EYM
supergravity with gauge group $U(1)^n\times SU(N)$. The Lagrangian
$\mathcal{L}_{11}$ has been written down and discussed in Section
\ref{Bulk}, while $\mathcal{L}_7^{(n)}$ has been dealt with in
Section \ref{reduction}. The final piece, $\mathcal{L}_{SU(N)}$, is
discussed in Appendix \ref{bigEYM}, where we provide the reader with a
general review of seven-dimensional supergravity theories. Crucial to
our construction is the way in which we identify the fields in the
supergravity and $U(1)^n$ gauge multiplets of the latter two
Lagrangians with the truncated bulk fields. Let us recall the
structure of this identification which has been worked out
in Section~\ref{reduction}. The bulk fields truncated to
seven dimensions form a $d=7$ gravity multiplet and $n$ $U(1)$
vector multiplets, where $n=1$ for the general $\mathbb{Z}_N$ orbifold
with $N>2$ and $n=3$ for the $\mathbb{Z}_2$ orbifold. The gravity multiplet
contains the purely seven-dimensional parts of the 11-dimensional
metric, gravitino and three-form, that is, $g_{\mu\nu}$, $\Psi_{\mu
i}$ and $C_{\mu\nu\rho}$, along with three vectors from $C_{\mu AB}$,
a spinor from $\Psi_{A\B{\imath}}$ and the scalar
$\mathrm{det}(e_A^{\ph{A}\underline{B}})$. Meanwhile, the vector
multiplets contain the remaining vectors from $C_{\mu AB}$, the
remaining spinors from $\Psi_{A\B{\imath}}$ and the scalars contained
in
$v_A^{\ph{A}\underline{B}}:=\mathrm{det}(e_A^{\ph{A}\underline{B}})^{-1/4}e_A^{\ph{A}\underline{B}}$,
the unit-determinant part of $e_A^{\ph{A}\underline{B}}$. The gravity
and $U(1)$ vector fields naturally combine together into a single
entity $A_\mu^I$, $I=1,\ldots(n+3)$, where the index $I$ transforms
tensorially under a global $SO(3,n)$ symmetry. Meanwhile, the vector
multiplet scalars naturally combine into a single $(3+n)\times (3+n)$
matrix $\ell$ which parameterizes the coset $SO(3,n)/SO(3)\times SO(n)$.
The precise mathematical form of these identifications is given in
equations \eqref{id1}-\eqref{idn2} for the general $\mathbb{Z}_N$
orbifold with $N>2$, and equations \eqref{id1}-\eqref{idn} and
\eqref{idn3}-\eqref{idn4} for the $\mathbb{Z}_2$ orbifold.

In addition to those states which arise from projecting bulk states to
the orbifold fixed plane the Lagrangian ${\cal L}_{SU(N)}$ also
contains a seven-dimensional $SU(N)$ vector multiplet, which is
genuinely located on the orbifold plane. It consists of gauge fields
$A_\mu^a$ with field strengths $F^a=\mathcal{D}A^a$, gauginos
$\la_i^a$, and $SU(2)$ triplets of scalars
$\phi_{a\ph{i}j}^{\ph{a}i}$. These fields are in the adjoint of
$SU(N)$ and we use $a,b,\ldots=4,\ldots,(N^2+2)$ for $su(N)$ Lie
algebra indices. It is important to write ${\cal L}_{SU(N)}$ in a form
where the $SU(N)$ states and the gravity/$U(1)^n$ states are
disentangled, since the latter must be identified with truncated bulk
states, as described above. For most of the fields appearing in
$\mathcal{L}_{SU(N)}$, this is just a trivial matter of using the
appropriate notation. For example, the vector fields in
$\mathcal{L}_{SU(N)}$ which naturally combine into a single entity
$A_\mu^{\ti{I}}$, where $\ti{I}=1,\ldots,(3+n+N^2-1)$, and transforms
as a vector under the global $SO(3,n+N^2-1)$ symmetry, can simply be
decomposed as $A_\mu^{\ti{I}}=(A_\mu^I,A_\mu^a)$, where $A_\mu^I$
refers to the three vector fields in the gravity multiplet and the
$U(1)^n$ vector fields and $A_\mu^a$ denotes the $SU(N)$ vector
fields. For gauge group $U(1)^n\times SU(N)$, the associated scalar
fields parameterize the coset $SO(3,n+N^2-1)/SO(3)\times SO(n+N^2-1)$.
We obtain representatives $L$ for this coset by expanding around
the bulk scalar coset $SO(3,n)/SO(3)\times SO(n)$, represented by matrices
$\ell$, to second order in the $SU(N)$ scalars $\Phi\equiv (\phi_a^{\ph{a}u})$.
For the details see Appendix~\ref{bigEYM2}. This leads to
\begin{equation}
L = \left( \begin{array}{ccc}
\ell+\frac{1}{2}h^2\ell\Phi^T\Phi & m & h\ell\Phi^T \\
h\Phi & 0 & \boldsymbol{1}_{N^2-1}+\frac{1}{2}h^2\Phi\Phi^T \\
\end{array} \right)\, .
\end{equation}
We note that the neglected $\Phi$ terms are of order $h^3$ and higher 
and, since we are aiming to construct the action only up to terms of order
$h^2$, are, therefore, not relevant in the present context. 
 
We are now ready to write down our final action.
As discussed in Section \ref{construct}, to order $h^2\sim g_{\mathrm{YM}}^{-2}$, it is given by
\begin{equation}\label{schematic_action}
S_{11-7}=\int_{\mathcal{M}^N_{11}} \mathrm{d}^{11}x \left[ \mathcal{L}_{11} +
         \delta^{(4)}(y^A)\mathcal{L}_{\mathrm{brane}} \right],
\end{equation}
where
\begin{equation}
\mathcal{L}_{\mathrm{brane}}=\mathcal{L}_{\mathrm{SU(N)}}-\mathcal{L}_7^{(n)},
\end{equation}
and $n=3$ for the $\mathbb{Z}_2$ orbifold and $n=1$ for $\mathbb{Z}_N$
with $N>2$. The bulk contribution, $\mathcal{L}_{11}$, is given in
equation \eqref{m-theory}, with bulk fields subject to the orbifold
constraints~\eqref{cond1}--\eqref{condn}. On the orbifold fixed plane,
$\mathcal{L}_7^{(n)}$ acts to cancel all the terms in
$\mathcal{L}_{\mathrm{SU(N)}}$ that only depend on bulk fields projected
to the orbifold plane. Thus none of the bulk gravity terms are replicated on the
orbifold place. To find $\mathcal{L}_{\mathrm{brane}}$ explicitly we need
to expand ${\cal L}_{SU(N)}$ in powers of $h$, using, in particular, the
above expressions for the gauge fields $A_\mu^{\tilde{I}}$ and the coset
matrices $L$, and extract the terms of order $h^2$. The further details 
of this calculation are provided in Appendix \ref{bigEYM}. The result is
\bea \label{braneaction}
\mathcal{L}_{\mathrm{brane}}\!\!\! &=&\!\!\! \frac{1}{g_{\mathrm{YM}}^2}
 \sqrt{-\ti{g}} \left\{ -\frac{1}{4}e^{-2\si}F^a_{\mu\nu}F_a^{\mu\nu}
-\frac{1}{2}\hat{\mathcal{D}}_\mu\phi_{a\ph{i}j}^{\ph{a}i}
\hat{\mathcal{D}}^\mu\phi_{\ph{aj}i}^{aj}-\frac{1}{2}\B{\la}^{ai}\Up^\mu\hat{\mathcal{D}}_\mu\la_{ai}
 -e^{-2\si}\ell_{I\ph{i}j}^{\ph{I}i}\phi_{a\ph{j}i}^{\ph{a}j}F^I_{\mu\nu}F^{a\mu\nu}\right.  \nn \\
&&\hspace{1.8cm} -\frac{1}{2}e^{-2\si}\ell_{I\ph{i}j}^{\ph{I}i}
{\phi^{aj}}_i\ell_{J\ph{k}l}^{\ph{J}k}\phi_{a\ph{l}k}^{\ph{a}l}
F^I_{\mu\nu}F^{J\mu\nu}-\frac{1}{2}p_{\mu\alpha\ph{i}j}^{\ph{\mu\alpha}i}
\phi_{a\ph{j}i}^{\ph{a}j}p^{\mu\alpha k}_{\ph{\mu\alpha k}l}\phi^{al}_{\ph{al}k} \nn \\
&&\hspace{1.8cm}+\frac{1}{4}\phi_{a\ph{i}k}^{\ph{a}i}
\hat{\mathcal{D}}_\mu\phi_{\ph{ak}j}^{ak}\left( \B{\psi}_\nu^j
\Up^{\nu\mu\rho}\psi_{\rho i} + \B{\chi}^j\Up^\mu\chi_i+ \B{\la}^{\alpha j}
\Up^\mu\la_{\alpha i} \right) \nn \\
&&\hspace{1.8cm}-\frac{1}{2\sqrt{2}}\left( \bar{\lambda}^{\alpha i}
\Upsilon ^{\mu \nu }\psi _{\mu j}+\bar{\lambda}^{\alpha i}\psi _{j}^{\nu }\right)
\phi_{a\ph{j}i}^{\ph{a}j}\phi^{ak}_{\ph{ak}l} p_{\nu\alpha \phantom{l}k}^{%
\phantom{\nu\alpha}l}-\frac{1}{\sqrt{2}}\left( \bar{\lambda}^{ai}
\Upsilon ^{\mu \nu }\psi_{\mu j}  +\bar{\lambda}^{ai}\psi_{j}^{\nu }\right)
\hat{\mathcal{D}}_\nu \phi_{ a\phantom{j}i}^{\phantom{ a}j} \nn \\
&& \hspace{1.8cm}+\frac{1}{192}e^{2\si }\ti{G}_{\mu \nu \rho
\sigma }\bar{\lambda}^{ai}\Upsilon ^{\mu
\nu \rho \sigma }\lambda_{ai}+\frac{i}{4\sqrt{2}}e^{-\si}F^I_{\mu\nu}
\ell_{I\ph{j}i}^{\ph{I}j}\B{\la}^{ai}\Up^{\mu\nu}\la_{aj}  \notag \\
&&\hspace{1.8cm}-\frac{i}{2}e^{-\si}\left(F_{\mu\nu}^I
\ell_{I\ph{k}l}^{\ph{I}k}\phi^{al}_{\ph{al}k}\phi^{\ph{a}j}_{a\ph{j}i}+
2F^a_{\mu\nu}\phi_{a\ph{j}i}^{\ph{a}j}\right)\left[ \frac{1}{4\sqrt{2}}\left( \bar{\psi}_{\rho
}^{i}\Upsilon ^{\mu \nu \rho \sigma }\psi _{\sigma j}+2\bar{\psi}^{\mu
i}\psi _{j}^{\nu }\right) \right.  \notag \\
&&\hspace{3.2cm}\left. +\frac{3}{20\sqrt{2}}\bar{\chi}^{i}\Upsilon ^{\mu \nu
}\chi _{j} -\frac{1}{4\sqrt{2}}\bar{\lambda}^{\alpha i}\Upsilon ^{\mu \nu
}\lambda _{\alpha j} +\frac{1}{2\sqrt{10}}\left( \bar{\chi}^{i}\Upsilon
^{\mu \nu \rho }\psi _{\rho j}-2\bar{\chi}^{i}\Upsilon ^{\mu }\psi _{j}^{\nu
}\right) \right]  \nn
\end{eqnarray}
\begin{eqnarray}
&&\hspace{1.8cm} +e^{-\si}F_{a\mu\nu}\left[ \frac{1}{4}\left( 2\bar{\lambda}^{ai}
\Upsilon ^{\mu }\psi _{i}^{\nu }-\bar{%
\lambda}^{ai}\Upsilon ^{\mu \nu \rho }\psi _{\rho i}\right) +\frac{1}{2%
\sqrt{5}}\bar{\lambda}^{ai}\Upsilon ^{\mu \nu }\chi _{i}\right] \nn \\
&&\hspace{1.8cm}+\frac{1}{4}e^{2\si}f_{bc}^{\ph{bc}a}f_{dea}
\phi^{bi}_{\ph{bi}k}\phi^{ck}_{\ph{ck}j}\phi^{dj}_{\ph{dj}l}
\phi^{el}_{\ph{el}i} -\frac{1}{2}e^{\si}f_{abc}\phi^{bi}_{\ph{bi}k}\phi^{ck}_{\ph{ck}j}
\left(\bar{\psi}_{\mu}^{j}\Upsilon ^{\mu }\lambda_{i}^a +\frac{2}{\sqrt{5}}\bar{\chi}^{j}\lambda _{i}^{%
\phantom{i}a}\right) \notag \\
&&  \hspace{1.8cm}-\frac{i}{\sqrt{2}}e^{\si }f_{ab}^{\ph{ab}c}\phi_{c\ph{i}j}^{\ph{c}i}
\bar{\lambda}^{aj}\lambda _{i}^b+\frac{i}{60\sqrt{2}}e^{\si }f_{ab}^{\ph{ab}c}
\phi^{al}_{\ph{al}k}\phi^{bj}_{\ph{bj}l}\phi_{c\ph{k}j}^{\ph{c}k}\left(
5\bar{\psi}_{\mu }^{i}\Upsilon ^{\mu \nu }\psi _{\nu i}+2\sqrt{5}\bar{\psi}%
_{\mu }^{i}\Upsilon ^{\mu }\chi _{i}\right.  \notag \\
&& \hspace{5.7cm}\left.\left. +3\bar{\chi}^{i}\chi _{i}-5\B{\la}^{\alpha i}
\la_{\alpha i}\right)-\frac{1}{96}\epsilon ^{\mu \nu \rho \sigma \kappa \lambda
 \tau }\ti{C}_{\mu\nu \rho} F_{\si \kappa}^aF_{a\la \tau }\right\}.
\eea
Here $f_{ab}^{\ph{ab}c}$ are the structure constants of $SU(N)$.
The covariant derivatives that appear are given by
\bea
\mathcal{D}_\mu A_{\nu a}&=&\partial_\mu A_{\nu a} -\ti{\Gamma}^\rho_{\mu\nu}A_{\rho a}+
f_{ab}^{\phantom{ab}c}A_\mu^bA_\nu^c, \\
\hat{\mathcal{D}} _{\mu }\la_{ai}&=&\partial _{\mu }\la_{ai}+\frac{1}{2}
q_{\mu i}^{\phantom{\mu i}j}\la_{aj}+\frac{1}{4}\ti{\omega} _{\mu }^{\phantom{\mu}%
\underline{\mu }\underline{\nu }}\Upsilon _{\underline{\mu }\underline{\nu }%
}\la_{ai}+ f_{ab}^{\phantom{ab}c}A_\mu^b\la_{ci}, \\
\hat{\mathcal{D}}_\mu\phi_{a\ph{i}j}^{\ph{a}i}&=&\partial_\mu \phi_{a\ph{i}j}^{\ph{a}i}
 -q_{\mu \ph{i}j\ph{k}l}^{\ph{\mu}i\ph{j}k}\phi_{a\ph{l}k}^{\ph{a}l}+
f_{ab}^{\phantom{ab}c}A_\mu^b\phi_{c\ph{i}j}^{\ph{c}i},
\eea
with the Christoffel and spin connections $\ti{\Gamma}$ and
$\ti{\omega}$ taken in the seven-dimensional Einstein frame, (with
respect to the metric $\ti{g}$). Finally, the quantities $p$ and $q$
are the Maurer-Cartan forms of the bulk scalar coset matrix
$\ell_I^{\ph{I}\underline{J}}$ as given by equations
\eqref{Maurer1}--\eqref{Maurern}. Once again, the identities for
relating the seven-dimensional gravity and $U(1)$ vector multiplet fields to
11-dimensional bulk fields are given in equations
\eqref{id1}--\eqref{idn2} for the generic $\mathbb{Z}_N$ orbifold with $N>2$,
and equations \eqref{id1}--\eqref{idn} and \eqref{idn3}--\eqref{idn4} for
the $\mathbb{Z}_2$ orbifold. We stress that these identifications are
part of the definition of the theory.

The leading order brane corrections to the supersymmetry transformation laws
\eqref{m-theory-var} of the bulk fields are computed using equations
\eqref{susy1} and \eqref{susy2}. They are given by
\bea
\delta^{\mathrm{brane}}\psi_{\mu i}&=& \frac{\kappa_7^2}{g_{\mathrm{YM}}^2}
                                       \left\{ \frac{1}{2}\left( \phi_{ak}^{\ph{ak}j}
\hat{\mathcal{D}}_\mu\phi^{a\ph{i}k}_{\ph{a}i}-\phi^{a\ph{i}k}_{\ph{a}i}
\hat{\mathcal{D}}_\mu\phi_{ak}^{\ph{ak}j}\right)\ve_j -\frac{i}{15\sqrt{2}}
\Upsilon _{\mu }\varepsilon _{i}f_{ab}^{\ph{ab}c}\phi^{al}_{\ph{al}k}\phi^{bj}_{\ph{bj}l}
\phi_{c\ph{k}j}^{\ph{c}k}e^{\si} \right. \nn \\
&& \left. \hspace{1.1cm} +\frac{i}{10\sqrt{2}}\left(
\Upsilon _{\mu }^{\ph{\mu}\nu\rho }-8\delta _{\mu }^{\nu }\Upsilon ^{\rho
}\right) \varepsilon _{j}\left( F_{\nu\rho }^{I}\ell_{I\phantom{k}l}^{%
\phantom{I}k}\phi^{al}_{\ph{al}k}\phi^{\ph{a}j}_{a\ph{j}i}+2F^a_{\nu\rho}
\phi^{\ph{a}j}_{a\ph{j}i}\right)e^{-\si} \right\}, \nn \\
\delta^{\mathrm{brane}}\chi_i&=&\frac{\kappa_7^2}{g_{\mathrm{YM}}^2}
\left\{ -\frac{i}{2\sqrt{10}}\Up^{\mu\nu}\ve_j\left( F_{\mu\nu}^I
\ell_{I\ph{k}l}^{\ph{I}k}\phi^{al}_{\ph{al}k}\phi^{\ph{a}j}_{a\ph{j}i}+
2F^a_{\mu\nu}\phi^{\ph{a}j}_{a\ph{j}i}\right)e^{-\si} \right. \nn \\ 
&&\hspace{1.1cm}\left.+\frac{i}{3\sqrt{10}}\varepsilon _{i}f_{ab}^{\ph{ab}c}
\phi^{al}_{\ph{al}k}\phi^{bj}_{\ph{bj}l}\phi_{c\ph{k}j}^{\ph{c}k}e^{\si} \right\}, \nn \\
\ell_{I\ph{i}j}^{\ph{I}i}\delta^{\mathrm{brane}}A^I_\mu&=&
\frac{\kappa_7^2}{g_{\mathrm{YM}}^2}\left\{ \left( \frac{i}{\sqrt{2}}
 \B{\psi }_{\mu }^{k}\varepsilon _{l} -\frac{i}{%
\sqrt{10}}\B{\chi }^{k}\Upsilon _{\mu }\varepsilon _{l} \right)
\phi^{al}_{\ph{al}k}\phi^{\ph{a}i}_{a\ph{i}j} e^{\si}-\B{\ve}^k
\Up_\mu\la^a_k\phi_{a\ph{i}j}^{\ph{a}i}e^\si \right\}, \label{bulksusycorr}
\end{eqnarray}
\begin{eqnarray}
\ell_{I}^{\ph{I}\alpha}\delta^{\mathrm{brane}}A^I_\mu&=&0, \nn \\
\delta^{\mathrm{brane}}\ell_{I\ph{i}j}^{\ph{I}i}&=&
\frac{\kappa_7^2}{g_{\mathrm{YM}}^2}\left\{ \frac{i}{\sqrt{2}}
\left[ \B{\ve}^k\la_{\alpha l}\phi^{al}_{\ph{al}k}
\phi^{\ph{a}i}_{a\ph{i}j}\ell_I^{\ph{I}\alpha}+\B{\ve}^l
\la_{ak}\phi^{ai}_{\ph{ai}j}\ell_{I\ph{k}l}^{\ph{I}k}-
\left( \B{\ve}^i\la_{aj}-\frac{1}{2}\delta^i_j\B{\ve}^m\la_{am}\right)
 \phi^{al}_{\ph{al}k}\ell_{I\ph{k}l}^{\ph{I}k} \right] \right\},\nn \\
\delta^{\mathrm{brane}}\ell_{I}^{\ph{I}\alpha}&=&
\frac{\kappa_7^2}{g_{\mathrm{YM}}^2}\left\{ -\frac{i}{\sqrt{2}}\B{\ve}^i\la^\alpha_j
\phi^{aj}_{\ph{aj}i}\phi^{\ph{a}l}_{a\ph{l}k}\ell_{I\ph{k}l}^{\ph{I}k}\right\}, \nn \\
\delta^{\mathrm{brane}}\la_i^\alpha &=&\frac{\kappa_7^2}{g_{\mathrm{YM}}^2}
\left\{ \frac{i}{\sqrt{2}}\Up^\mu\ve_j\phi_{ai}^{\ph{ai}j}
p_{\mu \ph{\alpha k}l}^{\ph{\mu}\alpha k}\phi^{al}_{\ph{al}k}\right\}, \nn
\eea
where $\ve_i$ is the 11-dimensional supersymmetry spinor $\eta$
projected onto the orbifold plane, as given in \eqref{spinorrel}. We note that
not all of the bulk fields receive corrections to their
supersymmetry transformation laws. The leading order supersymmetry
transformation laws of the $SU(N)$ multiplet fields are found using equation
\eqref{susy2} and take the form
\bea
\delta A_\mu^a&=& \B{\ve}^i\Up_\mu\la_i^ae^\si-\left( i\sqrt{2}\psi_\mu^i\ve_j
 -\frac{2i}{\sqrt{10}}\B{\chi}^i\Up_\mu\ve_j \right)\phi^{aj}_{\ph{aj}i}e^\si,\nn \\
\delta \phi_{a\ph{i}j}^{\ph{a}i}&=&-i\sqrt{2}\left( \B{\ve}^i\la_{aj}-
\frac{1}{2}\delta^i_j\B{\ve}^k\la_{ak}\right), \label{susybranexmfn} \\
\delta\la^a_i&=&-\frac{1}{2}\Up^{\mu\nu}\ve_i\left( F^I_{\mu\nu}
\ell^{\ph{I}j}_{I\ph{j}k}\phi^{ak}_{\ph{ak}j}+F^a_{\mu\nu}\right)e^{-\si}-i\sqrt{2}
\Up^\mu\ve_j\hat{\mathcal{D}}_\mu\phi^{a\ph{i}j}_{\ph{a}i}-i\ve_j
f^a_{\ph{a}bc}\phi^{bj}_{\ph{bj}k}\phi^{ck}_{\ph{ck}i}. \nn 
\eea

To make some of the properties of our result more transparent, it is
helpful to extract the bosonic part of the action. This bosonic part
will also be sufficient for many practical applications. We recall
that the full Lagrangian \eqref{braneaction} is written in the
seven-dimensional Einstein frame to avoid the appearance of
$\sigma$--dependent pre-factors in many terms. The bosonic part,
however, can be conveniently formulated in terms of $g_{\mu\nu}$, the
seven-dimensional part of the 11-dimensional bulk metric
$g_{MN}$. This requires performing the Weyl-rescaling~\eqref{Weyl}.
It also simplifies the notation if we rescale the scalar $\si$ as
$\tau=10\si/3$, and drop the tilde from the three-form
$\tilde{C}_{\mu\nu\rho}$ and its field strength
$\tilde{G}_{\mu\nu\rho\sigma}$, which exactly coincide with the purely
seven-dimensional components of their 11-dimensional counterparts.
Let us now write down the purely bosonic part of our action, subject
to these small modifications. We find
\begin{equation}
\mathcal{S}_{11-7,{\rm bos}}=\mathcal{S}_{11,{\rm bos}} + \mathcal{S}_{7,{\rm bos}}\, ,
\end{equation}
where $\mathcal{S}_{11,{\rm bos}}$ is the bosonic part of 11-dimensional
supergravity~\eqref{m-theory}, with fields subject to the orbifold constraints
\eqref{cond1}--\eqref{cond6}. Further, $\mathcal{S}_{7,{\rm bos}}$ is the bosonic
part of Eq.~\eqref{braneaction}, subject to the above modifications, for which
we obtain
\bea \label{boseaction}
\mathcal{S}_{7,{\rm bos}}&=& \frac{1}{g_{\mathrm{YM}}^2} \int_{y=0} \mathrm{d}^7x \sqrt{-g}
\left( -\frac{1}{4}H_{ab}F^a_{\mu\nu}F^{b\mu\nu} -\frac{1}{2}H_{aI}F^a_{\mu\nu}
F^{I\mu\nu} -\frac{1}{4}(\delta H)_{IJ}F^I_{\mu\nu}F^{J\mu\nu}  \right. \nn \\
&&\hspace{3.5cm} \left. -\frac{1}{2}e^\tau\hat{\mathcal{D}}_\mu
\phi_{a\ph{i}j}^{\ph{a}i}\hat{\mathcal{D}}^\mu\phi_{\ph{aj}i}^{aj}
-\frac{1}{2}\left( \delta K \right)^{\alpha j \ph{i} \beta l}_{\ph{\alpha j}i
\ph{\beta l} k}p_{\mu \alpha \ph{i} j}^{\ph{\mu\alpha}i\ph{j}}
p^{\mu \ph{\beta}k}_{\ph{\mu}\beta\ph{k}l}+\frac{1}{4}D^{ai}_{\ph{ai}j}D^{\ph{a}j}_{a\ph{j}i}\right) \nn \\
&&-\frac{1}{4g_{\mathrm{YM}}^2}\int_{y=0} C\wedge F^a \wedge F_a,
\eea
where
\bea
H_{ab}&=&\delta_{ab}, \label{H1} \\
H_{aI}&=&2\ell_{I\ph{i}j}^{\ph{I}i}\phi_{a\ph{j}i}^{\ph{a}j}, \label{H2} \\
(\delta H)_{IJ}&=&2\ell_{I\ph{i}j}^{\ph{I}i}\phi_{a\ph{j}i}^{\ph{a}j}
\ell_{J\ph{k}l}^{\ph{J}k}\phi_{a\ph{l}k}^{\ph{a}l},  \label{H3} \\
\left( \delta K \right)^{\alpha j \ph{i} \beta l}_{\ph{\alpha j}i\ph{\beta l} k}&=&
 e^\tau \delta^{\alpha\beta} \phi_{a\ph{j}i}^{\ph{a}j}\phi^{al}_{\ph{al}k}, \\
D^{ai}_{\ph{ai}j}&=&e^\tau f^a_{\ph{a}bc}\phi^{bi}_{\ph{bi}k}\phi^{ck}_{\ph{ck}j}.
\eea
The gauge covariant derivative is denoted by $\mathcal{D}$, while $\hat{\mathcal{D}}$ is given by
\begin{equation}
\hat{\mathcal{D}}_\mu\phi_{a\ph{i}j}^{\ph{a}i}=\mathcal{D}_\mu
 \phi_{a\ph{i}j}^{\ph{a}i} -q_{\mu \ph{i}j\ph{k}l}^{\ph{\mu}i\ph{j}k}\phi_{a\ph{l}k}^{\ph{a}l}\, .
\end{equation}
The Maurer-Cartan forms $p$ and $q$ of the matrix of scalars $\ell$ are defined by
\bea 
p_{\mu\alpha \ph{i}j}^{\ph{\mu\alpha }i}&=&\ell^{I}_{\ph{I}\alpha}\partial_\mu
 \ell_{I\ph{i}j}^{\ph{\mu}i}, \\
q_{\mu \ph{i}j\ph{k}l}^{\ph{\mu}i\ph{j}k}&=&\ell^{Ii}_{\ph{Ii}j}\partial_\mu
 \ell_{I\ph{k}l}^{\ph{\mu}k}.
\eea
The bosonic fields localized on the orbifold plane are the $SU(N)$
gauge vectors $F^a=\mathcal{D}A^a$ and the $SU(2)$ triplets of scalars
$\phi_{a\ph{i}j}^{\ph{a}i}$. All other fields are projected from the
bulk onto the orbifold plane, and there are algebraic equations relating
them to the 11-dimensional fields in $\mathcal{S}_{11}$.
As discussed above, these relations are trivial for the metric $g_{\mu\nu}$
and the three-form $C_{\mu\nu\rho}$, while the scalar $\tau$ is given by
\begin{equation} \label{rename1}
\tau = \frac{1}{2}\ln\det g_{AB}\, ,
\end{equation}
and can be interpreted as an overall scale factor of the orbifold
$\mathbb{C}^2/\mathbb{Z}_N$. For the remaining fields, the ``gravi-photons''
$F^I_{\mu\nu}$ and the ``orbifold moduli'' $\ell_I^{\ph{I}\underline{J}}$,
we have to distinguish between the generic $\mathbb{Z}_N$ orbifold with
$N>2$ and the $\mathbb{Z}_2$
orbifold. For $\mathbb{Z}_N$ with $N>2$ we have four $U(1)$ gauge fields,
so that $I=1,\ldots ,4$, and $\ell_I^{\ph{I}\underline{J}}$ parameterizes
the coset $SO(3,1)/SO(3)$. They are identified with 11-dimensional fields through
\bea \label{rename2}
F^I_{\mu\nu}&=&-\frac{i}{2}\mathrm{tr}\left( \si^IG_{\mu\nu}\right), \\
\ell_I^{\ph{I}\underline{J}}&=&\frac{1}{2}\mathrm{tr}\left( \B{\si}_I v \si^J v^{\dagger} \right),
\eea
where $G_{\mu\nu}\equiv (G_{\mu\nu p\B{q}})$,
$v\equiv (e^{\tau/4}e^{\B{p}}_{\ph{p}\B{\underline{q}}})$ and $\si^I$
are the $SO(3,1)$ Pauli matrices as given in Appendix \ref{Pauli}.
For the $\mathbb{Z}_2$ case, we have six $U(1)$ vector fields, so that
$I=1,\ldots ,6$, and $\ell_I^{\ph{I}\underline{J}}$ parameterizes the
coset $SO(3,3)/SO(3)^2$. The field identifications now read
\bea
F^I_{\mu\nu}&=&-\frac{1}{4}\mathrm{tr}\left( T^IG_{\mu\nu}\right), \\
\ell_I^{\ph{I}\underline{J}}&=&\frac{1}{4}\mathrm{tr}\left( \B{T}_I v T^J v^{T} \right),
\eea
where this time $G_{\mu\nu}\equiv (G_{\mu\nu AB})$,
$v\equiv (e^{\tau/4}e^{A}_{\ph{A}\underline{B}})$, and $T^I$
are the generators of $SO(4)$, as given in Appendix \ref{Pauli}.\\

Let us discuss a few elementary properties of the bosonic
action~\eqref{boseaction} on the orbifold plane, starting with the
gauge-kinetic functions \eqref{H1}--\eqref{H3}. The first observation
is, that the gauge-kinetic function for the $SU(N)$ vector fields is
trivial (to the order we have calculated), which confirms the result
of Ref.~\cite{Friedmann:2002ty}. On the other hand, we find non-trivial gauge
kinetic terms between the $SU(N)$ vectors and the gravi-photons, as
well as between the gravi-photons. We also note the appearance of the
Chern-Simons term $C\wedge F^a\wedge F_a$, which has been
predicted~\cite{Witten:2001uq} from anomaly cancellation in configurations
which involve additional matter fields on conical singularities, but,
in our case, simply follows from the structure of seven-dimensional
supergravity without any further assumption. We note that, while there
is no seven-dimensional scalar field term which depends only on
orbifold moduli, the scalar field kinetic terms in~\eqref{boseaction}
constitute a complicated sigma model which mixes the orbifold moduli
and the scalars in the $SU(N)$ vector multiplets. A further interesting
feature is the presence of the seven-dimensional D-term potential in
Eq.~\eqref{boseaction}. Introducing the matrices $\phi_a\equiv
(\phi^{\ph{a}i}_{a\ph{i}j})$ and $D^a\equiv (D^{ai}_{\ph{ai}j})$
this potential can be written as
\begin{equation}
 V=\frac{1}{4g_{\rm YM}^2}{\rm tr}\left( D^aD_a\right)\, ,
\end{equation}
where
\begin{equation}
D^a=\frac{1}{2}e^\tau f^a_{\ph{a}bc}[\phi^b,\phi^c]\; .
\end{equation}
The flat directions, $D^a=0$, of this potential, which correspond to
unbroken supersymmetry as can be seen from Eq.~\eqref{susybranexmfn},
can be written as
\begin{equation}
\phi^a=v^a\sigma^3
\end{equation}
with vacuum expectation values $v^a$. The $v^a$ correspond to elements
in the Lie algebra of $SU(N)$ which can be diagonalized into the
Cartan sub-algebra. Generic such diagonal matrices break $SU(N)$
to $U(1)^{N-1}$, while larger unbroken groups are possible for non-generic
choices. Looking at the scalar field masses induced from the D-term in such a
generic situation, we have one massless scalar for each
of the non-Abelian gauge fields which is absorbed as their longitudinal
degree of freedom. For each of the $N-1$ unbroken Abelian gauge fields,
we have all three associated scalars massless, as must be the case from
supersymmetry. This situation corresponds exactly to what happens when
the orbifold singularity is blown up. We can, therefore, see that within
our supergravity construction blowing-up is encoded by the D-term.
Further, the Abelian gauge fields in $SU(N)$ correspond to (a truncated version of)
the massless vector fields which arise from zero modes of the M-theory three-form
on a blown-up orbifold, while the $3(N-1)$ scalars in the Abelian vector fields
correspond to the blow-up moduli.


\section{Discussion and outlook}

In this Chapter, we have constructed the effective supergravity action for
M-theory on the orbifold $\mathbb{C}^2/\mathbb{Z}_N\times\mathbb{R}^{1,6}$,
by coupling 11-dimensional supergravity, constrained in accordance with
the orbifolding, to $SU(N)$ super-Yang-Mills theory located on the
seven-dimensional fixed plane of the orbifold. We have found that the
orbifold-constrained fields of 11-dimensional supergravity, when restricted
to the orbifold plane, fill out a seven-dimensional supergravity multiplet
plus a single $U(1)$ vector multiplet for $N>2$ and three $U(1)$ vector
multiplets for $N=2$. The seven-dimensional action on the orbifold plane,
which has to be added to 11-dimensional supergravity, couples these bulk
degrees of freedom to genuine seven-dimensional states in the $SU(N)$
multiplet. We have obtained this action on the orbifold plane by ``up-lifting''
information from the known action of ${\cal N}=1$ Einstein-Yang-Mills
supergravity and identifying 11- and 7-dimensional degrees of freedom appropriately.
The resulting 11-/7-dimensional theory is given as an expansion in the
parameter $h=\kappa^{5/9}/g_{\rm YM}$, where $\kappa$ is the 11-dimensional
Newton constant and $g_{\rm YM}$ is the seven-dimensional $SU(N)$ coupling.
The bulk theory appears at zeroth order in $h$, and we have determined the
complete set of leading terms on the orbifold plane which
are of order $h^2$. At order $h^4$ we encounter a singularity due to
a delta function square, similar to what happens in Ho\v{r}ava-Witten
theory~\cite{Horava:1996ma}. As in Ref.~\cite{Horava:1996ma}, we assume that this singularity
will be resolved in full M-theory, when the finite thickness of the
orbifold plane is taken into account, and that it does not invalidate the
results at order $h^2$. 

While we have focused on the A-type orbifolds
$\mathbb{C}^2/\mathbb{Z}_N$, we expect our construction to work
analogously for the other four-dimensional orbifolds of ADE type. Our
result represents the proper starting point for compactifications of
M-theory on $G_2$ spaces with singularities of the type
$\mathbb{C}^2/\mathbb{Z}_N\times B$, where $B$ is a three-dimensional
manifold. We consider this to be the first step in a programme, aiming
at developing an explicit supergravity framework for ``phenomenological''
compactifications of M-theory on singular $G_2$ spaces.

\chapter{Four-dimensional Effective M-theory on a Singular $G_2$ manifold}\label{c:G2Ch2}

\section{Introduction} 
In this Chapter, we shall investigate an M-theory compactification on $G_2$ spaces with $A$-type singularities, which leads to four-dimensional $SU(N)$ gauge multiplets. As we observed in Chapter \ref{G2Intro}, such a construction is an important step towards generating realistic particle phenomenology from M-theory. The associated effective theory for M-theory on spaces of the form
$\mathcal{M}_7\times\mathbb{C}^2/\mathbb{Z}_N$, where $\mathcal{M}_7$ is a smooth
seven-dimensional space with Minkowski signature, was derived in the previous Chapter (and \cite{Anderson:2006pb}). With this action available it is now possible to carry out an explicit reduction to four-dimensions of M-theory on a $G_2$ space containing
$\mathbb{C}^2/\mathbb{Z}_N$ singularities. 

The main objective of this Chapter is to study such compactifications of
M-theory, using the action derived in the previous chapter, \eref{schematic_action} and \eref{braneaction}, to explicitly describe
the non-Abelian gauge fields which arise at the singularities.  The
relevant $G_2$ spaces are constructed by dividing a seven-torus $\mathcal{T}^7$
by a discrete symmetry group $\Gamma$, such that the resulting
singularities are of co-dimension four and of A-type. We will then
perform the reduction to four dimensions on these spaces. This
includes the reduction of the seven-dimensional $\mathrm{SU}(N)$ gauge
theories on the three-dimensional singular loci within the $G_2$
space. For the orbifold examples considered in this chapter, the
singular loci will always be three-tori, $\mathcal{T}^3$. Hence, while the full
four-dimensional theory is $\mathcal{N}=1$ supersymmetric, the gauge
sub-sectors associated to each singularity have enhanced
$\mathcal{N}=4$ supersymmetry. Let us split the four-dimensional field
content into ``bulk fields'' which descend from 11-dimensional
supergravity and ``matter fields'' which descend from the
seven-dimensional super-Yang-Mills theories at the singularities. The
bulk fields correspond to the moduli of the $G_2$ space (plus
associated axions from the M-theory three-form). For the $G_2$ spaces
we consider, the only geometrical parameters that survive the
orbifolding are the seven radii of the torus. Hence we have seven
moduli $\tilde{T}^A$ in the reduced theory. At an orbifold singularity these
moduli can be divided up into $\tilde{T}^0$ and $\tilde{T}^{mi}$, where $m=1,2,3$ and
$i=1,2$. The modulus $\tilde{T}^0$ can be thought of as corresponding to the
volume of the three-torus locus of the singularity, while the $\tilde{T}^{mi}$
depend on one radius $R^m$ of the singular locus and two radii of $\mathcal{T}^7$
transverse to the singularity. For each singularity, the matter fields
consist of $\mathcal{N}=1$ vector multiplets with gauge group
$\mathrm{SU}(N)$, plus three chiral multiplets $\mathcal{C}^{am}$ per
vector multiplet, where $m=1,2,3$ and $a$ is an adjoint index of the
gauge group. This is indeed the field content of the $\mathcal{N}=4$
theory in $\mathcal{N}=1$ language.

Let us now summarize our main results. If we focus on one particular singularity,
the K\"ahler potential, gauge-kinetic function and superpotential of the
four-dimensional effective theory are given by
\bea
K&=& \frac{7}{\kappa_4^2}\ln 2 - \frac{1}{\kappa_4^2}\sum_{A=0}^6 \ln (\tilde{T}^A
+ {\bar{\ti{T}}}^A ) +
\frac{1}{4\lambda_4^2}\sum_{m=1}^3\frac{(\mathcal{C}_a^m+\bar{\mathcal{C}}_a^m)
(\mathcal{C}^{am}+\bar{\mathcal{C}}^{am})}{(\tilde{T}^{m1}+\bar{\tilde{T}}^{m1})
(\tilde{T}^{m2}+\bar{\tilde{T}}^{m2})}\,, \label{kahlerintro}\\
f_{ab}&=&\frac{1}{\lambda_4^2}\tilde{T}^0\delta_{ab}\, , \label{gaugeintro} \\ W
&=& \frac{\kappa_4^2}{24\lambda_4^2}f_{abc}\sum_{m,n,p=1}^3
\epsilon_{mnp}\mathcal{C}^{am}\mathcal{C}^{bn}\mathcal{C}^{cp}\,. \label{supotintro}
\eea
These expressions constitute the leading order terms of series
expanded in terms of the parameter $h_4=\kappa_4/\lambda_4$, that is, the
gravitational coupling divided by the gauge coupling. Complete
expressions for the above quantities involve a sum over all
singularities and are given in Section~\ref{s:reduction}.

It is interesting to compare these results with those found by
compactification on the associated smooth $G_2$ space, obtained by
blowing-up the singularities. The construction of smooth $G_2$ spaces
in this way has been pioneered by Joyce~\cite{Joyce} and the
associated four-dimensional effective theories for M-theory on such
$G_2$ spaces have been computed in Refs.~\cite{Lukas:2003dn,Barrett:2004tc}.  For an
explicit comparison, it is useful to recall that the geometrical
procedure of blowing-up can be described within the $\mathrm{SU}(N)$ gauge
theory as a Higgs effect, whereby VEVs are assigned to the (real parts
of the) Abelian matter fields $\mathcal{C}^{im}$ along the D-flat
directions~\cite{Anderson:2006pb}.  This generically breaks $\mathrm{SU}(N)$ to its
maximal Abelian subgroup $\mathrm{U}(1)^{N-1}$ and leaves only the $3(N-1)$ chiral
multiplets $\mathcal{C}^{im}$ massless. This field content corresponds
precisely to the zero modes of M-theory on the blown-up geometry, with
the Abelian gauge fields arising as zero modes of the M-theory
three-form on the $N-1$ two-spheres of the blow-up and the chiral
multiplets corresponding to its moduli.  With this interpretation of
the blowing-up procedure we can compare the above K\"ahler potential,
restricted to the Abelian matter fields $\mathcal{C}^{im}$, to
the smooth result. We find that subject to a suitable embedding of
the Abelian group $\mathrm{U}(1)^{N-1}$ into $\mathrm{SU}(N)$ they are indeed exactly
the same, provided the Abelian matter fields $\mathcal{C}^{im}$ are
identified with the blow-up moduli. Also note that, when restricted
to the Abelian fields $\mathcal{C}^{im}$, the above superpotential~\eqref{supotintro}
vanishes (consistent with the result one obtains in the smooth case).

We  also  show that  the  superpotential~\eqref{supotintro}  can  be
obtained from a Gukov-type formula which involves the integration
of  the  complexified   Chern-Simons  form  of  the  seven-dimensional
$\mathrm{SU}(N)$  gauge theory  over the  internal three-torus  $\mathcal{T}^3$.  In
addition, we show explicitly  that the superpotential for Abelian flux
of the $\mathrm{SU}(N)$  gauge fields on $\mathcal{T}^3$ is correctly obtained from
the same  Gukov formula, an observation first made in Ref.~\cite{Acharya:2002kv}.  This
result  provides a further  confirmation for the matching  between the
singular  and smooth theories  as the flux  superpotential in  both limits
is of the same form if the field identification suggested by the
comparison of K\"ahler potentials is used.

Finally, we consider one of the $\mathcal{N}=4$ sectors of our action with gravity
switched off. We point out that singular and blown-up geometries correspond
to the conformal phase and the Coulomb phase of this  $\mathcal{N}=4$ theory, respectively.
Additionally, its S-duality symmetry translates into a T-duality on the singular $\mathcal{T}^3$ locus.
We speculate about a possible extension of this S-duality to the full supergravity
theory and analyze some of the resulting consequences.  Various technical details are left until the Appendices. In Appendix \ref{A}, the construction, geometry, and
topology of orbifold based $G_2$ manifolds is discussed. There is also
a list of sixteen possible orbifold groups that lead to singular $G_2$
manifolds for which the results of this chapter are directly
applicable. In Appendix \ref{appred}  we reduce M-theory on a
smooth Joyce $G_2$ manifold, and then presents the K\"ahler potential for
M-theory on the blown-up orbifolds.  

Let us state the index conventions we shall use for the various spaces we
consider. We take 11-dimensional space-time to have mostly positive
signature, that is $(-++\ldots+)$, and use indices
$M,N,\ldots=0,1,\ldots,10$ to label 11-dimensional coordinates
$(x^M)$.  Four-dimensional coordinates on $\mathbb{R}^{1,3}$ are
labeled by $(x^\mu)$, where $\mu,\nu,\ldots=0,1,2,\ldots,3$, while
points on the internal $G_2$ space $\mathcal{Y}$ are labeled by
coordinates $(x^A)$, where $A,B,\ldots = 4,\ldots ,10$.  The
coordinates of co-dimension four singularities in the internal space
will be denoted by $(y^{\hat{A}})$, $\hat{A},\hat{B}\ldots=7,8,9,10$,
while the complementary three-dimensional singular $\mathcal{T}^3$ locus has
coordinates $x^m$, with $m,n,\ldots =4,5,6$. To describe the
seven-dimensional gauge theories it is also useful to introduce
coordinates $(x^{\hat{\mu}})$,
$\hat{\mu},\hat{\nu},\ldots=0,1,\ldots,6$ on the locus of the
singularity. Underlined versions of all the above index types denote
the associated tangent space indices. We frequently use the term
``bulk'' to refer to the full 11-dimensional space-time.

The basic structure of the $11$-dimensional theory that we derived in the previous Chapter is given in  \eref{schematic_action}. This form of the
action is valid for a single singularity of the type $\mathbb{C}^2/\mathbb{Z}_N$.
Below, in the context of $G_2$ compactifications, we will need a trivial
extension of this action incorporating a number of similar seven-dimensional
actions $\mathcal{S}_7$, one for each singularity. Considering just the bosonic part of the theory, at the orbifold fixed plane $11$-dimensional supergravity is enhanced by
\bea \label{boseaction2}
\mathcal{S}_{7}^{(0)}&=&  \int_{\mathbb{R}^{1,3}\times B}
     \mathrm{d}^7x \sqrt{-\hat{g}}\left(
     -\frac{1}{4}H_{ab}F^a_{\hat{\mu}\hat{\nu}}F^{b\hat{\mu}\hat{\nu}}
     -\frac{1}{2}H_{aI}F^a_{\hat{\mu}\hat{\nu}}F^{I\hat{\mu}\hat{\nu}}
     -\frac{1}{4}(\delta H)_{IJ}F^I_{\hat{\mu}\hat{\nu}}F^{J\hat{\mu}\hat{\nu}}      \right. \nn \\
&&\hspace{4.0cm} \left. -\frac{1}{2}e^\tau\hat{\mathcal{D}}_{\hat{\mu}}
     \phi_{au}\hat{\mathcal{D}}^{\hat{\mu}}\phi^{au}-\frac{1}{2}\left( \delta K      \right)_{I u J v}\partial_{\hat{\mu}}\ell^{Iu}\partial^{\hat{\mu}}\ell^{Jv}
     +\frac{1}{4}D^{au}D_{au}\right) \nn \\
&&-\frac{1}{4}\int_{\mathbb{R}^{1,3}\times B} C\wedge F^a \wedge F_a
     \, ,
\eea
where
\bea
H_{ab}&=&\delta_{ab}\, , \label{relabelH1} \\
H_{aI}&=&2\,\phi_{au}{\ell_{I}}^u\, , \label{relabelH2} \\
(\delta H)_{IJ}&=&2\,\phi_{au}{\phi^a}_v\ell_{I}^{\ph{I}u}\ell_{J}^{\ph{J}v}\, ,
     \label{relabelH3} \\
\left( \delta K \right)_{I u J v}&=&e^\tau  \phi_{au}\phi^{a}_{\ph{a}v}
     \delta_{\alpha\beta}{\ell_{I}}^\alpha{\ell_J}^\beta\, , \\
D_{au}&=&\frac{i}{\sqrt{2}}\epsilon_{uvw}e^\tau f_{abc}\phi^{bv}\phi^{cw}\, .
\eea
With this in hand, we turn now to a compactification of this theory on a $G_2$ space with $A$-type singularities


\section{The four-dimensional effective action on a $G_2$  orbifold} \label{s:reduction}
In this section, we will calculate the four-dimensional effective
theory for M-theory on a $G_2$ orbifold, using the
action~\eqref{m-theory}, \eref{schematic_action}, \eqref{boseaction}, \eref{boseaction2} for
11-dimensional supergravity coupled to seven-dimensional
super-Yang-Mills theory. 

We assume the $G_2$ orbifold $\mathcal{Y}$ takes the
form $\mathcal{T}^7/\Gamma$, where $\mathcal{T}^7$ is a seven-torus and $\Gamma$ is a
discrete group of symmetries of $\mathcal{T}^7$. We assume further that the
fixed points of the orbifold group $\Gamma$ are all of co-dimension
four, and also that points on the torus that are fixed by one
generator of $\Gamma$ are not fixed by other generators. A class of
such orbifolds was constructed in Ref.~\cite{joyce1,joyce2,Joyce,Barrett:2004tc}, and are
described in Appendix~\ref{A}. According to the ADE classification,
the singularities of $\mathcal{Y}$ are then all of type $A_{N-1}=\mathbb{Z}_N$,
for some $N$ \cite{Barrett:2004tc}. Furthermore, the approximate form of $\mathcal{Y}$
near a singularity is $\mathbb{C}^2/\mathbb{Z}_{N}\times \mathcal{T}^3$, where
$\mathcal{T}^3$ is a three-torus. The coordinates of the underlying torus $\mathcal{T}^7$
are denoted by $(x^A)$, where we change the range of indices to
be $A,B,\ldots = 1,\ldots ,7$, for convenience. For most of this
section, we will focus on one such singularity for simplicity. In the
neighborhood of a singularity, M-theory is described by the
seven-dimensional action reviewed in the previous section. Without
loss of generality, we consider this singularity of $\mathcal{Y}$ to be located
at $x^4=x^5=x^6=x^7=0$ and we split coordinates according to
\begin{equation} \label{7coordid}
(x^A)\to (x^m,y^{\hat{A}})\; ,
\end{equation}
where $m,n,\ldots =1,2,3$ and $\hat{A},\hat{B},\ldots = 4,5,6,7$.
The generator~\eqref{R} of the $\mathbb{Z}_N$ symmetry then acts
on the coordinates $(y^{\hat{A}})$.

For the purpose of our calculation, it is convenient to work with the orbifold
rather than any partially blown-up version thereof. However, we will see
later that the possibility of blowing up some of the singularities can,
in fact, be effectively described within the low-energy four-dimensional
gauge theories associated with the singularities.

\subsection{Background solution and zero modes}
Let us now discuss the M-theory background on
$\mathcal{M}_{11}=\mathbb{R}^{1,3}\times \mathcal{Y}$. Throughout our
calculations we take the expectation values of fermions to vanish. This means
that we only need concern ourselves with the bosonic equations of motion. For
the bulk 11-dimensional supergravity these are
\bea
\mathrm{d}G&=&0\, , \label{bulkeom1} \\
\mathrm{d}\ast G &=& -\frac{1}{2}G\wedge G\, ,  \label{bulkeom2}\\
\hat{R}_{MN}&=&\frac{1}{12}\left( G_{MPQR}{G_N}^{PQR}-\frac{1}{12}\hat{g}_{MN}G_{PQRS}G^{PQRS} \right)\, . \label{bulkeomn}
\eea
We take the background metric $\langle\hat{g}\rangle$ to be general
Ricci flat, while for the three-form, we choose vanishing background
$\langle C \rangle=0$. For $\mathcal{Y}$ being a toroidal $G_2$
orbifold, the Ricci flatness condition implies $\langle\hat{g}\rangle$
has constant components. In addition, these components should be
constrained in accordance with the orbifold symmetry. Truncating these
11-dimensional fields to our particular singularity, this background
leads to constant seven-dimensional fields $\langle \tau \rangle$ and
$\langle {\ell_I}^{\ul{J}}\rangle$, and vanishing $\langle
A_{\hat{\mu}}^I \rangle$, according to the identifications
\eqref{id1}, \eqref{idn}. Substituting this background into the field
equations for the localized fields gives
\bea
\mathcal{D}F&=&0\, , \\
\mathcal{D}_{\hat{\mu}}F^{a\hat{\mu}\hat{\nu}}&=&e^{\langle\tau\rangle}
     {f^a}_{bc}\phi^{b}_u\hat{\mathcal{D}}^{\hat{\nu}}\phi^{cu}\, ,\\
\hat{\mathcal{D}}_{\hat{\mu}}\hat{\mathcal{D}}^{\hat{\mu}}\phi_{au}&=&
     -\frac{i\sqrt{2}}{2}\epsilon_{uvw}f_{abc}\phi^{bv}
     D^{cw}\, .
\eea
A valid background is thus obtained by setting the genuine
seven-dimensional fields to zero, that is, $\langle A_{\hat{\mu}}^a\rangle=0$,
and $\langle \phi_{au} \rangle =0$. With these fields switched off, the singularity
causes no modification to the background for the bulk fields.

We now discuss supersymmetry of the background. Substitution of our background
into the fermionic supersymmetry transformation laws
\eqref{gravitino_var}, \eqref{bulksusycorr}, \eqref{susybranexmfn} makes every term vanish
except for the $\nabla_M \eta$ term in the variation of the gravitino. Hence,
the existence of precisely one Killing spinor on a $G_2$ space guarantees that
our background is supersymmetric, with $\mathcal{N}=1$ supersymmetry from a
four-dimensional point of view. 

Let us now discuss the zero modes of these background solutions, both for the
bulk and the localized fields. We begin with the bulk zero modes.
All the orbifold examples discussed in Appendix~\ref{A} restrict the internal metric
to be diagonal (and do not allow any invariant two-forms) and we will focus
on examples of this type in what follows. Hence, the 11-dimensional metric can be written as
\begin{equation} \label{metricA}
\mathrm{d}s^2=\left( \prod_{A=1}^7 R^A \right)^{-1} g_{\mu\nu}dx^\mu dx^\nu +
              \sum_{A=1}^{7}\left(R^A\mathrm{d}x^A\right)^2\, .
\end{equation}
Here the $R^A$ are precisely the seven radii of the underlying
seven-torus. The factor in front of the first part has been
chosen so that $g_{\mu\nu}$ is the four-dimensional metric in the
Einstein frame. There exists a $G_2$ structure $\varphi$, a harmonic
three-form, associated with each Ricci flat metric. For the seven-dimensional
part of the above metric and an appropriate choice of coordinates it is given by
\begin{eqnarray} \label{structure3}
\varphi & = & R^1R^2R^3\mathrm{d}x^1\wedge\mathrm{d}x^2\wedge\mathrm{d}x^3+
R^1R^4R^5\mathrm{d}x^1\wedge\mathrm{d}x^4\wedge\mathrm{d}x^5-
R^1R^6R^7\mathrm{d}x^1\wedge\mathrm{d}x^6\wedge\mathrm{d}x^7 \nonumber \\
 & &+R^2R^4R^6\mathrm{d}x^2\wedge\mathrm{d}x^4\wedge\mathrm{d}x^6 +
R^2R^5R^7\mathrm{d}x^2\wedge\mathrm{d}x^5\wedge\mathrm{d}x^7+
R^3R^4R^7\mathrm{d}x^3\wedge\mathrm{d}x^4\wedge\mathrm{d}x^7 \nonumber \\
 & & -R^3R^5R^6\mathrm{d}x^3\wedge\mathrm{d}x^5\wedge\mathrm{d}x^6.
\end{eqnarray}
It is the coefficients of $\varphi$ that define the metric moduli $a^A$, where
$A=0,\ldots ,6$, in the reduced theory, and these become the real bosonic parts of chiral
superfields. We thus set
\begin{equation} \label{period2}
\left. \begin{array}{cccc}
a^0=R^1R^2R^3, & a^{1}=R^1R^4R^5, & a^{2}=R^1R^6R^7, & a^{3}=R^2R^4R^6, \\
a^{4}=R^2R^5R^7, & a^{5}=R^3R^4R^7, & a^{6}=R^3R^5R^6. &  \, \\
\end{array} \right.
\end{equation}
Since there are no one-forms on a $G_2$ space, and our assumption
above states that there are no two-forms on $\mathcal{Y}$, the three-form field $C$ expands
purely in terms of three-forms, and takes the same form as $\varphi$, thus
\bea \label{Cexp1}
C&=& \nu^0\mathrm{d}x^1\wedge\mathrm{d}x^2\wedge\mathrm{d}x^3+
\nu^{1}\mathrm{d}x^1\wedge\mathrm{d}x^4\wedge\mathrm{d}x^5-
\nu^{2}\mathrm{d}x^1\wedge\mathrm{d}x^6\wedge\mathrm{d}x^7 +
 \nu^{3}\mathrm{d}x^2\wedge\mathrm{d}x^4\wedge\mathrm{d}x^6 \nn \\
 && + \nu^{4}\mathrm{d}x^2\wedge\mathrm{d}x^5\wedge\mathrm{d}x^7+
\nu^{5}\mathrm{d}x^3\wedge\mathrm{d}x^4\wedge\mathrm{d}x^7-
\nu^{6}\mathrm{d}x^3\wedge\mathrm{d}x^5\wedge\mathrm{d}x^6.
\eea
The $\nu^A$ become axions in the reduced theory, and pair up with the metric moduli
to form the superfields
\begin{equation}
\label{TA}
T^A=a^A+i\nu^A.
\end{equation}
In general, not all of the $T^A$ are independent. In fact, a simple
procedure determines which of the $T^A$ are constrained to be
equal. Each generator $\alpha_\tau$ of the orbifold group $\Gamma$
acts by simultaneous rotations in two planes, corresponding to index pairs
$(A,B)$ and $(C,D)$ say. If the order of $\alpha_\tau$ is greater than two, then we
identify $R^A$ with $R^B$ and $R^C$ with $R^D$.  We go through this
process for all generators of $\Gamma$ and then use \eqref{period2} to
determine which of the $a^A$, and hence $T^A$, are equal.

We now discuss a convenient relabeling of the metric moduli, adapted to the
structure of the singularity. Under the identification of coordinates \eqref{7coordid}, the
metric modulus $a^0$ can be viewed as the volume modulus of the three-torus locus
$\mathcal{T}^3$ of the singularity. The other moduli, meanwhile, are each a product
of one radius of the torus $\mathcal{T}^3$ with two radii of $\mathcal{Y}$ transverse to
the singularity. It is sometimes useful to change the notation for
these moduli to the form $a^{mi}$ where $m=1,2,3$ labels the radius on
$\mathcal{T}^3$ that $a^{mi}$ depends on, and $i=1,2$. Thus
\begin{equation}
\begin{array}{cccccc}
a^{11}=a^1, & a^{12}=a^2, & a^{21}=a^3, & a^{22}=a^4 & a^{31}=a^5,& a^{32}=a^6.
\end{array}
\end{equation}
We will also sometimes make the analogous change of notation for $\nu^A$
and $T^A$.

Having listed the bulk moduli, we now turn to the zero modes
associated with the singularity. The decomposition of
seven-dimensional fields works as follows. We take the straightforward
basis $(\mathrm{d}x^m)$ of harmonic one-forms on the three-torus, so
$A^a_{\hat{\mu}}$ simply decomposes into a four-dimensional vector $A^a_{\mu}$ plus the
three scalar fields $A^a_m$ under the reduction. The seven-dimensional
scalars $\phi_{au}$ simply become four-dimensional scalars. Setting
\bea
{b_a}^m&=&-A_{ma}, \\
{\rho_a}^1&=&\sqrt{a^{11}a^{12}}{\phi_a}^3, \\
{\rho_a}^2&=&-\sqrt{a^{21}a^{22}}{\phi_a}^2, \\
{\rho_a}^3&=&\sqrt{a^{31}a^{32}}{\phi_a}^1,
\eea
we can define the complex fields
\begin{equation} \label{cdefn}
{\mathcal{C}_a}^m={\rho_a}^m + i {b_a}^m\; .
\end{equation}
As we will see, the fields ${\mathcal{C}_a}^m$ are indeed the correct
four-dimensional chiral matter superfields.

The moduli in the above background solutions are promoted to four-dimensional fields,
as usual, and we will call the corresponding bulk fields $\hat{g}^{(0)}$ and $C^{(0)}$,
in the following. In a pure bulk theory, this would be a standard procedure and
the reduction to four dimension would proceed without further complication.
However, in the presence of localized fields there is a subtlety which we
will now discuss. Allowing the moduli to fluctuate introduces localized
stress energy on the seven-dimensional orbifold plane and this excites the
heavy modes of the theory which we would like to truncate in the reduction.
This phenomenon is well-known from Ho\v rava-Witten theory and can be dealt
with by explicitly integrating out the heavy modes, thereby generating
higher-order corrections to the effective theory~\cite{Lukas:1998ew}.
As we will now argue, in our case these corrections are always of higher
order. More precisely, we will compute the four-dimensional effective theory
up to second order in derivatives and up to order $\kappa_{11}^{4/3}$, relative
to the leading gravitational terms. Let  $\hat{g}^{(1)}$ and $C^{(1)}$
be the first order corrections to the metric and the three-form which originate from
integrating out the localized stress energy on the orbifold plane, so that
we can write for the corrected fields
\bea
\hat{g}^{(\mathrm{B})}&=&\hat{g}^{(0)}+ \kappa_{11}^{4/3} \hat{g}^{(1)}\, ,
\label{backrfields1}\\
C^{(\mathrm{B})}&=&C^{(0)}+\kappa_{11}^{4/3} C^{(1)}\,
.\label{backrfields2} \eea We note that these corrections are already
suppressed by $\kappa_{11}^{4/3}$ relative to the pure background
fields. Therefore, when inserted into the orbifold
action~\eqref{boseaction2}, the resulting corrections are of order
$\kappa_{11}^{8/3}$ or higher and will, hence, be neglected.
Inserted into the bulk action, the fields~\eqref{backrfields1} and
\eqref{backrfields2} lead to order $\kappa_{11}^{4/3}$ corrections
which can be written as
\begin{equation}
\delta\mathcal{S}_7=h_7^2\hat{g}_{MN}^{(1)}\left. \frac{\delta \mathcal{S}_{11}}
{\delta \hat{g}_{MN}}\right|_{\hat{g}=\hat{g}^{(0)},\, C=C^{(0)}} +
h_7^2C_{MNP}^{(1)}\left. \frac{\delta \mathcal{S}_{11}}{\delta C_{MNP}}
\right|_{\hat{g}=\hat{g}^{(0)},\, C=C^{(0)}}\, .
\label{dS}
\end{equation}
Let us analyze the properties of the terms contained in this
expression. The functional derivatives in the above expression
vanish for constant moduli fields since the background
configurations $\hat{g}^{(0)}$ and $\hat{C}^{(0)}$ are exact
solutions of the 11-dimensional bulk equations in this case. Hence,
allowing the moduli fields to be functions of the external
coordinates, the functional derivatives must contain at least two
four-dimensional derivatives. All terms in $\hat{g}^{(1)}$ and
$\hat{C}^{(1)}$ with four-dimensional derivatives will, therefore,
generate higher-dimensional derivative terms in four dimensions and
can be neglected. The only terms which are not of this type arise
from the D-term potential and covariant derivatives on the orbifold
plane and they appear within $\hat{g}^{(1)}$. These terms are of
order $\kappa^{4/3}$ and should in principle be kept. However, they
are of fourth order in the matter fields $\mathcal{C}^{am}$ and
contain two four-dimensional derivatives acting on bulk moduli. They
can, therefore, be thought of as corrections to the moduli kinetic
terms. As we will see, the K\"ahler potential of the
four-dimensional theory can be uniquely fixed without knowing these
correction terms explicitly.

\subsection{Calculation of the four-dimensional effective theory}
We will now reduce our theory to four dimensions starting with the
lowest order in the $\kappa$ expansion, that is, with the bulk theory.
The reduction of the bulk theory leads to a well-defined
four-dimensional supergravity theory in its own right. We shall keep
technical discussion to a minimum, and refer the reader to Appendix
\ref{appred} for further details of the method of reduction. The superpotential
and D-term vanish when one reduces 11-dimensional supergravity on a $G_2$
space~\cite{Papadopoulos:1995da,Beasley:2002db}. Also, we have no gauge fields to consider since our $G_2$
orbifolds do not admit two-forms. Thus we need only specify the K\"ahler potential to
determine the four-dimensional effective theory. To compute this we
use the formula
\begin{equation} \label{Kformula}
K=-\frac{3}{\kappa_4^2}\ln \left( \frac{V}{v_7} \right)\, ,
\end{equation}
 given in Ref.~\cite{Beasley:2002db}. Here, $V$ is the volume of the $G_2$ space $\mathcal{Y}$ as
measured by the internal part $g$ of the metric~\eqref{metricA}, and $v_7$ is
a reference volume,
\begin{equation}
V=\int_{\mathcal{Y}}\mathrm{d}^7x\sqrt{ \det g}\, , \hspace{0.5cm}
v_7=\int_{\mathcal{Y}}\mathrm{d}^7x \, .
\end{equation}
Here the four-dimensional Newton constant $\kappa_4$ is related to its
11-dimensional counterpart by
\begin{equation} \label{kappa114}
\kappa_{11}^2=\kappa_4^2v_7\, .
\end{equation}
For the $G_2$ orbifolds considered the volume is proportional to the
product of the seven radii $R^A$ in Eq.~\eqref{metricA}. The precise
form of the K\"ahler potential for 11-dimensional supergravity on
$\mathcal{Y}=\mathcal{T}^7/\Gamma$ is then given by
\begin{equation} \label{kahler1}
K_0=-\frac{1}{\kappa_4^2}\sum_{A=0}^6 \ln \left( T^A + \bar{T}^A
\right)+\frac{7}{\kappa_4^2}\ln 2\, .
\end{equation}

In order to perform the reduction of the seven-dimensional
Yang-Mills theory on the singular locus $\mathcal{T}^3$ to four
dimensions, we need to express the truncated bulk fields
$F_{\hat{\mu}\hat{\nu}}^I$, ${\ell_I}^{\ul{J}}$ and $\tau$ in terms
of the bulk metric moduli $a^A$ and the bulk axions $\nu^A$. We do
this by using the formulae \eqref{metricA}, \eqref{period2},
\eqref{Cexp1} for the 11-dimensional fields in terms of $a^A$ and
$\nu^A$, together with the field identifications
\eqref{id1}-\eqref{idn} between 11-dimensional and seven-dimensional
fields. We find that the only non-vanishing components of
$F_{\hat{\mu}\hat{\nu}}^I$ are some of the mixed components
$F^I_{\mu m}$, and these are given by
\bea
F^3_{\mu 1} =
\frac{1}{2}\left( -\partial_\mu \nu^{11} - \partial_\mu \nu^{12}
\right),
 && F^4_{\mu 1} = \frac{1}{2}\left( -\partial_\mu \nu^{11} + \partial_\mu \nu^{12} \right),
 \nn \\
F^2_{\mu 2} = \frac{1}{2}\left( \partial_\mu \nu^{21} + \partial_\mu
\nu^{22} \right),
 && F^5_{\mu 2} = \frac{1}{2}\left( \partial_\mu \nu^{21} - \partial_\mu \nu^{22} \right),
 \nn \\
F^1_{\mu 3} = \frac{1}{2}\left( -\partial_\mu \nu^{31} -
\partial_\mu \nu^{32} \right),
 && F^6_{\mu 3} = \frac{1}{2}\left( \partial_\mu \nu^{31} - \partial_\mu \nu^{32} \right).
\eea
For the coset matrix $\ell$, which is symmetric, we find the non-zero components
\bea
{\ell_1}^1={\ell_6}^6=\frac{a^{31}+a^{32}}{2\sqrt{a^{31}{a^{32}}}}, && {\ell_1}^6={\ell_6}^1=\frac{a^{31}-a^{32}}{2\sqrt{a^{31}{a^{32}}}}, \nn \\
{\ell_2}^2={\ell_5}^5=\frac{a^{21}+a^{22}}{2\sqrt{a^{21}{a^{22}}}}, && {\ell_2}^5={\ell_5}^2=\frac{-a^{21}+a^{22}}{2\sqrt{a^{21}{a^{22}}}}, \\
{\ell_3}^3={\ell_4}^4=\frac{a^{11}+a^{12}}{2\sqrt{a^{11}{a^{12}}}}, && {\ell_3}^4={\ell_4}^3=\frac{-a^{11}+a^{12}}{2\sqrt{a^{11}{a^{12}}}}. \nn
\eea
Finally, we have the following relation for the orbifold scale factor $\tau$:
\begin{equation}
e^{\tau} = (a^0)^{-2/3}\prod_{m=1}^3\left(a^{m1}a^{m2}\right)^{1/3}.
\end{equation}
We now present the results of our reduction of bosonic terms at the
singularity. We neglect terms of the form $\mathcal{C}^n(\partial
T)^2$, where $n\geq 2$, and thus neglect the back-reaction term
$\delta\mathcal{S}_7$ in Eq.~\eqref{dS} completely. From the
seven-dimensional action $\mathcal{S}_7^{(0)}$ in
Eq.~\eqref{boseaction2} we get the following terms, divided up into
scalar kinetic terms, gauge-kinetic terms and scalar potential:
\bea
\mathcal{L}_{4,\mathrm{kin}} &=&
 -\frac{1}{2{\lambda_4}^2}\sqrt{-g}\sum_{m=1}^3\left\{ \frac{1}{a^{m1}a^{m2}}
 \left(\mathcal{D}_\mu\rho_a^m\mathcal{D}^\mu\rho^{am}
 +\mathcal{D}_\mu b_a^m\mathcal{D}^\mu b^{am}\right)\right. \nn \\
 && \hspace{3.0cm}- \frac{1}{3} \sum_{A=0}^6 \frac{1}{a^{m1}a^{m2}a^A}
\partial_\mu a^A\left( \rho_a^m\mathcal{D}^\mu\rho^{am}+b_a^m\mathcal{D}^\mu b^{am}
 \right) \nn \\
&&  \hspace{3.0cm}- \frac{1}{(a^{m1})^2a^{m2}}\rho_{a}^m\left( \partial_\mu
    \nu^{m1}\mathcal{D}^\mu b^{am} + \partial_\mu a^{m1} \mathcal{D}^\mu \rho^{am} \right)
    \nn \\
&& \hspace{3.0cm}\left. -\frac{1}{a^{m1}(a^{m2})^2}\rho_a^m\left(
    \partial_\mu \nu^{m2}\mathcal{D}^\mu b^{am} + \partial_\mu a^{m2}
    \mathcal{D}^\mu \rho^{am} \right) \right\},\label{kinactual} \\
\mathcal{L}_{4,\mathrm{gauge}}&=&-\frac{1}{4\lambda_4^2}\sqrt{-g}\left(a^0
    F_{\mu\nu}^aF^{\mu\nu}_a-\frac{1}{2}\nu^0\epsilon^{\mu\nu\rho\si}
    F^a_{\mu\nu}F_{a\rho\si}\right), \label{gauge}\\
\mathcal{V}&=&\frac{1}{4\lambda_4^2a^0}\sqrt{-g}{f^a}_{bc}f_{ade}
    \sum_{m,n,p=1}^3 \epsilon_{mnp}\frac{1}{a^{n1}a^{n2}a^{p1}a^{p2}}
    \left( \rho^{bn}\rho^{dn}\rho^{cp}\rho^{ep}+\rho^{bn}\rho^{dn}b^{cp}b^{ep}\right. \nn \\
    &&\hspace{7.3cm}\left. +b^{bn}b^{dn}\rho^{cp}\rho^{ep}+b^{bn}b^{dn}b^{cp}b^{ep}\right).
    \label{scalarpot}
\eea
The four-dimensional gauge coupling $\lambda_4$ is related to the seven-dimensional analogue by
\begin{equation} \label{la74}
\lambda_4^{-2}=v_3\lambda_7^{-2}\, ,\qquad v_3=\int_{\mathcal{T}^3}\mathrm{d}^3x\, ,
\end{equation}
where $v_3$ is the reference volume for the three-torus.
Note that the above matter field action is suppressed relative to the gravitational
action by a factor $h_4^2=\kappa_4^2/\lambda_4^2\sim \kappa_{11}^{4/3}$, as mentioned earlier.
\\

\subsection{Finding the superpotential and K\"ahler potential}
The above reduced action must be the bosonic part of a four-dimensional
$\mathcal{N}=1$ supergravity and we would now like to determine the
associated K\"ahler potential and superpotential.
We start by combining the information from the expression \eqref{kahler1}
for the bulk K\"ahler potential $K_0$ descending from 11-dimensional supergravity
with the matter field terms \eqref{kinactual}, \eqref{gauge} and \eqref{scalarpot}
descending from the singularity to obtain the full K\"ahler potential. In general,
one cannot expect the definition~\eqref{TA} of the moduli in terms of
the underlying geometrical fields to remain unchanged in the presence of
additional matter fields. We, therefore, begin by writing the most general
form for the correct superfield $\tilde{T}^A$ in the
presence of matter fields as
\begin{equation}
\ti{T}^A=T^A + F^A\left( T^B, \bar{T}^B, \mathcal{C}_m^a, \bar{\mathcal{C}}_m^a \right)
 \,.\label{Tcorr}
\end{equation}
Analogously, the most general form of the K\"ahler potential in the presence
of matter can be written as
\begin{equation}
K=K_0+K_1\left( T^A, \bar{T}^A, \mathcal{C}_m^a, \bar{\mathcal{C}}_m^a \right)\,. \label{Kcorr}
\end{equation}
Given this general form for the superfields and the K\"ahler potential, we can work
out the resulting matter field kinetic terms by taking second derivatives of $K$ with
respect to $\ti{T}^A$ and $\mathcal{C}_a^m$. Neglecting terms of order $C^n(\partial T)^2$,
as we have done in the reduction to four dimensions, we find
\bea \label{kingeneric}
\mathcal{L}_{4,\mathrm{kin}}&=&-\sqrt{-g}\left\{ \sum_{m,n=1}^3
\frac{\partial^2K_1}{\partial \mathcal{C}_a^m \partial \bar{\mathcal{C}}_b^n}
\mathcal{D}_\mu\mathcal{C}_a^m\mathcal{D}^\mu\bar{\mathcal{C}}_b^n+
\left( 2\sum_{m=1}^3\sum_{A=0}^6\frac{\partial^2K_1}{\partial \mathcal{C}_a^m
 \partial \bar{T}^{A}}\mathcal{D}_\mu\mathcal{C}_a^m\partial^\mu \bar{T}^A+
 \mathrm{c.c.} \right)\right. \nn \\
&& \hspace{1.5cm} + \left. \left( \sum_{A,B=0}^6 \frac{\partial^2K_0}{\partial T^A
   \partial \bar{T}^B}\partial_\mu T^A \partial^\mu \bar{F}^B + \mathrm{c.c.} \right)\right\}\, .
\eea
By matching kinetic terms \eqref{kinactual} from the reduction with
the kinetic terms in the above equation \eqref{kingeneric} we can
uniquely determine the expressions for the superfields $\ti{T}^A$
and the K\"ahler potential. They are given respectively by
\bea
\ti{T}^A &=& T^A -\frac{1}{24\lambda_4^2}\left( T^A+\bar{T}^A \right) \sum_{m=1}^3
 \frac{\mathcal{C}_a^m\bar{\mathcal{C}}^{am}}{(T^{m1}+\bar{T}^{m1})(T^{m2}+\bar{T}^{m2})}\,,\\
K&=& \frac{7}{\kappa_4^2}\ln 2 - \frac{1}{\kappa_4^2}\sum_{A=0}^6
\ln (\ti{T}^A + \bar{\ti{T}}^A ) + \frac{1}{4\lambda_4^2}
\sum_{m=1}^3\frac{(\mathcal{C}_a^m+\bar{\mathcal{C}}_a^m)(\mathcal{C}^{am}+
\bar{\mathcal{C}}^{am})}{(\ti{T}^{m1}+\bar{\ti{T}}^{m1})(\ti{T}^{m2}+\bar{\ti{T}}^{m2})}\,.
\label{1singkahler}
\eea

We now come to the computation of the gauge-kinetic function $f_{ab}$
and the superpotential $W$. The former is straightforward to read off
from the gauge-kinetic part \eqref{gauge} of the reduced action and is given by
\begin{equation}
f_{ab}=\frac{1}{\lambda_4^2}\ti{T}^0\delta_{ab}.
\end{equation}
To find the superpotential, we can compare the scalar potential \eqref{scalarpot} of
the reduced theory to the standard supergravity formula \cite{Wess} for the scalar potential
\begin{equation}
\mathcal{V}=\frac{1}{\kappa_4^4}\sqrt{-g}e^{\kappa_4^2K}\left( K^{X\bar{Y}}\mathcal{D}_X W
\mathcal{D}_{\bar{Y}} \bar{W} - 3 \kappa_4^2 \lvert W \rvert^2\right) +
\sqrt{-g}\frac{1}{2\kappa_4^4}(\mathrm{Re}f)^{-1ab}D_a D_b\, ,
\end{equation}
taking into account the above results for the K\"ahler potential and the gauge
kinetic function. This leads to the superpotential and D-terms
\bea
W &=& \frac{\kappa_4^2}{24\lambda_4^2}f_{abc}\sum_{m,n,p=1}^3
\epsilon_{mnp}\mathcal{C}^{am}\mathcal{C}^{bn}\mathcal{C}^{cp},
\label{1singsuperpot} \\
D_a &=& \frac{2i\kappa_4^2}{\lambda_4^2}f_{abc}\sum_{m=1}^{3} \frac{\mathcal{C}^{bm}\bar{\mathcal{C}}^{cm}}{(\ti{T}^{m1}+\bar{\ti{T}}^{m1}) (\ti{T}^{m2}+\bar{\ti{T}}^{m2})}.
\eea
It can be checked that these D-terms are consistent with the gauging of
the $\mathrm{SU}(N)$ K\"ahler potential isometries, as they should be.

We are now ready to write down our formulae for the quantities that
specify the four-dimensional effective supergravity for M-theory on
$\mathcal{Y}=\mathcal{T}^7/\Gamma$, including the contribution from all singularities.
To do this we simply introduce a sum over the singularities.

Let us present the notation we need to write down these results. We
introduce a label $(\tau,s)$ for each singularity, the index $\tau$
labeling the generators of the orbifold group, and $s$ labeling
the $M_\tau$ fixed points associated with the generator
$\alpha_\tau$. We write $N_\tau$ for the order of the generator
$\alpha_\tau$. For the 16 types of orbifolds, these integer numbers
can be computed from the information provided in Appendix~\ref{A}.
Thus, near a singular point, $\mathcal{Y}$ takes the approximate
form $\mathcal{T}^3_{(\tau,s)}\times
\mathbb{C}^2/\mathbb{Z}_{N_\tau}$, where $\mathcal{T}^3_{(\tau,s)}$
is a three-torus. The matter fields at the singularities we denote
by $(\mathcal{C}^{(\tau,s)})_a^m$ and it is understood that the
index $a$ transforms in the adjoint of $\mathrm{SU}(N_\tau)$. The
gauge couplings depend only on the type of singularity, that is on
the index $\tau$, and are denoted by $\la_{(\tau)}$. M-theory
determines the values of these gauge couplings, and they can be
derived using equations \eqref{couplerel}, \eqref{kappa114} and
\eqref{la74}. We find
\begin{equation}
\la_{(\tau)}^2=\left( 4\pi \right)^{4/3}
\frac{v_7^{1/3}}{v_3^{(\tau)}}\kappa_4^{2/3}\,
\end{equation}
in terms of the reference volumes $v_7$ for $\mathcal{Y}$ and
$v_3^{(\tau)}$ for $\mathcal{T}^3_{(\tau,s)}$.

The respective formulae for the moduli superfields, K\"ahler potential, superpotential and D-term potential are
\begin{equation} \label{superfield}
\ti{T}^A = T^A -\left( T^A+\bar{T}^A \right) \sum_{\tau,s,m}
\frac{1}{24\la_{(\tau)}^2}\frac{(\mathcal{C}^{(\tau,s)})_a^m(\bar{\mathcal{C}}^{(\tau,s)})^{am}}{(T^{B(\tau,m)}+\bar{T}^{B(\tau,m)})(T^{C(\tau,m)}+\bar{T}^{C(\tau,m)})}\,,
\end{equation}
\begin{equation} \label{fullkahler}
K= - \frac{1}{\kappa_4^2}\sum_{A=0}^6 \ln (\ti{T}^A + \bar{\ti{T}}^A ) +
\sum_{\tau,s,m}\frac{1}{4\lambda_{(\tau)}^2}\frac{\left[(\mathcal{C}^{(\tau,s)})_a^m+(\bar{\mathcal{C}}^{(\tau,s)})_a^m\right]\left[(\mathcal{C}^{(\tau,s)})^{am}+(\bar{\mathcal{C}}^{(\tau,s)})^{am}\right]}{(\ti{T}^{B(\tau,m)}+\bar{\ti{T}}^{B(\tau,m)})(\ti{T}^{C(\tau,m)}+\bar{\ti{T}}^{C(\tau,m)})}
+\frac{7}{\kappa_4^2}\ln 2 \,,
\end{equation}
\begin{equation} \label{fullsuperpot}
W = \frac{1}{24}\sum_{\tau,s,m,n,p}\frac{\kappa_4^2}{\lambda_{(\tau)}^2}f_{abc} \epsilon_{mnp}(\mathcal{C}^{(\tau,s)})^{am}(\mathcal{C}^{(\tau,s)})^{bn}(\mathcal{C}^{(\tau,s)})^{cp},
\end{equation}
\begin{equation} \label{fullD}
D_a = 2i\sum_{\tau,s,m}\frac{\kappa_4^2}{\lambda_{(\tau)}^2}f_{abc} \frac{(\mathcal{C}^{(\tau,s)})^{bm}(\bar{\mathcal{C}}^{(\tau,s)})^{cm}}{(\ti{T}^{B(\tau,m)}+\bar{\ti{T}}^{B(\tau,m)})(\ti{T}^{C(\tau,m)}+\bar{\ti{T}}^{C(\tau,m)})}.
\end{equation}
The index functions $B(\tau,m)$, $C(\tau,m)\in \{0,\ldots,6\}$ indicate
by which two of the seven moduli the matter fields are divided by in
equations \eqref{superfield}, \eqref{fullsuperpot} and
\eqref{fullD}. Their values depend only on the generator index $\tau$
and the R-symmetry index $m$. They may be calculated from the formula
\begin{equation}
a^{B(\tau,m)}a^{C(\tau,m)}=\frac{\left( R^m_{(\tau)} \right) ^2
\prod_A R^A}{\prod_n R^n_{(\tau)}}\, ,
\end{equation}
where $R^m_{(\tau)}$ denote the radii of the three-torus
$\mathcal{T}^3_{(\tau,s)}$. The possible
values of the index functions are given in Table~1.
\begin{table}[t]
\begin{center}
\begin{tabular}{|c|c|c|c|c|}
\hline
$\mathrm{Fixed \, directions\, of\, } \alpha_\tau$ &
$(B(\tau,1),C(\tau,1))$ & $(B(\tau,2),C(\tau,2))$ & $(B(\tau,3),C(\tau,3))$ & $h(\tau)$ \\
\hline
(1,2,3) & (1,2) & (3,4) & (5,6) & 0\\
\hline
(1,4,5) & (0,2) & (3,5) & (4,6) & 1\\
\hline
(1,6,7) & (0,1) & (3,6) & (4,5) & 2\\
\hline
(2,4,6) & (0,4) & (1,5) & (2,6) & 3\\
\hline
(2,5,7) & (0,3) & (1,6) & (2,5) & 4\\
\hline
(3,4,7) & (0,6) & (1,3) & (2,4) & 5\\
\hline
(3,5,6) & (0,5) & (1,4) & (2,3) & 6\\
\hline
\end{tabular}
\caption{\emph{Values of the index functions $(B(\tau,m),C(\tau,m))$ and
$h(\tau)$ that appear in the superfield definitions,
K\"ahler potential, D-term potential and gauge-kinetic functions.}}
\end{center}
\end{table}

There is a universal gauge-kinetic function for each $\mathrm{SU}(N_\tau )$
gauge theory given by
\begin{equation}
f_{(\tau)}=\frac{1}{\la_{(\tau)}^2}\ti{T}^{h(\tau)},
\end{equation}
where $\ti{T}^{h(\tau)}$ is the modulus that corresponds to the volume
of the fixed three-torus $\mathcal{T}^3_{(\tau,s)}$ of the symmetry
$\alpha_\tau$. The value of $h(\tau)$ in terms of the fixed directions
of $\alpha_\tau$ is given in Table~1.
\\

\subsection{Comparison with results for smooth $G_2$ spaces}
As mentioned in the introduction, one can construct a smooth $G_2$
manifold $\mathcal{Y}^{\mathrm{S}}$ by blowing up the singularities of the $G_2$
orbifold $\mathcal{Y}$ \cite{Joyce,Barrett:2004tc,Lukas:2003dn}. The moduli K\"ahler potential
for M-theory on this space has been computed in
Refs.~\cite{Barrett:2004tc,Lukas:2003dn}. An outline of this calculation, together with the
full result is given in Appendix~\ref{appred}.
Here, we focus on the contribution from a single singularity, which gives
\begin{equation} \label{Ksing}
K =  -\frac{1}{\kappa_4^2}\sum_{A=0}^{6}\ln (T^A+\bar{T}^A) +
     \frac{2}{N c_\Gamma\kappa_4^2}\sum_m\frac{\sum_{i\leq j}
     \left(\sum_{k=i}^{j}(U^{km}+\bar{U}^{km})\right)^2}{(T^{m1}+
     \bar{T}^{m1})(T^{m2}+\bar{T}^{m2})} + \frac{7}{\kappa_4^2}\ln2\, .
\end{equation}
Here, as usual, $T^A$ are the bulk moduli and $U^{im}$ are the blow-up
moduli. As for the formula for the singular manifold, the index $m$
can be thought of as labelling the directions on the three-torus
transverse to the particular blow-up. Each blow-up modulus is
associated with a two-cycle within the blow-up of a given singularity,
and the index $i,j,\ldots =1,\ldots (N-1)$ labels these. Finally,
$c_\Gamma$ is a constant, dependent on the orbifold group.

In computing the K\"ahler potential \eqref{Ksing}, the M-theory action
was taken to be 11-dimensional supergravity, and so the result is
valid when all the moduli, including blow-up moduli, are large
compared to the Planck length. Therefore, the above result for the
K\"ahler potential cannot be applied to the orbifold limit, where
$\mathrm{Re}(U^{im})\rightarrow 0$. However, the corresponding singular result
\eqref{1singkahler} can be used to consider the case of
small blow-up moduli. As discussed in the introduction, the Abelian
components of the matter fields $\mathcal{C}^{im}$ correspond to
moduli associated with the blow-up of the singularity, while the
non-Abelian components correspond to membrane states that are massless
only in the singular limit. Blowing up the singularity is, from this
point of view, described by turning on VEVs for (the real parts of)
the Abelian fields $\mathcal{C}^{im}$ along the D-flat
directions. This, generically, breaks the gauge group $\mathrm{SU}(N)$
to $\mathrm{U}(1)^{N-1}$ and only leaves the $3(N-1)$ matter fields
$\mathcal{C}^{im}$ massless. This massless field content matches
exactly the zero modes of the blown-up geometry. Therefore, by
switching off the non-Abelian components of $\mathcal{C}^{am}$ in
equation \eqref{1singkahler}, one obtains a formula for the moduli
K\"ahler potential for M-theory on $\mathcal{Y}^{\mathrm{S}}$ with small blow-up
moduli. At first glance this is slightly different from the
smooth result \eqref{Ksing} which contains a double-sum over the Abelian
gauge directions. However, we can show that they are actually equivalent.
First, we identify the bulk moduli $T^A$ in \eqref{Ksing} with $\ti{T}^A$ in \eqref{1singkahler}.
One obvious way of introducing a double-sum into the singular result
\eqref{1singkahler} is to introduce a non-standard basis $X_i$ for the Cartan
sub-algebra of $\mathrm{SU}(N)$, which introduces a metric
\begin{equation}
 \label{kij}
 \kappa_{ij}=\mathrm{tr}(X_iX_j)\, .
\end{equation}
Neglecting an overall rescaling of the fields, identification of the smooth
and singular results for $K$ then requires the identity
\begin{equation} \label{identity}
\sum_{i,j}\kappa_{ij}(\mathcal{C}^{im}+\bar{\mathcal{C}}^{im})(\mathcal{C}^{jm}+
\bar{\mathcal{C}}^{jm})=\sum_{i\leq j}\left(\sum_{k=i}^{j}(U^{(k,m)}+\bar{U}^{(k,m)})\right)^2
\end{equation}
to hold. So far we have been assuming the canonical choice $\kappa_{ij}=\delta_{ij}$,
which is realized by the standard generators
\begin{equation}
X_1= \frac{1}{\sqrt{2}}\mathrm{diag}(1,-1,0,\ldots,0)\,, \hspace{0.3cm} X_2= \frac{1}{\sqrt{6}}\mathrm{diag}(1,1,-2,0,\ldots,0)\, , \hspace{0.3cm}  \cdots \, , \nn
\end{equation}
\begin{equation}
X_{N-1}= \frac{1}{\sqrt{N(N-1)}}\mathrm{diag}(1,1,\ldots,1,-(N-1))\,.
\end{equation}
Clearly, the relation~\eqref{identity} cannot be satisfied with a holomorphic
relation between fields for this choice of generators. Instead, from the
RHS of Eq.~\eqref{identity} we need the metric $\kappa_{ij}$ to be
\begin{equation} \label{killingmetric}
\kappa_{ij}= \left\{ \begin{array}{c}
 (N-j)i\, , \hspace{0.3cm} i\leq j\, ,\\
 (N-i)j\, , \hspace{0.3cm} i>j\, .
\end{array} \right.
\end{equation}
From Eq.~\eqref{identity} this particular metric $\kappa_{ij}$ is
positive definite and, hence, there is always a choice of generators
$X_i$ which reproduces this metric via Eq.~\eqref{kij}. For the simplest case $N=2$,
there is only one generator $X_1$ and the above statement becomes trivial.
For the $N=3$ case, a possible choice for the two generators $X_1$ and $X_2$ is
\begin{equation}
X_1=\mathrm{diag}(0,-1,1)\,, \hspace{0.3cm} X_2=\mathrm{diag}(1,0,-1)\,.
\end{equation}
Physically, these specific choices of generators tell us how the Abelian
group $\mathrm{U}(1)^{N-1}$ which appears in the smooth case is embedded into the
$\mathrm{SU}(N)$ group which is present in the singular limit.


\section{Symmetry Breaking and Discussions}

\label{wilflux}

In this section, we will consider more general background configurations than
those discussed in Section 3. We can investigate the effects of such phenomena
as flux vacua and Wilson lines and in addition, we can study how
gauge symmetry is broken in these configurations. In particular, we would like
to examine the explicit symmetry breaking patterns obtained through Wilson
lines, and the effects of $G$- and $F$-flux on our four-dimensional theory.
We will also briefly explore how to express the super-Yang-Mills sector in the
language of $\mathcal{N}=4$ supersymmetry, a rephrasing that will yield new insight into
the structure of our theory close to a singular point in the $G_2$ space.

\subsection{Wilson lines}

We would now like to discuss breaking of the $\mathrm{SU}(N)$ gauge symmetry
through inclusion of Wilson lines in the internal three-torus $\mathcal{T}^3$.
Let us briefly recall the main features of Wilson-line
breaking~\cite{Wess, gsw, Witten:1985xc, Breit:1985ud, Cvetic:1995dt}. A Wilson line is a configuration
of the (internal) gauge field  $A^{a}$ with vanishing associated field
strength. For a non-trivial Wilson-line to be possible, the first fundamental group, $\pi_1$,
of the internal space needs to be non-trivial, a condition satisfied in our case,
as $\pi_1(\mathcal{T}^3)=\mathbb{Z}^3$. Practically, a Wilson line around a non-contractible loop $\gamma$ can be
described by
\begin{equation}
U_{\gamma}=P\exp\left(  -i\oint_{\gamma}X_{a}A^{a}{}_{m}dx^{m}\right)
\end{equation}
where $X_{a}$ are the generators of the Lie algebra of the gauge group, $G$.
This expression induces a group homomorphism, $\gamma\rightarrow U_{\gamma}$,
between the fundamental group and the gauge group of our theory.

We can explicitly determine the possible symmetry breaking patterns by
examining particular embeddings (that is, choices of representation) of
the fundamental group into the gauge group. For convenience, we will
focus on gauge groups $\mathrm{SU}(N)$, where $N=2,3,4,6$, since these are
the gauge groups known to arise from explicit constructions of $G_2$
orbifolds~\cite{Barrett:2004tc}. For example, we may choose a
representation for $\pi _{1}(\mathcal{T}^3)=\mathbb{Z}^{3}$ in the following
way. Let a generic group element of $\mathbb{Z}^{3}$ be given by a
triple of integers (taking addition as the group multiplication),
\begin{equation}
g=\left(  n,m,p\right).
\end{equation}
Then we may embed this in $\mathrm{SU}(4)$ as
\begin{equation}
g=\left(
\begin{array}
[c]{ccc}%
e^{in}{\bf 1}_{2\times2} &  & \\
& 1 & \\
&  & e^{-2 in}%
\end{array}
\right)
\end{equation}
which will clearly break the symmetry to $\mathrm{SU}(2)\times \mathrm{U}(1)\times \mathrm{U}(1)$. There
is, however, a great deal of redundancy in these choices of embedding and the
homomorphisms we define are clearly not unique. For example, we could have
alternately chosen the map so as to take $g$ to an element in the subgroup
$\mathrm{SU}(2)\times \mathrm{SU}(2)\times \mathrm{U}(1)$, say
\begin{equation}
g=\left(
\begin{array}
[c]{ccc}%
(-1)^{n}{\bf 1}_{2\times2} &  & \\
& e^{ im} & \\
&  & e^{- im}%
\end{array}
\right)\, ,
\end{equation}
which would also break $\mathrm{SU}(4)$ to $\mathrm{SU}(2)\times \mathrm{U}(1)\times \mathrm{U}(1)$. A nice
example of the types of reduced symmetry possible with Wilson lines is given
by the following embedding of $\mathbb{Z}^{3}$ into $\mathrm{SU}(6)$~:
\begin{equation}
\left(
\begin{array}
[c]{ccc}%
e^{in}{\bf 1}_{2\times2} &  & \\
& e^{\frac{-2in}{3}}{\bf 1}_{3\times3} & \\
&  & 1
\end{array}
\right)\, .
\end{equation}
This breaks $\mathrm{SU}(6)$ to the subgroup $\mathrm{SU}(3)\times
\mathrm{SU}(2)\times \mathrm{U}(1)\times \mathrm{U}(1)$, which
contains the symmetry group of the Standard Model. (Though even in
this case, our theory does not contain the particle content of the
Standard Model.)

Having given a number of examples, we can now classify in general, which unbroken
subgroups of $\mathrm{SU}(N)$ are possible (using the group-theoretical tools
provided in Ref.~\cite{Slansky:1981yr}). Clearly, the generic unbroken subgroup is
$\mathrm{U}(1)^{N-1}$, however, certain choices of embedding leave a larger symmetry group intact.
These special choices are of particular interest, but we may smoothly deform
from such a choice to a generic solution by varying a parameter in our
embedding. For example, let the mapping of a group element $(n,m,p)$ in
$\mathbb{Z}^{3}$ into $\mathrm{SU}(3)$ be given by
\begin{equation}
g=\left(
\begin{array}
[c]{ccc}%
e^{i\alpha m+ip} &  & \\
& e^{i\alpha n+ip} & \\
&  & e^{-i\alpha(n+m)-2ip}%
\end{array}
\right)
\end{equation}
where the parameter $\alpha$ may be freely varied. For general values of
$\alpha$ this embedding breaks to $\mathrm{U}(1)^{2}$, however for $\alpha=0$ we may
break to the larger group, $\mathrm{SU}(2)\times \mathrm{U}(1)$.

We find the Wilson lines can
break the $\mathrm{SU}(N)$ symmetry group to any subgroup with the rank $N-1$,
with the generic choice being the Cartan algebra itself. In addition, by introducing
a parameter, as in the above $\mathrm{SU}(3)$ example, any of the possible breakings
can be continuously deformed to the generic breaking.
The results for all possible unbroken gauge groups are summarized in Table 2. \begin{table}[ptb]
\begin{center}%
\begin{tabular}
[c]{|c|c|}\hline
Gauge Group & Residual Gauge Groups from Wilson lines\\\hline
$\mathrm{SU}_{2}$ & $\mathrm{U}_1$\\\hline
$\mathrm{SU}_{3}$ & $\mathrm{SU}_{2}\times \mathrm{U}_1$, $\mathrm{U}_1^{2}$\\\hline
$\mathrm{SU}_{4}$ & $\mathrm{SU}_{3}\times \mathrm{U}_1$, $\mathrm{SU}_{2}\times \mathrm{U}_1^{2}$, $\mathrm{SU}_{2}^{2}\times
\mathrm{U}_1$, $\mathrm{U}_1^{3}$\\\hline
$\mathrm{SU}_{6}$ & $\mathrm{SU}_{5}\times \mathrm{U}_1$, $\mathrm{SU}_{4}\times \mathrm{U}_1^{2}$, $\mathrm{SU}_{2}\times
\mathrm{SU}_{3}\times \mathrm{U}_1^{2}$, $\mathrm{SU}_{2}^{2}\times \mathrm{U}_1^{3}$, $\mathrm{SU}_{2}\times \mathrm{U}_1%
^{4}$,\\
& $\mathrm{SU}_{3}\times \mathrm{U}_1^{3}$, $\mathrm{SU}_{2}\times \mathrm{SU}_{4}\times \mathrm{U}_1$, $\mathrm{SU}_{2}%
^{3}\times \mathrm{U}_1^{2}$, $\mathrm{SU}_{3}^{2}\times \mathrm{U}_1$, $\mathrm{U}_1^{5}$\\\hline
\end{tabular}
\end{center}
\caption{\emph{The symmetry group reductions in the presence of Wilson lines}}%
\end{table}Note that the Cartan subgroups are included as the last entries for
each of the gauge groups. (For other examples of Wilson lines in $G_2$ spaces see, for example, Refs.~\cite{Friedmann:2002ct},~\cite{Friedmann:2002ty}.) 

It is worth noting briefly that we can view this symmetry breaking
by Wilson lines in an alternate light in four-dimensions. Rather
than consider a seven-dimensional compactification and Wilson lines,
we could obtain the same results by turning on VEVs for certain
directions of the scalar fields in our four-dimensional
theory\footnote{In fact, the scalars which directly correspond to
Wilson lines in seven dimensions are the axionic, Abelian parts of
the fields ${\cal C}^{am}$.}. For example, if we give generic VEVs
to all the Abelian directions of the scalar fields in
Eq.~(\ref{scalarpot}) we can break the symmetry to a purely Abelian
gauge group. This corresponds to a generic embedding in the Wilson
line picture. Likewise, we can obtain the larger symmetry groups
listed in Table 2 by giving non-generic VEVs to the scalar fields.

\subsection{$\boldsymbol{G}$- and $\boldsymbol{F}$-Flux}\label{Ch2_flux}

The previous discussion of Wilson lines can be thought of as describing
non-trivial background configurations for which we still maintain the
condition $F=0$ on the field strength. However, to gain a better understanding
of the possible vacua and their effects, we need to consider the
contributions of flux both from bulk and seven-dimensional field strengths.
Let us start with a bulk flux $G_{\mathcal{Y}}$ for the internal
part~\footnote{We do not discuss flux in the external part of $G$.} of the M-theory
four-form field strength $G$. For M-theory compactifications on smooth $G_2$
spaces this was discussed in Ref.~\cite{Beasley:2002db}. In our case, all we have
to do is modify this discussion to include possible effects of the
singularities and their associated seven-dimensional gauge theories.
However, inspection of the seven-dimensional gauge field
action~\eqref{boseaction2} shows that a non-vanishing internal $G_{\mathcal{Y}}$ will not
generate any additional contributions to the four-dimensional scalar potential,
apart from the ones descending from the bulk. Hence, we can use the
standard formula~\cite{Beasley:2002db}
\begin{equation}
W=\frac{1}{4}\int_{\mathcal{Y}}\left(  \frac{1}{2}C+i\varphi\right)  \wedge G_{\mathcal{Y}}\; ,
\label{Wfluxgen}
\end{equation}
where $\mathcal{Y}$ is a general seven-dimensional manifold of
$G_{2}$ holonomy, $C$
is the 3-form of 11-dimensional supergravity and $\varphi$ is the $G_{2}%
$-structure of $X$. For a completely singular $G_2$ space, where the torus moduli
$T^A$ are the only bulk moduli, this formula leads to a flux superpotential
\begin{equation}
 W\sim n_AT^A\; ,
\end{equation}
with flux parameters $n_A$, which has to be added to the ``matter field''
superpotential~\eqref{1singsuperpot}. If some of the singularities are
blown up we also have blow-up moduli $U^{im}$ and the flux superpotential contains
additional terms, thus
\begin{equation}
 W\sim n_AT^A + n_{im}U^{im} \; . \label{Wflux}
\end{equation}

We now turn to a discussion of the seven-dimensional SYM theory at the singularity.
First, it is natural to ask whether the matter field superpotential~\eqref{1singsuperpot}
can also be obtained from a Gukov-type formula, analogous to Eq.~\eqref{Wfluxgen},
but with an integration over the three-dimensional internal space on which
the gauge theory is compactified. To this end, we begin by defining the complexified
internal gauge field
\begin{equation}
\mathcal{C}_{a}=\rho_{am}\mathrm{d}x^{m}+ib_{am}\mathrm{d}x^{m}\; .
\end{equation}
It is worthwhile to note at this stage, that writing the real parts
of these fields (which are scalar fields in the original
seven-dimensional theory) as forms is, in fact, an example of the
procedure referred to as
`twisting'~\cite{Bershadsky:1995sp,Acharya:1998pm}. In this particular
case, the twisting amounts to identifying the $R$-symmetry index
($m=1,2,3$) of our original seven-dimensional supergravity with the
tangent space indices of the three-dimensional compact space,
$\mathcal{T}^{3}$. A plausible guess for the Gukov-formula for the
seven-dimensional gauge theory is an expression proportional to the
integral of the complexified Chern-Simons form
\begin{equation}
\omega_{CS}=\left(  \mathcal{F}^{a}\wedge\mathcal{C}_{a}-\frac{1}{3}%
f_{abc}\mathcal{C}^{a}\wedge\mathcal{C}^{b}\wedge\mathcal{C}^{c}\right)
\end{equation}
over the three-dimensional internal space \cite{Acharya:2002kv}. Here, $\mathcal{F}$ is the
complexified field strength
\begin{equation}
\mathcal{F}^{a}=\mathrm{d}\mathcal{C}^{a}+f^{a}{}_{bc}\mathcal{C}^{b}\wedge
\mathcal{C}^{c}.
\end{equation}
Indeed, if we specialize to the case of vanishing flux, that is $\mathrm{d}\mathcal{C}^a=0$,
our matter field superpotential~\eqref{1singsuperpot} is exactly reproduced by the
formula
\begin{equation}
W=\frac{\kappa_{4}^{2}}{16\lambda_{4}^{2}}\frac{1}{v_{3}}\int_{\mathcal{T}^{3}}%
\omega_{CS}\; . \label{NewGukov}%
\end{equation}

To see that Eq.~\eqref{NewGukov} also correctly incorporates the contributions of
$F$-flux, we can look at the following simple example of an Abelian $F$-flux.
Let the Abelian parts of the gauge field strength, $F^i$, be expanded in a basis of the harmonic
two-forms, $\omega_{m}=\frac{1}{2}\epsilon_{mnp}\mathrm{d}x^n\wedge \mathrm{d}x^p$, on the internal
three-torus $\mathcal{T}^3$, as
\begin{equation}
F^i=f^{im}\omega_m\; ,\label{fluxAnsatz}
\end{equation}
where $f^{im}$ are flux parameters.  Substituting this expression into the seven-dimensional
bosonic action~\eqref{boseaction2} and performing a compactification on $\mathcal{T}^3$ we find
a scalar potential which, taking into account the K\"ahler potential~\eqref{1singkahler},
can be reproduced from the superpotential
\begin{equation}
W=\frac{\kappa_{4}^{2}}{8\lambda_{4}^{2}}f_{im}\mathcal{C}^{im}\; . \label{Wfluxabelian}
\end{equation}
This superpotential is exactly reproduced by the Gukov-type
formula~\eqref{NewGukov} which, after substituting the flux
Ansatz~\eqref{fluxAnsatz}, specializes to its Abelian part. Hence, the
formula~\eqref{NewGukov} correctly reproduces the matter field
superpotential as well as the superpotential for Abelian $F$-flux.
The explicit Gukov formula for multiple singularities
is analogous to Eq.~\eqref{NewGukov}, with an additional sum to run over all
singularities as in Eq.~\eqref{fullsuperpot}. We also note that the
$F$-flux superpotential~\eqref{Wfluxabelian} is consistent with the
blow-up part of the $G$-flux superpotential~\eqref{Wflux} when the identification
of the Abelian scalar fields $\mathcal{C}^{im}$ with the blow-up moduli
$U^{im}$ is taken into account.

\subsection{Relation to $\mathcal{N}$=4 Supersymmetric Yang-Mills Theory}

In the previous sections we have explored aspects and modifications of
the four-dimensional $\mathcal{N}=1$ effective theory. We shall now
take a step back and look at the theory without flux, rephrasing it in order to
provide us with several new insights. The M-theory compactification discussed
in the previous sections is clearly $\mathcal{N}=1$ supersymmetric, by
virtue of our choice to compactify on a $G_{2}$ holonomy
space. However, if we neglect the gravity sector (that is, in
particular hold constant the moduli $T^A$) the remaining theory is
$\mathcal{N}=4$ super-Yang-Mills theory, an expected outcome since we
are compactifying the seven-dimensional SYM theory on a three-torus.
We will now make this connection more explicit by matching the
Yang-Mills part of our four-dimensional effective theory with
$\mathcal{N}=4$ SYM theory in its standard form. This connection is of
particular interest since $\mathcal{N}=4$ SYM theory is of central
importance in many current aspects of string theory, particularly in
the context of the AdS/CFT conjecture \cite{Maldacena:1997re}. We will begin
with a brief review of the central features of $\mathcal{N}=4$
Yang-Mills theory itself before identifying this structure in our
M-theory compactification.

In addition to a non-Abelian gauge symmetry (given in our case by $\mathrm{SU}(N)$),
the $\mathcal{N}=4$ SYM Lagrangian in
four-dimensions is equipped with an internal $\mathrm{O}(6)\sim \mathrm{SU}(4)$ R-symmetry.
In terms of $\mathcal{N}=1$ language its field content consists of
Yang-Mills multiplets  $(A_{\mu}^{a},\lambda^{a})$, where $a$ is a gauge index,
and a triplet of chiral multiplets $(A^a_m+iB^a_m,\chi^a_m)$ per gauge multiplet,
where $A^a_m$ and $B^a_m$ are real scalars, $\chi^a_m$ are Weyl fermions and
$m,n,\ldots = 1,2,3$. All we require in order to identify fields is the bosonic part of
the $\mathcal{N}=4$ Lagrangian which is given by \cite{Gliozzi:1976jf,Brink:1976bc,D'Hoker:2002aw}
\begin{align}
\mathcal{L}_{\mathcal{N}=4} &  = -\frac{1}{4g^{2}}G_{\mu\nu}^{a}G_{a}^{\mu\nu
}+\frac{\theta}{64\pi^{2}}\epsilon^{\mu\nu\rho\sigma}G_{\mu\nu}^{a}%
G_{a\rho\sigma}-\frac{1}{2}\left(  \mathcal{D}_{\mu}A_{m}^{a}\mathcal{D}^{\mu}A_{a}^{m}%
-\frac{1}{2}\mathcal{D}_{\mu}B_{m}^{a}\mathcal{D}^{\mu}B_{a}^{m}\right)  \nonumber\\
&  \hspace{0.8cm}+\frac{g^{2}}{4}\mathrm{tr}\left( [A_{m},A_{n}][A^{m}%
,A^{n}]+[B_{m},B_{n}][B^{m},B^{n}]+2[A_{m},B_{n}][A^{m},B^{n}]\right)\; .  \label{N4YM}%
\end{align}
With these $\mathcal{N}=4$ definitions in mind, we turn now to the
four-dimensional effective theory~\eqref{kinactual}--\eqref{scalarpot}
derived in the previous sections and consider the case
where the gravity sector is neglected and the geometric moduli are held
constant. This is the situation when we are in the neighborhood of a singular
point on the $G_{2}$ space and we are neglecting all bulk contributions. By
inspection, we need the following field identifications
\begin{align}
A_{a}^{m} &  =\frac{1}{\lambda_{4}\sqrt{a^{m1}a^{m2}}}\rho_{a}^{m}\, ,\label{Aa}\\
B_{a}^{m} &  =\frac{1}{\lambda_{4}\sqrt{a^{m1}a^{m2}}}b_{a}^{m}\, ,\\
G_{\mu\nu}^{a} &  =F_{\mu\nu}^{a}\; .
\end{align}
The $\mathcal{N}=4$ coupling constants are related to the $\mathcal{N}=1$
constants by%
\begin{equation}
g^{2} =\frac{\lambda_{4}^{2}}{a^{0}}\, ,\qquad
\theta =\frac{8\pi^{2}\nu^{0}}{\lambda_{4}^{2}}.\label{4couplings}%
\end{equation}
With these identifications, Eqs.~(\ref{kinactual})-(\ref{scalarpot}) exactly
reproduce Eq.~(\ref{N4YM}).

We can now consider the Montonen-Olive and S-duality conjecture~\cite{Montonen:1977sn} in the
context of our theory. This duality acts on the complex coupling
\begin{equation}
\tau\equiv\frac{\theta}{2\pi}-\frac{4\pi i}{g^{2}}%
\end{equation}
by the standard $\mathrm{SL}(2,\mathbb{Z})$ transformation
\begin{equation} \label{mod}
\tau\rightarrow\frac{a\tau+b}{c\tau+d}%
\end{equation}
with $ad-bc=1$ and $a,b,c,d\in$ $\mathbb{Z}$. Note that these transformations
contain in particular $\tau\rightarrow -\frac{1}{\tau}$, an interchange of strong and weak coupling.
Specifically, the S-duality conjecture is the statement that a $\mathcal{N}=4$
Yang-Mills theory with parameter $\tau$ as defined above and gauge group $G$,
is identical to the theory with coupling parameter transformed as in \eqref{mod} and the
dual gauge group, $\widehat{G}$. Note that here ``dual group'' refers to the
Langlands dual group, (which for $G=\mathrm{SU}(N)$, is given by $\widehat
{G}=\mathrm{SU}(N)/\mathbb{Z}_{N}$)~\cite{Goddard:1981rv}.

When we consider the above transformations within the context of our theory,
several interesting features emerge immediately. With the field
identifications in Eq.~(\ref{Aa})--(\ref{4couplings}) we have
\begin{equation}
\tau=-\frac{4\pi i\ti{T}^{0}}{\lambda_{4}^{2}}.
\end{equation}
Therefore, the shift symmetry $\tau\rightarrow\tau+b$ is equivalent
to an axionic shift of $\ti{T}^0$ and
$\tau\rightarrow-\frac{1}{\tau}$ is given by
$\ti{T}^0\rightarrow\frac{1}{\ti{T}^0}$. Since
$\mathrm{Re}(T^0)=a^{0}$ describes the volume of the torus,
$\mathcal{T}^{3}$, S-duality in the present context is really a form
of T-duality.

Bearing in mind this behavior in the Yang-Mills sector, we turn now to the
gravity sector. In a toroidal compactification of M-theory the T-duality
transformation of $\ti{T}^0$ would be part of the U-duality group \cite{Hull:1994ys}
and would, therefore, be an exact symmetry. One may speculate that this
is still the case for our compactification on a $G_2$ orbifold and we proceed
to analyze the implications of such an assumption. Examining the
structure of our four-dimensional effective theory~\eqref{kinactual}--\eqref{scalarpot}
we see that the expressions for $K$, $W$ and
$D$ are indeed invariant under axionic shifts of $\ti{T}^0$. However, it is not
so clear what happens for $\ti{T}^0\rightarrow\frac{1}{\ti{T}^0}$.
An initial inspection of Eqs.~\eqref{kinactual}--\eqref{scalarpot}
shows that while the K\"ahler potential changes by%
\begin{equation}
\delta K\sim\ln\left(  \ti{T}^0\bar{\ti{T}}^{0}\right)\; ,
\end{equation}
the kinetic terms and superpotential will remain unchanged.
In order for the whole supergravity theory to be invariant
we need the supergravity function $\mathcal{G}=K+\ln |W|^2$
to be invariant. However as stands, with the $\ti{T}^0$ independent
superpotential~\eqref{1singsuperpot} this is clearly not the case.
One should, however, keep in mind that this superpotential is valid only
in the large radius limit and can, therefore, in principle be subject to
modifications for small $\mathrm{Re}(T^0)$. Such a possible modification which
would make the supergravity function $\mathcal{G}$ invariant and reproduce the
large-radius result~\eqref{1singsuperpot} for large $\mathrm{Re}(T^0)$ is given by
\begin{equation}
W\rightarrow h(\ti{T}^0)W,
\end{equation}
where
\begin{equation}
h(  \ti{T}^0)  = \frac{1}{\eta^{2}(i\ti{T}^0)\left(
j(i\ti{T}^0)-744\right)  ^{1/12}}\,
\end{equation}
and $\eta$ and $j$ are the usual Dedekind $\eta$-function and Jacobi $j$-function.
For large $\mathrm{Re}(T^0)$ the function $h$ can be expanded as
\begin{equation}
 h(\ti{T}^0)=1+2e^{-2\pi\ti{T}^0}+\dots\, .
\end{equation}
Recalling that $\mathrm{Re}(T^0)$ measures the volume of the singular
locus $\mathcal{T}^3$, the above expansion suggests that the function $h$
may arise from membrane instantons wrapping this three-torus.
It would be interesting to verify this by an explicit membrane instanton
calculation along the lines of Ref.~\cite{Harvey:1999as}.

It is well known that there are two dynamical phases in $\mathcal{N}=4$
Yang-Mills theory in four-dimensions \cite{D'Hoker:2002aw}. A supersymmetric ground
state of the $\mathcal{N}=4$ theory is attained when the full scalar
potential in Eq.~\eqref{N4YM} vanishes. This is equivalent to the condition
\begin{equation}
\left[  Z^{am},Z^{bn}\right]  =0 \label{comm}
\end{equation}
with $Z^{am}=A^{am}+iB^{am}$. There are two classes of solutions to
this equation.  The first, the ``superconformal phase'', corresponds
to the case where $\left\langle Z^{am}\right\rangle =0$ for all
$a,m$. The gauge symmetry is unbroken for this regime, as is the
superconformal symmetry. In the present context, this phase
corresponds to the neighborhood of a $\mathbb{C}^{2}/\mathbb{Z}_{N}$
singularity in which the full $\mathrm{SU}(N)$ symmetry is present.
As a result of the $\mathcal{N}=4$ supersymmetry in the gauge sector,
this phase will not be destabilized by low-energy gauge dynamics and,
hence, the theory will not be driven away from the orbifold point by
such effects. However, one can also expect a non-perturbative moduli
superpotential from membrane instantons~\cite{Harvey:1999as} whose
precise form for small blow-up cycles is unknown. It would be
interesting to investigate whether such membrane instanton corrections
can stabilize the system at the orbifold point or whether they drive
it away towards the smooth limit.

The second phase, called the ``Coulomb phase'' (or spontaneously
broken phase) corresponds to the flat directions of the potential
where Eq.~\eqref{comm} is satisfied for $\left\langle
  Z^{am}\right\rangle \neq0$.  The dynamics depend upon the amount of
unbroken symmetry. For generic breaking, $\mathrm{SU}(N)$ is reduced
to $\mathrm{U}(1)^{N-1}$. If this breaking is achieved through
non-trivial VEVs in the $A^{am}$ directions it corresponds,
geometrically, to blowing up the singularity in the internal $G_2$
space.


\section{Conclusion and outlook}

In this chapter we have constructed, for the first time, the explicit
four-dimensional effective supergravity action for M-theory on a
singular $G_2$ manifold. The class of $G_2$ manifolds for which our
results are valid consists of quotients of seven-tori by discrete
symmetry groups that lead to co-dimension four singularities, around
which the manifold has the structure
$\mathcal{T}^3\times\mathbb{C}^2/\mathbb{Z}_N$.  Breaking the
$\mathrm{SU}(N)$ gauge theory, generically to $\mathrm{U}(1)^{N-1}$,
by assigning VEVs to (the real parts) of the chiral multiplets along
D-flat directions can be interpreted as an effective four-dimensional
description of blowing up the orbifold. We have used this
interpretation to compare our result for the $G_2$ orbifold with its
smooth counterpart obtained in earlier papers~\cite{Lukas:2003dn,Barrett:2004tc}. We
find that, subject to choosing the correct embedding of the Abelian
group $\mathrm{U}(1)^{N-1}$ into $\mathrm{SU}(N)$, the results for the
K\"ahler potentials match exactly. This result seems somewhat
surprising given that there does not seem to be a general reason why
the smooth K\"ahler potential should not receive corrections for small
blow-up moduli when the supergravity approximation breaks down. At any
rate, our result allows us to deal with M-theory compactifications
close to and at the singular limit of co-dimension four A-type
singularities. This opens up a whole range of applications, for
example, in the context of wrapped branes and their associated
low-energy physics.

An interesting extension of the work presented here would be to
attempt a more general compactification of M-theory on a $G_2$
manifold whose singular loci are different from $\mathcal{T}^3$. This would
allow a reduction of the $\mathcal{N}=4$ supersymmetry in the gauge
theory sub-sector to $\mathcal{N}=1$, giving rise to richer
infrared gauge dynamics.

In continuing a programme for M-theory phenomenology, we aim
to explicitly include conical singularities into these models,
thereby incorporating charged chiral matter. This problem is
currently under investigation.

\chapter{Heterotic compactifications on Calabi-Yau three-folds}\label{HetIntro}

\section{Model building in $E_8 \times E_8$ heterotic string theory}
Compactification of the $E_8\times E_8$ heterotic string on Calabi-Yau
three-folds~\cite{Candelas:1985en, Greene:1986ar} is one of the oldest approaches to
particle phenomenology from string theory. Heterotic models have a
number of phenomenologically attractive features typically not shared
by alternative string constructions. Most notably,
gauge unification is ``automatic'' and standard model families
originate from an underlying spinor-representation of $\rm{SO}(10)$. 
Since the formulation of the  heterotic string \cite{Gross:1985fr,Gross:1984dd}, attempts have been made to develop realistic particle physics, namely the symmetries and spectra of the Standard Model, within a heterotic model. These include
compactification of the $10$-dimensional heterotic string theory on Calabi-Yau $3$-manifolds \cite{Candelas:1985en, Greene:1986ar,Greene:1986bm,Greene:1986jb}, theories constructed on toroidal orbifolds \cite{Dixon:1985jw, Dixon:1986jc} (which are singular limits of Calabi-Yau manifolds) and various other constructions \cite{Gepner:1987qi,Gepner:1987vz,Kawai:1986ah}. In the following chapters we shall explore the first of these, heterotic string compactifcations on a space of the form $M_{4}\times X$ where $M_4$ is Minkowski four-space and $X$ is a (real) six-dimensional Calabi-Yau manifold. Furthermore, we shall be interested in compactifications which produce $N=1$ supersymmetric theories in four-dimensions with symmetries and particle content consistent with the Standard Model.

In the following Chapters, we shall discuss in detail a class of models with broadly desirable phenomenological features and outline techniques
for systematically and algorithmically scanning a set of possible heterotic vacua. Before we begin these investigations, however, it will be useful to briefly review the fundamentals of heterotic string compactifications on Calabi-Yau manifolds. For a more complete treatment of heterotic constructions we recommend to the reader the classic text by Green, Schwarz and Witten \cite{gsw}, the comprehensive  `Bestiary' of Calabi-Yau manifolds and compactifications by Hubsch \cite{hubsch} and a number of useful review articles \cite{Distler:1987ee,Quevedo:1997uy,Uranga,Kachru:1997pc,Braun:2005ux}.
\section{The $10$-dimensional effective heterotic lagrangian and supersymmetry}
The field content of $N=1$, $10$-dimensional supergravity coupled to Yang-Mills theory is as follows: A Yang-Mills supermultiplet $({A^a}_{M}, \chi^a)$ consisting of a vector potential ${A^a}_{M}$ and a gaugino $\chi^a$ and the supergravity multiplet: $({e^A}_{M}, B_{MN}, \phi, \psi_M, \lambda)$ where ${e^A}_{M}$ is the vielbein, $B_{MN}$ an anti-symmetric NS $2$-form, $\phi$ the dilaton, $\psi_M$ the gravitino, and a spinor $\lambda$ (the dilatino). Following the notation of \cite{gsw}, the $10$-dimensional effective theory is described by the lagrangian,
\beq \label{heterotic_lagrangian}
S_{het} = \frac{1}{2\kappa^2} \int d^{10}x (-G)^{1/2} [R -\partial_{M}\phi \partial^{M}\phi - \frac{3\kappa^{2}}{8g^{4}\phi^2} \mid{H_3}\mid^2 - \frac{\kappa^2}{4g^{2}\phi}Tr(\mid F_{2}\mid^2)+ \ldots]
\eeq
where $H_3 = dB_2 - \omega_3$ is the field strength associated to the $2$-form and $F$ is the Yang-Mills field strength. Here $\omega_3$ is the Chern-Simons $3$-form $\omega_{3}=A_{a}F^{a} - \frac{1}{3}gf_{abc}A^{a}A^{b}A^{c}$. The index $a$ runs over the degrees of freedom of $E_{8}\times E_{8}$. The parameter in the perturbative expansion, $\alpha'$ is given by $\kappa^2 \sim g^{2}\alpha'$.

It is our goal to investigate compactifications of this theory which will result in a phenomenologically relevant $N=1$ supersymmetric theory in four dimensions. Specifically, we are seeking vacuum solutions to the $10$-dimensional theory compactified on $M_4 \times X$ where $M_4$ is a maximally symmetric four-dimensional space (de Sitter space, anti de Sitter space or Minkowski space) and $X$ is a compact, $6$-dimensional ``internal" space. To this end, we will decompose all the fields in \eref{heterotic_lagrangian} explicitly into external ($M_4$) and internal ($X$) components. If we decompose the Lorentz group $SO(1,9)$ over this product manifold as $SO(1,3) \times SO(6) \approx SO(1,3) \times SU(4)$, then a ${\bf 16}$ component Majorana-Weyl spinor of $SO(1,9)$ will satisfy ${\bf 16} \approx ({\bf 2},{\bf4}) \oplus ({\bf 2'}, {\bf \bar{4}})$.  In this work, we will be interested in Riemannian candidates for the internal space $X$. Such manifolds are partially classified by their holonomy groups \cite{berger}. For a $6$-dimensional, irreducible\footnote{Irreducible manifolds are those that cannot be written as a direct product of sub-manifolds. It can be shown that it is difficult to combine a reducible internal space with the presence of chiral fermions in $M_4$ and such choices may be phenomenologically disfavored. See e.g. \cite{hubsch}} Riemannian manifold, the available holonomy groups are $SO(6)$, $U(3)$ and $SU(3)$. In this chapter, we shall investigate the choice of $SU(3)$ holonomy - namely Calabi-Yau $3$-folds. Finally, we will write the bundle $U$ associated with the $E_8 \times E_8$ field strength $F$ as a direct product bundle, $U \approx V \times \tilde{V}$, where each of $V, \tilde{V}$ is a bundle with structure group $\subseteq E_8$. 

We recall that finding an unbroken supersymmetry at tree level amounts to finding a supersymmetry transformation such that the variation $\delta\psi$ vanishes for every fermionic field $\psi$ \cite{gsw}. In \eref{heterotic_lagrangian} the only elementary fermions are the gravitino, $\psi_M$, the spin $1/2$ `dilatino' $\lambda$ and the gauginos $\chi^a$. The supersymmetry variations of these fermions in terms of the SUSY parameterizing spinor $\eta$ are
\bea \label{fermion_variation}
\delta\psi_{M} & = & \frac{1}{\kappa}D_{M}\eta + \frac{\kappa}{32g^{2}\phi}({\Gamma_M}^{NPQ}-9{\delta^N}_{M}\Gamma^{PQ})\eta H_{NPQ} + \ldots \\ 
\delta\chi^{a}  & = &  -\frac{1}{4g\surd{\phi}}\Gamma^{MN}{F^a}_{MN}\eta +\ldots \\ \nn
\delta\lambda & = & -\frac{1}{\surd{2}{\phi}}(\Gamma\cdot\partial\phi)\eta + \frac{\kappa}{8\surd{2}g^{2}\phi}\Gamma^{MNP}{\eta}H_{MNP} +\ldots \nn
\eea
where we have omitted $(\text{fermi})^2$ and higher terms. 

In the following sections, we will be searching for solutions of the system $\delta\psi_{i}=\delta\lambda=\delta\chi=0$. In specifying such a solution, we are free to specify the space-time metric and the dilaton, $\phi$. However, there exists a constraint on $F$ and $H$ in the form of the Bianchi identities. In the minimal $10$-dimensional theory, the Bianchi identity for $H$ is just $dH=-tr(F\wedge F)$. Within the context of string theory, however, this is corrected by constraints placed on the model by anomaly cancellation. The Green-Schwarz mechanism \cite{gsw} requires that
\beq \label{bianchi}
dH=tr R\wedge R - tr F\wedge F.
\eeq

Next, we note that in order for the theory to have $N=1$ supersymmetry in four dimensions, we must consider \eref{fermion_variation} restricted to the internal space. Letting the indices, $M,N=0,1 \ldots 9$ decompose as those ranging over $M_4$: $\alpha,\beta=0,1 \ldots 3$ and those over the compact space $X$: $i,j=4, \ldots 9$, \eref{fermion_variation} becomes
\bea \label{fermion_variation_cy}
0 = \delta\psi_{i} & = & \frac{1}{\kappa}D_{i}\eta + \frac{\kappa}{32g^{2}\phi}({\Gamma_i}^{jkl}-9{\delta^j}_{i}\Gamma^{kl})\eta H_{jkl} + \ldots \\
0 = \delta\chi^{a} & = & -\frac{1}{4g\surd{\phi}}\Gamma^{ij}{F^a}_{ij}\eta +\ldots \nn \\ 
0= \delta\lambda & = & -\frac{1}{\surd{2}{\phi}}(\Gamma\cdot\partial\phi)\eta + \frac{\kappa}{8\surd{2}g^{2}\phi}\Gamma^{ijk}{\eta}H_{ijk} +\ldots \nn
\eea
Before we consider specific solutions to \eref{fermion_variation_cy}, we make several general observations.

\section{Calabi-Yau three-folds}\label{s:CY}
To reduce $10$ dimensional heterotic string theory to four-dimensions, we must compactify on a $6$-dimensional manifold. In this work, we will take the compact space, $X$, to be a Calabi-Yau manifold. There are many good reviews of Calabi-Yau manifolds available in the literature \cite{gsw,hubsch, Uranga}, and we will not attempt a comprehensive discussion here. Rather, we shall simply provide an overview of the basic properties of such spaces and the results important to the constructions presented in the following Chapters.

A $2n$-dimensional Calabi-Yau manifold is a K\"ahler manifold obeying the toplogical property $c_{1}(TX)=0$. For such a manifold, a powerful theorem due to Yau \cite{calabi, yau, yau2}, guarantees the existence of a Ricci-flat metric with $SU(n)$ holonomy. Thus, we are assured of the existence of a Ricci-flat metric and are not obliged to explicitly construct it. For this work, we shall consider Calabi-Yau $3$-folds, that is $3$-complex dimensional compact manifolds admitting metrics of $SU(3)$ holonomy. 

A real $2n$-dimensional manifold can be regarded as a $n$-dimensional complex manifold only if it admits a complex structure.  That is, it must admit a globally defined tensor, $J^{i}_{j}$ satisfying \cite{gsw, hubsch}
\beq\label{niejenhuis}
J^{i}_{j}J^{k}_{i} = - \delta^{k}_{j} ~~\text{and}~~ N^{k}_{ij}=\partial_{[j}J^{k}_{i]}-J^{p}_{[i}J^{q}_{j]}\partial_{q}J^{k}_{p}=0
\eeq
where $N^{k}_{ij}$ is called the Niejenhuis tensor \cite{hubsch}. A tensor $J^{i}_{j}$ is called an `almost complex structure' if it satisfies only the first condition in \eref{niejenhuis} and a `complex structure' if, in addition, the Niejenhuis tensor vanishes. A real manifold, in principle, may admit many complex structures.

For a complex manifold, cohomology and homology classes may be constructed according to their Dolbeaut cohomology (i.e. written in terms of the number of holomorphic and antiholomorphic indices). That is, the cohomology groups decompose as
\beq
H^{p}(TX) = \bigoplus _{r+s=p}H^{r,s}(TX)
\eeq
where $H^{r,s}$ corresponds to forms with $r$ holomorphic and $s$ antiholomorphic indices. The dimensions of $H^{r,s}$ are denoted by $h^{r,s}$ and are known as Hodge numbers. These numbers are topological invariants of $X$ and do not depend on the choice of complex structure \cite{Joyce}.

A hermitian metric on a complex  manifold is called K\"ahler if it can be written in the form $ds^2=g_{a\bar{b}}dz^{a}d\bar{z}^{\bar{b}}$ (with $a,b$ as complex coordinates on $X$) and in addition, the associated $(1,1)$ form
\beq
J = g_{a\bar{b}}dz^{a}\wedge d\bar{z}^{\bar{b}}
\eeq
is closed, i.e. $dJ=0$ \cite{hubsch}. This $(1,1)$ form corresponds to lowering an index on the complex structure $J^{i}_{j}$ with the hermitian metric and is known as the K\"ahler form. A manifold which admits a K\"ahler form is called a K\"ahler manifold. This is a topological property of a space since the K\"ahler form $J$ defines a non-trivial\footnote{The class is non-trivial since $\int_{X} J\wedge J \wedge J = Vol(X)$.} cohomology class in $H^{1,1}(TX)$. 

To specify a Calabi-Yau $3$-fold, we must specify both the complex and K\"ahler structures. The set of parameters that span the space of complex structures is called the complex structure moduli space and has dimension $h^{2,1}(TX)$. Likewise, the set of parameters which define the K\"ahler class is known as the K\"ahler moduli space and $h^{1,1}$ counts the deformations of the K\"ahler structure. Locally, the complete moduli space of a Calabi-Yau space is a direct product of these two spaces \cite{Candelas:1989bb,Candelas:1990pi}.

As a result of the above, a Calabi-Yau manifold is characterized by a simple set of topological information. The Hodge numbers form the so-called ``hodge diamond" structure: $h^{i,j}$, $i+j \leq 3$ with all Hodge numbers fixed\footnote{$h^{3,0}=h^{0,3}=1,h^{1,0}=h^{0,1}=0,h^{0,2}=h^{2,0}=0$.} except for $h^{2,1}$ and $h^{1,1}$. Thus, by the index theorem \cite{AG2} the Euler number of $X$ is just $\frac{1}{2}\chi=h^{1,1}-h^{2,1}$.

Expanding $J$ in a set of basis forms, $J= t^{r} J_{r}$ with $r =1, \ldots h^{1,1}(TX)$ we call the set of parameters $t^{r}$ the \emph{K\"ahler cone} of $X$. For each such $(1,1)$-form $J_{r}$, there is a dual 2-cycle, $C_{s}$ in homology $h_{2}(TX)$, with duality defined by
\beq
\int_{C_{s}} J_r = \delta_{rs}~.
\eeq
The set of all $C_r$ is known as the dual-cone to the K\"ahler cone, or the \emph{Mori Cone} \cite{Hosono:1994ax}. It follows that the set of $C_r$, $[W]$, is an \emph{effective} class of curves. That is, the class $[W]$ has a holomorphic representative $C$.
\section{Anomaly Cancellation}
Regardless of the vacuum configurations chosen for $H$, $G$ and $F$, there is a further observation that we can make about the anomaly cancellation of \eref{bianchi}. We note that $trF \wedge F$ and $trR \wedge R$ represent the second Chern characters of the $E_8 \times E_8$ bundle, $U$,  and of the tangent bundle, $TX$, of $X$, respectively and that $dH$ is trivial in cohomology.  Thus, it follows that in order for \eref{bianchi} to have a solution, $V$ and $\tilde{V}$ must be holomorphic vector bundles satisfying the following topological identities:
\bea
c_{1}(V)=c_{1}(\tilde{V})=0~~\text{mod} 2 \\
ch_{2}(TX)-ch_{2}(V)-ch_{2}(\tilde{V})=0
\eea
where $c_1$ is the first Chern class and $ch_{2} = \frac{1}{2}c^{2}_{1}-c_{2}$ is the second Chern character\footnote{As stated in Section \ref{s:CY}, for a Calabi-Yau space $c_{1}(TX)=0$ so $ch_{2}(TX)= -c_{2}(TX)$.}.

This expression is corrected still further if we allow for the existence of $5$-branes in the theory. Allowing for either NS $5$-branes in the weakly coupled theory or $M5$-branes in the strongly coupled picture of Ho\v{r}ava-Witten theory (see Section \ref{s:hor_witt_review}), the anomaly condition is naturally modified. Locally, this changes the Bianchi identity and its associated topological condition to  
\begin{equation}
ch_{2}(TX)-ch_{2}(V)-ch_{2}(\tilde{V})=W_{5}  \label{anomaly}
\end{equation}
where $W_{5}$ is the class of two-cycles in $X$ which we allow $5$-branes to wrap. In this work we will maintain a general viewpoint and allow for the existence of $5$-branes. Further, we require the two-cycles to be well-behaved in the sense that they are defined by holomorphic curves. Hence, the five-brane class $W_{5}$ must be chosen such that it indeed has a holomorphic curve representative $C$, with $W=[C]$. Classes $W\in H_2(X,\mathbb{Z})$ with this property are called {\em effective} (see \cite{AG1} for details). We shall refer to \eref{anomaly} as the \textit{anomaly cancellation condition}.

Note that for this work, we will construct all the dynamics of the theory entirely within one $E_8$ factor and treat the other $E_8$ bundle as a `hidden sector', (which is coupled to the physical theory only through gravity). That is, we will not specify a hidden bundle and instead demand that
\beq
ch_{2}(TX)-ch_{2}(V)~~\text{is an effective class on}~X
\eeq 
So, for example, the anomaly cancellation condition would be satisfied for a trivial bundle $\tilde{V}$ and a $5$-brane class $W_{5} =ch_{2}(TX)-ch_{2}(V)$~\footnote{Of course, there may be other choices  which involve a non-trivial hidden bundle $\tilde{V}$. Since we are mostly interested in the observable sector at this stage the important point for now is the existence of a viable hidden sector.}.
Thus, in the following sections, we will investigate only the internal space $X$ and a single $E_8$ bundle, $V$ to  solve  \eref{fermion_variation} and \eref{bianchi}. We will make one more general observation about \eref{fermion_variation} before looking at specific solutions.
\section{Supersymmetry and vector bundle stability}\label{stability}
We begin our analysis of the fermion variations in \eref{fermion_variation_cy} by considering the gaugino variation 
$\delta\chi^{a} =0$, which leads to $\Gamma^{ij}{F^a}_{ij}=0$. Then, re-written in terms of holomorphic indices over $X$, the vanishing of the gaugino variation implies that the $E_{8}\times E_{8}$ gauge connection, $A$, must satisfy the hermitian Yang-Mills equations:
\beq\label{HYME}
F_{ab}=F_{\overline{a}\overline{b}}=g^{a\overline{b}}F_{\overline{b}a}=0
\eeq
where $F$ is the field strength of $A$. The first two expressions in \eref{HYME}, $F_{ab}=F_{\overline{a}\overline{b}}=0$ are simply the condition that the vector bundle be holomorphic. To address the final equality, $g^{a\overline{b}}F_{\overline{b}a}=0$, once again, we ignore the $E_{8}$
factor of the ``hidden'' sector (or alternatively take the vector bundle ${\tilde{V}}$ to have a trivial vacuum) and focus on a single ``visible'' $E_{8}$ connection only. In order to preserve $4$-dimensional supersymmetry, we must find a bundle $V$ whose field strength satisfies \eref{HYME}.

In general \eref{HYME} is a set of very complicated differential equations for $A$ and no generic solution techniques are known. However, for Calabi-Yau manifolds $X$ there exists a powerful way of transforming this question into an alternate one in algebraic geometry. For a Calabi-Yau manifold, the Donaldson-Uhlenbeck-Yau theorem \cite{duy1,duy2} states that on each \textit{poly-stable} holomorphic vector bundle $V$, there exists a unique connection satisfying the Hermitian Yang-Mills equation \eref{HYME}. Thus, to verify that our vector bundle is consistent with supersymmetry on a Calabi-Yau compactification,
we need only to verify that it possesses the property of poly-stablility. 

We will not review the technical details of the theorem here and the reader is referred
to \cite{duy1,duy2} for the details. For our purposes, it suffices to
consider {\em stable holomorphic vector bundles}.

Stability can be examined by defining a number called the \textit{slope} of a bundle (or sheaf):
\begin{equation}\label{slope}
\mu (V)\equiv \frac{1}{\rk(V)}\int_{X}c_{1}(V)\wedge J^{\dim(X)-1}  
\end{equation}
where $J$ is the K\"{a}hler form on $X$ and ${\rm rk}(F)$ and $c_1(F)$ are the rank and the first Chern class of the sheaf, respectively. A bundle $V$ is now called stable (resp.~semi-stable)
if for all sub-sheafs $F\subset V$ with $0<{\rm rk}(F)<{\rm rk}(V)$ the
slope satisfies $\mu (F)<\mu (V)$ (resp.~$\mu (F)\leq \mu (V)$).  A bundle is poly-stable if it can be decomposed into a direct sum of stable bundles, all with the same slope. It follows that stable $\rightarrow$ poly-stable $\rightarrow$ semi-stable. Unfortunately, although the presence of a stable vector bundle $V$ guarantees a solution to \eref{HYME}, we are faced with ``conservation of misery" since the property of stability is notoriously difficult to prove in algebraic geometry. It is worth mentioning here that a bundle $V$ is stable if and only if its dual, $V^*$ is stable, and that any ``twisting" (i.e. tensoring by a line bundle) of a stable bundle is stable.

With these general constraints in hand, we will now discuss full solutions to our supersymmetry conditions \eref{fermion_variation_cy}. We will begin by reviewing what was historically the earliest and simplest solution to the equations in \eref{fermion_variation_cy}, the so-called ``Standard Embedding".

\section{The Standard Embedding}
In the search for supersymmetric compactifications of the heterotic lagrangian, the simplest solution to  \eref{fermion_variation} begins with the assumption that the three-form $H$ vanishes and that the dilaton, $\phi$ is a constant. With these assumption, finding a supersymmetric vacuum reduces to solving
\bea
0 & = &\delta\psi_{i}=D_{i}\eta \\
0 & = &\delta\chi^{a}=\Gamma^{ij}{F^a}_{ij}\eta
\eea

The first of these equations is simply the condition that there exists a single covariantly constant spinor on $X$. In six dimensions, according to Berger's classification \cite{berger}, it is precisely $SU(3)$ holonomy that guarantees the existence of such a spinor. Thus, for Calabi-Yau manifolds $X$, $\delta\psi_{i}=0$ is satisfied and the spinor field $\eta$ necessarily obeys an integrability condition $[D_{i},D_{j}]\eta=0$ which in turn implies that 
\beq
\Gamma^{k}R_{ik}\eta =0
\eeq
Thus, our $SU(3)$ holonomy metric must be Ricci-flat as discussed in Section \ref{s:CY}. With the choice of a Ricci-flat Calabi-Yau manifold as the internal space, we still have to satisfy two additional conditions for supersymmetry. First, we need a bundle $V$ over $X$ which satisfies the Hermitian Yang-Mills equation $g^{a\overline{b}}F_{\overline{b}a}$. Further, with $H=0$ the anomaly condition (in the absence of $5$-branes), requires $tr R\wedge R =tr F\wedge F$.

The simplest non-trivial solution to this is to set the background spin connection equal to the background Yang-Mills connection, $A$. That is, we will take the bundle $V$ to be the tangent bundle of the Calabi-Yau. Such a choice satisfies the Bianchi identity exactly in field space. Finally, with $X$ a Ricci-flat K\"ahler manifold, it is easy to see that the tangent bundle is stable. That is, \eref{HYME} is satisfied. With this choice for the bundle $V$, it is clear that over the internal space, the structure group is no longer $E_8$, but rather $SU(3)$. That is, we have broken $E_8$ to the subgroup $E_8 \rightarrow E_6 \times SU(3)$.\footnote{It is clear that such breaking is necessary for our goal of producing realistic $4$-dimensional physics. Since $E_8$ has no complex representations it is not suitable for Grand Unified model building in four dimensions. Instead, we will be interested in the GUT symmetry groups $E_6$, $SO(10)$ and $SU(5)$.}

In general, if we solve \eref{bianchi} by using only a subset of the $E_8$ group indices (setting $F=0$ for the rest), then the $E_8$ bundle breaks into a product bundle with structure group $H \times G$ where $G,H \in E_8$, $H$ is the Yang-Mills group of physical space ($M_4$) and $G$ is the structure group of $V$, the bundle over the internal space. So, for a given choice of bundle and structure group, $G$, we can determine the maximal $H$ such that $H \times G \in E_8$. $H$ is called the ``commutant" of G in $E_8$. For the above choice of $G=SU(3)$, we have $H=E_6$. 

The relevant fermionic fields in the low-energy four-dimensional theory arise from the decomposition of the spinor (gaugino) in the vector supermultiplet which transforms in the ${\bf 248}$ of $E_8$.  We will discuss the particle spectra that arises in such compactifications more generally in the next sections, but for now we simply state one result. The chiral asymmetry (i.e. the number of generations of quarks/leptons) is given by the Euler number, $\chi$ of $X$.
 \beq\label{standard_gen}
 N_{gen}= \frac{1}{2}\chi=h^{1,1}-h^{2,1}
 \eeq
 This property was seen as a potentially attractive feature of early heterotic string compactifications, since it implies that the the topology of space itself determines the number of particle families. The complete spectra of particles, and other features such as Yukawa couplings \cite{Candelas:1987se} can be determined for the Standard Embedding. 
 
 Therefore, in the Standard Embedding, a compact Calabi-Yau manifold, $X$ is the sole ingredient necessary to determine the particle spectra of the four-dimensional effective theory. By choosing $X$ we have determined an $E_6$ Grand Unified Theory (GUT) in four-dimensions. Despite its appealing simplicity, the Standard Embedding is of limited usefulness from a phenomenological perspective since very few known Calabi-Yau $3$-folds have the right features for string phenomenology. In particular, those with Hodge numbers satisfying \eref{standard_gen} are rare and even when they do exist, tend to exhibit the wrong particle spectra \cite{hubsch}. As well, in the standard embedding, we are clearly limited to $E_6$ GUT theories. In our search for realistic particle physics from string theory, it is clearly of interest to ask if there are more general solutions to the constraints \eref{anomaly}, \eref{HYME}.
\section{General Embeddings}

Having obtained an $E_6$ GUT theory from the Standard Embedding, it is a natural question to ask whether one could obtain other GUT groups such as $SO(10)$ and $SU(5)$ through Calabi-Yau compactification?  It was realized in \cite{Witten:1985bz} that such constructions are indeed possible by choosing $V$ not as the tangent bundle, $TX$, as in the Standard Embedding above, but as a more general holomorphic vector bundle over $X$ with structure group $G$. As discussed above, we must break the $E_8$ symmetry in order to have realistic physics in four-dimensions. We will take the structure group of $10$-dimensional bundle to decompose into a product, $H \times G \in E_{8}$. For a general structure group $G$, the four-dimensional symmetry group, $H$, will be the commutant of $G$ in $E_8$. For instance, if we are interested in obtaining a $SO(10)$ theory in $4$-dimensions, we must select $G$ to be $SU(4)$. Similarly, a bundle $V$ with $SU(5)$ symmetry will lead to an $SU(5)$ GUT theory. Such choices of holomorphic vector bundles are known as \emph {General Embeddings} and will be the subject of the rest of this thesis.

However, we must ask whether such general vector bundles satisfy the conditions for supersymmetric vacua \eref{fermion_variation_cy}? We begin by trying to satisfy \eref{bianchi}, \eref{fermion_variation_cy} by again taking $H=d\phi=0$, but this time without embedding the spin connection in the gauge group. The vanishing of the gravitino variation in \eref{fermion_variation_cy} once again determines $X$ as a manifold of $SU(3)$ holonomy. So, we will consider these ``General Embeddings" to be bundles over Calabi-Yau manifolds. With this solution, $\delta\lambda=0$ is obeyed identically. Finally, we have already addressed the issue of the gaugino variation $\delta\chi=0$ in the previous section. If $V$ is a general holomorphic vector bundle (with structure group $G$) over a Calabi-Yau $3$-fold then supersymmetry demands that the bundle satisfy the Hermitian Yang-Mills equation \eref{HYME}. That is, $V$ must be a {\textit stable} bundle. For a $SU(n)$ bundle, the first Chern Class vanishes ($c_1(V)=0$) and thus we must have
\beq
\mu(\mathcal{F})\leq 0.
\eeq
for all coherent sub-sheaves, $\mathcal{F} \in V$. It is a useful result that 
\beq\label{H0=0}
H^0(X,V)=H^3(X,V)=0
\eeq
 for a stable $SU(n)$ bundle, $V$. While not sufficient to prove stability, the above condition on cohomology is a useful and non-trivial check. 

The only remaining concern is the anomaly cancellation condition arising from the Bianchi identity \eref{bianchi}. We first note that for $SU(n)$ bundles $c_1(V)=0$, so the anomaly cancellation condition \eref{anomaly} takes the form $c_{2}(TX)-c_{2}(V)-c_{2}(\tilde{V})=W_{5}$. The condition $dH=trR\wedge R -trF \wedge F$ is obeyed with $H=0$ only if the spin connection is embedded in the gauge group. We must now consider the fact that \eref{heterotic_lagrangian}, \eref{fermion_variation_cy}, and \eref{bianchi} are actually expansions in $\alpha'$ and in a parameter $r$ related to the radius of the compact Calabi-Yau, $X$ \cite{Witten:1986kg}. One normally works at lowest order in this expansion. This corresponds to working at low energies, in the supergravity limit, and is a good approximation if the curvature length scale of $X$ is large compared with the string length. It can be shown that it is possible to modify the base manifold $X$ and bundle $V$ away from the standard embedding so that \eref{fermion_variation_cy},\eref{bianchi} will be satisfied for a general embedding. The details of this argument\footnote{For a generalization to include $5$-branes see \cite{Witten:1996mz}.} are too lengthy to be reviewed here (see \cite{Witten:1986kg,Witten:1985bz} for details). However, we may state the result clearly. In a general embedding heterotic compactification we choose a stable holomorphic vector bundle $V$ over a Calabi-Yau manifold $X$ and solve the conditions for supersymmetry order by order in a perturbative expansion in $\alpha'/r$. At first order, the configuration will solve \eref{fermion_variation_cy} over a Calabi-Yau manifold, though we will have perturbed away from the flat metric ($g=g_{0} + \alpha' \tilde{g}+ \ldots$). It can be shown that while $X$ will no-longer have $SU(3)$ holonomy, it will still be Calabi-Yau (i.e. a K\"ahler space with $c_{1}(TX)=0$) \cite{Witten:1986kg}.

Starting with the pioneering work in~\cite{Witten:1985bz,Distler:1987ee}, 
there has been continuing
activity on Calabi-Yau based non-standard embedding models
over the years. Recently, there has been significant
progress both from the mathematical and the model-building viewpoint,
leading to models edging closer and closer towards the standard
model~\cite{Braun:2005nv, Braun:2005bw, Bouchard:2005ag}. 
Two types of constructions, one based on
elliptically fibered Calabi-Yau spaces with bundles of the
Friedman-Morgan-Witten type~\cite{Friedman:1997yq,Friedman:1997ih,Dongai:1997mg} and
generalizations~\cite{Braun:2005ux, Braun:2005zv},
\cite{Thomas:1999iz, Braun:2006ae}, the other based on complete
intersection Calabi-Yau spaces with monad
bundles~\cite{Distler:1987ee, Blumenhagen:2006wj, Blumenhagen:1997vt, Kachru:1995em, Douglas:2004yv},
have been pursued in the literature. It is the last of these, the Monad construction, that we will study in detail in the following chapters. As a final comment, we mention here that the standard embedding is included as a special case of the general embedding. That is, it is simply one choice of a general $SU(3)$ holomorphic vector bundle.

\subsection{Wilson Lines and Symmetry Breaking}
We note that this discussion was motivated by asking which choices of vector bundle, $V$, would produce realistic GUT symmetries in $4$-dimensions. We have argued above that it is possible to produce GUT theories in four-dimensions. However, clearly once we have obtained the GUT group \cite{Ross:2007az}, we must address the task of breaking the symmetry to something containing the symmetry $SU(3) \times SU(2) \times U(1)$, of the Standard Model of particle physics. In Heterotic model-building, this is accomplished by a ``two-stage" symmetry breaking. The first stage (described above) reduces $E_8$ to a GUT group ($E_6$, $SO(10)$ or $SU(5)$). In the second stage, Wilson lines are introduced to break the GUT symmetry to the Standard Model symmetry (plus possibly $U(1)$ factors) \cite{gsw}.

In order to introduce non-trivial Wilson lines, the manifold $X$ must not be a simply connected space (i.e. it must have non-trivial fundamental group, $\pi_{1}(X)$). Starting with a simply connected Calabi-Yau manifold, we can obtain a smooth, non-simply connected one by dividing the manifold by a freely acting discrete symmetry
\beq
X \rightarrow \frac{X}{\Gamma}
\eeq
where $\Gamma$ is a discrete group of finite order, $|\Gamma|$. The fundamental group of the quotient $X/\Gamma$ is $\Gamma$. For example, with a non-trivial fundamental group, $\Gamma = \IZ_2$ one can break $SU(5)$ down to $SU(3) \times SU(2) \times U(1)$ and $\Gamma = \IZ_3 \times \IZ_3$
can break $SO(10)$ down to $SU(3) \times SU(2) \times U(1)^2$ \cite{gsw}. A complete heterotic string model must thus include a Calabi-Yau manifold and vector bundle complete with discrete symmetries that will reduce the GUT symmetry to that of the Standard Model.

Finally, we might ask, is it possible to choose a vector bundle, $V$, to break $E_8$ directly to something containing Standard Model symmetry, thus by-passing GUT theories altogether? This may indeed be accomplished by choosing $G$ to be a unitary group $U(N)$ rather than special unitary, $SU(N)$. Such constructions have been explored in the literature \cite{Blumenhagen:2006ux, Blumenhagen:2006wj}. These constructions form an interesting field in their own right, but will not be considered here.

In this work, we shall require our bundles to have structure group $SU(n)$ with $n=3,4,5$, leading to GUT groups $E_6$, $SO(10)$ and $SU(5)$, respectively (followed by the standard ``two-stage" breaking by Wilson lines).

\section{Particle Spectra in Heterotic Calabi-Yau Compactifications}\label{het_spectra}
Having defined the area of our interest to be the General Embedding of holomorphic vector bundles over Calabi-Yau $3$-folds, we now turn to the general structure of the low-energy
particle spectrum. In addition to the dilaton, $h^{1,1}(X)$ K\"ahler
moduli and $h^{2,1}(X)$ complex structure moduli of the Calabi-Yau
space, each of the $E_8$ gauge theories as well as the five-branes
give rise to a sector of particles in the low-energy theory. 

As described in the previous section, the low-energy gauge group $H$ in the observable sector is given by the commutant of the structure group $G$ within $E_8$. For
$G={\rm SU}(3)$, ${\rm SU}(4)$, ${\rm SU}(5)$ this implies the standard
grand unified groups $H=E_6$, ${\rm SO}(10)$, ${\rm SU}(5)$, respectively.
In order to find the matter field representations, we have to
decompose the adjoint ${\bf 248}$ of $E_8$ under $G\times H$. In
general, this decomposition can be written as
\begin{equation}
248\rightarrow (1,\mbox{Ad}(H))\oplus \bigoplus_{i}(R_{i},r_{i})  
\label{adjoint}
\end{equation}
where $\mbox{Ad}(H)$ denotes the adjoint representation of $H$ and
$\{(R_{i},r_{i})\}$ is a set of representations of $G\times H$.
The adjoint representation of $H$ corresponds to the low-energy
gauge fields while the low-energy matter fields transform in the representations
$r_i$ of $H$. For the three relevant structure groups these matter
field representations are explicitly listed in Table \ref{spec}.

To be observable at low energy, the fermion fields transforming under the GUT symmetry must be massless modes of the Dirac operator on $X$. It can be shown that the number of massless modes for a given representation equals the dimension of a certain bundle-valued cohomology group \cite{gsw}. The number of supermultiplets occuring in the low energy theory for
each representation $r_i$ is given by
$n_{r_{i}}=h^{1}(X,V_{R_{i}})$, where $r_{i}, R_{i}$ are defined by the decomposition~\eqref{adjoint}.
For $G={\rm SU}(n)$, the relevant representations $R_i$ can be obtained
by appropriate tensor products of the fundamental representation. The relevant cohomology groups and hence the number of low-energy representations can then be computed as
summarized in Table \ref{spec}.
\begin{table}[h]
\begin{center}
\begin{tabular}{|l|l|l|}
\hline
$G\times H$ & Breaking Pattern:  
${\bf 248}\rightarrow $ 
& Particle Spectrum\\  
\hline\hline
{\small $\rm{SU}(3)\times E_{6}$} & {\small $({\bf 1},{\bf
  78})\oplus ({\bf 3},{\bf 27})\oplus 
(\overline{\bf 3},\overline{\bf 27})\oplus ({\bf 8},{\bf 1})$ }
&
$
\ba{rcl}
n_{27}&=&h^{1}(V)\\ 
n_{\overline{27}}&=&h^{1}(V^\star)=h^{2}(V)\\
n_{1}&=&h^{1}(V\otimes V^\star)
\ea
$
\\  \hline
{\small $\rm{SU}(4)\times\rm{SO}(10)$} &{\small $({\bf 1},{\bf
  45})\oplus ({\bf 4},{\bf 16}) 
\oplus (\overline{\bf 4},\overline{\bf 16})\oplus ({\bf 6},{\bf
  10})\oplus ({\bf 15},{\bf 1})$ } 
&
$
\ba{rcl}
n_{16}&=&h^{1}(V)\\
n_{\overline{16}}&=&h^{1}(V^\star)=h^2(V)\\
n_{10}&=&h^{1}(\wedge ^{2}V)\\
n_{1}&=&h^{1}(V\otimes V^\star)
\ea
$
\\  \hline
{\small $\rm{SU}(5)\times\rm{SU}(5)$} &{\small $({\bf 1},{\bf 24})\oplus
({\bf 5},{\bf 10})\oplus (\overline{\bf 5},\overline{\bf 10})\oplus
({\bf 10},\overline{\bf 5})\oplus 
(\overline{\bf 10},{\bf 5})\oplus ({\bf 24},{\bf 1})$}
&
$
\ba{rcl}
n_{10}&=&h^{1}(V)\\ 
n_{\overline{10}}&=&h^{1}(V^\star)=h^2(V)\\ 
n_{5}&=&h^{1}(\wedge^{2}V^\star)\\
n_{\overline{5}}&=&h^{1}(\wedge ^{2}V)\\
n_{1}&=&h^{1}(V\otimes V^\star)
\ea
$
\\ \hline
\end{tabular}
\caption{{\em A vector bundle $V$ with structure group $G$ can break the $E_8$
  gauge group of the heterotic string into a GUT group $H$. The low-energy representation are found from the branching of the ${\bf 248}$ adjoint of $E_8$ under $G\times H$ and the low-energy spectrum is obtained by computing the indicated bundle cohomology groups.}}\label{spec}
\end{center}
\end{table}
Further, the Atiyah-Singer index theorem \cite{AG1,AG2}, applied to the case
$c_1(TX)=c_1(V)=0$, tells us that the index of $V$ can be expressed as 
\begin{equation}
\ind(V)=\sum_{p=0}^3 (-1)^{p}\, h^{p}(X,V)=\frac{1}{2}\int_{X}c_3(V) \ ,
\label{index}
\end{equation}
where $c_3(V)$ is the third Chern class of $V$. For a stable $SU(n)$ bundle
we have $h^0(X,V)=h^3(X,V)=0$ and comparison with Table \ref{spec} shows
that, in this case, the index counts the chiral asymmetry, that is,
the difference of
the number of generations and anti-generations. The index is usually easier
to compute than individual cohomologies and is useful to impose a physical constraint on the chiral asymmetry.

We note here that Serre duality \eref{serre}, a consequence of stability \eref{H0=0},
and the Atiyah-Singer index theorem \eref{AS}
reduce our necessary particle spectra computations by half. The cohomology
of $V$ and $V^*$ are related as
\beq\label{indV}
- h^{1}(X,V) + h^{1}(X,V^*) = \ind(V) = \frac12 \int_X  c_3(V) \ .
\eeq
Hence, we only need to compute one of the two groups on the
left of \eref{indV}. Similarly, provided that $H^0(X,\Lambda^2 V)=H^0(X,\Lambda^2 V^*)=0$, as will indeed be the case for the stable bundles we consider,
the index theorem applied to $\Lambda^2 V$ together with the relation $c_3(\Lambda^2 V)=(n-4)c_3(V)$ leads to
\beq
(n-4) \ind(V) = - h^{1}(X,\wedge^2V) + h^{1}(X,\wedge^2 V^*) \ . \label{indtheorem2}
\eeq

\subsection{Net Families of Particles}
A physical constraint arises from the requirement that the particle spectra generated by the heterotic model produces three generations of quarks/leptons. This condition will be satisfied so long as the topological index of $V$, 
\begin{equation}
\ind(V)=\frac{1}{2}\int_{X}c_{3}(V)
\end{equation} is an integer multiple of $3$. We make this choice, rather than requiring that
the index be exactly $3$, with the intention of eventually adding discrete symmetries and Wilson lines on the Calabi-Yau manifold which would reduce this number of generations further (depending on the choice of discrete symmetry).  With Wilson lines in mind, we see that the number of generations is  constrained by the limited choice of discrete symmetries. Suppose there is a discrete symmetry group $\Gamma$ which acts freely on the manifold $X$. Then the Euler number of $Y=X/\Gamma$ is simply that of $X$ divided by group order $|\Gamma|$\footnote{Note, the individual Hodge numbers do not divide in this way.}. Therefore $|\Gamma|$ must by construction divide the Euler number of $X$.

Under this $\Gamma$-action,
the index of $V$, given above, also descends by division of $\Gamma$. Therefore, combining this with the conclusion of the previous paragraph, we require that
\begin{equation}\label{k-div}
\frac13\left( \frac{1}{2}\int_{X}c_{3}(V) \right) \mbox{ divides the
Euler number of } X
\end{equation}

\section{Heterotic Compactification and Physical Constraints}\label{s:het}
In the following chapters, we will construct large classes of heterotic models consisting of holomorphic, $SU(n)$-bundles over Calabi-Yau $3$-folds. We will use the results of this chapter to impose physical constraints on the constructions and to select which models are worthy of more detailed study. 

Because we will make repeated use of these constraints, we shall briefly summarize here the key observations made in the preceding sections. (For a deeper discussion of these constraints and their origins see for example \cite{Candelas:1985en,gsw,Donagi:2004ia,Donagi:2004su,Donagi:2004ub}).

\paragraph{The Ingredients of a Heterotic Compactification:}
In addition to a Calabi-Yau three-fold $X$ with tangent bundle, $TX$,
we need two holomorphic vector bundles $V$ and $\tilde{V}$ with
associated structure groups which are sub-groups of $E_8$. In the
present context, we will be interested in bundles with rank $n=3,4,5$
and corresponding structure group $G=\rm{SU}(n)$. In general,
heterotic vacua can also contain five-branes. For a supersymmetric compactification, the five-branes have to wrap a holomorphic curve in the Calabi-Yau space $X$, whose second homology
class we denote by $W\in H_2(X,{\mathbb Z})$ and take to be effective.

\paragraph{The Physical Constraints:}
\begin{itemize}
\item {\bf Structure group ${\bf SU(n)}$} 

We shall take $V$ to be a holomorphic $SU=(n)$ bundle with $n=3,4,5$ in order to produce $E_6$, $SO(10)$ and $SU(5)$ GUT theories in $4$-dimensions. Thus, we impose on our holomorphic bundles the restriction that
\beq
c_1(V)=0
\eeq

\item {\bf Stability} 

To preserve $N=1$ supersymmetry in $4$-dimensions, we require $V$ and $\tilde{V}$ to be stable, holomorphic vector bundles. For $SU(n)$ bundles this implies $h^0(X,V)=h^3(X,V)=0$ (see Chapter \ref{StabCh}).

\item {\bf Anomaly Cancellation}

To guarantee anomaly cancellation, our choice of Calabi-Yau manifold, $X$, and $SU(n)$ vector bundle, $V$ must satisfy a topological constraint $c_2(TX)-c_2(V) =W$ where $W\in H_2(X,\mathbb{Z})$ is {\em effective}\footnote{In practice, this will reduce to the condition that the coefficient of $J^2$ in $c_2(TX)-c_2(V)$ is non-negative.}. We shall denote this schematically as
\beq
c_2(TX)-c_2(V) \geq 0\; . \label{anomaly_final}
\eeq

\item {\bf Three Generations}

Requiring three generations of particles places a constraint on the index of the bundle, relating it to the index of the Calabi-Yau $3$-fold $X$.
\begin{equation}\label{k-div_final}
\frac13\left( \frac{1}{2}\int_{X}c_{3}(V) \right) \mbox{ divides the
Euler number of } X
\end{equation}
\end{itemize}

We now turn to building a new class of heterotic models, subject to the physical constraints described above.

\chapter{An algorithmic approach to heterotic compactification: Monad bundles and Cyclic Calabi-Yau manifolds}\label{MonCh1}

\section{Introduction}
 While the $E_8 \times E_8$ heterotic string theory has many phenomenologically attractive features, heterotic model building is still a long way from one of its major goals: finding an example which does not merely have standard model spectrum but reproduces the standard model exactly,
including detailed properties such as, for example, Yukawa couplings.

One of the main obstacles in achieving this goal is the inherent
mathematical difficulty of heterotic models. In addition to a
Calabi-Yau three-fold $X$, heterotic models require two holomorphic
(semi)-stable vector bundles $V$ and $\tilde{V}$ on $X$. Except for
the simple case of standard embedding, where $V$ is taken to be the
tangent bundle $TX$ of the Calabi-Yau space and $\tilde{V}$ is
trivial, construction of these vector bundles is often not
straightforward and the computation of their properties is usually
involved.  For example, stability of these bundles, an essential
property if the model is to preserve supersymmetry, is notoriously
difficult to prove. In addition, when searching for realistic particle
physics from heterotic string theory, these mathematical obstacles
have to be resolved for a large number of Calabi-Yau spaces and
associated bundles, as every single model (or even a small number of
models) is highly likely to fail when confronted with the detailed
structure of the standard model.  One of the main purposes of this thesis is to
present an algorithmic approach to this problem by presenting a large new data set of heterotic models and the toolkit  from algebraic geometry necessary to analyze them.  By an algorithmic approach we mean a set of techniques which allow us to construct classes of vector
bundles on (certain) Calabi-Yau spaces systematically, prove their
stability and compute the resulting low-energy particle spectra
completely. In the following chapters we will focus on developing the details of the monad construction of vector bundles and the necessary
analytic and computer algebra methods by concentrating on relatively explicit constructions of Calabi-Yau manifolds which can be obtained as complete intersections in products of projective spaces.  A generalization of these methods to more general
complete intersection Calabi-Yau manifolds and a detailed analysis of
the particle physics properties of these models will be the subject of
Chapter \ref{c:posCICY}.

The heterotic models considered in this chapter will be constructed
following the process described in Chapter \ref{HetIntro}. After choosing a Calabi-Yau space $X$
(which we will take to be one of the five Calabi-Yau spaces realized
as intersections in a single ordinary projective space), we will scan
over a certain, well-defined class of (monad) bundles, $V$, on $X$.  We
will think of these bundles as bundles in the observable sector and
take the hidden bundle $\tilde{V}$ to be trivial. The anomaly
condition~\eqref{anomaly} can then be satisfied by including
five-branes as long as $c_2(TX)-c_2(V)$ is an effective class on $X$.
This is precisely what we will require. In addition, we will only
consider bundles $V$ whose index is a (non-zero) multiple of three as in \eref{k-div}.
Only such bundles have a chance, after dividing out by a discrete symmetry,
of producing a model with chiral asymmetry three. We also require
stability for all such bundles and the ability to compute their complete low-energy spectrum.

In this chapter, we will construct all positive monad bundles of rank 3,
4 and 5 on the five complete intersection Calabi-Yau spaces in a
single projective space subject to the physical constraints of Section \ref{s:het}.
(The bundles should be such that heterotic anomaly cancellation can be
accomplished and their chiral asymmetry should be a (non-zero)
multiple of three). We find $37$ examples in total. We then prove
stability for all these bundles using a variant of a simple criterion
due to Hoppe~\cite{hoppe}. Recently, this criterion has been
used~\cite{maria}, although in a slightly different way from the
present chapter, to prove stability for a class of positive bundles on
the quintic~\cite{Douglas:2004yv}. Further, we compute the complete
spectrum for all
bundles, including gauge singlet fields. It turns out that a common
feature of our models is that they only lead to generations but no
anti-generations. While the present chapter deals with a relatively
small number of examples, we have shown that the relevant methods can
be applied in a systematic and algorithmic way. A
significantly larger class of complete intersection Calabi-Yau spaces
and bundles on them can be treated in a similar way (see
\cite{Distler:2007av} for a recent constraint on classifying bundles
in general). This
generalization and the analysis of the particle physics of the
resulting models will be the subject of the following Chapters.

The plan of the chapter is as follows. In the following sections, we will
construct vector bundles from monad sequences and constrain them with the 
physical requirements of Section \ref{s:het}. In Section \ref{s:monad}, we discuss the monad construction,
its main properties and prove a number of general results for such
bundles. In Section \ref{s:eg}, we classify the positive monad bundles on our
five Calabi-Yau spaces, prove their stability and compute the
spectra. The appendices to this chapter include Appendix \ref{appA} which provides a
short summary
of the relevant tools in commutative algebra and how they are applied
in the context of the Macaulay computer algebra package~\cite{m2}. Appendix \ref{proofs} contains several useful technical results.
%
%
\section{Monad Construction of Vector Bundles}\label{s:monad}
To begin our systematic construction of vector bundles for heterotic
compactifications, we will make use of a standard and powerful
technique for defining bundles, known as the {\it monad
  construction}. On complex projective varieties, this method of
constructing vector bundles dates back to the early works on $\IP^4$
by \cite{HM} and systematic approaches by \cite{beilinson,rosa,maruyama}. This construction defines a vast class of vector
bundles; in fact, every bundle on $\IP^n$ can be expressed as a monad
\cite{barth,HM}. Bundles defined as monads have been widely used in
the mathematics and physics literature. The reader is referred to
\cite{monadbook} for the most general construction of monads and their
properties.  In this work we will use a restricted form prevalent in
the physics literature.
\subsection{The Calabi-Yau Spaces}
Our monad bundles will be constructed on complete intersection
Calabi-Yau manifolds, $X$, which are defined in a single projective ambient space
${\cal A}=\IP^m$. There are five such Calabi-Yau manifolds~\cite{hubsch}
and their properties are summarized in Table 3.
\begin{table}
\begin{center}
\begin{tabular}{|c|c|c|c|c|c|c|c|}\hline
Intersection&${\cal A}$ &Configuration& $\chi (X)$ & $h^{1,1}(X)$ &
 $h^{2,1}(X)$ & $d(X)$ & $\tilde{c}_2(TX)$ \\ \hline
Quintic &$\IP^4$& $[4|5]$ & $-200$ & $1$ & $101$&$5$ & $10$ \\
Quadric and quartic& $\IP^5$ & $[5|2 \ 4]$ & $-176$ & $1$ & $89$ &$8$&
  $7$ \\
Two cubics&$\IP^5$ & $[5|3 \ 3]$ &$-144$ & $1$ &$73$ &$9$& $6$ \\
Cubic and 2 quadrics&$\IP^6$ & $[6|3 \ 2 \ 2]$ & $-144$ & $1$ &
  $73$ & $12$ & $5$ \\
Four quadrics&$\IP^7$ & $[7|2 \ 2 \ 2 \ 2]$ & $-128$ & $1$ & $65$ & $16$&
  $4$ \\ \hline
\end{tabular}
\caption{\em The five complete intersection Calabi-Yau manifolds in a single
projective space. Here, $\chi (X)$ is the Euler number, $h^{1,1}(X)$ and $h^{2,1}(X)$
are the Hodge numbers, $d(X)$ is the intersection number and $c_2(TX)=\tilde{c}_2(TX)J^2$
is the second Chern class. The normalization of the K\"ahler form $J$ is defined in
the main text.\label{t:cy}} 
\end{center}
\end{table}
They are most conveniently described by the configurations
$[m|q_1,\ldots ,q_K]$ listed in the Table, where $m$ refers to the
dimension of the ambient space $\mathbb{P}^m$ and the numbers $q_a$
indicate the degree of the defining polynomials. In this notation the
Calabi-Yau condition $c_1(TX)=0$ translates to $\sum_{a=1}^Kq_a=m+1$.
Furthermore, note that $h^{1,1}(X)=1$ for all five
cases. Hence, these manifolds have their Picard group,
${\rm Pic}(X)$, being isomorphic to ${\mathbb Z}$. 
Such manifolds are called {\it cyclic} \cite{jardim}. 
The K\"ahler form $J$
descends from the the ambient space $\IP^n$ and is normalized
as
\begin{equation}
 \int_{\IP^n}J^m=1\; .
\end{equation}
Integrals over $X$ of any three-form $w$, defined on ${\cal A}=\IP^m$,
can be reduced to integrals over the ambient space using the formula
\begin{equation}
 \int_Xw=d(X)\int_{\IP^m}w\wedge J^{m-3}\; ,
\end{equation}
where $d(X)$ are the intersection numbers listed in Table 3.
The second homology $H_2(X,\mathbb{Z})$ is dual to the integer multiples of
$J\wedge J$ and the Mori cone of $X$ corresponds to all positive multiples
of $J\wedge J$~\cite{Hosono:1994ax}.

For our subsequent analysis it is useful to discuss some properties of line
bundles on the above Calabi-Yau manifolds. We denote by $\cO (k)$ the $k^{\rm th}$
power of the hyperplane bundle, $\cO(1)$, on the ambient space $\IP^m$ and by
$\cO_X (k)$ its restriction to the Calabi-Yau space $X$. The normal bundle $\cN$
of $X$ in the ambient space is then given by
\beq
 \cN=\bigoplus_{a=1}^K\cO (q_a)\; . \label{normal}
\eeq
In general, one finds, for the Chern characters of line bundles on $X$,
\bea
{\rm ch}_1(\cO_X(k))&=&c_1(\cO_X(k))=kJ \ ,\\
{\rm ch}_2(\cO_X(k))&=&\frac{1}{2}k^2J^2 \ ,\\
{\rm ch}_3(\cO_X(k))&=&\frac{1}{6}k^3J^3\; .
\eea
From the Atiyah-Singer index theorem the index of $\cO_X(k)$ is given by
\bea
 {\rm ind}(\cO_X(k))&\equiv&\sum_{q=0}^3(-1)^qh^q(X,\cO_X(k))\nn \\
  &=&\int_X\left[{\rm ch}_3(\cO_X(k))+\frac{1}{12}c_2(TX)\wedge
   c_1(\cO_X(k))\right]\nn \\ 
  &=&\frac{d(X)k}{6}\left(k^2+\frac{1}{2}\tilde{c}_2(TX)\right)\; , 
\label{indline}
\eea
where the numbers $\tilde{c}_2(TX)$ characterize the second Chern class of $X$
and $d(X)$ are the intersection numbers. The values for these quantities
can be read off from Table 3.

We recall that the Kodaira vanishing theorem \cite{AG1} states that
on a K\"ahler manifold $X$, $H^q(X, L \otimes K_X)$ vanishes for $q>0$
and $L$ a positive line bundle. Here, $K_X$ is the canonical bundle on
$X$. For Calabi-Yau manifolds $K_X$ is of course trivial and, hence,
the only non-vanishing cohomology for positive line
bundles on Calabi-Yau manifolds is $H^0$. The dimension of this
cohomology group can then be computed from the index theorem.
In fact, inserting the values for the intersection numbers and
the second Chern class from Table 3 into Eq.~\eqref{indline} we 
explicitly find, for the five Calabi-Yau spaces and for line bundles
$\cO_X(k)$ with $k>0$, that
\bea
 h^0([4|5],\cO_X(k))&=&\frac{5}{6}(k^3+5k) \ , \label{cy1}\\
 h^0([5|2\,4],\cO_X(k))&=&\frac{2}{3}(2k^3+7k) \ , \\
 h^0([5|3\,3],\cO_X(k))&=&\frac{3}{2}(k^3+3k) \ , \\
 h^0([6|3\,2\,2],\cO_X(k))&=&2k^3+5k \ , \\
 h^0([7|2\,2\,2\,2],\cO_X(k))&=&\frac{8}{3}(k^3+2k)\; .\label{cy5}
\eea
For negative line bundles $L=\cO_X (-k)$, where $k>0$, it follows
from Serre duality on the Calabi-Yau three-fold $X$, 
$h^q(X,L)=h^{3-q}(X,L^*)$, that only
$H^3(L,X)$ can be non-zero and that its dimension
$h^3(X,\cO_X(-k))=h^0(X,\cO_X(k))$ is given by one of the explicit
expressions~\eqref{cy1}--\eqref{cy5}. Finally, we have
\beq
h^0(X,\cO_X)=h^3(X,\cO_X)=1\; ,\qquad h^1(X,\cO_X)=h^2(X,\cO_X)=0\; .
\label{h0}
\eeq
Now we explicitly know the cohomology for all line bundles on
the five Calabi-Yau manifolds under consideration.
In particular, we conclude that $h^0(X,\cO_X(k))>0$ precisely for $k\geq
0$ and, hence, that only the line bundles $\cO_X(k)$ with $k\geq 0$
have a non-trivial section. This is one of the underlying conditions
for the validity of Hoppe's criterion which will play a central role
in the stability proof for our bundles.

\subsection{Constructing the Monad} 
Having discussed the manifold $X$ and line bundles thereon, we now
construct the requisite vector bundles $V$.
Our construction proceeds as follows. On a Calabi-Yau manifold $X$, a
monad bundle $V$ 
is defined by the short exact sequence 
\beq\label{monad}
0 \to V \stackrel{f}{\longrightarrow} B \stackrel{g}{\longrightarrow}
C \to 0 \; ,
\eeq
where $B$ and $C$ are bundles on $X$. It is standard to take
$B$ and $C$ to be direct sums of line bundles over $X$, that is
\beq
 B=\bigoplus\limits_{i=1}^{r_B} \cO_X(b_i)\; ,\qquad
 C=\bigoplus\limits_{i=1}^{r_C} \cO_X(c_i)\; .
\eeq
Here, $r_B$ and $r_C$ are the ranks of the bundles $B$ and $C$, respectively.
The exactness of \eref{monad} implies that $\ker(g) = \im(f)$ and $\ker(f)=0$,
so that the bundle $V$ can be expressed as
\[
V = \ker(g) \ .
\]
The map $g$ is a morphism between bundles and can be defined as a $r_C
\times r_B$ matrix whose entries, $(i,j)$, are sections of $\cO_X (c_i-b_j)$.
As we have seen in the previous subsection, such sections exist iff
$c_i\geq b_j$ and so this is what we should require. In fact,
if $c_i=b_j$ for an index pair $(i,j)$ the two corresponding line bundles
can simply be dropped from $B$ and $C$ without changing the resulting bundle $V$.
In the following, we will, therefore, assume the stronger condition $c_i>b_j$
for all $i$ and $j$. 

The Calabi-Yau manifolds discussed in this chapter are complete
intersections in a single projective space $\IP^m$. We can, therefore, write down
an analogous short exact sequence
\beq\label{monada}
0 \to \cV \stackrel{\tilde{f}}{\longrightarrow}\cB \stackrel{\tilde{g}}
{\longrightarrow} \cC \to 0 \; ,
\eeq 
on the ambient space where
\beq
 \cB=\bigoplus\limits_{i=1}^{r_B} \cO(b_i)\; ,\qquad
 \cC=\bigoplus\limits_{i=1}^{r_C} \cO(c_i)\; .
\eeq
The map $\tilde{g}$ can be viewed as a  $r_C \times r_B$ matrix
whose entries, $(i,j)$, are homogeneous polynomials of degree $c_i-b_j$.
This sequence defines a vector bundle $\cV$ on the ambient space whose
restriction to $X$ is $V$. Further, the map $g$ can be seen as the
restriction of its ambient space counterpart $\tilde{g}$ to $X$.
Unless explicitly stated otherwise, we will assume throughout that this map is generic.

It is natural to enquire whether $V$ thus defined is always a bona fide
bundle rather than a sheaf. We are assured on this point by the
following theorem \cite{fulton}.
\begin{theorem}
Over any smooth variety $X$, if $g : B \to C$ is a morphism between 
locally free sheaves $B$ and $C$, then $\ker(g)$ is locally free.
\end{theorem}
Now, by definition, a locally free sheaf of constant rank is a vector
bundle. Therefore, by the above theorem, it only remains to check
whether $\ker(g)$ has constant rank on $X$. Indeed, $g$ could be
less than maximal rank on a singular (sometimes called `degeneracy')
locus. We note that exactness of the sequence, that is ${\rm coker}(g)=0$,
is equivalent to this degeneracy locus being empty.

To show that the degeneracy locus is empty for our bundles, it turns
out to be convenient to consider the dual bundle $V^*$ defined by the
dual sequence
\beq\label{monaddual}
0 \to C^* \stackrel{g^T}{\longrightarrow} B^* 
\longrightarrow V^* \to 0 \; ,
\eeq
where
\beq\label{defV^*}
V^* = \coker(g^T) \ .
\eeq
We can now apply the following theorem~\cite{maria,lazarsfeld}.
\begin{theorem}\label{dualtheorem}
Let $\phi: E \to F$ be a morphism of vector bundles on a
variety of dimension $N$ and let $e = \rk(E)$, $f = \rk(F)$ and $e \le
f$. If $E^* \otimes F$ is globally generated and $f - e + 1 > N$, then
generic maps $\phi$ have a vanishing degeneracy locus.
\end{theorem}
Therefore, take $\phi = g^T$, $E = C^*$ and $F = B^*$.  For all our
bundles of interest, $N=3$ and $e < f$. In fact, $f-e$ is the rank of
$V$, which is 3, 4, or 5 for the bundles of interest in heterotic
compactifications. Finally, $E^* \otimes F$ is globally generated
because $B$ and $C$ are direct sums of line bundles with $c_i > b_j$
for all $i,j$. Hence, all the conditions in the theorem are obeyed and
we see that the degeneracy locus of $g^T$, and hence the one for $g$,
is vanishing for the bundles of interest on the Calabi-Yau. However,
one should note that this criterion will not always be satisfied when
writing monad sequences on the higher dimensional ambient spaces, as in
Eq.~\eqref{monada}. (Such
issues will be discussed further in section 4.4). For more on the
degeneracy locus of bundle maps, and why Theorem \ref{dualtheorem}
guarantees its vanishing in the dual monad, see e.g.~\cite{Munoz:math0107226,laytimi}.)

For later reference we present the formulae for the Chern classes of
$V$ (see Ref.~\cite{hubsch}). Simplifying the expressions for $c_2(V)$ and
$c_3(V)$ by imposing the vanishing of the first Chern class, we have
\bea\label{Vcherns}
 \rk(V) &=& r_B - r_C \ , \\ 
 c_1(V) &=& \left(\sum\limits_{i=1}^{r_B} b_i - \sum\limits_{i=1}^{r_C}c_i\right)
  J \equiv 0 \ , \label{c1} \\
 c_2(V) &=& -\frac12 \left(\sum\limits_{i=1}^{r_B} b_i^2 - 
  \sum\limits_{i=1}^{r_C} c_i^2\right) J^2 \ , \label{c2}\\
 c_3(V) &=& \frac13 \left(\sum\limits_{i=1}^{r_B} b_i^3 - 
  \sum\limits_{i=1}^{r_C} c_i^3\right) J^3 \ .
\eea
Hence, from Eq.~\eqref{index} and the above expression for the third Chern class,
the index of $V$ is explicitly given by
\begin{equation}
 {\rm ind}(V)= \sum_{p=0}^3(-1)^p\,h^p(X,V)=
 \frac{d(X)}{6}\left(\sum\limits_{i=1}^{r_B} b_i^3 - 
  \sum\limits_{i=1}^{r_C} c_i^3\right)\; . \label{relabel_indV}
\end{equation}
Within this chapter, we will make extensive use of the computer algebra system
\cite{m2} in analyzing the monads in \eref{monad}. Utilizing this
powerful tool we are able to catalog efficiently bundle cohomologies
previously too difficult to be calculated. Indeed, computing particle
spectra, that is, 
sheaf cohomology, is ordinarily a tremendous task even for a single bundle,
and it would be unthinkable to attempt to calculate by hand the
hundreds of such cohomologies necessary in a systematic study of
monad bundles. However, the recent advances in algorithmic
algebraic geometry allow us to explicitly and efficiently compute the
requisite cohomology groups for a certain class of bundles.
For the first time, we describe in detail how to use this technology
in the context of string compactification.

With this approach in mind, we recall that in computational algebraic
geometry \cite{schenck}, sheaves are expressed in the language of
graded modules over polynomial rings. If $X$ is embedded in $\IP^m$
with homogeneous coordinates $[x_0:x_1:\ldots:x_m]$, we can let $R$ be
the coordinate ring $\IC[x_0, x_1, \ldots, x_m] / (X)$ where $(X)$ is
the ideal associated with $X$. The bundles $B$ and $C$ are then
described by free-modules of $R$ with appropriate degrees
(grading). We leave to the Appendix a detailed tutorial of the
sheaf-module correspondence and the construction and relevant
computation of monad bundles using computer algebra.

\subsection{Stability of Monad Bundles}
\label{sec:stability}
As mentioned in the previous section, (semi-)stability of the vector
bundle is of central
importance to heterotic compactifications. In general, proving
stability is an overwhelming technical obstacle and a systematic analysis
has so far been elusive.
However, for a class of manifolds, a sufficient but by no means
necessary condition is of great utility; this is the so-called Hoppe's
criterion \cite{hoppe,huybrechts}:
\begin{theorem}\label{hoppe}
[Hoppe's Criterion]
Over a projective manifold $X$ with Picard group $Pic(X) \simeq \IZ$
(i.e., $X$ is cyclic),
let $V$ be a vector bundle with $c_1(V)=0$. If $H^0(X, \bigwedge^p V) = 0$
for all $p = 1, 2, \ldots, \rk(V) - 1$, then $V$ is stable.
\end{theorem}
We also recall that for the Calabi-Yau manifolds used in this chapter all
positive line bundles have a section, an underlying assumption for the
validity of Hoppe's theorem which is, hence, satisfied.

The strategy is therefore clear. To prove stability for the monad
bundles \eref{monad} over cyclic manifolds $X$ using
Hoppe's criterion, we need to show the vanishing of $H^0(X, \wedge^p
V)$ for $p=1,\ldots,\rk(V)-1$. In the following paragraphs, we will
outline the basis for this stability proof and make note of certain
results and properties that are of particular use.

One additional assumption which we will make is that all line bundles
involved in the definition of the bundles $V$ are positive, that is,
for all $i$,
\beq\label{positive}
b_i>0 \quad \mbox{and }c_i>0 \; .
\eeq
We will refer to this property as ``positivity'' of the bundle $V$.
While this is not required for a consistent definition of the bundle
or the associated heterotic model, it turns out to be
a crucial technical assumption which facilitates the stability
proof. The essential point is that  positivity of $V$ allows one to use Kodaira
vanishing when applying Hoppe's criterion to the dual bundle $V^*$.
To see how this works, recall that the dual bundle is defined by
the sequence  $0 \to C^* \longrightarrow B^* \longrightarrow V^* \to 0$
and that its stability is equivalent to that of $V$. The associated
long exact sequence in cohomology is
\bea
0&\to& H^0(X, C^*) \to H^0(X, B^*) \to \fbox{\mbox{$H^0(X, V^*)$}}\nn\\
 &\to& H^1(X, C^*) \to H^1(X, B^*)\to H^1(X,V^*)\nn\\
 &\to&H^2(X,C^* )\to  H^2(X, B^*)\to H^2(X,V^*)\nn\\
 &\to& H^3(X,C^*) \to H^3(X, B^*) \to H^3(X, V^*) \to 0 \ .
\label{dual-seq}
\eea
Given that we are dealing with positive bundles $V$, it follows that
$B^*$ and $C^*$ are sums of negative line bundles and, hence,
$H^0(X, B^*)$ and
$H^1(X, C^*)$ in the above sequence are zero due to Kodaira vanishing.
This means the ``boxed'' cohomology $H^0(X, V^*)$ also vanishes.
(For later considerations we note that Kodaira vanishing also
implies $H^1(X,B^*)=H^2(X,C^*)=0$ and, hence, $H^1(X,V^* )\simeq H^2(X,V)=0$.)
In order to prove stability of $V^*$ by applying Hoppe's criterion we
have to show that $H^0(X, \wedge^p V^*)=0$ for $p=1,\ldots, \rk(V)-1$
and we have just completed the first step for $p=1$.

Next, we need to compute the cohomologies $H^0(X, \wedge^p V^*)$ for $p>1$.
However, a further simplification occurs because we are dealing with
unitary bundles. In fact, for an $SU(n)$ bundle $V$, we have
\beq
\wedge^{n-1} V^* \simeq V \label{wedge}
\eeq
(see, for example Ref.~\cite{FH}) .
Therefore, to cover the case $p=n-1$, the highest exterior power
relevant to Hoppe's criterion, we only need to show that $H^0(X,V) =
0$.  This is indeed the case for all bundles considered in this chapter
and the explicit proof, which is somewhat lengthy, is presented in the Appendix of
\cite{Anderson:2007nc}\footnote{In the interests of space we have omitted this appendix from the current work since its content is discussed in Chapter \ref{StabCh}.}. This completes the stability proof for the
rank $3$ bundles. 

For rank $4$ and $5$ bundles we have to look at further exterior powers of
$V^*$, namely $\Lambda^pV^*$ for $p = 2, \ldots, \rk(V)-2$. To deal
with those we consider the standard long exact (``exterior power'') 
sequence~\cite{maria,AG1} for $\Lambda^pV^*$
\bea
0 &\to& S^p C^* \to S^{p-1} C^* \otimes B^* \to S^{p-2} C^* \otimes
\wedge^2 B^* \to \ldots\nn\\
& \to& A \otimes \wedge^{p-1} B^* \to \wedge^p B^* \to \wedge^p V^* \to 0 \ ,
\label{wedge-seq1}
\eea
which is induced by the short exact sequence \eqref{monaddual}.
Here $S^i$ is the $i$-th symmetrized tensor power of a bundle.
Such a sequence does not itself induce a long exact sequence in
cohomology; we need to slice it up into groups of three. In other
words, we introduce co-kernels $K_i$ such that \eref{wedge-seq1} becomes
the following set of short exact sequences
\bea
0 &\to& S^p C^* \to S^{p-1} C^* \otimes B^* \to K_1 \to 0 \ , \nn\\
0& \to& K_1 \to S^{p-2} C^* \otimes \wedge^2 B^* \to K_2 \to 0 \ , \nn \\
&&\qquad\qquad\qquad\qquad\vdots \nn \\
 0 &\to& K_{p-1} \to \wedge^p B^* \to \wedge^p V^* \to 0 \ .
\label{wedge-seq}
\eea
Each of the above now induces a long exact sequence in cohomology in
analogy to \eref{dual-seq}:
\beq\label{wedge-hom}
{\hspace{-1cm}
\ba{rcl}
&&0 \to H^0(X, S^p C^*) \to H^0(X, S^{p-1} C^* \otimes B^*) \to H^0(X,
K_1) \to H^1(X, S^p C^*) \to \ldots \to 0 \ , \\
&&0 \to H^0(X, K_1) \to H^0(X, S^{p-2} C^* \otimes \wedge^2 B^*) 
\to H^0(X, K_2) \to H^1(X, K_1) \to \ldots \to 0 \ , \\
&& \hspace{6cm} \vdots\\
&&0 \to H^0(X,  K_{p-1}) \to H^0(X, \wedge^p B^*) \to 
\fbox{\mbox{$H^0(X, \wedge^p V^*)$}}
\to H^1(X,  K_{p-1}) \to \ldots \to 0 \ . \\
\ea
}
\eeq
The term we need is boxed and we need to trace through the various
sequences, using the readily computed cohomologies of the
symmetric and antisymmetric powers of $B^*$ and $C^*$, to arrive at the
answer. Let us now do this explicitly for the case $p=2$, that is, 
$H^0(X,\Lambda^2V^*)$.
The long exact sequence~\eqref{wedge-seq1} then specializes to
\begin{equation}
 0\to S^2C^*\to C^*\otimes B^*\to\Lambda^2B^*\to\Lambda^2
 V^*\to 0\; ,
\end{equation}
which needs to be broken up into the two short exact sequences
\bea
 &0&\to S^2C^*\to C^*\otimes B^*\to K\to 0\label{short1}\\
&0&\to K\to \Lambda^2B^*\to\Lambda^2 V^*\to 0\; .\label{short2}
\eea
From the first of these we have the long exact sequence
\bea
 0&\to&H^0(X,S^2C^* )\to H^0(X,C^*\otimes B^* )\to H^0(X,K)\nn\\
  &\to&H^1(X,S^2C^* )\to H^1(X,C^*\otimes B^* )\to H^1(X,K)\nn\\
  &\to&H^1(X,S^2C^* )\to\ldots\; .
  \label{seq1}
\eea
Since $B^*$ and $C^*$ are sums negative line bundles, so are their
various tensor products which appear in the above sequences. From Kodaira
vanishing all cohomologies of such bundles vanish except for the
third. 
Applying this
to \eqref{seq1} we immediately deduce that $H^0(X,K)=H^1(X,K)=0$. Using this
information in the long exact sequence
\beq
0\to H^0(X,K)\to H^0(X,\Lambda^2B^* )\to H^0(X,\Lambda^2V^* )
\to H^1(X,K)\to\ldots
\eeq
which follows from \eqref{short2} we find $H^0(X,\Lambda^2V^* )=0$,
as desired. This completes the stability proof for rank $4$
bundles~\footnote{Together with $H^0(X,V^* )=0$, which we have shown earlier, it
also provides an independent argument for the stability of rank $3$ bundles.}.

Finally, for rank $5$ bundles, we still need to compute $H^0(X,\Lambda^3V^*)$.
Repeating the above steps for this case one finds that Kodaira vanishing on $X$
alone does not quite provide sufficient information to conclude that
$H^0(X,\Lambda^3V^*)=0$. In this case, we need to employ the additional
technique of  \textit{Koszul sequences} \cite{hubsch,AG1} which rely
on the embedding of the Calabi-Yau manifold in an ambient space ${\cal A}$.
Specifically, for a vector bundle $\cW$ on ${\cal A}$ the Koszul sequence reads
\begin{equation}\label{koszul}
0\rightarrow \wedge ^{K}\cN^{* }\otimes \cW\rightarrow ...\rightarrow \wedge
^{2}\cN^{* }\otimes \cW\rightarrow \cN^{* }\otimes \cW\rightarrow \cW\overset{
\rho }{\rightarrow }\cW|_{X}\rightarrow 0 \ ,
\end{equation}
where $\cW|_X$ denotes the restriction of $\cW$ to $X$ and $\rho$ is the
associated restriction map. Here $\cN^*$ is the dual of the Calabi-Yau
normal bundle, defined in Eq.~\eqref{normal}. As will be shown in the
next section, the Koszul sequence can be used to compute the relevant
cohomologies directly from the ambient space. This will allow us to
complete the stability proof for rank 5 bundles.

\section{Classification and Examples}\label{s:eg}
Armed with the general information about the five Calabi-Yau manifolds
and monad bundles we can now proceed to classify such bundles, prove their
stability and compute their spectrum.
\subsection{Classification of Configurations}\label{s:class}
For the monad bundles defined by the short exact
sequence~\eqref{monad}, we can immediately formulate a classification
scheme. Recall that, taking the bundles $B$ and $C$ to be direct sums
of line-bundles over the manifold $X$, we have
\beq\label{monad2}
0 \to V \to \bigoplus\limits_{i=1}^{r_B} \cO_X(b_i) 
\stackrel{g}{\longrightarrow}
\bigoplus\limits_{i=1}^{r_C} \cO_X(c_i) \to 0 \ ,
\qquad
V \simeq \ker(g) \ .
\eeq
From our discussion so far these bundles are subject to a number
of physical and mathematical constraints which can be summarized
as follows:
\begin{enumerate}
\item As discussed earlier we require all $b_i$ and $c_i$
  to be positive; this is a technical assumption which will
  significantly simplify our
  computations.
\item We furthermore require that $b_i < c_j$ for all $i$ and $j$;
  this is to ensure that the map $g$, which consists of sections of
  $\cO_X(c_j-b_i)$, has no zero entries. Further, we require the map
  $g$ to be generic. Then, all conditions of Theorem~\eqref{dualtheorem}
  are met and we are guaranteed that $V$, as defined
  by the sequence~\eqref{monad2}, is indeed a bundle.
\item Since we are dealing with special unitary bundles we impose $c_1(V) = 0$.
\item For a given Calabi-Yau space $X$ and a bundle $V$ we need to ensure
      that the anomaly condition~\eqref{anomaly} can be satisfied. 
      To do this we impose the condition that $c_2(TX) - c_2(V)$ must be effective.
      Then, we can choose a trivial hidden bundle $\tilde{V}$
      and a five-brane wrapping a holomorphic curve with
      homology class  $c_2(TX) - c_2(V)$. In practice, this condition simply means that
      the coefficient of $J^2$ in  $c_2(TX) - c_2(V)$ must be non-negative~\footnote{However,
      for a given example there may well be other ways to satisfy the anomaly
      condition which involve a non-trivial hidden bundle $\tilde{V}$.} .
\item We require that the index of $V$ is a non-zero multiple of three.
      Only such models may lead to three generations after dividing by a discrete
      symmetry.
\item Since we are interested in low-energy grand unified groups we consider bundles
      $V$ with structure group ${\rm SU}(n)$, where $n={\rm rk}(V)=3,4,5$.
\end{enumerate}
Therefore, an integer partitioning problem immediately presents
itself to us: find partitions $\{b_i\}_{i=1,\ldots ,r_C+n}$ and
$\{c_j\}_{i=1,\ldots ,r_C}$ of positive integers $b_i>0$, $c_i>0$ satisfying
$b_i<c_j$ for all $i$, $j$ and subject to the condition
$\sum\limits_{i=1}^{r_B} b_i - \sum\limits_{i=1}^{r_C} c_i = 0$ 
for vanishing first Chern class of $V$ (see Eq.~\eqref{c1}). Further,
we demand that the index of $V$, Eq.~\eqref{relabel_indV}, is non-zero and divisible
by three and that the coefficient of $J^2$ in $c_2(TX) - c_2(V)$ be non-negative,
in order to ensure the existence of a holomorphic five-brane curve. From
Eq.~\eqref{c2} the last constraint can be explicitly written as
\beq
0 \le -\frac12 (
\sum\limits_{i=1}^{r_C + n} b_i - \sum\limits_{i=1}^{r_C} c_i)
\le \tilde{c}_2(TX) \ , \label{c2cons}
\eeq
where the numbers $\tilde{c}_2(TX)$ for the second Chern class of $X$
are given in Table 3. Since $b_i<c_j$ for all $i$, $j$ it is clear
that this constraint implies an upper bound on $b_i$ and
$c_j$ and, hence, that the number of vector bundles in our class is
finite~\footnote{The constraint~\eqref{c2cons} arises because we require $N=1$
supersymmetry in four dimensions. If we relaxed this condition and allowed for
anti-five branes there would be no immediate bound on the number of vector
bundles. However, in this case, the stability of such non-supersymmetric models
has to be analyzed carefully~\cite{Gray:2007zza,Gray:2007qy}.}.
To derive this bound explicitly we slightly modify an argument from
Appendix B of Ref.~\cite{Douglas:2004yv}. Define the quantity
\beq
S = \sum\limits_{i=1}^{r_C + n} b_i = \sum\limits_{i=1}^{r_C} c_i \ ,
\eeq
and consider the following chain of inequalities
\bean
2\,\tilde{c}_2(TX) \ge
\sum\limits_{i=1}^{r_C} c_i^2 - \sum\limits_{i=1}^{r_C + n}
b_i^2
& \ge & (b_{max}+1)\sum\limits_{i=1}^{r_C} c_i - 
 \sum\limits_{i=1}^{r_C + n} b_i^2 \\
& = & S + \sum\limits_{i=1}^{r_C + n} b_{max} b_i -
 \sum\limits_{i=1}^{r_C + n} b_i^2 \ge S \ .
\eean
From Table 3, $\tilde{c}_2(TX)$ is at most $10$ and, hence, the sum
$S$ cannot exceed $20$, thereby placing an upper bound on our partitioning
problem.

Given the finiteness of the problem, the classification of all
positive monad bundles subject to the above constraints is now
easily computerizable. Given these conditions, we found $37$ bundles on the
five Calabi-Yau manifolds in question, $20$ for rank $3$, $10$ for rank $4$ and
$7$ for rank $5$. 
Had we relaxed the condition that $c_3$ should be
divisible by 3, we would have found 43, 15, 10, 6, and 3 bundles,
respectively on the 5 cyclic manifolds, for a total of 77.
A complete list of all such bundles for 
the five Calabi-Yau manifolds of concern is given in the Tables 4--8.
\begin{table}
\begin{center}
\begin{tabular}{|c|c|c|c|c|} \hline
Rank & $\{b_i\}$ & $\{c_i\}$ & $c_2(V)/J^2$ & $\ind(V)$ \\ \hline
3 & (2, 2, 1, 1, 1) & (4, 3) & 7 & -60 \\
3 & (2, 2, 2, 1, 1) & (5, 3) & 10 & -105\\ 
3 & (3, 2, 1, 1, 1) & (4, 4) & 8 & -75\\ 
3 & (1, 1, 1, 1, 1, 1) & (2, 2, 2) & 3 & -15\\ 
3 & (2, 2, 2, 1, 1, 1) & (3, 3, 3) & 6 & -45\\ 
3 & (3, 3, 3, 1, 1, 1) & (4, 4, 4) & 9 & -90 \\ 
3 & (2, 2, 2, 2, 2, 2, 2, 2) & (4, 3, 3, 3, 3) & 10 & -90\\ 
3 & (2, 2, 2, 2, 2, 2, 2, 2, 2) & (3, 3, 3, 3, 3, 3) & 9 & -75\\ 
4 & (2, 2, 1, 1, 1, 1) & (4, 4) & 10 & -90\\ 
4 & (1, 1, 1, 1, 1, 1, 1) & (3, 2, 2)  & 5 & -30\\ 
4 & (2, 2, 2, 1, 1, 1, 1) & (4, 3, 3) & 9 & -75\\ 
4 & (2, 2, 2, 2, 1, 1, 1, 1) & (3, 3, 3, 3) & 8 & -60\\ 
5 & (1, 1, 1, 1, 1, 1, 1, 1) & (3, 3, 2) & 7 & -45\\ 
5 & (1, 1, 1, 1, 1, 1, 1, 1) & (4, 2, 2) & 8 & -60\\ 
5 & (2, 2, 2, 2, 2, 1, 1, 1, 1, 1) & (3, 3, 3, 3, 3) & 10 & -75\\ \hline
\end{tabular}
\caption{\em Positive monad bundles on the quintic, $[4|5]$.}
\end{center}
\end{table}
\begin{table}
\begin{center}
\begin{tabular}{|c|c|c|c|c|} \hline
Rank & $\{b_i\}$ & $\{c_i\}$ & $c_2(V)/J^2$ & $\ind(V)$ \\ \hline
3& (2, 2, 1, 1, 1) & (4, 3) & 7 & -96\\ 
3& (1, 1, 1, 1, 1, 1) & (2, 2, 2) & 3 & -24\\ 
3& (2, 2, 2, 1, 1, 1) & (3, 3, 3) & 6 & -72\\ 
4& (1, 1, 1, 1, 1, 1, 1) & (3, 2, 2) & 5 & -48\\ 
5& (1, 1, 1, 1, 1, 1, 1, 1) & (3, 3, 2) & 7 & -72 \\ \hline
\end{tabular}
\caption{\em Positive monad bundles on $[5|2 \ 4]$.}
\end{center}
\end{table}
\begin{table}
\begin{center}
\begin{tabular}{|c|c|c|c|c|} \hline
Rank & $\{b_i\}$ &$ \{c_i\}$ & $c_2(V)/J^2$ & $\ind(V)$ \\ \hline
3& (1, 1, 1, 1) & (4) & 6 & -90\\ 
3& (1, 1, 1, 1, 1) & (3, 2) & 4 & -45\\ 
3& (2, 1, 1, 1, 1) & (3, 3) & 5 & -63\\ 
3& (1, 1, 1, 1, 1, 1) & (2, 2, 2) & 3 & -27 \\ 
3& (2, 2, 2, 1, 1, 1) & (3, 3, 3) & 6 & -81\\ 
4& (1, 1, 1, 1, 1, 1) & (3, 3) & 6 & -72\\ 
4& (1, 1, 1, 1, 1, 1, 1) & (3, 2, 2) & 5 & -54\\
4& (1, 1, 1, 1, 1, 1, 1, 1) & (2, 2, 2, 2) & 4 & -36\\ 
5& (1, 1, 1, 1, 1, 1, 1, 1, 1) & (3, 2, 2, 2) & 6 & -63\\ 
5& (1, 1, 1, 1, 1, 1, 1, 1, 1, 1) & (2, 2, 2, 2, 2) & 5 & -45\\ \hline
\end{tabular}
\caption{\em Positive monad bundles on $[5|3 \ 3]$.}
\end{center}
\end{table}
\begin{table}
\begin{center}
\begin{tabular}{|c|c|c|c|c|} \hline
Rank & $\{b_i\}$ & $\{c_i\}$ & $c_2(V)/J^2$ & $\ind(V)$ \\ \hline
3& (1, 1, 1, 1, 1) & (3, 2) & 4 & -60\\ 
3& (2, 1, 1, 1, 1) & (3, 3) & 5 & -84\\ 
3& (1, 1, 1, 1, 1, 1) & (2, 2, 2) & 3 & -36\\ 
4& (1, 1, 1, 1, 1, 1, 1) & (3, 2, 2) & 5 & -72\\ 
4& (1, 1, 1, 1, 1, 1, 1, 1) & (2, 2, 2, 2) & 4 & -48\\ 
5& (1, 1, 1, 1, 1, 1, 1, 1, 1, 1) & (2, 2, 2, 2, 2) & 5 & -60\\
\hline
\end{tabular}
\caption{\em Positive monad bundles on $[6|2 \ 2 \ 3]$.}
\end{center}
\end{table}
\begin{table}
\begin{center}
\begin{tabular}{|c|c|c|c|c|} \hline
Rank & $\{b_i\}$ & $\{c_i\}$ & $c_2(V)/J^2$ & $\ind(V)$ \\ \hline
3 & (1, 1, 1, 1, 1, 1) & (2, 2, 2) & 3 & -48 \\
\hline
\end{tabular}
\caption{\em Positive monad bundles on $[7|2 \ 2 \ 2 \ 2]$.}
\end{center}
\end{table}
\subsection{$E_6$-GUT Theories}
The first case we shall analyze is $E_6$-GUT theories which arise
from $SU(3)$ bundles. We have already seen in Section~\ref{sec:stability}
that all such bundles are indeed stable. This result has been explicitly
confirmed by a computer algebra computation of $H^0(X,V^*)$ and
$H^0(X,\Lambda^2V^*)$ along the lines described in Appendix~\ref{appA}.
We can, therefore, directly turn to a computation of their particle spectrum.
\subsubsection{Particle Content}
The number of ${\bf 27}$ and $\overline{\bf 27}$ representation of
$E_6$ is easy to obtain. Since $V$ is stable we already know that
$H^0(X,V)=H^3(X,V)=0$. From the long exact sequence~\eqref{dual-seq}
we have deduced earlier that $H^2(X,V)\simeq H^1(X,V^*)=0$
so that $H^1(X,V)$ is the only non-vanishing cohomology. Its
dimension can be directly computed from the index~\eqref{relabel_indV}, so that
\beq\label{E6-27}
n_{27} = h^1(X,V) = -\ind(V)\; ,\qquad n_{\overline{27}} = h^2(X,V)=0 \; .
\eeq
Therefore, for the rank 3 bundles in Tables 4--8, the (negative of
the) right-most column gives the number of ${\bf 27}$ representations.
This result also provides the first example of what is a general feature
of positive monad bundles, namely the absence of anti-generations.
The numbers $n_{27}$ have been independently verified by computer algebra.

What about the $E_6$ singlets? These correspond to the cohomology
$H^1(X, \text{ad}(V)) = H^1(X, V \otimes V^*)$. We begin by
tensoring the defining sequence \eqref{monaddual} for $V^*$ by $V$.
This leads to a new short exact sequence
\beq
0 \to C^* \otimes V \to B^* \otimes V \to V^* \otimes V \to 0 \ .
\eeq
One can produce two more short exact sequences by multiplying
\eqref{monaddual} with $B$ and $C$. Likewise, three short exact
sequences can be obtained by multiplying the original sequence~\eqref{monad}
for $V$ with $V^*$, $B^*$ and $C^*$. The resulting six
sequences can then be arranged into the following web of three horizontal
sequences $h_{I}$, $h_{II}$, $h_{III}$ and three vertical ones
$v_I$, $v_{II}$, $v_{III}$.
\beq\ba{cccccccccl}
&&0&&0&&0&&& \\
&&\downarrow&&\downarrow&&\downarrow&&& \\
0&\to& C^* \otimes V &\to& B^* \otimes V &\to& V^* \otimes V &\to&0
\qquad &h_{I} \\
&&\downarrow&&\downarrow&&\downarrow&&& \\
0&\to& C^* \otimes B &\to& B^* \otimes B &\to& V^* \otimes B &\to&0
\qquad &h_{II} \\
&&\downarrow&&\downarrow&&\downarrow&&& \\
0&\to& C^* \otimes C &\to& B^* \otimes C &\to& V^* \otimes C &\to&0
\qquad &h_{III}  \\
&&\downarrow&&\downarrow&&\downarrow&&& \\
&&0&&0&&0&&& \\
&&v_I&&v_{II}&&v_{III}&&& \\
\ea
\eeq
The long exact sequence in cohomology induced by $h_{I}$ reads
\bea\label{VVseq}
0 &\to& H^0(X,C^* \otimes V) \to H^0(X,B^* \otimes V)
\to H^0(X, V^* \otimes V)\nn\\
&\to&H^1(X,C^* \otimes V) \to
 H^1(X, B^* \otimes V)\to \fbox{\mbox{$H^1(X,V^* \otimes V)$}}\nn \\
&\to& H^2(X,C^* \otimes V) \to \ldots\label{h1}
\eea
and we have boxed the term which we would like to compute. We will also need the
long exact sequences which follow from $v_{I}$ and $v_{II}$. They
are given by
\bea
 0&\to&H^0(X,C^*\otimes V)\to H^0(X,C^*\otimes B)\to H^0(X,C^*\otimes C)\nn\\ 
  &\to&H^1(X,C^*\otimes V)\to H^1(X,C^*\otimes B)\to H^1(X,C^*\otimes C)\nn\\
 &\to&H^2(X,C^*\otimes V)\to H^2(X,C^*\otimes B)\to H^2(X,C^*\otimes C)
 \to\ldots\label{v1}\\[0.3cm]
 0&\to&H^0(X,B^*\otimes V)\to H^0(X,B^*\otimes B)\to H^0(X,B^*\otimes C)\nn\\ 
  &\to&H^1(X,B^*\otimes V)\to H^1(X,B^*\otimes B)\to H^1(X,B^*\otimes C)\nn\\
 &\to&H^2(X,B^*\otimes V)\to H^2(X,B^*\otimes B)\to H^2(X,B^*\otimes C)\to\ldots
 \label{v2}
 \eea
Now, because of the integers defining $B$ and $C$ satisfy  $b_i < c_j$,
the tensor product $C^* \otimes B$ is a direct
sum of negative line bundles and, hence, all its cohomology groups
vanish except the third. Further, the middle cohomologies
$H^1$ and $H^2$ of $B^*\otimes B$ and $C^*\otimes C$ vanish.
From the sequence~\eqref{v1} this implies
\beq
 H^0(X,C^*\otimes V)=H^2(X,C^*\otimes V)=0\; , \qquad 
 H^1(X,C^*\otimes V)=  H^0(X,C^*\otimes C)\; .
\eeq
Vanishing of $H^2(X,C^*\otimes V)$ means that the long exact sequence
\eqref{h1} breaks after the second line and we get
\beq
 h^1(X,V^*\otimes V)=h^1(X,B^*\otimes V)-h^1(X,C^*\otimes V)
 +h^0(X,V^*\otimes V)
 -h^0(X,B^*\otimes V)\; .
 \label{nsing0}
\eeq
Using the additional information
\beq
 h^1(X,B^*\otimes V)- h^0(X,B^*\otimes V)=h^0(X,B^*\otimes C)
 -h^0(X,B^*\otimes B)\; .\\
\eeq
which follows from the sequence \eqref{v2} and the fact that
$h^0(X,V^*\otimes V)=1$ (see Theorem B.1 of Ref.~\cite{hubsch})
Eq.~\eqref{nsing0} can be re-written as
\beq
 h^1(X,V^*\otimes V)=h^0(X,B^*\otimes C)-h^0(B^*\otimes B)
                        -h^0(C^*\otimes C)+1\; .\label{nsing}
\eeq
This equation, together with Eqs.~\eqref{cy1}--\eqref{cy5} and
\eref{h0}, allows us to directly compute the number $n_1$ of
$E_6$-singlets and the results are given in \tref{t:E6}. For
reference, we have also included the number of ${\bf
  27}$-representations (the number of $\overline{\bf 27}$ particles,
we recall, is zero). In addition, the results for $h^1(X,V^*\otimes
V)$ have been independently confirmed using Macaulay~\cite{m2},
following the procedure outlined in Appendix~\ref{appA}. We note
that the above derivation of Eq.~\eqref{nsing} is independent
of the rank of the vector bundle $V$ and, hence, it remains
valid for rank $4$ and $5$ bundles.
\begin{table}[h]
\begin{center}
\begin{tabular}{|c|c|c|c|c|} \hline
$X$ & $\{b_i\}$ & $\{c_i\}$ & $n_{27}$ &$ n_1$ 
\\ \hline
$[4|5]$ & (2, 2, 1, 1, 1) & (4, 3) & 60 & 141 
\\
& (2,2,2,1,1) & (5, 3) & 105 & 231 
\\
& (3, 2, 1, 1, 1) & (4, 4) & 75 & 171 
\\
& (1, 1, 1, 1, 1, 1) & (2, 2, 2) & 15 & 46 
\\
& (2, 2, 2, 1, 1, 1) & (3, 3, 3) & 45 &109 
\\
& (3, 3, 3, 1, 1, 1) & (4, 4, 4) & 90 & 199 
\\
& (2, 2, 2, 2, 2, 2, 2, 2) & (4, 3, 3, 3, 3) & 90 & 180 
\\
& (2, 2, 2, 2, 2, 2, 2, 2, 2) & (3, 3, 3, 3, 3, 3)& 75 & 154 
\\
\hline
$[5|2 \ 4]$ & (2, 2, 1, 1, 1) & (4, 3) & 96 & 206 
\\
& (1, 1, 1, 1, 1, 1) & (2, 2, 2) & 24 & 64 
\\
& (2, 2, 2, 1, 1, 1) & (3, 3, 3) & 72 & 154 
\\
\hline
$[5|3 \ 3]$& (1, 1, 1, 1) & (4) & 90 & 200 
\\
& (1, 1, 1, 1, 1) & (3, 2) & 45 & 103 
\\
& (2, 1, 1, 1, 1) & (3, 3) & 63 & 136 
\\
& (1, 1, 1, 1, 1, 1) & (2, 2, 2) & 27 & 64 
\\
& (2, 2, 2, 1, 1, 1) & (3, 3, 3) & 81 & 163 
\\
\hline
$[6|2 \ 2 \ 3]$& (1, 1, 1, 1, 1) & (3, 2) & 60 & 132 
\\
& (2, 1, 1, 1, 1) & (3, 3) & 84 & 174 
\\ 
& (1, 1, 1, 1, 1, 1) & (2, 2, 2) & 36 & 82 
\\
\hline
$[7|2 \ 2 \ 2 \ 2]$ & (1, 1, 1, 1, 1, 1) & (2, 2, 2) &48 & 100 
\\
\hline
\end{tabular}
\caption{\emph{The particle content for the $E_6$-GUT theories arising from
  our classification of stable, positive $SU(3)$ monad bundles $V$ on the
  Calabi-Yau threefold $X$. The number $n_{\overline{27}}$ of
  anti-generations vanishes.}}
\label{t:E6}
\end{center}
\end{table}
\subsection{$SO(10)$-GUT Theories}
Grand Unified theories with gauge group ${\rm SO}(10)$ are obtained
from rank $4$ bundles with structure group ${\rm SU}(4)$. We have
already shown the stability of positive rank $4$ monad bundles $V$ in
Section~\ref{sec:stability}. As before, we have explicitly confirmed this general
result for the rank $4$ bundles in our classification with Macaulay~\cite{m2},
by showing that $H^0(X,\Lambda^pV^*)$ for $p=1,2,3$ vanishes.
We proceed to analyze the particle content of ${\rm SO}(10)$ GUT
theories.
\subsubsection{Particle Content}
Recall from Table 2, that for $SO(10)$-GUT theories we need to
compute $n_{16}=h^{1}(X,V)$, $n_{\overline{16}}=h^1(X,V^*) = h^{2}(X,V)$,
$n_{10}=h^{1}(X, \wedge ^{2}V)$ and $n_{1}=h^{1}(X, V\otimes V^*)$.

Let us begin with the generations and anti-generations in ${\bf 16}$
and $\overline{\bf 16}$. As in the case of rank $3$
bundles, stability implies that $H^0(X,V)=H^3(X,V)=0$ and, further, from
the sequence~\eqref{dual-seq}, also $H^2(X,V)=H^1(X,V^*)$ is zero.
Hence, as before, the number of anti-generations vanishes and the
number of generations can be computed from the index, so that
\beq
n_{16} =  h^1(X,V) = -\ind(V) \; , \qquad n_{\overline{16}} = 0 \ .
\eeq
Thus, for the rank $4$ bundles in Tables 4--8, the (negative) of
the right-most column gives the number of ${\bf 16}$ representations.

Next, we need to compute the Higgs content which is given by
$n_{10}=h^{1}(X, \wedge ^{2}V)$. It can be shown in general that
for generic maps $g:B\to C$ the number of ${\bf 10}$ representations
always vanishes, that is
\beq
 n_{10}=0\; .
\eeq
The proof is somewhat technical and can be found in Appendix~\ref{SO(10)proof}.
Again, this result can be readily verified using computer algebra.

Finally, we need to compute the number $n_1$ of ${\rm SO}(10)$ singlets
which is easily obtained from Eq.~\eqref{nsing}. The results for
the spectrum from rank $4$ bundles are summarized in \tref{t:SO10}.

A vanishing number, $n_{10}$, of Higgs particles is not desirable
from a particle physics viewpoint. One might, therefore, wonder whether
more specific choices of the map $g$ in \eqref{monad} could produce a
non-zero value for $n_{10}$. This problem has been encountered
in Ref.~\cite{Bouchard:2005ag,Donagi:2004qk,Donagi:2004ia} where the
spectrum of compactification was shown to depend on the region of
moduli space. 
Specifically, it was shown that the spectrum takes
a generic form with possible enhancements in special regions of the
moduli space; this was dubbed the ``jumping phenomenon'' in
\cite{Donagi:2004qk,Donagi:2004ia}.

To see that a similar phenomenon can arise for monad bundles, let is consider the
following $SU(4)$ bundle on the quintic, $[4|5]$.
\beq\label{5eg-1}
0 \to V \to \cO_X^{\oplus 2}(2) \oplus  \cO_X^{\oplus 4}(1)
\stackrel{g}{\longrightarrow} \cO_X^{\oplus 2}(4) \to 0 \; .
\eeq
This bundle and its particle content for a generic map $g$ is given in
the first line of \tref{t:SO10}. Now we explicitly define the map
$g$ by
\beq\label{5eg-2}
g = \left(
\begin{matrix}
4x_{3}^2& 9x_{0}^2 + x_{2}^2& 8x_{2}^3& 2x_{3}^3&
4x_{1}^3& 9x_{1}^3 \cr
x_{0}^2 + 10x_{2}^2& x_{1}^2& 9x_{2}^3&
7x_{3}^3& 9x_{1}^3 + x_{2}^3& x_{1}^3 + 7x_{4}^3
\end{matrix}
\right) \ .
\eeq
where $x_0,\ldots ,x_4$ are the homogeneous coordinates of $\IP^4$.
This choice for $g$ is no longer completely generic,
although the sequence~\eqref{5eg-1} is still exact.
Following the steps in Appendix \ref{ap:eg}, we can use Macaulay
to calculate the spectrum for this case. We find
\beq
n_{16} =  90\; , \qquad n_{\overline{16}}=0\; ,\qquad
n_{10} = 13\; ,\qquad n_1=277 \; .
\eeq
This is identical to the generic result in Table~\ref{t:SO10},
except for the number of ${\bf 10}$ representations which has
changed from $0$ to $13$.
\begin{table}
\begin{center}
\begin{tabular}{|c|c|c|c|c|c|} \hline
$X$ & $\{b_i\}$ & $\{c_i\}$ & $n_{16}$& $n_1$\\ \hline
$[4|5]$ & (2, 2, 1, 1, 1, 1) & (4, 4) & 90 & 277 
\\
&  (1, 1, 1, 1, 1, 1, 1) & (3, 2, 2) & 30  & 112 
\\
& (2, 2, 2, 1, 1, 1, 1) & (4, 3, 3) & 75 &  236 
\\
& (2, 2, 2, 2, 1, 1, 1, 1) & (3, 3, 3, 3) & 60 &  193
\\
\hline
$[5|2 \ 4]$ & (1, 1, 1, 1, 1, 1, 1) & (3, 2, 2) &48 & 159
\\
\hline
$[5|3 \ 3]$ & (1, 1, 1, 1, 1, 1) & (3, 3) & 72 &  213
\\ 
& (1, 1, 1, 1, 1, 1, 1) & (3, 2, 2) & 54 &  166 
\\
& (1, 1, 1, 1, 1, 1, 1, 1) & (2, 2, 2, 2) & 36 &  113 
\\
\hline
$[6|2 \ 2 \ 3]$ & (1, 1, 1, 1, 1, 1, 1) & (3, 2, 2) &72& 213
\\ 
& (1, 1, 1, 1, 1, 1, 1, 1) & (2, 2, 2, 2) & 48 & 145
\\
\hline
\end{tabular}
\caption{\emph{The particle content for the $SO(10)$-GUT theories arising from
  our classification of stable, positive, $SU(4)$ monad bundles $V$ on the
  Calabi-Yau threefold $X$. The number $n_{\overline{16}}$ of anti-generations
  vanishes. The number $n_{10}$ vanishes for generic choices of the
  map $g$ in the monad sequence~\eqref{monad}, but can be made
  non-vanishing with particular choices of $g$.}}
\label{t:SO10}
\end{center}
\end{table}

\subsection{$SU(5)$-GUT Theories}
Finally, we should consider ${\rm SU}(5)$ GUT theories which originate
from rank $5$ bundles with structure group ${\rm SU}(5)$. To demonstrate
their stability from Hoppe's criterion we have to show that
$H^0(X,\Lambda^pV^*)$ for $p=1,2,3,4$ vanish. For $p=1,2,4$ this
has already been accomplished in Section~\ref{sec:stability}, so it
remains to deal with the case $p=3$.

Unfortunately, for $p=3$ the long exterior power sequences~\eqref{wedge-hom}
together with Kodaira vanishing are not quite sufficient to prove that
$H^0(X,\Lambda^3V^*)=0$. Indeed, writing down~\eqref{wedge-seq} for
$p=3$ we find
\bea
\nn&& 0 \to S^3 C^* \to S^2 C^* \otimes B^* \to K_1 \to 0 \ , \\
&& 0 \to K_1 \to C^* \otimes \wedge^2 B^* \to K_2 \to 0 \ , \label{wedge3split}\\
&& 
0 \to K_{2} \to \wedge^3 B^* \to \wedge^3 V^* \to 0 \ .
\nn
\eea
Now, using the 3 intertwined long exact sequences in cohomology
induced by the above 3 sequences, together with Kodaira vanishing
for the negative bundles formed from the symmetric and
anti-symmetric powers of $B^*$ and $C^*$, we can only conclude that
\beq\label{h0Xwedge3}
H^0(X,\wedge^3 V^*) \simeq H^2(X,K_1) \ .
\eeq
We will now show that the stability proof can be completed by applying Koszul
resolutions to our rank $5$ bundles. This technique makes explicit use of the
embedding in the ambient space ${\cal A} =\mathbb{P}^m$ and its complexity grows with
the number of co-dimensions of the Calabi-Yau manifold $X$ in ${\cal A}$. We, therefore,
start with the quintic, $X=[4|5]$, the only co-dimension one example among the
five Calabi-Yau manifolds under consideration, before we proceed to the more
complicated examples.
\subsubsection{Stability for Rank 5 Bundles on the Quintic}
For the quintic, the normal bundle is simply given by $\cN=\mathcal{O}(5)$
and the Koszul sequence~\eref{koszul}, applied to $\cW=\Lambda^3\cV^*$, explicitly reads
\begin{equation}\label{koszul1}
0\rightarrow \cN^{* }\otimes \wedge ^{3}\cV^{* }\rightarrow \wedge
^{3}\cV^{* }\rightarrow  \wedge ^{3}V^{* }\rightarrow
0 \ .
\end{equation}
From this, we have the long exact sequence in cohomology,
\beq
0\rightarrow H^{0}(\mathcal{A},\cN^{* }\otimes \wedge ^{3}\cV^{*
})\rightarrow H^{0}(\mathcal{A},\wedge ^{3}\cV^{* })\rightarrow
H^{0}(X,\wedge ^{3}V^{* })
\rightarrow H^{1}(\mathcal{A},\cN^{* }\otimes
\wedge ^{3}\cV^{* })\rightarrow ...  
\label{ext}
\eeq
Thus, if we knew $H^{0}(\mathcal{A},\wedge ^{3}\cV^{* })$ and $H^{1}(%
\mathcal{A},N^{* }\otimes \wedge ^{3}\cV^{* })$, we could hope to
determine $H^{0}(X,\wedge ^{3}V^{* })$ itself. In fact, we can show that $%
H^{0}(\mathcal{A},\wedge ^{3}\cV^{* })=H^{1}(\mathcal{A},N^{* }\otimes
\wedge ^{3}\cV^{* })=0$ by writing down the ambient space version of the
exterior power sequences~\eqref{wedge3split} tensored by $\cN^*$.
\begin{eqnarray}
0 &\rightarrow &\cN^{* }\otimes S^{3}\cC^{* }\overset{h}{\rightarrow }
\cN^{* }\otimes S^{2}\cC^{* }\otimes \cB^{* }\rightarrow
\cK_{1}\rightarrow 0  \ , \nn \\
0 &\rightarrow &\cN^{* }\otimes \cK_{1}\rightarrow \cN^{* }\otimes
\cC^{* }\otimes \wedge ^{2}\cB^{* }\rightarrow\cK_{2}\rightarrow 0 
 \ , \\
0 &\rightarrow &\cK_{2}\rightarrow \cN^{* }\otimes \wedge ^{3}\cB^{*
}\rightarrow \cN^{* }\otimes \wedge ^{3}\cV^{* }\rightarrow 0 \
.
\nn
\end{eqnarray}
Since $\cB^{* }$, $\cC^{* }$ and $\cN^{* }$ are all negative bundles, it
follows that $H^{0}(\mathcal{A},\wedge ^{3}\cV^{* })=0$ and 
$h^{1}(\mathcal{A},\cN^{* }\otimes \wedge ^{3}\cV^{* })= h^{3}(\mathcal{A}
,\cK_{1})=\ker (h^{\prime })$, where 
$h^{\prime }:$ $H^{4}({\cal A},\cN^{* }\otimes S^{3}\cC^{* })\rightarrow
H^{4}({\cal A},\cN^{* }\otimes S^{2}\cC^{* }\otimes \cB^{* })$ is the map induced
from $h$ above. Now, we note that since the ranks of the maps in the
defining monads were chosen, by construction, to be maximal rank, 
it follows that the
induced map $h$ in the exterior power sequence is also maximal rank. To
proceed further, we finally observe that for any generic, maximal rank
map $h:{\cal U}\to {\cal W}$ between two ambient space bundles ${\cal U}$ and ${\cal W}$
the induced map $\tilde{h}:H^0({\cal A},{\cal U})\to H^0({\cal A},{\cal W})$
is also maximal rank (see Appendix \ref{ap:gen}).
Since the sequences above are all defined over the ambient space and $h$
is maximal rank, it follows from the above argument that $h^{\prime }$ is
maximal rank and $\ker (h^{\prime })=0$. Therefore, 
\begin{equation}
h^{1}(\mathcal{A},\cN^{* }\otimes \wedge ^{3}\cV^{* })=0.
\end{equation}
Thus, returning to \eref{ext}, we find that $H^{0}(X,\wedge ^{3}V^{* })=0$
and by Hoppe's criterion, \textit{all generic, positive }$SU(5)$\textit{\ bundles are
stable on the quintic}.

\subsubsection{The Co-dimension 2 and 3 Manifolds}

The stability proof for our remaining rank $5$ bundles is similar in
approach, but slightly more lengthy than that given in the previous
subsection. In the
interests of space, we will only give an overview of it here. 
We recall from Subsection \ref{s:class} that the remaining
Calabi-Yau manifolds with rank $5$ bundles are defined by two and three
constraints in $\mathbb{P}^{5}$ and $\mathbb{P}^{6}$ respectively. We first look
at the co-dimension two case.

For co-dimension two, the normal bundle takes the form $N=\mathcal{O}
(q_1)\oplus \mathcal{O}(q_2)$ with $q_1,q_2>0$. This time the Koszul sequence
\eref{koszul} is no longer short-exact, but reads
\begin{equation}
0\rightarrow \wedge ^{2}\cN^{* }\otimes \wedge ^{3}\cV^{* }\rightarrow
\cN^{* }\otimes \wedge ^{3}\cV^{* }\rightarrow \wedge ^{3}\cV^{* }\overset%
{\rho }{\rightarrow }\wedge ^{3}V^{* }\rightarrow 0 \ .
\end{equation}
It can be split into two short exact sequences,
\begin{eqnarray}
0 &\rightarrow &\wedge ^{2}\cN^{* }\otimes \wedge ^{3}\cV^{* }\rightarrow
\cN^{* }\otimes \wedge ^{3}\cV^{* }\rightarrow \cK
\rightarrow 0 \ , \notag \\
0 &\rightarrow &\cK\rightarrow \wedge ^{3}\cV^{* }\overset{
\rho }{\rightarrow }\wedge ^{3}V^{* }\rightarrow 0 \ .
\end{eqnarray}
From the long cohomology sequences of these two resolutions, we find that $
H^{0}(X,\wedge ^{3}V^{* })\simeq$
$H^{2}(\mathcal{A},\wedge ^{2}\cN^{*}\otimes \wedge ^{3}\cV^{* })$
(since $H^{0}(\mathcal{A},\wedge ^{3}\cV^{*
})=H^{0}(\mathcal{A},\cN^{* }\otimes \wedge ^{3}\cV^{* })=0$ by the same
arguments as before). 
Next, the exterior power sequence~\eqref{wedge-seq1},
multiplied by $\Lambda^2\cN^*$ and written over $\mathbb{P}^{5}$ yields,
\begin{eqnarray}
0 &\rightarrow &\wedge ^{2}\cN^{* }\otimes S^{3}\cC^{* }\overset{h}{
\rightarrow }\wedge ^{2}\cN^{* }\otimes S^{2}\cC^{* }\otimes \cB^{*
}\rightarrow \cK_{1}\rightarrow 0  \ , \notag \\
0 &\rightarrow &\cK_{1}\rightarrow \wedge
^{2}\cN^{* }\otimes \cC^{* }\otimes \wedge ^{2}\cB^{* }\rightarrow
\cK_{2}\rightarrow 0  \ , \notag \\
0 &\rightarrow &\cK_{2}\rightarrow \wedge ^{2}\cN^{* }\otimes \wedge
^{3}\cB^{* }\rightarrow \wedge ^{2}\cN^{* }\otimes \wedge ^{3}\cV^{*
}\rightarrow 0 \ .
\end{eqnarray}
Once again, we find that $H^{2}(\mathcal{A},\wedge ^{2}\cN^{* }\otimes
\wedge ^{3}\cV^{* })\simeq H^{4}(\mathcal{A},\cK_{1})$ and $h^{4}(
\mathcal{A},\cK_{1})=\ker (h^{\prime })$ where $h^{\prime }:$ $
H^{5}({\cal A},\wedge ^{2}\cN^{* }\otimes S^{3}\cC^{* })\rightarrow
H^{5}({\cal A},\wedge^{2}\cN^{* }\otimes S^{2}\cC^{* }\otimes \cB^{* })$.
As before, it follows from our definition of the monad that $h^{\prime
}$ is maximal rank
and $\ker (h^{\prime })=0$. Therefore, all positive rank $5$ bundles
on the manifolds $[5|2\;4]$ and $[5|3\;3]$ are stable.

With this analysis complete, we are left with only one rank $5$ bundle
on the co-dimension 3 manifold, $[6|2\;2\;3]$, to consider. In this
case, we could directly apply the Koszul resolution techniques as
above, with a normal bundle, $\mathcal{N}=\mathcal{O}(2) \oplus
\mathcal{O}(2) \oplus \mathcal{O}(3)$, and higher antisymmetric powers
in the Koszul resolution \eref{koszul}. Note, however, that in this
case we are not assured that the dual sequence \eref{monaddual} is
well defined on the ambient space, since the numeric criteria in
Theorem \ref{dualtheorem} are not satisfied on
$\mathbb{P}^6$. However, we can still compute the cohomology of the
relevant sheaves on $\mathbb{P}^6$.  The calculation is lengthy, but
straightforward. 

It is worth noting that there is an alternative approach to this
case. Instead of viewing the Koszul resolution as describing the
restriction of objects on $\mathbb{P}^m$ to the Calabi-Yau, we may
view $X=[6|2\;2\;3]$ as a sub-variety in the 4-fold 
$Y=[6|2\;2]$. Then we may apply the Koszul techniques exactly as before,
viewing the normal bundle to the Calabi-Yau as a line bundle,
$\mathcal{O}_Y(3)$ in $[6|2\;2]$. The analysis then reduces to that
described for the
co-dimension 1 case \eref{koszul1} (that is, that of the rank 5 bundles on the
quintic). A straightforward calculation shows that $H^{0}(X,\wedge
^{3}V^{* })=0$ and the final rank 5 bundle is stable.

\subsubsection{Particle Content}
\begin{table}
\begin{center}
\begin{tabular}{|c|c|c|c|c|c|} \hline
$X$ & $\{b_i\}$ & $\{c_i\}$ & $n_{\overbar{10}}$ & 
 $n_1$\\ \hline
$[4|5]$ 
 & (1, 1, 1, 1, 1, 1, 1, 1) & (3, 3, 2) & 45 &  202
\\ 
& (1, 1, 1, 1, 1, 1, 1, 1) & (4, 2, 2) &60 & 262
\\ 
& (2, 2, 2, 2, 2, 1, 1, 1, 1, 1) & (3, 3, 3, 3, 3) &75 &301
\\ \hline
$[5|2 \ 4]$
& (1, 1, 1, 1, 1, 1, 1, 1) & (3, 3, 2) & 72 & 288 
\\
\hline
$[5|3 \ 3]$
&(1, 1, 1, 1, 1, 1, 1, 1, 1) & (3, 2, 2, 2) &63 &243
\\ 
&(1, 1, 1, 1, 1, 1, 1, 1, 1, 1) & (2, 2, 2, 2, 2) &45 &176
\\
\hline
$[6|2 \ 2 \ 3]$ & 
(1, 1, 1, 1, 1, 1, 1, 1, 1, 1) & (2, 2, 2, 2, 2) &60 &226
\\
\hline
\end{tabular}
\caption{\emph{The particle content for the $SU(5)$-GUT theories arising from
  our classification of stable, positive, $SU(5)$ monad bundles $V$ on the
  Calabi-Yau threefold $X$.
  The number of $n_{10}$'s, vanishes. Further,
  $n_5=n_{\overline{10}}$. Moreover, $n_{\overline{5}}=0$ for generic choices of the map
  $g$ in Eq.~\eqref{monad}, and can be made non-vanishing in special
  regions of moduli space.}}
\label{t:SU5}
\end{center}
\end{table}
We have shown, using the Koszul sequence, that all positive rank 5
bundles in our classification are stable. Let us now analyze their
particle spectrum. From Table 2, we need to compute $n_{10}=h^{1}(X,V^*)= h^{2}(X,V)$,
$n_{\overline{10}}=h^1(X,V)$, 
$n_{5}=h^{1}(X,\wedge ^{2}V)$, $n_{\overline{5}}=
h^{1}(X,\wedge ^{2}V^*) = h^2(X, \wedge^2V)$, and 
$n_{1}=h^{1}(X,V\otimes V^*)$. As for rank 4 and 5 bundles, we have
$h^0(X,V)=h^3(X,V)=0$ from stability and $h^2(X,V)=0$ from the
sequence~\eqref{dual-seq}. Consequently, we find
\beq
n_{10} = h^1(X,V^*) , \qquad n_{\overline{10}} = h^{1}(X,V)=-ind(V) \ .
\eeq
As before, we have no anti-generations and the (negative) of the index,
listed in right-most column of Tables 4--7, gives the number $n_{10}$
for all rank $5$ bundles. We include these in \tref{t:SU5} for reference.

Next, we need to compute the $H^1(X,\wedge ^{2}V)$ and $H^2(X,\wedge ^{2}V)$.
From the above arguments we know that $V$ is stable; hence $\wedge
^{2}V$ is also stable and thus $H^0(X,\wedge ^{2}V)$ and $H^3(X,\wedge ^{2}V)$
both vanish (recall that we have already shown explicitly that
$H^0(X,\wedge ^{2}V^*) = H^3(X,\wedge^{2}V)$ vanishes). Therefore, applying
the index theorem~\eqref{index} to $\Lambda^2V$ we have
\beq\label{indwedge2}
-h^1(X,\wedge ^{2}V)+h^2(X,\wedge ^{2}V) = \ind(\wedge ^{2}V) =
\frac12 \int_X c_3(\wedge ^{2}V) \ .
\eeq
For $SU(n)$ bundles one has (see Eq.~(339) of Ref.~\cite{Donagi:2004ia}),
\beq\label{c3wedge2}
c_3(\wedge ^{2}V) = (n-4) c_3(V) \ .
\eeq
Hence, combining \eref{indwedge2} and \eref{c3wedge2}, we find the relation
\beq\label{n5-n5bar}
-n_5 + n_{\overline{5}} = \ind(V) = -n_{\overbar{10}} \ .
\eeq
We still need to compute one of the numbers $n_5$ and
$n_{\overline{5}}$. Macaulay~\cite{m2} can very easily
calculate $n_{\overline{5}}= h^{1}(X,\wedge ^{2}V^*) = h^2(X,
\wedge^2V)$. It turns out that
\begin{equation}
 n_{\overline 5}=0
\end{equation}
for all rank $5$ bundles and generic choices~\footnote{Presumably, $n_{\overline 5}$ can
be different from zero for non-generic choices of the map $g$, similar to the case
of $n_{\overline 10}$ for rank $4$ bundles.} of the map $g$. From Eq.~\eqref{n5-n5bar}
this implies
\begin{equation}
 n_5=n_{\overbar{10}}\; ,
\end{equation}
and, hence, the complete spectrum is determined by $n_{10}$ and $n_1$. We have listed
these numbers in~\tref{t:SU5}.

\section{Conclusion}
In this chapter, we have presented a classification of positive ${\rm SU}(n)$ monad bundles
on the five Calabi-Yau manifolds defined by complete intersections in a single
projective space. We have required that these bundles can be incorporated
into a consistent heterotic compactification where the heterotic anomaly cancellation
condition can be satisfied by including an appropriately wrapped five-brane.
In addition, we have imposed two ``physical'' conditions, namely that the
rank of bundle be $n=3,4,5$ (in order to obtain a suitable grand unification group)
and that the index of the bundle (that is, the chiral asymmetry) is a non-zero
multiple of three. Given these conditions, we found $37$ bundles on the
five Calabi-Yau manifolds in question, $20$ for rank $3$, $10$ for rank $4$ and
$7$ for rank $5$. 
Using a simple criterion due to Hoppe, we have shown that all
these bundles are stable and, hence, lead to supersymmetric compactifications.
We have also computed the full particle spectrum for all $37$ cases, including
the number of gauge singlets. A generic feature of all our bundles is that
the number of anti-generations vanishes.

These results show that a combination of analytic computations and computer algebra
can be used to analyze a class of models algorithmically. In particular, we have seen
that the notoriously difficult problem of proving stability can be addressed
systematically and that the full particle spectra can be obtained for all cases.
Although the final number of models is still relatively small we expect that
these methods can be extended to much larger classes of Calabi-Yau manifolds, such
as complete intersections in products of projective spaces and in weighted
projective spaces. Such a large-scale analysis which leads to a substantial number of examples with broadly the right physical properties is the subject of the next Chapter.
\chapter{Compactifying on Complete Intersection Calabi-Yau Manifolds}\label{c:posCICY}

\section{Introduction}
As we saw in the previous chapter, the monad construction of vector bundles can be used to algorithmically construct heterotic models. We can produce  four-dimensional effective theories with the gauge groups of grand unified theories (GUTs) and under suitable symmetry breaking (i.e. Wilson lines, etc) they can contain the symmetry of the standard model. While the cyclic Calabi-Yau manifolds provide a good test of our methodology, it is our ultimate goal in this work to construct a large class of heterotic models and present the detailed techniques for analyzing them. To this end, we now greatly extend our class of bundles and manifolds by generalizing the techniques to build bundles over  much larger data set: the $7890$ complete intersection Calabi-Yau manifolds (CICYs).

It is our hope that by formulating a systematic construction of a large class of vector bundles over an explicit and relatively simple set of Calabi-Yau manifolds, we can build a substantial number of heterotic models which can be thoroughly scanned for physically relevant properties. In the following sections, we begin by reviewing the constructions of complete 
intersection Calabi-Yau manifolds and the monad construction of vector bundles 
over these spaces. 

Of the CICYs, $4515$ are the so called ``favorable" manifolds which possess a simple K\"ahler structure: the K\"ahler forms, $J$ of the manifold $X$ descend directly from those in the ambient projective space. That is, for a favorable CICY defined in $\cA = \mathbb{P}^{n_1}\times \ldots \mathbb{P}^{n_m}$, $h^{1,1}(TX)=m$. We shall use these manifolds as our starting set. And construct \emph{positive} monad bundles over them. In particular, we review how bundles can be constructed as the kernels and cokernals of maps between direct sums of line bundles and how the topological data (chern classes, cohomology groups, etc) for such bundles can be obtained. 

For such positive monads,  we once again prove that the number of physically relevant bundles is finite. We show that of the over $4000$ manifolds at our disposal, only $36$ admit realistic positive monad models. Over these $36$ manifolds we find $7118$ bundles. For these models we can compute the spectra using general techniques. As in the cyclic case, we find no anti-generations and the Higgs particle content depends on the bundle moduli. While we do not yet prove stability of these bundles in this chapter, we show that for all bundles $H^0(X,V)=H^0(X,V^*)=0$, a non-trivial check of stability for $SU(n)$ bundles.

In chapter \ref{MonCh1}, a computational framework was developed for the 5 \textit{cyclic} Calabi-Yau manifolds using a combination of analytic results and computational algebraic geometry computer packages. However, since existing computer packages do not yet handle manifolds as complicated as the CICYs; for this work it is clear that we must develop new analytic techniques and computer applications. In this chapter and the following chapters, we will continue to develop our algorithmic approach to heterotic model building by addressing the mathematical obstacles to computing particle spectra and proving bundle stability.
%
\section{Complete Intersection Calabi-Yau Threefolds}
To begin our construction of vector bundles for heterotic models, we turn first to the compact Calabi-Yau spaces. Ever since the realization that Calabi-Yau three-folds played a central role in superstring compactification \cite{Candelas:1985en}, constructions of so-called ``complete intersection Calabi-Yaus'' (CICYs) \cite{Candelas:1987kf,Candelas:1987du} have been a topic of interest. Indeed, this method of Calabi-Yau construction was used in one of the first attempts to systematically study families of Calabi-Yau manifolds. Subsequent work, especially in
light of mirror symmetry, was carried out in explicit mathematical
detail \cite{Green:1987cr,He:1990pg,Green:1987zp,Gagnon:1994ek,Hosono:1994ax} for
half a decade, culminating in the comprehensive review \cite{hubsch} 
on the subject. The manifolds in \cite{Anderson:2007nc}, used to
illustrate a new algorithmic approach in heterotic compactification,
are special cases of these CICYs.

Unfortunately, much of the original data was stored on computer media,
such as magnetic tapes at CERN, which have
been rendered obsolete by progress. Partial results,
including, the list of the CICY threefolds themselves, 
can be found on
the Calabi-Yau Homepage \cite{cypage}. In this section, we shall
resurrect some of the useful facts concerning the CICY threefolds,
which will be of importance to our bundle constructions later.
We will present only the essentials, leaving most of the details to Appendix
\ref{a:cicy}.

\subsection{Configuration Matrices and Classification}
We are interested in manifolds $X$ which can be described as algebraic varieties, that is, as intersections of the zero loci of $K$ polynomials $\{p_j\}_{j=1,\ldots ,K}$ in an ambient space $\cA$. For our purpose, we will consider ambient spaces $\cA =\IP^{n_1} \times \ldots \times \IP^{n_m}$ given by a product of $m$ ordinary projective spaces with dimensions $n_r$. We denote the projective coordinates of each factor $\IP^{n_r}$ by $[x_0^{(r)}:x_1^{(r)}:\ldots:x_{n_r}^{(r)}]$, its Kahler form by $J_r$ and the $k^{\rm th}$ power of the hyperplane bundle by $\cO_{\IP^{n_r}}(k)$. The Kahler forms are normalised such that
\begin{equation}
 \int_{P^{n_r}}J_r^{n_r}=1\; .
\end{equation}
The manifold $X$ is called a {\it complete intersection} if the dimension of $X$ is equal the dimension of $\cA$ minus the number of polynomials. This is, in a sense, the optimal way in which polynomials can intersect. To obtain threefolds $X$  from complete intersections we then need
\beq\label{ci}
\sum_{r=1}^m n_r - K = 3 \ .
\eeq
Each of the defining homogeneous polynomials $p_j$ can be characterised by its multi-degree ${\bf q}_j=(q_j^1,\ldots , q_j^m)$, where $q_j^r$ specifies the degree of $p_j$ in the coordinates ${\bf x}^{(r)}$ of the factor $\IP^{n_r}$ in $\cA$.  A convenient way to encode this information is by a {\it configuration matrix}
\beq\label{cy-config}
\left[\ba{c|cccc}
\IP^{n_1} & q_{1}^{1} & q_{2}^{1} & \ldots & q_{K}^{1} \\
\IP^{n_2} & q_{1}^{2} & q_{2}^{2} & \ldots & q_{K}^{2} \\
\vdots & \vdots & \vdots & \ddots & \vdots \\
\IP^{n_m} & q_{1}^{m} & q_{2}^{m} & \ldots & q_{K}^{m} \\
\ea\right]_{m \times K}\; .
\eeq
Note that the $j^{\rm th}$ column of this matrix contains the multi-degree of the polynomial $p_j$.
In order that the resulting manifold be Calabi-Yau, the condition
\beq\label{cy-deg}
\sum_{j=1}^K q^{r}_{j} = n_r + 1 \qquad \forall r=1, \ldots, m
\eeq
needs to imposed (essentially to guarantee that $c_1(TX)$ vanishes).
Henceforth, a CICY shall mean a Calabi-Yau threefold, specified by the
configuration matrix \eref{cy-config}, satisfying the conditions
\eref{ci} and \eref{cy-deg}. In fact, the condition \eref{cy-deg} even
obviates the need for the first column $\IP^{n_1} \ldots \IP^{n_m}$ in the
configuration matrix. Subsequently, we will frequently need the normal bundle $\cN$ of $X$ in $\cA$ which is given by
\begin{equation}
 \cN = \bigoplus_{j=1}^K\cO_\cA ({\bf q}_j)\; . \label{normalbundle}
\end{equation}
Here and in the following we employ the short-hand notation $\cO_\cA ({\bf k})=\cO_{\IP^{n_1}}(k^1)\otimes\dots\otimes\cO_{\IP^{n_r}}(k^r)$ for line bundles on the ambient space $\cA$. 

As an archetypal example, the famous quintic in $\IP^4$ is simply
denoted as ``$[4 | 5]$''. One might
immediately ask how many possible non-isomorphic (one obvious
isomorphism being row and column permutations) configurations could
there be. This question was nicely settled in
\cite{Candelas:1987kf,He:1990pg} and the number is, remarkably,
finite. A total of 7890 is found and can be accessed at \cite{cypage}. We have compiled an electronic list of these CICYs which contains all the essential information including configuration matrices, Euler numbers $\chi (X)$, second Chern classes $c_2(TX)$, Hodge numbers $h^{1,1}(X)$ and $h^{2,1}(X)$ and allows for easy calculation of triple intersection numbers. It also contains previously unknown information, in particular about redundancies within the CICY list. This data underlies many of the subsequent calculations for monad bundles on CICYs. For more details on this ``legacy" subject see Appendix~\ref{a:cicy}.
\subsection{Favorable Configurations}
Our choice of complete intersection Calabi-Yau manifolds is motivated largely by the explicit and relatively simple nature of the constructions. Perhaps the most valuable advantage of the presence of the ambient space $\cA$ is the existence of relatively straightforward methods to identify discrete symmetries, a crucial step for the implementation of Wilson line breaking. To take maximal advantage of the presence of the ambient space we will focus on CICYs for which this explicit embedding is particularly useful. For some CICYs, the second cohomology $H^2(X)$ is not entirely spanned by the restrictions of the ambient space Kahler forms $J_r$. For example, in the case of
the well-known Tian-Yau manifold, $X=\left[
\begin{array}
[c]{c}%
3\\
3
\end{array}
\left|
\begin{array}
[c]{ccc}%
3 & 0 & 1\\
0 & 3 & 1
\end{array}
\right.  \right]  $, there are two Kahler forms descending from the two $\mathbb{P}^{3}$'s, but $h^{1,1}(X)=14$. Here, we would like to focus on CICYs $X$ for which the second cohomology is entirely spanned by the ambient space Kahler forms and which are, hence, characterised by
\[
h^{1,1}(X)=m=\#\text{ of }\mathbb{P}^{n}\text{'s.}
\]
We shall call manifolds with this property \emph{favourable}. Such favourable CICYs offer a number of considerable practical advantages. The Kahler cone, that is the set of allowed Kahler forms $J$ on $X$, is simply given by $\{J=t^rJ_r\,|\,t^r\geq 0\}$, where $t^r$ are the Kahler moduli. Further, the set of all line bundles on $X$, the Picard group $\mbox{Pic}(X)$, is isomorphic to $\mathbb{Z}^{m}$, so  line bundles on $X$ can be characterised by an integer vector ${\bf k}=(k^1,\ldots ,k^m)$. We denote these line bundles by ${\cal O}_X({\bf k})$ and they can be obtained by restricting their ambient space counterparts $\cO_\cA ({\bf k})$ to $X$. We can also introduce a dual basis $\{\nu^r \}$ of $H^4(X,\mathbb{Z})$, satisfying
\begin{equation}
 \int_X\nu_r\wedge J_s=\delta^r_s\; ,
\end{equation} 
and, via Poincar\'e duality $H^4(X,\mathbb{Z})\simeq H_2(X,\mathbb{Z})$, we can use this basis to describe the second integer homology of $X$. The effective classes $W\in H_2(X,\mathbb{Z})$ then correspond precisely to the  positive integer linear combinations of $\nu^r$, that is $w_r\nu^r$ with $w_r\geq 0$. This property makes checking our version of the anomaly cancellation condition~\eqref{anomaly} very simple. If we expand second Chern classes in the basis $\{\nu^r \}$, writing $c_2(U)=c_{2r}(U)\nu^r$ for any bundle $U$, then the condition~\eqref{anomaly} simply turns into the inequalities
\begin{equation}
 c_{2r}(V)\leq c_{2r}(TX)\; .\label{effcond1}
\end{equation} 
The triple intersection numbers $d_{rst}$ of favourable CICYs $X$ can simply be obtained by integration
\begin{equation}
 d_{rst}=\int_XJ_r\wedge J_s\wedge J_t \label{drst}
\end{equation}
of the Kahler forms over $X$.  
In practical calculations of the second Chern class (or the second Chern characters), one usually arrives at an expansion of the form $c_2(U)=c_2^{rs}(U)J_r\wedge J_s$. It is useful to note that the conversion into the basis $\nu^r$ involves a contraction with the intersection numbers \eref{intersection}, that is, 
\begin{equation}
c_{2r}(U)=d_{rst}c_2^{st}(U)\; .
\end{equation}

Scanning through the CICY data, we find that  there is a total of 4515 CICYs which are favourable. This is still a large dataset and we shall henceforth restrict our attention to these.

\section{Line bundles on CICYs}\label{s:line}
As we will see, line bundles on CICYs are the main building blocks of the monad bundles considered in this Chapter, so we need to know their detailed properties. In particular we need to be able to fully determine the cohomology of line bundles on CICYs. We will return to this problem shortly after briefly reviewing a few more elementary properties. For an ambient space $\cA$ with $m$ projective factors, we consider a generic line bundle $L=\cO_X({\bf k})$ on a CICY $X$, where ${\bf k}=(k^1,\ldots ,k^m)$ is an $m$-dimensional integer vector. The Chern characters of such a line bundle are given by 
\beq\ba{rl}
\ch_{1}(L) & = c_1(L) = k^rJ_r\\
\ch_{2}(L) & =\frac{1}{2}k^rk^sJ_r\wedge J_s\\
\ch_{3}(L) & =\frac{1}{6}k^rk^sk^tJ_r\wedge J_s\wedge J_t \ ,
\ea\eeq
with implicit summation in $r,s,t = 1, \ldots, m$. Note that every line bundle on a CY $3$-fold is uniquely classified by its first Chern class, as can be seen explicitly from the above expression for $\ch_1$. The dual of the line bundle $L$ is simply given by $L^*=\cO_X(-{\bf k})$. Using the Atiyah-Singer index theorem \cite{AG1}, the index of $L$ can be written as
\begin{align}\label{ind}
\ind(L)  &\equiv\sum_{q=0}^{3}(-1)^{q}h^{q}(X,L) = \int_X
\ch(L)\wedge\td(X)
=\int_{X}\left[  \ch_{3}(L)+\frac{1}{12}\ch_{2}(TX)\wedge
c_{1}(L)\right]  \nonumber\\
& =\frac{1}{6}\left( d_{rst} k^rk^sk^t+\frac{1}{2}k^rc_{2r}
(TX)\right)\; .
\end{align}

A special class of line bundles are the so-called {\em positive line bundles} which, in the present case, are the line bundles $L=\cO_X({\bf k})$ with all $k^r>0$.  The Kodaira vanishing theorem \eref{kodaira} applies to such positive bundles and (given the canonical bundle $K_X$ of a Calabi-Yau manifold is trivial) it implies that $h^{q}(X,L)=0$ for all $q\neq 0$. This means that  $h^{0}(X,L)$ is the only non-vanishing cohomology and it can, hence, be easily calculated from the index \eref{ind} since $h^0(X,L) = \ind(L)$. The situation is just as simple for {\em negative line bundles} $L$, that is line bundles for which $L^*$ is positive. In our case, the negative line bundles $L=\cO_X({\bf k})$ are of course the ones with all $k^r<0$. Applying the Kodaira vanishing theorem to $L^* =\cO (-{\bf k})$ and then using Serre duality it follows that $h^3(X,L)$ is the only non-vanishing cohomology of a negative line bundle. Again, it can be computed from the index using $h^3(X,L)=-{\rm ind}(V)$. These result for positive and negative line bundles can also be checked using the techniques of spectral sequences. In this case, the dimension of the single non-zero cohomology can be computed without explicitly knowledge of the Leray maps $d_{i}$ (see Chapter \ref{c:line_bundle_coho}) between cohomologies.  
One more general statement can be made. It turns out that semi-positive line bundles, that is line bundles  $L=\cO_X({\bf k})$, where $k^r\geq 0$ for all $r$, always have at least one section, so $h^0(X,L)>0$. One might be tempted to conclude that the line bundles with sections are precisely the semi-positive ones. While this is indeed the case for some CICYs it is by no means always true and for some CICYs the class of line bundles with a section is genuinely larger than the class of semi-positive line bundles.

Further quantitative statements about the cohomology of line bundles $L=\cO ({\bf k})$ containing ``mixed" or zero entries $k^{r}$ are not so easily obtained. For a general line bundle with mixed sign or zero entries, computing the dimensions $h^{q}(X,\mathcal{O}_{X}({\bf k}))$ does require explicit information about  the dimensions of kernels and ranks of the Leray maps $d_i$ in \eref{leray_iterate}. Fortunately, this information can be obtained based on a computational variation of the Bott-Borel-Weil theorem. In this way, we are able to calculate all line bundle cohomologies on favourable CICYs explicitly. The details are technical and are explained in Chapter \ref{c:line_bundle_coho}. The general result involves a large number of case distinctions, analogous to but significantly more complex than the Bott formula \eref{bott} for line bundle cohomology over $\IP^n$.

As an illustration, we provide a ``generalised Bott formula" for mixed line bundles of the form $\cO_X(-k,m)$ with $k \geq 1$, and $m \geq 0$ on the manifold $X=\left[
\begin{array}
[c]{c}%
1\\
3
\end{array}
\left|
\begin{array}
[c]{ccc}%
2 \\
4
\end{array}
\right.  \right]  $. We find that
\beq\label{example_coho}
h^{q}(X, \mathcal{O}_{X}(-k,m))=\left\{
\begin{array}
[c]{ll}%
(k+1)\binom{m}{3}-(k-1)\binom{m+3}{3} & q=0\quad k<\frac{(1+2m)(6+m+m^2)}{3(2+3m(1-m))}\\
(k-1)\binom{m+3}{3}- (k+1)\binom{m}{3} & q=1\quad k>\frac{(1+2m)(6+m+m^2)}{3(2+3m(1-m))}\\
0 & \mbox{otherwise}
\end{array}
\right. \ .
\eeq
where $\binom{n}{m}$ is the usual binomial coefficient with the convention that $\binom{0}{m}=1$.

It should be clear from the above formula that the analytic cohomology of a line bundle with mixed positive/negative entries on an arbitrary CICY is a complicated object in general. We present the outline of our algorithm for computing the cohomology of an arbitrary line bundle and its computer implementation in Chapter \ref{c:line_bundle_coho}.
\section{The monad construction on CICYs}
As was discussed in Ref.~\cite{Anderson:2007nc},
large classes of vector bundles can be
constructed over projective varieties using a variant of Horrock's monad construction \cite{monadbook}. Vector bundles defined through the monad short exact sequences can be thought
of as kernels of maps between direct sums of line bundles. For reviews of this construction and some of its applications, see Ref.~\cite{monadbook,Blumenhagen:2006ux}.  The {\em monad bundles} $V$ considered in this Chapter are defined through the short exact sequence
\bea
\nn &&0 \to V \to B \stackrel{f}{\longrightarrow} C \to 0\ ,
\mbox{ where} \\
B &=& \bigoplus_{i=1}^{r_B} \cO_X({\bf b}_i) \ , \quad
C = \bigoplus_{j=1}^{r_C} \cO_X({\bf c}_j) \ .
\label{defV}
\eea
are sums of line bundles with ranks $r_B$ and $r_C$, respectively. From the exactness of \eref{defV}, it follows that the bundle $V$ is defined as
\beq
V=\ker(f) \ .
\eeq
The rank $n$ of $V$ is easily seen, by exactness of \eref{defV}, to be
\beq
n = \rk(V) = r_B - r_C \ .
\eeq
Because the Calabi-Yau manifolds discussed in this work are defined as
complete intersection hypersurfaces in a product of projective spaces,
we can write a short exact sequence analogous to \eref{defV} but over the
ambient space, $\cA$.
\bea\label{ambmonad}
\nn &&0 \to \cV \to \cB \stackrel{\tilde{f}} {\longrightarrow} \cC \to 0\ ,\mbox{ where} \\
\cB &=& \bigoplus_{i=1}^{r_B} \cO_\cA({\bf b}_i) \ , \quad
\cC = \bigoplus_{j=1}^{r_C} \cO_\cA({\bf c}_j) \ .
\eea
Here, the map $\tilde{f}$ is a matrix whose entries are homogeneous polynomials of (multi-)degree ${\bf c}_j -{\bf b}_i$. The sequence \eref{ambmonad} defines a coherent sheaf $\cV$ on $\cA$ whose restriction to $X$ is $V$ (and additionally the map $f$ can be viewed as the restriction of $\tilde{f}$).\\

\noindent A number of mathematical constraints should be imposed on the above monad construction.
\paragraph{Bundleness: } It is not a priori obvious that the exact sequence~\eqref{monad} indeed defines a bundle rather than a coherent sheaf. However, thanks to a theorem of Fulton and Lazarsfeld~\cite{lazarsfeld} this is the case provided two conditions are satisfied (see also~\cite{Anderson:2007nc,maria}). First, all line bundles in $C$ should be greater or equal than all line bundles in $B$. By this we mean that $c^r_j\geq b^r_i$ for all $r$, $i$ and $j$. Second, the map $f:B\rightarrow C$ should be sufficiently generic~\footnote{The actual condition of Fulton and Lazarfeld's theorem \cite{lazarsfeld}, apart from genericity of $f$, is that $C^*\otimes B$ is globally generated so has at least $r_B r_C$ sections. This is indeed the case if $c^r_j\geq b^r_i$ for all $r$, $i$ and $j$ since, in this case, the line bundles $\cO_X({\bf c}_j-{\bf b}_i)$ which make up $C^*\otimes B$ are semi-positive so have at least one section each. On some CICYs the line bundles with sections extend beyond the semi-positive ones, as discussed earlier, and for those CICYs one can likely allow monads where some of the entries in $C$ are smaller than the ones in $B$ and still preserve ``bundleness" of $V$. In the present Chapter, we will not pursue this very case-dependent possibility further.}. Phrased in terms of ambient space language this means that the map $\tilde{f}:\cB\rightarrow\cC$ should be made up from sufficiently generic homogeneous polynomials  of degree ${\bf c}_j-{\bf b}_i$. We will henceforth require these two conditions. An immediate consequence of $V$ being a bundle is that \eqref{defV} can be dualized to the short exact sequence
\beq\label{dualV}
0 \to C^* \stackrel{f^T} {\longrightarrow} B^* \to V^* \to 0 \ ,
\eeq
so that the dual bundle $V^*$ is given by
\beq
V^*=\coker(f^T)\, .
\eeq
\paragraph{Non-triviliaty: } The above constraint on the integers $c^r_j$ and $b^r_i$ can be slightly strengthened. Suppose that a monad bundle $V$ is defined by the short exact sequence
\beq
0 \to V \to B \oplus R \stackrel{f'}{\longrightarrow} C \oplus R \to 0 \ ,
\eeq
where the repeated summand $R$ is a line bundle or direct sum of line bundles. The so-defined bundle $V$ is indeed equivalent to the one defined by the sequence~\eqref{monad}, so the common summand $R$ is, in fact, irrelevant \cite{monadbook}. To exclude common line bundles in $B$ and $C$ we should demand that all line bundles in $C$ are strictly greater than all line bundles in $B$. By this we mean that $c^r_j\geq b^r_i$ for all $r$, $i$ and $j$ and, in addition, that for all $i$ and $j$ strict inequality, $c^r_j>b^r_i$, holds for at least one $r$ (which can depend on $i$ and $j$).
\paragraph{Positivity: } We require that all line bundles in $B$ and $C$ are positive, that is $b^r_i>0$ and $c^r_j>0$ for all $i$, $j$ and $r$. Monads discussed in the physics literature~\cite{Distler:1987ee,Blumenhagen:2006ux,Blumenhagen:2006wj,Kachru:1995em} have typically been of this type and we will refer to them as {\em positive monads}. The reasons for this constraint are mainly of a practical nature. We have seen in our discussion of line bundles on CICYs that the cohomology of positive line bundles is particularly simple and easy to calculate from the index theorem. 
This fact significantly simplifies the analysis of positive monads. Furthermore, experience seems to indicate that non-positive bundles are "more likely" to be unstable. As an extreme case, one can show (see Chapter \ref{StabCh} that monads constructed only from negative line bundles are unstable. Of course we are not implying that all non-positive monads are unstable. In fact, in a forthcoming paper we investigate such non-positive monads~\cite{semi-positives} and we will show that allowing zero entries can still be consistent with stability. However, from the point of view of stability, starting with positive monads seems the safest bet, and we will focus on this class in the present work. 

In addition to the constraints of a more mathematical nature above we should consider physical constraints (see Chapter \ref{HetIntro}). To formulate them we need explicit expressions for the Chern classes of monad bundles. One finds
\bea
\nn \rk(V) &=& r_B - r_C = n  \ , \\
\nn c_1^r(V) &=& \sum_{i=1}^{r_B} b^r_i - \sum_{j=1}^{r_C} c^r_j \ ,
\\
c_{2r}(V &=& \frac12  d_{rst} 
   \left(\sum_{j=1}^{r_C} c^s_j c^t_j- 
   \sum_{i=1}^{r_B} b^s_i b^t_i \right) \ , 
\label{chernV} \\
\nn c_3(V) &=& \frac13 d_{rst} 
   \left(\sum_{i=1}^{r_B} b^r_i b^s_i b^t_i - \sum_{j=1}^{r_C} c^r_j
   c^s_j c^t_j \right) \ ,
\eea
where $d_{rst}$ are the triple intersection numbers~\eqref{drst} on $X$ and the relations for $c_{2r}(V)$ and $c_3(V)$ have been simplified assuming that $c_1^r(V)=0$. Then we need to impose two physics constraints.
\paragraph{Correct structure group:} To have bundles with structure group ${\rm SU}(n)$ where $n=3,4,5$ we first of all need that $n=r_B-r_C=3,4,5$. In addition,  the first Chern class of $V$ needs to vanish which, from the second Eq.~\eqref{chernV}, can be expressed as
\beq
S^r := \sum_{i=1}^{r_C+n} b^r_i = \sum_{j=1}^{r_C} c^r_j  \qquad \forall r=1,
   \ldots, k \ .
\label{cons1}
\eeq
We have defined the quantities $S^r$ which represent the first Chern classes of $B$ and $C$ and will be useful for the classification of positive monads below.

\paragraph{Anomaly cancellation/effectiveness:} As we have seen this condition can be stated in the simple form~\eqref{anomaly_final}. Inserting the above expression for the second Chern class gives
\beq
d_{rst} \left( \sum_{j=1}^{r_C} c^s_j c^t_j - \sum_{i=1}^{r_B} b^s_i b^t_i \right) \leq 2c_{2r}(TX) \qquad \forall r \ . \label{effcond2}
\eeq

In addition, we should of course prove stability of positive monads, a task which will be systematically dealt with in Chapter \ref{StabCh}. This completes the set-up of monads bundles. To summarise, we will consider monad bundles $V$ of rank $3$, $4$ or $5$, defined by the short exact sequence~\eqref{defV} with positive line bundles only. In addition, all line bundles in $C$ must be strictly greater than all line bundles in $B$ and the two constraints~\eqref{cons1} and \eqref{effcond2} must be satisfied.
\section{Classification of positive monads on CICYs}\label{s:positive}
An obvious question is whether the class of monads defined in the previous section is finite. In this section, we show that this is indeed the case and subsequently classify all such monads.

We start by stating the classification problem in a more formal way. For any favourable CICY manifold $X$ with second Chern class $c_{2r}(TX)$ and triple intersection numbers $d_{rst}$, defined in a product of $m$ projective spaces, and for any $n=3,4,5$, we wish to find all sets of integers $b^r_i$ and $c^r_j$, where $r=1,\ldots ,m$, $i=1,\ldots ,r_B=r_C+n$ and $j=1,\ldots , r_C$ satisfying the conditions
\bea
\nn 1.&&b^r_i\geq 1\; ,\quad c^r_j\geq 1\; ,\quad \forall i,j,r;\\
\nn 2.&&c^r_j \ge b^r_j \ , \quad \forall i,j,r;\\ 
\nn 3. &&\forall i,j \mbox{ there exists at least one } r \mbox{ such that }c^r_j>b^r_i;\\
 4.&&\sum_{i=1}^{r_B} b^r_i = \sum_{j=1}^{r_C} c^r_j = S^r \ , \quad
    \forall r; \label{class}\\
\nn 5.&&d_{rst} \left( \sum_{j=1}^{r_C} c^s_j c^t_j - \sum_{i=1}^{r_B} b^s_i b^t_i \right) \leq 2c_{2r}(TX) \qquad \forall r \, .
\eea
Our first task is to show that this defines a finite class. Although all that is involved are simple manipulations of inequalities it is not complete obvious at first which approach to take. We start by defining the maxima $b^r_{\rm max}={\rm max}_i\{b^r_i\}$, minima $c^r_{\rm min}={\rm min}_j\{c^r_j\}$ and their differences $\theta^r=c^r_{\rm min}-b^r_{\rm max}\geq 0$ which are of course positive for all $r$. Then we can write
\beq
b^r_i = b^r_{\rm max} - T^r_i, \qquad
c^r_j = c^r_{\rm min} + D^r_j \ , \label{bcmaxmin}
\eeq
where $T^r_i$ and $D^r_j$ are the deviations from the maximum and minimum values. It is also useful to introduce the sums
\begin{equation}
 T^r=\sum_{i=1}^{r_B} T^r_i\; ,\quad D^r=\sum_{j=1}^{r_C} D^r_j \label{TDdef}
\end{equation}
of these deviations. Given theses definitions, it is easy to see that
\begin{equation}
S^r = b^r_{\rm max} r_B - T^r\; ,\quad S^r = c^r_{\rm min} r_C + D^r\; .
\end{equation}
Subtracting these two equations and using $r_B=r_C+n$ it follows that
\beq\label{thetaid}
\theta^rr_C + (D^r + T^r) = n b^r_{\rm max} \ .
\eeq
We will use this identity shortly.  Next, from the definition \eqref{cons1}, and since all $c_s^j \ge 1$, we
obtain the two inequalities
\beq\label{ineq2}
S^r \ge \sumj \II^r = r_C \II^r \; , \quad S^r \le \sumi b^r_{\rm max} = b^r_{\rm max} r_B\; ,\quad \forall~r\; ,
\eeq
where $\II_s$ is a vector with all entries being 1. After this preparation, we come to the key part of the argument which involves working out the consequences of condition 5 in \eqref{class}.
\beq\label{boundS}
\ba{llll}
2c_{2r}(TX) 
&\ge& 
d_{rst} \left(
\sumj c^s_j c^t_j - \sumi b^s_i b^t_i 
\right) & \\
&=&
d_{rst}\left(
\sumj (c^s_{\rm min}+D^s_j) c^t_j - 
\sumi (b^s_{\rm max}-T^s_i) b^t_i 
\right) & \mbox{inserting }\eqref{bcmaxmin}\\
&=& d_{rst}\left(
(c^s_{\rm min} - b^s_{\rm max}) S^t + \sumj D^s_j c^t_j + 
\sumi T^s_i b^t_i 
\right) & \mbox{using }\eqref{cons1}\\
&\ge& d_{rst}\left(
\theta^s S^t + (D^s + T^s) \II_t
\right) & \mbox{since } c^t_j, b^t_i \ge 1\ ,\mbox{using }\eqref{TDdef}\\
&\ge& d_{rst}\left(
\theta^s (r_C \II^t) + (D^s + T^s) \II^t 
\right) & \mbox{by first inequality }\eref{ineq2}\\
&=& d_{rst}\left(
n b^s_{\rm max} \II^t
\right) & \mbox{from \eref{thetaid}}\\
&\ge& \frac{n}{r_B} 
d_{rst}\left( S^s \II^t
\right) & \mbox{by second inequality}
\ea
\eeq
From the second last line in the above chain of inequalities, we can also express this result as a bound in the variables $b^r_{\rm max}$ (the maximum entries the bundle $B$ can have in each projective space), resulting in
\beq\label{bsmaxineq}
2c_{2r}(TX) \ge n\sum_{s,t} d_{rst} b^s_{\rm max} \ .
\eeq
It turns out that the matrices $\sum_td_{rst}$ are always non-singular, so this equation provides an upper bound for $b^r_{\rm max}$. Moreover, since each $b^r_{\rm max} \in \IZ_{\ge 1}$, and since the
matrix $n \sum_t d_{rst}$ has entries in $\IZ_{\ge 0}$,
Eq.~\eref{bsmaxineq} may not have solutions for all manifolds. In fact, of
the 4515 favourable CICYs, Eq.~\eref{bsmaxineq} immediately 
eliminates all but 63 which include the 5 cyclic ones studied in Ref.~\cite{Anderson:2007nc}. 
One finds that the values for $b^r_{\rm max}$ are very small indeed and never exceed 6.
\comment{
{7643, 7668, 7707, 7708, 7725, 7727, 7728, 7758, 7759, 7779, 7789, 7797,
7799, 7806, 7807, 7808, 7809, 7816, 7817, 7819, 7821, 7822, 7823, 7831, 7833,
7834, 7836, 7840, 7844, 7845, 7853, 7854, 7858, 7859, 7861, 7862, 7863, 7865,
7866, 7867, 7868, 7869, 7870, 7871, 7872, 7873, 7874, 7875, 7876, 7877, 7878,
7879, 7880, 7881, 7882, 7883, 7884, 7885, 7886, 7887, 7888, 7889, 7890}
}

So far, we have bounded the maximal entries of the bundle $B$. What about $r_B$, the rank of $B$? It turns out there are various ways to derive an upper bound on $r_B$.  First note that, from the third condition in \eref{class}, for all $j\in \{1,\ldots ,r_C\}$, there exists a $\sigma\in \{1,\ldots, m\}$, call it $\sigma (j)$, such that
\begin{equation}
c^r_j-b^r_{\rm max} \geq \delta^{r \sigma(j)} \; .
\end{equation}
Introduce
\begin{equation}\label{def-nu}
\nu^r = \sum_{j=1}^{r_C}\delta^{r \sigma(j)} \; ,
\end{equation}
the number of line bundles in $C$ which are bigger than the ones in $B$ due to the $r$-th entry. 
Since all line bundles in $C$ are bigger than the ones in $B$ it follows that
 \begin{equation}
 \sum_{r=1}^m \nu^r = r_C = r_B + n\; .\label{nusum}
 \end{equation}
We conclude that
 \begin{equation}
 r_Bb^r_{\rm max}\geq\sum_{i=1}^{r_B}b^r_i=\sum_{j=1}^{r_C}c^r_j
 \geq\sum_{j=1}^{r_C}(b^r_{\rm max}+\delta^{r\sigma (j)})=r_cb^r_{\rm max}+\nu^r
 \end{equation}
and, hence, that $nb^r_{\rm max}\geq\nu^r$. Summing this result over $r$ one easily finds that
 \begin{equation}
  r_B\leq n\left(1+\sum_{r=1}^mb^r_{\rm max}\right)\; . \label{rBbound1}
 \end{equation}
Since we have already bounded $b^r_{\rm max}$ (independently of $r_B$) this provides an upper bound for $r_B$. This shows that our class of bundles is indeed finite. While the above bound is simple, for the practical purpose of classifying all bundles it often turns out to be too weak, and requires computationally expensive scanning of monads with large $r_B$ and, hence, a large number of integer entries. Based on Eq.~\eqref{rBbound1} alone, a classification on a desktop machine is likely impossible. Fortunately, one can derive other constraints on $r_B$ which in many cases turn out to be stronger. Using $nb^r_{\rm max}\geq\nu^r$ in Eq.~\eqref{bsmaxineq} leads to
 \begin{equation}
  \sum_{r,s}d_{rst}\nu^t\leq 2c_{2r}(TX)\; . \label{rBbound2}
\end{equation}
For each CICY, one can find all integer solutions $(\nu^r)$ (subject to the constraint
$\nu^r\geq 0$, of course) to this equation and then calculate the maximal possible value for $r_B$ from Eq.~\eqref{nusum}. Finally, starting again from condition 5 of \eref{class} we find
\beq\ba{rcl}
 2c_{2r}(TX)
&\geq& 
d_{rst}\left[
\sum\limits_{j=1}^{r_C}c^s_jc^t_j-\sum\limits_{i=1}^{r_B}b^s_ib^t_i\right]\\
&\geq& 
d_{rst}\left[\sum\limits_{j=1}^{r_C}(b^s_{\rm max}+\delta^{s\sigma (j)})
(b^t_{\rm max}+ \delta^{t\sigma (j)})- 
\sum\limits_{i=1}^{r_B}b^s_i b^t_i\right]\\
&=&
d_{rst}\left[\sum\limits_{j=1}^{r_C}b^s_{\rm max}b^t_{\rm max} -
\sum\limits_{i=1}^{r_B}b^s_i b^t_i+ 2\nu^s b^t_{\rm max} +
\delta_{s}^t \nu^t\right]\\
&\geq
&d_{rst}\left[ -n b^s_{\rm max} b^t_{\rm max} + 2\nu^s b^t_{\rm max}
+\delta_{s}^t\nu^t\right]\; .
\ea\eeq
Rewriting this as an system of linear inequalities for $\nu^s$, we have that
\begin{equation}\label{rBbound3}
\sum\limits_s\left(2\sum\limits_td_{rst}b^t_{\rm max}+d_{rss}\right)\nu^s\leq 2c_{2r}(TX)+nd_{rst}b^s_{\rm max}b^t_{\rm max}\; .
\end{equation}
Again, this equation can be solved for all non-negative integers $\nu^r$ since
$b^{max}_s$ is bounded from \eref{bsmaxineq} and, subsequently, we can compute the
maximal $r_B$ from Eq.~\eqref{nusum}. In practice, we evaluate all three bound~\eqref{rBbound1}, \eqref{rBbound2}, \eqref{rBbound3} for every CICY and use the minimum value obtained. In this way we find maximal values for $r_B$ ranging from $8$ to $22$ depending on the CICY. 

The explicit classification is now simply a matter of computer search. For each of the 63 CICYs with solutions to the inequality~\eqref{bsmaxineq} we scan over all allowed values of $n$, $r_B$ and over all values for $S^r$ subject to the last inequality in \eqref{boundS}. For each fixed set of these quantities we then generate all multi-partitions of entries $b^r_i$ and $c^r_j$ eliminating, of course, trivial redundancies due to permutations.
\comment{
Computing $r_B$ from the three constraints \eqref{cons1}, \eqref{cons2} and \eqref{cons3} and taking the minimal value I find for the
63 CICYs which pass the $b_{\rm max}$ test I find (in pairs (CICY\#,maximal rank of $B$)):
\begin{equation}
\begin{array}{llllllll}
(7643, 8)& (7668, 8)& ( 7707, 11)& (7708, 11)& (7725, 8)& (7727, 11)& (7728, 11)& (7758, 8)\\
( 7759, 8)& (7779, 12)& (7789, 12)& (7797, 11)& (7799, 8)& (7806, 14)& (7807, 8)& (7808, 8)\\
(7809, 8)& (7816, 11)& (7817, 11)& (7819, 11)& (7821, 8)& (7822, 11)& (7823, 11)& (7831, 12)\\
(7833, 10)& (7834, 11)& (7836, 11)& (7840, 14)& (7844, 9)& (7845, 12)& (7853, 9)& (7854, 11)\\
(7858, 14)& (7859, 14)& (7861, 7)& (7862,16)& (7863, 9)& (7865, 11)& (7866, 11)& (7867, 14)\\
(7868, 10)& (7869,14)& (7870, 13)& (7871, 11)& (7872, 12)& (7873, 14)& (7874, 14)& (7875,14)\\
(7876, 11)& (7877, 14)& (7878, 9)& (7879,8)& (7880, 16)& (7881, 14)& (7882, 16)& (7883, 10)\\
(7884, 12)& (7885, 22)& (7886, 22)& (7887, 20)& (7888, 22)& (7889, 9)& (7890, 12)&
\end{array}\nonumber
 \end{equation}
}
Upon performing this scan, we find that positive monad bundles only exist over 36 favourable CICYs (out of the 63 which passed the initial test). These 36 manifolds (tabulated in Appendix \ref{s:posCICY})together with the number of monad bundles over them, are listed in Table~\ref{t:pos-cicy}. 
\begin{table}
{\scriptsize
$\ba{|c|c||c|c||c|c||c|c|}\hline
\mbox{Config} & \mbox{No.Bundles} & \mbox{Config} & \mbox{No.Bundles} &
\mbox{Config} & \mbox{No.Bundles} & \mbox{Config} & \mbox{No.Bundles}
\\ \hline \hline
\tconf{5} & (20, 14, 9)
& \tconf{3 & 3 } & (5, 3, 2)
& \tconf{4 & 2 } & (7, 5, 3)
& \tconf{3 & 2 & 2 } & (3, 2, 1)
\\ \hline
\tconf{2 & 2 & 2 & 2 \cr} & (2, 1, 0)
& \tconf{ 2 \cr 4 \cr  } & (611, 308, 56)
& \tconf{ 3 \cr 3 \cr  } & (62, 43, 14)
& \tconf{0 & 2 \cr 2 & 3 \cr} & (80, 12, 0)
\\ \hline
\tconf{0 & 2 \cr 3 & 2 \cr  } & (12, 5,,0)
& \tconf{0 & 2 \cr 4 & 1 \cr  } & (126, 17, 0)
& \tconf{ 1 & 1 \cr 3 & 2 \cr  } & (15, 8, 0)
& \tconf{ 1 & 1 \cr 4 & 1 \cr  } & (153, 35, 19)
\\ \hline
\tconf{ 2 & 1 \cr 1 & 3 \cr  } & (3, 0, 0)
& \tconf{ 2 & 1 \cr 2 & 2 \cr  } & (5, 0, 0)
& \tconf{ 2 & 1 \cr 3 & 1 \cr  } & (13, 2, 0)
& \tconf{ 0 & 0 & 2 \cr 2 & 2 & 2 \cr  } & (5, 0, 0)
\\ \hline
\tconf{ 0 & 0 & 2 \cr 3 & 2 & 1 \cr  } & (5, 0, 0)
& \tconf{ 0 & 1 & 1 \cr 2 & 2 & 2 \cr  } & (5, 0, 0)
& \tconf{ 0 & 1 & 1 \cr 2 & 3 & 1 \cr  } & (12, 5, 0)
& \tconf{ 0 & 1 & 1 \cr 3 & 2 & 1 \cr  } & (8, 0, 0)
\\ \hline
\tconf{ 0 & 1 & 1 \cr 4 & 1 & 1 \cr  } & (126, 17, 0)
& \tconf{ 0 & 2 & 1 \cr 2 & 2 & 1 \cr  } & (2, 0, 0)
& \tconf{ 1 & 1 & 1 \cr 3 & 1 & 1 \cr  } & (2, 0, 0)
& \tconf{ 2 & 1 & 1 \cr 2 & 1 & 1 \cr  }  & (1, 0, 0)
\\ \hline
\tconf{0 & 0 & 1 & 1 \cr 2 & 2 & 2 & 1 \cr  } & (3, 0, 0)
& \tconf{0 & 0 & 1 & 1 \cr 3 & 2 & 1 & 1 \cr  } & (5, 0, 0)
& \tconf{ 2 \cr 2 \cr 3 \cr  } & (553, 232, 0)
& \tconf{ 0 & 2 \cr 1 & 2 \cr 1 & 2 \cr  } & (8, 0, 0)
\\ \hline
\tconf{ 1 & 1 \cr 0 & 2 \cr 1 & 3 \cr  } & (74, 0, 0)
& \tconf{ 1 & 1 \cr 0 & 2 \cr 2 & 2 \cr  } & (9, 0, 0)
& \tconf{ 1 & 1 \cr 1 & 1 \cr 1 & 3 \cr  } & (25, 0, 0)
& \tconf{ 1 & 1 \cr 1 & 1 \cr 2 & 2 \cr  } & (9, 0, 0)
\\ \hline
\tconf{ 1 & 1 \cr 1 & 2 \cr 0 & 3 \cr  } & (34, 0, 0)
& \tconf{ 1 & 1 \cr 2 & 1 \cr 2 & 1 \cr  } & (3, 0, 0)
& \tconf{1 & 1 & 0 \cr 1 & 0 & 1 \cr 3 & 1 & 1 \cr  } & (9, 0, 0)
& \tconf{2 \cr 2 \cr 2 \cr 2 \cr } & (3665, 625, 0)
\\
\hline
\ea$
}
\caption{{\em The 36 manifolds which admit positive monads. The No.Bundles column next
to each manifold is a triple, corresponding to the numbers of respectively ranks 3,4,
and 5 monads.}}
\label{t:pos-cicy}
\end{table}
In total, we find 7118 positive monad bundles. These include the 77 positive monad bundles on the 5 cyclic CICYs (these are the CICYs with $h^{1,1}(X)=1$) found in Ref.~\cite{Anderson:2007nc}.
Some explicit examples are listed in Table~\ref{t:monex}.
\begin{table}
\begin{center}
\beq
{\small
\ba{|c|c|c|c|c|c|}\hline
\mbox{CICY } X & B & C & \rk(V) &
\left[ \ba{c} c_2(TX) \\ c_2(V) \ea \right] & \mbox{ind}(V) = \frac12c_3(V)
\\ \hline \hline
\left[\begin{array}{c|c}
1 & 2\\
1 & 2\\
1 & 2\\
1 & 2\\
\end{array}\right]
&
\cO_X(1,1,1,1)^{\oplus 8} &
\ba{c}
\cO_X(5, 1, 1, 1) \\
\oplus \cO_X(1, 5, 1, 1) \\
\oplus \cO_X(1, 1, 5, 1) \\
\oplus \cO_X(1, 1, 1, 5)
\ea
& 4 & \left[ \ba{c} (24, 24, 24, 24) \\ (24, 24, 24, 24) \ea \right]
& -64 \\ \hline
\left[\begin{array}{c|cc}
1 & 1 & 1\\
2 & 2 & 1\\
2 & 2 & 1\\
\end{array}\right]
& \cO_X(1,1,1)^{\oplus 10} &
\ba{c}
\cO_X(1, 1, 2)^{\oplus 3} \\
\oplus \cO_X(1, 2,1)^{\oplus 3} \\
\oplus \cO_X(4, 1, 1)
\ea
& 3 & \left[ \ba{c} (24, 36, 36) \\ (24, 36, 36) \ea \right]
& -69 \\ \hline
\left[\begin{array}{c|c}
1 & 2 \\
3 & 4 \\
\end{array}\right]
& \cO_X(1,1)^{\oplus 11}
& \cO_X(6,1) \oplus \cO_X(1,2)^{\oplus 5}
& 5 & \left[ \ba{c} (24, 44) \\ (20, 30) \ea \right]
& -40 \\ \hline
[4|5] & \cO_X(1)^{\oplus 6} & \cO_X(2)^{\oplus 3}
& 3 & \left[ \ba{c} (50) \\ (15) \ea \right]
& -15 \\ \hline
\ea\nn
}
\eeq
\caption{\em Some examples from the 7118 positive monads on favourable CICYs.}
\label{t:monex}
\end{center}
\end{table}
Focusing on the different ranks of $V$ considered, we find 5680 bundles of rank 3, 1334 of rank 4, and 104 of
rank 5 on these 36 manifolds. To get an idea of the distribution, in part (a) of Fig.~\ref{f:c3pos} we have plotted the number of monads as a function of the index ${\rm ind}(V)$. It seems, at first glance, that the distribution is roughly Gaussian. For comparison, in part (b) of Fig.~\ref{f:c3pos}, we have plotted the number of monads which satisfy the 3-generation constraints~\eqref{k-div_final}. The same data, but split up into the three cases $n=3,4,5$ for the rank of $V$, is shown in Fig.~\ref{f:c3pos345}. The total numbers of bundles in all cases has been collected in Table~\ref{t:pos-monad}.
\begin{table}[h!!!]
{\small
\begin{center}
\begin{tabular}{|c|c|c|c|c|}  \hline
  & Bundles & ${\rm ind}(V) = 3k$ & \begin{tabular}{l} ${\rm ind}(V) = 3k$ \\ and $k$ divides $\chi(X)$ \end{tabular} 
  & \begin{tabular}{l} ${\rm ind}(V) = 3k$ \\ $|{\rm ind}(V)|<40$ \\
  and $k$ divides $\chi(X)$ \end{tabular} \\ 
\hline
rank 3 & 5680 & 3091 & 458 & 19\\
rank 4 & 1334 & 207 & 96 & 2 \\
rank 5 & 104 & 52 & 5 & 0 \\ \hline
Total & 7118 & 3350 & 559 & 21\\ \hline
\end{tabular}
\end{center}
\caption{{\em The number of positive monad bundles on favourable CICYs.
Imposing that the third Chern class is divisible by 3 reduces the number
and requiring in addition that the quotient of $c_3(V)$ by 3 divides the
Euler number of the corresponding CICY further reduces the number.}}
\label{t:pos-monad}
}
\end{table}
\begin{figure}[t]
\centerline{(a)\epsfxsize=3in\epsfbox{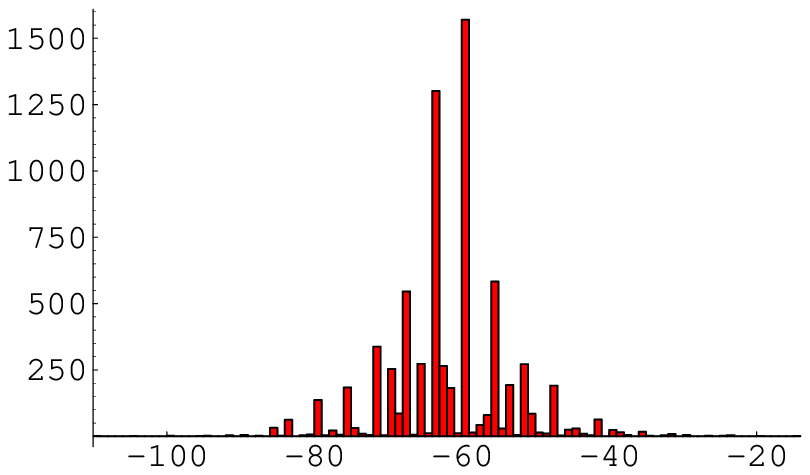}
(b)\epsfxsize=3in\epsfbox{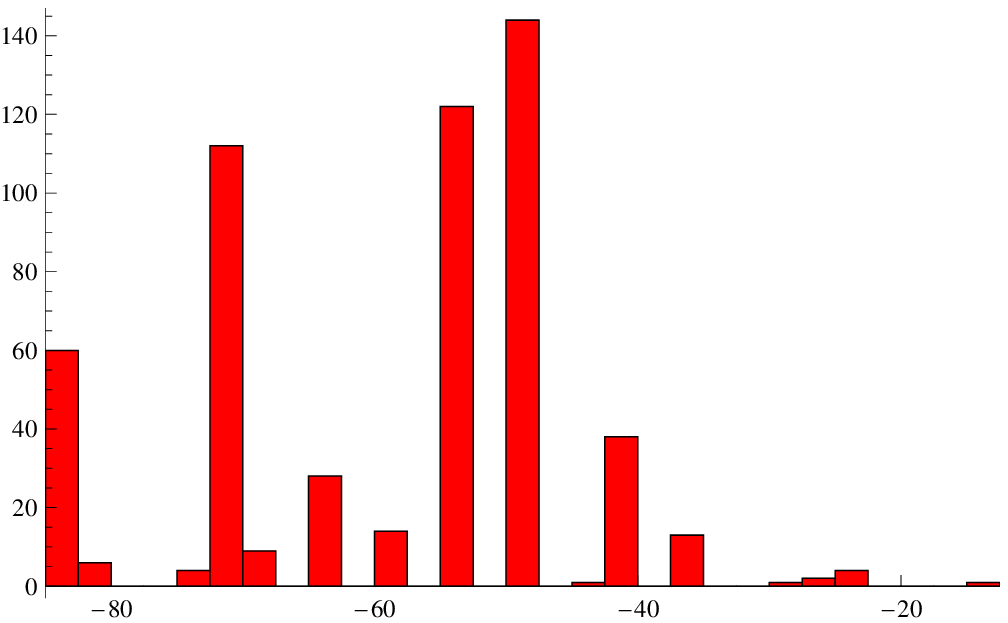}}
\caption{{\em (a) Histogram for the index,
${\rm ind}(V)$, of the 7118 positive monads
found over 36 favourable CICYs: the horizontal axis is ${\rm ind}(V)$ and the
vertical, the number of bundles;
(b) the same data set,
but only taking those monads which have ${\rm ind}(V) = 3k$ for some positive 
integer $k$ and such that $k$ divides the Euler number of the 
corresponding CICY.
}}
\label{f:c3pos}
\end{figure}
\begin{figure}[t]
\centerline{(a)\epsfxsize=3in\epsfbox{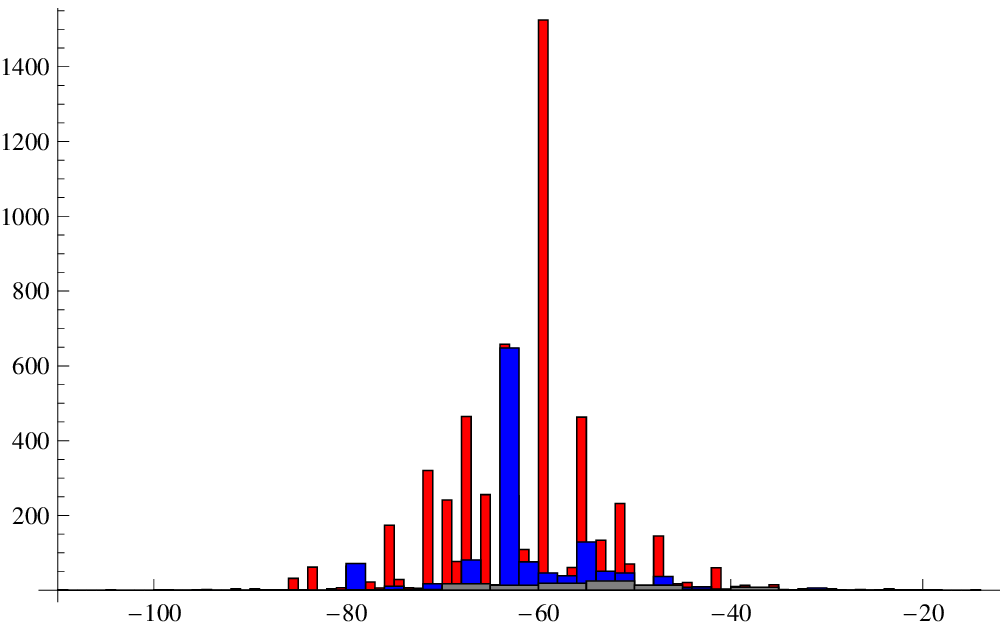}
(b)\epsfxsize=3in\epsfbox{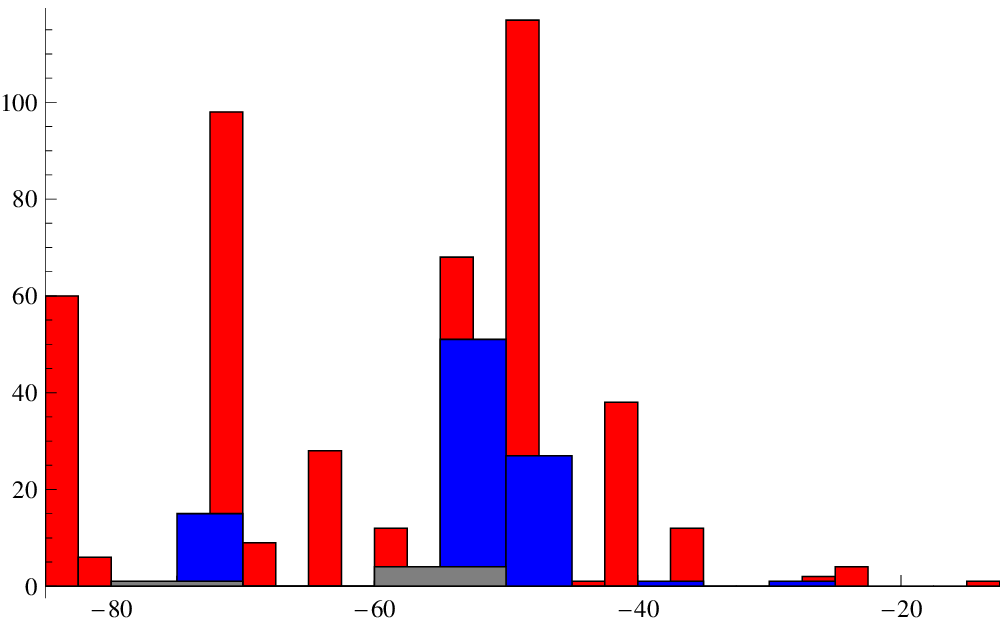}}
\caption{{\em (a) Histogram for the index, ${\rm ind}(V)$, of the
positive monads, 5680 of rank 3 (in red), 1334 of rank 4 (in blue), and 104 of
rank 5 (in gray), found over 36 favourable CICYs: the horizontal axis is ${\rm ind}(V)$ and the
vertical, the number of bundles; (b) the same data set, but only taking those monads which have ${\rm ind}(V) = 3k$ for some positive integer $k$ and such that $k$ divides the Euler number of the corresponding CICY.
}}
\label{f:c3pos345}
\end{figure}
It is clear from this table that even the two very rudimentary physical constraints~\eqref{k-div_final} lead to a very substantial reduction of the number of viable bundles. If these two constraints are combined with a ``sensible" limit on the index, for example ${\rm ind}(V)\leq 40$ (assuming that the discrete symmetries one is likely to find are of order $\leq 13$), then part (b) of the figures show that the number of remaining bundles is very small indeed, with only $21$ bundles less than $40$.

\section{Stability}\label{s:stable}

As mentioned in the introduction, a crucial property of vector bundles
in heterotic compactification is stability, one of the conditions for the low-energy theory to be supersymmetric. In general, it is a difficult task to prove stability of vector bundles and we will rigorously investigate this issue in Chapter \ref{StabCh}. Presently, we will be satisfied checking the necessary but generally not sufficient  cohomology conditions \eref{H0=0} as a first (and non-trivial) test of stability.
Establishing these results will also be very helpful for the calculation of the particle spectrum in the next Section.

To begin, consider the familiar short exact sequence defining the monad
\eref{defV}
\beq
0 \to V \to B \stackrel{f}{\longrightarrow} C \to 0 \ ,
\eeq
which induces the following long exact sequence in cohomology
\beq\label{HVBC}
\ba{lll}
0&\to& \fbox{$H^0(X,V)$} \to H^0(X,B) \to H^0(X,C) \to\\
 &\to& H^1(X,V) \to H^1(X,B) \to H^1(X,C)\to\\
 &\to& H^2(X,V) \to H^2(X,B) \to H^2(X,C)\to\\
 &\to& \fbox{$H^{3}(X,V)$} \to H^{3}(X,B) \to H^{3}(X,C) \to 0 \ .
\ea
\eeq
We have boxed the terms which we would like to show are vanishing. We begin with $H^3(X,V)$ which, due to Serre duality, is equivalent to $H^0(X,V^*)$. Fortunately, for the case of positive monads, it is easy to see that $H^0(X,V^*)$ vanishes. Consider the sequence dual to \eref{defV} which reads
\beq \label{H0dual}
\sseq{C^*}{B^*}{V^*}
\eeq
and its associated long exact sequence in cohomology
\beq\label{HC*B*V*}
\ba{lll}
0&\to& H^0(X,C^*) \to H^0(X,B^*) \to H^0(X,V^*) \to\\
 &\to& H^1(X,C^*) \to \ldots \to
\ea
\eeq
Now, $H^0(X,B^*)$ and $H^1(X,C^*)$ both vanish because $B^*$ and $C^*$ are both 
sums of negative line bundles on $X$. Hence, positioned in between those zero entries, $H^0(X,V^*)\simeq H^3(X,V)$ also vanishes.

Showing that $H^0(X,V)$ vanishes as well is not quite as simple. Let $X$ be a CICY of codimension $K$,
embedded in $\cA$, a product of
projective spaces, and $\cN^*$  the dual of the normal bundle \eref{normalbundle} of $X$.  To show that $H^{0}(X,V)=0$ we will use the correspondence of $SU(n)$ bundles: $V \equiv \wedge^{n-1}V^*$ where $rk(V)=n$ and prove the result for each rank separately. As an illustration, we shall provide here the proof for the rank $3$ bundles. The proof for $SU(4)$ and $SU(5)$ bundles is entirely analogous, but more lengthy since it requires the use of Koszul and Leray sequences. To begin, for an $SU(3)$ bundle $V \equiv \wedge^{2}V^*$ and the exterior power sequence \cite{AG1,AG2} arising from \eref{defV} gives us directly:
\beq
0 \to  S^{2}C^*\to  C^* \times B^* \to\wedge^{2}B^* \to \wedge^{2}V^* \to 0~;
\eeq
which can be decomposed into short exact sequences in the standard way:
\bea
0 \to S^{2}C^* \to C^* \times B^* \to K_1 \to 0 \\
0 \to K_1 \to \wedge^{2}B^* \to \wedge^{2}V^* \to 0
\eea
Then, since $B^*,C^*$ are direct sums of negative line bundles, by \eref{kodaira} it follows that the long-exact sequences in cohomology immediately yield that $H^0(X,\wedge^{2}V^*)=0$ if $H^0(X,\wedge^{2}B^*)=H^{1}(X,K_1)=0$. But since $B$ is a negative line bundle the first condition is clearly satisfied. Furthermore, the cohomology of the first sequence gives
\beq
{\small
\ba{l}
0 \to H^{2}(X,K_1) \to H^{3}(X, S^{2}C^*) \to H^{3}(X, C^*\times B^*) \to H^{3}(X, K_1) \to 0 \\
H^{1}(X,K_1)=0
\ea
}
\eeq
Thus, $H^0(X,\wedge^{2}V^*)=0$ and hence, $H^0(X,V)=0$ for $SU(3)$ bundles. For rank $4$ and $5$ bundles we shall provide explicit examples (including for example, the vanishing of $H^{0}(X, \wedge^{3}V^*)$) in Chapter \ref{StabCh}). We have explicitly calculated that for all positive monads, the conditions $H^0(X,V) = H^0(X,V^*) = 0$ are always satisfied.  
\section{Computing the Particle Spectrum}
\subsection{Bundle Cohomology}
While computing the full cohomology of monad bundles is generally a difficult task, it will become clear in the following that significant simplifications arise for positive monads. This computational advantage is of course one of the motivations to consider positive monads and it will lead to a number of general statements about their cohomology.

\subsubsection{The number of families and anti-families: $H^1(X,V)$ and $H^1(X,V^*)$}
The defining short exact sequence~\eqref{monad} of the monad bundle $V$ induces the long exact sequence
\beq
\ba{lllllll}
 0&\to&H^0(X,V)&\to&H^0(X,B)&\to&H^0(X,C)\\
&\to&H^1(X,V)&\to&H^1(X,B)&\to&H^1(X,C)\\
&\to&H^2(X,V)&\to&H^2(X,B)&\to&H^2(X,C)\\
&\to&H^3(X,V)&\to&H^3(X,B)&\to&H^3(X,C)\to 0
\ea
\eeq 
Since both $B$ and $C$ are sums of positive line bundles we know from Kodaira vanishing that the cohomologies $H^q(X,C)=H^q(X,B)=0$ for all $q>0$. The above long exact sequence then immediately implies that $H^2(X,V)=0$. In the previous Section we have already shown that $H^0(X,V)=H^3(X,V)=0$ always, so that the only non-vanishing cohomology of positive monads is $H^1(X,V)$. The dimension $h^1(X,V)$ of this first cohomology can then be calculated from the index theorem~\eqref{indV} or indeed the above long exact sequence. In summary, one finds
\begin{equation}
 h^1(X,V)=h^0(X,C)-h^0(X,B)=-{\rm ind}(V)\; ,\quad h^q(X,V)=0\mbox{ for } q\neq 1\; .
\end{equation} 
This means that the number of anti-families always vanishes and that the number of families can easily be obtained from the index which we have already computed in Figs.~\ref{f:c3pos} and \ref{f:c3pos345}.
The absence of vector-like pairs of families might be considered an attractive feature and is certainly a pre-requisite for compactifications with the exact standard model spectrum. We stress that this property is directly linked to the property of positivity and will not generally hold if we allowed zero or negative integer entries in the line bundles defining the monad.

For ${\rm SU}(3)$ bundles we have $V\simeq \Lambda^2 V^*$ and, hence,  the cohomology groups $H^1(X,\wedge^2 V)$ and $H^1(X, \wedge^2 V^*)$ contain no new information. However, for ${\rm SU}(4)$ and ${\rm SU}(5)$ this is not the case and we have to perform another calculation. In the case of rank four, $\wedge^2V\simeq \wedge^2V^*$, so that $H^1(X,\wedge^2 V)\simeq H^1(X, \wedge^2 V^*)$. For rank five the situation is less trivial, but from Eq.~\eqref{indtheorem2} we know that $h^1(X,\wedge^2 V)$ and $h^1(X,\wedge^2V^*)$ are related by the index, ${\rm ind}(V)$, of $V$. Hence, in both the rank four and five cases it is enough to compute one of $H^1(X,\wedge^2 V)$ and $H^1(X,\wedge^2V^*)$ and, in the following, we will opt for $H^1(X,\wedge^2V^*)$.

To calculate this cohomology, we can essentially proceed along the lines of Appendix B.3 of Ref.~\cite{Anderson:2007nc}. We start by writing down the Koszul resolution~\eref{koszul} for $\wedge^2V^*$ which is given by
\beq\label{koszul-cv2s}
0 \to \cv2s \otimes \wedge^K \cN^* \to \cv2s \otimes \wedge^{K-1} \cN^*
\to \ldots \to \cv2s \otimes \cN^* \to \cv2s \to \wedge^2V^* \to 0 \ .
\eeq
Recall that $K$ is the co-dimension of the CICY $X$ embedded in the ambient space $\cA$ and $\cN$ is the normal bundle~\eqref{normalbundle} of $X$ in $\cA$. As a first step we will now derive vanishing theorems for the cohomologies of the bundles $\wedge^2V^* \otimes \wedge^j\cN^*$ which appear in the above Koszul sequence. 
To do this, we start the exact sequence for antisymmetric products of bundles \cite{AG1,Anderson:2007nc}
\beq\label{wedge2V*}
0 \to S^2 \cC^* \to \cC^* \otimes \cB^* \to \wedge^2 \cB^* \to \cv2s \to 0 \ ,
\eeq
which is induced from the dual sequence 
\beq
\sseq{\cC^*}{\cB^*}{\cV^*}\; .
\eeq
We can then tensor \eref{wedge2V*} by $\wedge^j \cN^*$ for $j=0,\ldots, K$ and break the resulting 4-term exact sequence into two short exact sequences
\beq\ba{l}
\sseq{S^2 \cC^* \otimes \wedge^j \cN^*}
     {\cC^* \otimes \cB^* \otimes \wedge^j \cN^*}{Q_j} \ ;\\
\sseq{Q_j}{\wedge^2 \cB^* \otimes \wedge^j \cN^*}
     {\wedge^2 \cV^* \otimes \wedge^j \cN^*} \ ;
\ea \qquad
j = 0, \ldots, K \ ,
\eeq
where $Q_j$ are approriate (co)kernels. This induces two inter-related
long exact sequences in cohomology on $\cA$ which are given by

\beq\label{Hwedege2V*}
{\scriptsize
\ba{l}
\ba{llllllllll}
0
 &\to& \cancelto{0}{H^0(\cjn)} &\to& \cancelto{0}{H^0(\cbjn)} &\to& H^0(\cA, Q_j) &\to&\\
 &\to& \cancelto{0}{H^1(\cjn)} &\to& \cancelto{0}{H^1(\cbjn)} &\to& H^1(\cA, Q_j) &\to&\\
 &\to& &&\vdots &&  &\to& \\
 &\to& \cancelto{0}{H^{K+2}(\cjn)} &\to& \cancelto{0}{H^{K+2}(\cbjn)} &\to& H^{K+2}(\cA, Q_j) &\to&\\
 &&&&&&&&\\ 
 &\to& H^{K+3}(\cjn) &\to& H^{K+3}(\cbjn) &\to& H^{K+3}(\cA, Q_j) &\to& 0 \ ;\\ \\
\ea
\\
\ba{llllllllll}
0
 &\to& H^0(\cA, Q_j) &\to& \cancelto{0}{H^0(\bjn)} &\to&  {H^0(\cA, \cv2s\otimes \wedge^j \cN^*)} &\to&\\
 &\to& H^1(\cA, Q_j) &\to& \cancelto{0}{H^1(\bjn)} &\to&  {H^1(\cA, \cv2s\otimes \wedge^j \cN^*)} &\to&\\
 &\to& &&\vdots &&  &\to& \\
 &\to& H^{K+2}(\cA, Q_j) &\to& \cancelto{0}{H^{K+2}(\bjn)} &\to& 
 		{H^{K+2}(\cA,\cv2s\otimes \wedge^j \cN^* )} &\to& \\
 &&&&&&&&\\ 
 &\to& H^{K+3}(\cA, Q_j) &\to& H^{K+3}(\bjn) &\to& H^{K+3}(\cA,\cv2s\otimes \wedge^j \cN^* ) &\to&
0 \ .
\ea
\ea
}
\eeq
Note that since $X$ is of codimension $K$, the ambient space has
dimension $K+3$ and hence there are no non-vanishing cohomology groups
above $H^{K+3}$. Moreover,
the bundles $\cN^*$, $\cB^*$ and $\cC^*$ as well as their various tensor and wedge products are all negative and, 
hence, all their cohomologies except the highest one, namely $K+3$, vanish by Kodaira \eref{kodaira}; we have marked this explicitly
in \eref{Hwedege2V*}.

Therefore, the sequences~\eref{Hwedege2V*} immediately imply that for all $j$,
\beq\label{wedege2V*vanish}
\ba{lll}
H^i(\cA, Q_j) = 0 \ , && i=0, \ldots, K+1\ ; \\
H^i(\cA, \cv2s\otimes \wedge^j \cN^*) \simeq H^{i+1}(\cA, Q_j)  = 0 \ , && i=0, \ldots, K \ ; \\
H^{K+1}(\cA, \cv2s\otimes \wedge^j \cN^*) \simeq H^{K+2}(\cA, Q_j)  &&
\ea
\eeq
as well as two 4-term exact sequences:
\beq\label{posHwedege2V*}
{\small
\ba{l}
0 \to H^{K+2}(\cA, Q_j) \to H^{K+3}(\cjn) \stackrel{g}{\longrightarrow} H^{K+3}(\cbjn) \to H^{K+3}(\cA, Q_j) \to 0
\ ;
\\ \\
0 \to H^{K+2}(\cA, \cv2s\otimes \wedge^j \cN^*) \to H^{K+3}(\cA, Q_j) \to H^{K+3}(\bjn) \to H^{K+3}(\cA, \cv2s\otimes \wedge^j \cN^*) \to
0 \ .
\ea
}
\eeq
In \eref{posHwedege2V*} we have introduced a map $g$ which is induced from the defining map $f$ of the monad in Eq.~\eref{defV}.  
As in the previous subsection, $f$ is generic, and $g$ is so as well and thus has maximal rank. The top
sequence then implies that $H^{K+2}(\cA, Q_j)$ vanishes, and whence, by \eref{wedege2V*vanish},
$H^{K+1}(\cA, \cv2s\otimes \wedge^j \cN^*)$ vanishes as well. In other words, we have one more vanishing that
what the long exact sequences automatically guarantee.
To summarize then,  we find the vanishing cohomology groups
\beq\label{Hiwedgej}
H^{i}(\cA, \cv2s\otimes\wedge^j \cN^*)=0 \ , \qquad
\forall~i = 0, \ldots K+1, \ j = 0, \ldots, K \ .
\eeq

Equipped with these results, we can re-examine the Koszul sequence~\eref{koszul-cv2s}. It has $K+2$ terms
and we can break it up into $K$ short exact sequences, introducing (co)kernels much like we did above. Then, the vanishing of the cohomology groups
\beq\label{Hj+1wedgej}
H^{j+1}(\cA, \wedge^2\cV^* \otimes \wedge^j \cN^*) = 0 \ , \qquad
\forall~j=0, \ldots, K \ ,
\eeq
which represent a subset of the vanishing theorems~\eqref{Hiwedgej}, implies that
\beq\label{H1wedge=0}
H^1(X, \wedge^2 V^*) = 0 \ .
\eeq

We emphasise that the assumption of a generic map $f$ is crucial to arrive at this result. For rank four bundles with low-energy gauge group ${\rm SO}(10)$ it implies (see Table~\ref{spec}) that
\begin{equation}
 n_{10}=h^1(X,\wedge^2 V)=0\; ,
\end{equation}
and, hence, a vanishing number of Higgs multiplets. For rank five bundles with low-energy gauge group ${\rm SU}(5)$ we have  
\begin{equation}
 n_5=h^1(X,\wedge^2V^*)=0\; ,\quad n_{\bar 5}=-{\rm ind}(V)\; ,
\end{equation}
where Eq.~\eref{indtheorem2} has been used. This means the number of ${\bf 10}$ and $\bar{\bf 5}$ representations  is the same, forming an appropriate number of complete ${\rm SU}(5)$ families and there are no vector-like pairs of ${\bf 5}$ and $\bar{\bf 5}$ representations. The absence of Higgs multiplets in the ${\rm SO}(10)$ and ${\rm SU}(5)$ models is a phenomenologically problematic feature which was already observed in the previous Chapter (Ref.~\cite{Anderson:2007nc}). There, it was also shown that the number of Higgs multiplets can be non-zero once the assumption of a generic map $f$ is dropped. A similar situation was encountered in \cite{Bouchard:2005ag}. We expect a similar bundle-moduli dependence of the spectrum, (as first discussed in \cite{Donagi:2004qk}), for the more general class of models considered in this Chapter. 
It remains a matter of a more detailed analysis, focusing on physically promising models within our classification, to decide if a realistic particle spectrum can be obtained from such a mechanism.
\subsubsection{Singlets and $H^1(X, V \otimes V^*)$}
Finally, we need to calculate the number of gauge group singlets which correspond to the cohomology
$H^1(X, \text{ad}(V)) = H^1(X, V \otimes V^*)$. We begin by tensoring the defining sequence \eqref{dualV} for $V^*$ by $V$. This leads to a new short exact sequence
\beq
0 \to C^* \otimes V \to B^* \otimes V \to V^* \otimes V \to 0 \ .
\eeq
One can produce two more short exact sequences by multiplying
\eqref{dualV} with $B$ and $C$. Likewise, three short exact
sequences can be obtained by multiplying the original sequence~\eqref{defV}
for $V$ with $V^*$, $B^*$ and $C^*$. The resulting six
sequences can then be arranged into the following web of three horizontal
sequences $h_{I}$, $h_{II}$, $h_{III}$ and three vertical ones
$v_I$, $v_{II}$, $v_{III}$.
\beq
{\small
\ba{cccccccccl}
&&0&&0&&0&&& \\
&&\downarrow&&\downarrow&&\downarrow&&& \\
0&\to& C^* \otimes V &\to& B^* \otimes V &\to& V^* \otimes V &\to&0
\qquad &h_{I} \\
&&\downarrow&&\downarrow&&\downarrow&&& \\
0&\to& C^* \otimes B &\to& B^* \otimes B &\to& V^* \otimes B &\to&0
\qquad &h_{II} \\
&&\downarrow&&\downarrow&&\downarrow&&& \\
0&\to& C^* \otimes C &\to& B^* \otimes C &\to& V^* \otimes C &\to&0
\qquad &h_{III}  \\
&&\downarrow&&\downarrow&&\downarrow&&& \\
&&0&&0&&0&&& \\
&&v_I&&v_{II}&&v_{III}&&& \\
\ea
}
\eeq
The long exact sequence in cohomology induced by $h_{I}$ reads
\beq\label{VVseq_relabel}
{\small
\ba{lll}
0 &\to& H^0(X,C^* \otimes V) \to H^0(X,B^* \otimes V)
\to H^0(X, V^* \otimes V)\\
&\to&H^1(X,C^* \otimes V) \to
 H^1(X, B^* \otimes V)\to \fbox{\mbox{$H^1(X,V^* \otimes V)$}} \\
&\to& H^2(X,C^* \otimes V) \to \ldots \\
\ea
}
\eeq
and we have boxed the term which we would like to compute. We will also need the
long exact sequences which follow from $v_{I}$ and $v_{II}$. They
are given by
\beq
{\small
\ba{lll}
 0&\to&H^0(X,C^*\otimes V)\to H^0(X,C^*\otimes B)\to H^0(X,C^*\otimes C)\\
  &\to&H^1(X,C^*\otimes V)\to H^1(X,C^*\otimes B)\to H^1(X,C^*\otimes C)\\
 &\to&H^2(X,C^*\otimes V)\to H^2(X,C^*\otimes B)\to H^2(X,C^*\otimes C)\\
&\to&H^3(X,C^*\otimes V)\to H^3(X,C^*\otimes B)\to H^3(X,C^*\otimes C) \to 0 \\ \label{v1_relabel}
\\[0.1cm]
 0&\to&H^0(X,B^*\otimes V)\to H^0(X,B^*\otimes B)\to H^0(X,B^*\otimes C)\\ 
  &\to&H^1(X,B^*\otimes V)\to H^1(X,B^*\otimes B)\to H^1(X,B^*\otimes C)\\
 &\to&H^2(X,B^*\otimes V)\to H^2(X,B^*\otimes B)\to H^2(X,B^*\otimes C) \\
 &\to&H^3(X,B^*\otimes V)\to H^3(X,B^*\otimes B)\to H^3(X,B^*\otimes C)\to 0 \\ 
 \ea
 }
 \eeq
To make progress we need information about the cohomologies of $B^*\otimes B$, $C^*\otimes C$ and $C^*\otimes B$. For the general case this is difficult to determine since $B^*\otimes B$, $C^*\otimes C$ and $C^*\otimes B$ may contain ``mixed" line bundles with different sign or zero entries which may have non-vanishing middle cohomologies. This means in some cases there will not be sufficiently many zero entries in the above long exact sequences to compute $h^1(X,V\otimes V^*)$ without additional input, for example about the rank of maps. 

However, a general formula can be derived for all monads satisfying
\begin{equation}
 H^1(X,C^*\otimes C)=H^2(X,C^*\otimes B)=0\; . \label{vancond}
\end{equation} 
Since we can compute all line bundle cohomologies we can explicitly check for each given example whether these conditions are actually satisfied. Let us focus on cases where this is the case. Then the sequence~\eref{v1_relabel} implies that $H^2(X,C^*\otimes V)=0$ which means that \eqref{VVseq_relabel} breaks after the second line and this 6-term exact sequence implies:
 \beq
 h^1(X,V^*\otimes V)=h^1(X,B^*\otimes V)-h^1(X,C^*\otimes V)
 +h^0(X,V^*\otimes V) -h^0(X,B^*\otimes V) + h^0(X, C^* \otimes V) \; .
 \label{nsing0_relabel}
\eeq
In the above, we have used the fact that for any long exact sequence, however the number of terms, the total alternating sum of the dimensions of the terms vanishes.

We can apply a similar trick to the other 2 long exact sequences.
Using our assumption $H^1(X,B^*\otimes C)\simeq H^2(X,C^*\otimes B)=0$ in the second sequence in \eref{v1_relabel} and 
$H^1(X,C^*\otimes C)=0$ in the first sequence \eref{v1_relabel} gives the two relations
\beq
{\small
\ba{ccc}
 h^1(X,B^*\otimes V)- h^0(X,B^*\otimes V)&=&h^0(X,B^*\otimes C) -h^0(X,B^*\otimes B)+h^1(X,B^*\otimes B)\\
 h^0(X,C^*\otimes V)-h^1(C^*\otimes V)&=&h^0(X,C^*\otimes B)-h^0(X,C^*\otimes C)-h^1(X,C^*\otimes B)\; .
\ea
}
\eeq
Inserting these into Eq.~\eref{nsing0_relabel} and using the fact that for a stable $SU(n)$ bundle $V$,
$h^0(X,V \otimes V^*) = 1$ (cf. Section 4.2 of \cite{Anderson:2007nc}) gives the final result:
\beq
{\small
\ba{lll}
 n_1=h^1(X,V^*\otimes V)&=&h^0(X,B^*\otimes C)-h^0(X,B^*\otimes B)-h^0(X,C^*\otimes C)\\
 &&+h^0(X,C^*\otimes B)-h^1(X,C^*\otimes B)+h^1(X,B^*\otimes B)+1 \label{n1res}
\ea
}
\eeq
for the number of singlets. We emphasise that this is result is valid provided the monad satisfies the two conditions~\eref{vancond}. 
In this case, Eq.~\eref{n1res} allows an explicit calculation of the number of singlets from the known line bundle cohomologies.

As an example, we consider the manifold 
$\left[
\begin{array}
[c]{c}
1\\
3
\end{array}
\left|
\begin{array}
[c]{ccc}
2 \\
4
\end{array}
\right.  \right]  $, and the rank 4 monad bundle defined by
\beq
B=\cO_X(1,1)^{\oplus 6} \oplus \cO_X(2,1)^{\oplus 2}\; ,\quad C=\cO_X(2,3)^{\oplus2} \oplus \cO_X(3,1)^{\oplus2} \; .
\eeq
It can be checked from the known line bundle cohomologies that this bundle indeed satisfies the conditions~\eqref{vancond}. The number of singlets, calculated from Eq.~\eqref{n1res}, is then given by $n_1=241$.

For bundles which do not satisfy \eqref{vancond} other methods can be employed. In favourable cases, the cohomologies of $B^*\otimes B$, $C^*\otimes C$ and $C^*\otimes B$ may have a different pattern of zeros which still allows the derivation of a formula for $n_1$ analogous to Eq.~\eqref{n1res} by combining appropriate parts of the sequences~\eqref{h1}, \eqref{v1} and \eqref{v2}. If this is not possible one has to resort to ambient space methods and Koszul resolutions in combination with our results for the ranks of maps in Leray spectral sequences explained in Chapter \ref{c:line_bundle_coho}. Here, we will not present such a calculation which is likely to be complicated and, if required at all, should probably be only carried out for physically promising models. However, we stress that all the necessary technology is available so that the number of singlets can, not just in principle but in practice, be obtained for all positive monads on favourable CICYs.
\section{Conclusions and Prospects}
In this Chapter, we have analysed positive monad bundles with structure group ${\rm SU}(n)$ (where $n=3,4,5$) on favourable CICY manifolds in the context of $N=1$ supersymmetric compactifications of the $E_8\times E_8$ heterotic string. We have shown that the class of these bundles, subject to the heterotic anomaly condition, is finite and consists of $7118$ examples. More specifically, we find that these $7000$ monads are concentrated on only $36$ CICYs. All other of the $4500$ or so CICYs do not allow positive monads which satisfy the anomaly condition. As a non-trivial test for the stability of these bundles we have shown that $H^0(X,V)=H^3(X,V)=0$ for all examples. A systematic method of analyzing stability will be presented in Chapter \ref{StabCh}. We have also shown how to calculate the complete particle spectrum for these models. In particular, we found that the number of anti-families always vanishes so that there are no vector-like family anti-family pairs present in any of the models. For low-energy groups ${\rm SO}(10)$ and ${\rm SU}(5)$ ($n=4,5$) the number of Higgs fields vanishes at generic points in the bundle moduli space. However, as shown in Ref.~\cite{Anderson:2007nc}, for non-generic values of the bundle moduli, Higgs multiplets can arise. The details of this moduli-dependence of the spectrum have to be analysed for specific models, preferably focusing on physically promising examples. We have also shown that the number of gauge singlets can be calculated, in many cases in terms of a generic formula, or else by applying more elaborate methods\footnote{A crucial technical result, essential for many of these derivations, is the calculation of line bundle cohomology for all line bundles on favourable CICYs presented in Chapter \ref{c:line_bundle_coho}.}.  Based on the results for the particle spectrum, we have scanned the $7118$ bundles imposing two rudimentary physical conditions. First, the number of families should equal $3k$ for some non-zero integer $k$, so there is a chance to obtain three families after dividing by a discrete symmetry of order $k$. In addition, the Euler number of the Calabi-Yau space should be divisible by $k$. It turns out that only $559$ out of the $7118$ bundles pass this basic test. If, in addition, one demands that the order $k$ of the symmetry does not exceed $13$ one is left with only $21$ models. 

This drastic reduction of the number of viable models due to a few basic physical constraints is not uncharacteristic and has been observed in the context of other string constructions~\cite{Gmeiner:2005vz}. In our case, the main reason for this reduction is the relatively large values for the Euler characteristic of our models (roughly, a Gaussian distribution with a maximum at about $60$, see Fig.~\ref{f:c3pos}) in conjunction with the empirical fact that large discrete symmetries of Calabi-Yau manifolds are hard to find. In order to make this statement more precise a systematic analysis of discrete symmetries $\Gamma$ on CICYs $X$ (which lead to a smooth quotient $X/\Gamma$) has to be carried out and the results of this analysis have to be combined with the results of the present Chapter. We are planning to carry this out explicitly in the near future. However, even in the absence of such a classification of discrete symmetries we find it likely that the vast majority of positive monads will fail to produce three-family models on $X/\Gamma$ given the large number of families on the ``upstairs" manifold $X$. 

These large numbers are, of course, directly related to the property of positivity. An obvious course of action is, therefore, to relax this condition and also allow zero or even slightly negative integers $b_i^r$ and $c_j^r$ in the definition~\eqref{monad} of the monad. The number of these non-positive monads is vastly larger than the number of positive ones and it turns out the distribution of their Euler characteristics is peaked at smaller values, as expected. Crucially, as will be shown in Ref.~\cite{semi-positives,stability}, some of these non-positive monads are still stable and, hence, lead to supersymmetric models. We, therefore, believe that the generalisation to non-positive monads is a crucial step towards realistic models within this framework and work in this direction is underway~\cite{semi-positives}.

\chapter{Cohomology of Line Bundles on CICYs}\label{c:line_bundle_coho}

In order to use the monad construction of vector bundles to build heterotic models, it is essential to understand the properties of vector bundles on complete intersection Calabi-Yau spaces. One of the most critical properties for the computation of physical properties is the cohomology of $V$ and its dual and tensor powers (see sections \ref{het_spectra} and \ref{s:het}). For monad-defined bundles, \eref{monad}
the cohomology of $V$ is clearly determined by the cohomologies of its defining line bundles. So, in order to determine $H^*(X,V)$ and hence the physical particle spectra of the heterotic models we first must determine the cohomology of line bundles on CICYs. General results regarding properties of bundle cohomologies exist in the mathematics literature, but very few techniques have been developed for explicit computation. Since this is critical to the development of our program of heterotic model building, we will take this section to carefully review the existing approaches and add to these new tools for determining bundle cohomology.

\section{The Koszul Resolution}\label{koszul_rev}
The standard method of computing the cohomology of a vector bundle
$V= \cV|_X$  coming from the restriction of $\cV$ from an ambient space $\cA$ to
the variety $X$ is the so-called {\it Koszul Resolution} of
$V|_X$. In general, if $X$ is a smooth hypersurface of co-dimension $K$, which is the zero locus of a holomorphic section $s$ of the bundle $N$, then the following exact sequence exists \cite{AG1,AG2}:
\beq\label{koszulA}
0 \to \cV \otimes \wedge^K N_X^* \to \cV \otimes \wedge^{K-1} N_X^*
\to \ldots \to \cV \otimes N_X^* \to \cV \to \cV|_X \to 0 \ .
\eeq
Thus, if the cohomology of the bundles $\wedge^j N^* \otimes \cV$ are known on the ambient space, we can use the Koszul sequence to determine the cohomology of $V|_X$. Here, $\cN_X^*$ is the dual to the normal bundle. We recall that for a CICY, the normal bundle to the space is given by the configuration matrix \eref{cy-config}:
\beq\label{normal2}
\cN_X = \bigoplus_{j=1}^K \cO(q_j^1, \ldots, q_j^m) \ .
\eeq
In the above, we have generalized the standard notation that
$\cO_{\IP^n}(k)$ denotes the line-bundle over $\IP^n$ 
whose sections are degree $k$ polynomials in the coordinates of $\IP^n$; that is, $\cO(q^j_1,
\ldots, q^j_m)$ is the line-bundle over $\IP^{n_1} \times \ldots
\times \IP^{n_m}$ whose sections are polynomials of degree $q^j_1,
\ldots, q^j_m$ in the respective $\IP^{n_i}$-factors. Being a direct
sum, the rank of $\cN_X$ is $K$. 

We can break the sequence \eref{koszulA} into a series of short exact sequences as 
\bea
0 \to \cV \otimes \wedge^K N_X^* \to \cV \otimes \wedge^{K-1} N_X^*
\to \mathcal{K}_1 \to 0 \\
0 \to \mathcal{K}_1 \to \cV \otimes \wedge^{K-2} N_X^* \to \mathcal{K}_2 \to 0 \\
\ldots \\
0 \to \mathcal{K}_{K-1} \to \cV \to \cV|_X \to 0
\eea
and each of these short exact sequences will give rise to a long exact sequence in cohomology:
\bea \label{long_exact}
0&\to& H^0(\cA, \cV \otimes \wedge^K N_X^*) \to H^0(\cA, \cV \otimes \wedge^{K-1} N_X^*)
\to H^0(\cA, \mathcal{K}_1) \\
0 &\to& H^0(\cA,\mathcal{K}_1) \to H^0(\cA, \cV \otimes \wedge^{K-2} N_X^*) \to H^0(\cA, \mathcal{K}_2) \to \ldots \\
\ldots \\
0 &\to& H^0(\cA, \mathcal{K}_{K-1}) \to H^0(\cA, \cV) \to H^0(X, \cV|_X) \to \ldots
\eea
To find $H^*(X, \cV|_{X})$ we must determine the various cohomologies in \eref{long_exact}. It is easy to see that for higher co-dimensional spaces or tensor powers of bundles, this decomposition of sequences is a laborious process. Fortunately, the analysis of these arrays of exact sequences is dramatically simplified by the use of spectral sequences. Spectral sequences are completely equivalent to the collection of exact sequences described above, but designed for explicit cohomology computation. Since there are many good reviews of spectral sequence available in the literature \cite{hubsch,Distler:1987ee, AG1,AG2}, we will only discuss the essential features in the following paragraphs.
\section{The Spectral Sequence}\label{s:leray}
To obtain the necessary cohomology of $V|_X$ from \eref{koszulA}, we define a tableaux
\beq\label{leray}
E^{j,k}_{1}(V) := H^j(A,  V \otimes \wedge^{k} N_X^*), \qquad
k = 0, \ldots, K; \  j=0, \ldots, \dim(A) = \sum_{i=1}^m n_i \ .
\eeq
This forms the first term of a so-called {\it Leray spectral sequence} \cite{AG1,AG2}. A Leray sequence is a complex defined by differential maps $d_i : E^{j,k}_i \to
E^{j-i+1,k-i}_i$ for $j = 1,2,\ldots$ {\it ad infinitum} where $d_i \circ d_i =0$. The subsequent terms in the spectral sequence are defined by
\beq \label{leray_iterate}
E^{j,k}_{i+1}(V)= \frac{ker(d_{i}: E^{j,k}_{i}(V) \rightarrow E^{j-i+1,k-i}_{i}(V))}{Im(d_{i}: E^{j+i-1,k+i}_{i}(V) \rightarrow E^{j,k}_{i}(V))}
\eeq
Since the number of terms in the Koszul sequence  \eref{koszulA} is finite, there exists a limit to the spectral sequence. That is, the sequence of tableaux converge after a finite number of steps to $E^{j,k}_{\infty}(V)$. The actual cohomology of the bundle $V$ is constructed from this limit tableaux:
\beq\label{hodge_converge}
h^q(X, V|_X) = \sum^{K}_{m=0} \text{rank} E^{q+m,m}_{\infty}(V) \ .
\eeq
where $h^q(X, V|_X)= \text{dim} (H^q(X, V|_X))$. 

In practice, the tableaux $E^{p,q}_r$ converges fairly rapidly because many of its entries will turn out to be zero and the associated maps $d_i$, vanish; hence the spectral sequence converges after only a few steps. However, the reader will  have noticed that in general, all computations which involve long exact cohomology sequences \eref{long_exact} and Leray spectral sequences \eref{leray_iterate} rely upon the ability to discern the action of maps between cohomologies on the ambient space $\mathcal{A}$. In fortunate cases, the tableau are sufficiently sparse that is possible to determine the required cohomologies without knowing any maps explicitly. But in general the obstacle cannot be avoided. Fortunately, this otherwise impossible task of computing the rank and kernels of the Leray maps can be accomplished straightforwardly for some bundles using the coset representation of Flag spaces and the tensor algebra associated with representations of Lie groups. \cite{barton,hubsch}. 

\section{Cohomology of line bundles on CICYs}
Up to this point, our comments on bundle cohomology has been general. However, to proceed further we now focus on the monad bundles which form the basis of this work. As stated above, the cohomology of a monad bundle $V$ is determined by the cohomology of direct sums of line bundles in its defining exact sequence \eref{monad}. The question therefore becomes, is it possible to fully determine the cohomology of line bundles on CICYs? As we saw in section \ref{s:line}, it is possible to determine the cohomology of strictly positive (and negative) line bundles of the form $\cO_{X}(k_{1},k_{2},...,k_{m})$  with $k_{r} < 0$ using the techniques of Koszul and Leray spectral sequences. However, for a generic line bundle with $k_{r}$ positive, negative or zero, we need new techniques. The most important of these is a computational variation on the Bott-Borel-Weil theorem \cite{hubsch} which we turn to next.

\subsection{Flag Spaces the Bott-Borel-Weil Theorem}

It can be shown that every simply connected compact homogeneous complex space is homeomorphic to a torus-bundle over a product of certain coset spaces $G/H$, where $G$ is a compact simple Lie group and $H$ is a regular semi-simple subgroup. Such spaces are known as C-spaces or 'generalized flag varieties' \cite{hubsch}. In fact, the simplest example of this is $\mathbb{P}^{n}= (\frac{U(n+1)}{U(1) \times U(n)})$. Viewing $\mathbb{P}^n$ in this way will prove useful to us since it can be shown that homogeneous holomorphic vector bundles over such flag varieties, $\mathbb{F} = (G_{\mathbb(C)}/H)$, are labeled by representations of $H$ (for our applications, $H=U(1) \times U(n)$). This will provide us with a powerful new tool to investigate bundle cohomology on CICYs.

Recalling that a representation can be written as a direct sum of irreducible ones, we can focus on irreducible homogeneous holomorphic vector bundles. Further, we know that such representations are uniquely labeled by their highest weight, so we have a convenient notation for such bundles. For this purpose, we will use the well-known Young tableaux (see e.g. \cite{georgi}. We will be dealing strictly with unitary groups and will adopt the following conventions. To denote a bundle, we write $(a_{1}, \ldots, a_{n})$, where $a_r \leq a_{r+1}$ is the number of boxes in the rth row of the tableau. For $a_r >0$ ($<0$) the boxes are arrayed to the right (left) of the 'spine'. Therefore, in the standard tensorial notation, $(-1,0, \ldots 0)$ denotes a covariant vector $v_{\mu}$ while $(0, \ldots ,0,1)$ labels the contravariant vector $v^{\mu}$. All other representations can be obtained from these by multiplication and then decomposition into a direct sum of irreducible components through symmetrizing, anti-symmetrizing and taking traces with the invariant tensor (${\delta}^{\mu}_{\nu}$). A tensor product of representations of factor $U(n_{f})$'s can be written as the Young tableau,
\beq
(a_{1}, \ldots a_{n_1} | b_{1}, \ldots , b_{n_2} | \dots |d_{1}, \ldots , d_{n_F})
\eeq
or for a more condensed notation, we can stack the partitions vertically on top of each other. 

For the case of line bundles, we recall that we may view $\mathbb{P}^n$ as the space of all lines $L \approx \mathbb{C}^1$ through the origin of $\mathbb{C}^{n+1}$. Each line is defined as the zero set of some linear polynomial $l(x)$ over $\mathbb{C}^{n+1}$. Now, from the definition of the hyperplane bundle $\cO(1)$ on $\mathbb{P}^n$ as the line bundle whose (global holomorphic) sections are linear polynomials we may formulate a line bundle in the language of flag spaces above. Viewing $\mathbb{P}^n$ as a quotient of unitary groups and a bundle over it as a representation of $U(1) \times U(n)$, a little thought reveals that we may denote $\cO(1)$ as $(-1|0, \ldots 0)$ (and similarly, its dual bundle $\cO(-1)$ is written $(1|0, \ldots 0)$). 

With this notation in hand, let $\mathbb{F}= \frac{U(N)}{U(n_1) \times \ldots \times U(n_F)}$ (with $N = \sum_{f} n_{f}$) be a flag space as above and $V$ be a holomorphic homogeneous vector bundle over it. Then
\begin{theorem}{Bott-Borel-Weil} \label{BBW}

(1) Homogeneous vector bundles $V$ over $\mathbb{F}$ are in 1-1 correspondence with the $U(n_1) \times ...\times U(n_F)$ representations.

(2) The cohomology $H^{i}(\mathcal(A,V)$ is non-zero for at most one value of $i$, in which case it provides an irreducible representation of $U(N)$, $H^{i}(\mathbb{F},V) \approx (c_1,...,c_N)\mathcal{C}^N$.

(3) The bundle, $(a_1,...,a_{n_1}|...|b_1,...,b_{n_F})$, determines the cohomology group $(c_1,...,c_N)$, according to the following algorithm:

1. Add the sequence $1..., N$ to the entries in $(a_1,...,a_{n_1}|...|b_1,...,b_{n_F}) $.

2. If any two entries in the result of step 1 are equal, all cohomology vanishes; otherwise proceed.

3. swap the minimum number ($=i$) of neighboring entries required to produce a strictly increasing sequence.

4. Subtract the sequence $1,...N$ from the result of $3$, to obtain $(c_{1},c_{2},...,c_{N})$.
\end{theorem}

Using this algorithm, it is straightforward to reproduce the Bott-formula \cite{AG1,AG2,hubsch} for cohomology of line bundles on single projective spaces by simply counting the dimensions of the the associated Young tableau $(c_{1},c_{2},...,c_{N})$ of the unitary representations. The result is 

\beq\label{bottformula}
h^{q}(\mathbb{P}^{n}, \mathcal{O}_{\IP^n}(k))=\left\{
\begin{array}
[c]{ll}%
\binom{k+n}{n} & q=0\quad k>-1\\
1 & q=n\quad k=-n-1\\
\binom{-k-1}{-k-n-1} & q=n\quad k<-n-1\\
0 & \mbox{otherwise}
\end{array}
\right. \ .
\eeq

where the binomial coefficients arise from the dimensions of Young tableau\footnote{See \cite{georgi} for a review of the hook-length formulas.}. 

The computation of line bundle cohomology described by the Bott-Borel-Weil theorem is easily generalized to products of projective space using the K\"unneth formula \cite{AG1,AG2} which gives the cohomology of bundles over a direct product of
spaces. For products of projective spaces it states that:
\beq\label{kunneth}
H^n(\IP^{n_1} \times \ldots \times \IP^{n_m}, \cO(q_1, \ldots, q_m)) =
\bigoplus_{k_1+\ldots+k_m = n} H^{k_1}(\IP^{n_1},\cO(q_i)) \times
\ldots \times H^{k_m}(\IP^{n_m},\cO(q_m)) \ ,
\eeq

With this in hand, we can compute the cohomology of line bundles over the ambient space. For example, in the notation of flag varieties, the line bundle $l = \mathcal{O}(k_{1},-k_{2})$ on $\mathbb{P}^{n_1} \times \mathbb{P}^{n_2}$ (with $k_{2} \geq n_{2}+1$) can be denoted by a product of irreps of $(U(1) \times U(n_{1})) \times (U(1) \times U(n_{2}))$:
\beq
l \sim \binom{-k_{1}|0,\ldots ,0}{{~k_2}|0, \ldots, 0}
\eeq
where there are $n_1$ zeroes in the first row and $n_2$ zeroes in the second. Using Bott-Borel-Weil and the Kunneth formula then, the cohomology of this line bundle on the ambient space would be described by 
\beq \label{sample_coho}
h^{n_{2}}(\cA , l) \sim \binom{-k_{1},0,\ldots ,0}{1, \ldots, 1,({k_2}-n_{2})}
\eeq
where $(-k_{1},0,\ldots ,0)$ denotes the Young Tableau of a irreducible representation of $U(n_{1} +1)$,  $(1, \ldots, 1,({k_2}-n_{2}))$ is the Young tableau of a $U(n_{2}+1)$ irrep and the Kunneth product of the restricted cohomologies is denoted by the vertical stacking of tableau. We recall that the dimension of a Young tableau may be easily computed from the hook-length formula (see \cite{georgi}, for example). For instance, the dimension of $(-k_{1},0,\ldots ,0)$ is just the degrees of freedom in a totally symmetric tensor in $(n_{1}+1)$ variables, namely $\binom{k_{1} + n_{1}}{n_{1}} $. In counting the degrees of freedom in the tableau $(1, \ldots, 1,({k_2}-n_{2}))$, it is useful to recall that the totally anti-symmetric tensor, $\epsilon^{[a,...,b]}$ is a singlet under $U(n)$. Thus we can strip a Levi-Civita tensor from the tableau $(1, \ldots, 1,({k_2}-n_{2}))= (1, \ldots, 1) \otimes (0, \ldots , 0,({k_2}-n_{2}-1))$ and just consider the dimension of $(0, \ldots , 0,k{_2}-n_{2}-1)$ which is yet another symmetrized tensor whose degrees of freedom may be counted as before. Therefore, the total cohomology/tableau $\binom{-k_{1},0,\ldots ,0}{1, \ldots, 1,({k_2}-n_{2})}$ has dimension $\binom{k_{1} + n_{1}}{n_{1}} \times \binom{k_{2}-1}{n_{2}}$. 

To simplify future calculations, we introduce a short-hand notation for the dimension of a space of symmetric tensors in a product of projective spaces (${\mathbb{P}}^{n_1} \times \ldots {\mathbb{P}}^{n_F}$). If we list the numbers of symmetrized indices in each $\mathbb{P}^{n_j}$ as a vector $(r, \ldots, s)$ then the dimension of the tableau shall be denoted by
\beq\label{dims}
[r, \ldots, s] \equiv \binom{r+n_1}{ n_1} \times \dots \times \binom{s+n_F}{ n_F}
\eeq
In this notation, the dimension of the cohomology \eref{sample_coho} can be written as $[k_{1}, k_{2} -n_{2}]$

In summary, by using the Bott-Borel-Weil theorem we are able to represent the cohomologies of line bundles over the ambient space $\mathcal{A}$ as irreducible representations of unitary groups (and readily compute their dimensions). Returning to the task of computing the line bundle cohomology on the Calabi-Yau $3$-fold, $X$, we note that this technique will dramatically simplify the Leray spectral sequence calculations of the previous section by  providing a simple representation for the ambient space cohomology groups involved. We will reduce the abstract task of determining the properties of maps between line bundle cohomology groups to the more straightforward one of studying maps between irreps of unitary groups. 
\subsection{Computing the Ranks and Kernels of Leray maps}
 We return now to the Leray tableau of Section \ref{s:leray} and the task of computing the kernels and ranks of the maps in \eref{leray_iterate}. When computing the cohomology of line bundles on CICYs, it is easy to see that the spectral sequence converges after only two steps (i.e. $E_{2}^{j,k} = E_{\infty}^{j,k} $) and in order to explicitly determine the cohomology, we must consider maps of the form
 \beq \label{main_map}
{d_1}^k : E^{j,k}_{1}(V) \to E^{j,k-1}_{1}(V)
\eeq
 where $E^{j,k}_1 =H^j(\cA , \wedge^k N^* \otimes V)$ can be written as a direct sum of products of irreps of unitary groups \ref{BBW}. 
 
 Considering \eref{main_map} for $V$ a direct sum of line bundles, we find that the tensor representations involved are simply of the form $(-m,0, \ldots ,0)$ (or $(0, \ldots ,0,m)$). That is, a line bundle $\cO(k) $ ($k>0$)over $\mathbb{P}^n$ generates a cohomology group that can be represented as a totally symmetrized contravariant (or covariant) tensor, $\lambda ^{(a, \ldots b)}$ with $k$ indices (where $a,b=1 \ldots ,n+1$) and the round brackets $(a,...,b)$ denoting symmetrization.  As above, the degrees of freedom in $\lambda$ are given by $[k]=\binom{k+n}{n}$
 
 For a line bundle $\cO(k_{1}, \ldots,k_{n_{F}})$ over $\mathbb{P}^{n_1} \times \dots \times \mathbb{P}^{n_F}$ with $k_{i} >0$ its cohomology will be given by the symmetric tensor
\beq
\lambda_{(a, \ldots, b) \dots (\alpha, \ldots, \beta)}
\eeq
where each index-type runs over the coordinate range of a single $\mathbb{P}^{n_i}$ and there are a total of $k_1$ $a,b$-type indices running ($1 \dots n_{1}+1$), and $k_{n_F}$ $\alpha , \beta$-type indices, etc. While our results hold for arbitrary products of projective space $\mathbb{P}^{n_1} \times \ldots \mathbb{P}^{n_2}$, to simplify notation in the following sections, we will present our results in a product of two such spaces $\mathbb{P}^{n_1}\times \mathbb{P}^{n_2}$. To generalize the formulae and discussion to higher product ambient spaces simply replace the words ``bi-degree" with ``multi-degree'' and dimensions $[r,s]$, with $[r,s,u,\ldots]$, as appropriate.

\subsubsection{Individual Maps}
To begin our analysis of cohomology maps, consider a map between the cohomology groups of two individual line-bundles, $f: V_1 \rightarrow V_2$, with an ambient space, $\cA = \mathbb{P}^{n_1} \times \mathbb{P}^{n_2}$. Written as a map between tensors this is
\beq \label{tensor_map}
{f _{(a, \ldots , b)}}^{(\alpha, \ldots , \beta)} \lambda{^{(a, \ldots,b,c \ldots d)}}_{(\alpha, \ldots, \beta ,\sigma, \ldots \gamma)}={\tau^{(c, \ldots, d)}}_{(\sigma, \ldots , \gamma)}
\eeq

where $\lambda \in V_{1}$ and $\tau \in V_{2}$. 

To proceed, we are interested in computing $ker(f)$:
\beq \label{ker}
{f _{(a, \ldots , b)}}^{(\alpha, \ldots , \beta)} \lambda{^{(a, \ldots,b,c \ldots d)}}_{(\alpha, \ldots, \beta ,\sigma, \ldots \gamma)}=0
\eeq
By writing these tensors explicitly, we have reduced the problem to an exercise in linear algebra. 

In this work, we will consider the maps $f$ to be {\emph{generic}} and view \eref{ker} as a linear system determining components of $\lambda$. This linear system is then $dim(V_{2})$ equations in $dim(V_{1})$ unknowns. Therefore,
\beq
ker(f) = dim(V_{1}) - dim(V_{2})
\eeq 
We now make an observation about the representations involved. In each map $f$ of the form above, $f$ is itself a tensor with totally symmetric indices (of several different types) that acts on other symmetric tensors. 

There is an easy way to think about symmetric tensors. Note that a symmetric tensor $\lambda_{(a \ldots b)}$ with $k$ indices running $1 \dots n$ is in one-to-one correspondence with a homogeneous polynomial of degree $k$ in n variables. Simply stated, $\lambda$ can be viewed as the ``coefficients" of the polynomial. 
\beq
\lambda_{(a \ldots b)}x^{a} \dots x^{b}
\eeq
That is, the number of independent homogeneous polynomials of degree ``k" is equal to the degrees of freedom in $\lambda$. For a tensor involving several Young Tableau such as $\lambda_{(a, \ldots, b) \dots (\alpha, \ldots, \beta)}$ this is equivalent to a polynomial of a certain multi-degree over several different variable types (i.e. $\lambda_{(a, \ldots, b) \dots (\alpha, \ldots, \beta)}x^{a} \dots x^{b}y^{\alpha} \dots y^{\beta}$). Using this notation, we can specify the number of indices in $\lambda$ or the degree of the polynomial as a vector $(r_1, \ldots, r_m)$. 

Instead of considering a mapping between two tableau we can then consider maps between homogeneous polynomials. For instance, the mapping in $\mathbb{P}^{n_1} \times \mathbb{P}^{n_2}$
\beq\label{toref1}
f ^{(a, \ldots , b (\alpha, \ldots , \beta} \lambda^{c \ldots d) \sigma, \ldots \gamma)}= \tau^{(a, \ldots, d)(\alpha, \ldots , \gamma)}
\eeq
where the map, $f$, has $k_1$ a-type indices and $k_2$ $\alpha$-type indices, $\lambda$ has $(r_{1},r_{2})$ $a,\alpha$ indices respectively and $\tau$ has $(r_{1} + k_{1}, r_{2} + k_{2})$ indices can be written as
\beq\label{toref2}
P_{f}P_{\lambda} = P_{\tau}
\eeq
where $P_{\lambda}$ is a homogeneous polynomial of bi-degree $(r_{1},r_{2})$, $P_{\tau}$ is the polynomial of bi-degree $(r_{1} + k_{1}, r_{2} + k_{2})$ and even the map between cohomologies, $P_f$ itself, is a homogeneous polynomial of bi-degree $(k_{1},k_{2})$.

Finally, note that in the language of polynomials described above, index contraction as in \eref{tensor_map} can be thought of as polynomials containing derivative operators. That is ${f _{(a, \ldots , b)}} \lambda{^{(a, \ldots,b,c \ldots d)}}$ corresponds to $P_{f}P_{\lambda}$ where
\beq\label{derivs}
P_{f} = f^{(a\ldots b)}\partial x_{a} \dots \partial x_{b}
\eeq
acting on the regular polynomial $\lambda^{(a, \ldots,b,c \ldots d)}x^{a} \ldots x^{b}$. 

In the calculation of Leray tableau and line bundle cohomology, we note that the cohomology groups of line bundles on $\cA$ will always be symmetric polynomials of given bi-degree in the coordinates of $\cA$. The maps $d_i$ given in \eref{leray_iterate} however, will generally be polynomials in both coordinates and derivative operators.

\subsection{Symmetric Tensors and Polynomials}
We shall begin by introducing some notion for the ranks of the maps described above. Let us denote by $(k,m)$ the bi-degree of a map, where negative numbers refer to derivative operators and positive entries to polynomial multiplication. Through the following cases we introduce the notation for the dimension of a space of symmetric tensors in a product of two projective spaces (${\mathbb{P}}^{n_1} \oplus {\mathbb{P}}^{n_2}$) of bi-degree $(r,s)$
\beq
[r,s]= \binom{r+n_1}{ n_1} \binom{s+n_2}{ n_2}
\eeq
We turn now to several distinct distinct varieties of map.

\subsubsection{The Maps}\label{big_hairy_maps}
Below we distinguish the sub-cases necessary to analyze a general Leray tableau for line bundle cohomology.
\paragraph{Case 1: Mapping one space of polynomials to another}Consider $f: V_1 \rightarrow V_2$ where $V_1$ is a space of polynomials of bi-degree $(l - k, p - m)$, $V_2$ of bi-degree $(l,p)$ and $f$ a polynomial map of degree $(k, m)$. In this case,
\beq
ker(f) = Max[0, [l-k,p-m] -[l,p]]
\eeq
where each of $k,m$ can be positive, negative or zero. This is simply the relationship $ker(f)=$ ``number of unknowns - number of equations". In this case, a single polynomial map is always injective and a single derivative map is always surjective.

\paragraph{Case 2:  ``Many-to-one"}
Consider the case where $F: (V_1 \oplus V_2 ....\oplus V_N) \rightarrow V$ and where the map $f$ is decomposable into component maps, $f_i : V_i \rightarrow V$. Let $V$ be composed of polynomials of bi-degree (r,s), then $dim(V)=[r,s]$ in the notation above. Let the component maps have bi-degree $(k_i,m_i)$. Then, as in the above case, $ker(f_i)= Max[0, dim(V_i)-dim(V)]$. There are two sub-cases to consider here.
We will investigate the image of the map $F$ above. To begin, we can make several observations simply from linear algebra. First, we observe that if $f_j$ is surjective for any $j$, then the total map, $F$ is surjective. That is if $ker(f_j)>0$ for any $j$ then 
\beq\label{surj}
ker(F) = Max[0, \sum_i dim(V_{i}) -dim(V)] =Max[0, \sum_i dim(V_{i}) -[r,s]]
\eeq
Further, we have that the dimension, $im(F)$, of the image, $Im(F)$, is given by
\beq\label{image}
im(F) = {\sum_i}^{N}im(f_i) - \displaystyle\sum_{i<j}im(f_i)\cap im(f_{j}) + \displaystyle\sum_{i<j<l}im(f_i)\cap im(f_j)\cap im(f_l)  \ldots 
\eeq
That is, the dimension of image in $V$ of the composite map $F$ is simply the sum of the images of the individual maps $f_i$, subtracting off the overcounting of `overlapping' pairs of images (and triples, quadruples, etc.) Thus, to compute Im(F) we need only understand how to count the intersection of two such images, i.e.
\beq
im(f_i)\cap im(f_{j})
\eeq 
For the case of $F: (V_1 \oplus V_2 ....\oplus V_N) \rightarrow V$, we introduce the notation $dim(V)=[r,s]$ and the map $f_i$ of bi-degree $(k_i,m_i)$ (where $k_i,m_i$ can be positive, negative or zero according to the tensor structure in \eref{tensor_map}). We observe that if $f_{i}$ (and/or $f_j$) is surjective then, $im(f_i)\cap im(f_{j})=im(f_{j})$. That is, if at least one of the maps is surjective then it must follow that the intersection of images yields the smaller of the two. But what if both maps are injective? (i.e. $ker(f_i)=ker(f_j)=0$). In this case, respresenting the problem in terms of polynomials, the intersection is given by
\beq
P_{f_{i}}P_{v_{i}}=P_{f_{j}}P_{v_{j}}
\eeq
But by inspection, a polynomial equation of this type has a solution:
\beq
P_{v_{i}}=P_{f_{j}}P_{{\sigma}_{ij}} \quad P_{v_{j}}=P_{f_{i}}P_{{\sigma}_{ij}}
\eeq
where $P_{\sigma_{ij}}$ is a polynomial of bi-degree $(r-k_{i}-k_{j},s-m_{i}-m_{j})$ (and hence $P_{\sigma_{ij}}=P_{\sigma_{ji}}$). Note that $P_{\sigma_{ji}}$ is a reasonable answer so long as $[r-k_{i}-k_{j},s-m_{i}-m_{j}] \leq Min[im(f_{i}),im(f_{j})]$. Thus, we can consolidate these observations, and state that this pair-wise intersection takes the form
\beq\label{im_intersec}
im(f_i)\cap im(f_{j})=\left\{
\ba{ll}
~Min[im(f_{i}),im(f_{j})]=\tau & ~\mbox{if at least one of}~f_{i}, f_{j}~ \mbox{surjective} \\
~Min[\tau,[r-k_{i}-k_{j},s-m_{i}-m_{j}] ]&~\mbox{both}~f_{i},f_{j}~\mbox{injective}
\ea
\right. 
\eeq

More generally for triple and higher intersections of images, we find that a polynomial solution to $im(f_i)\cap im(f_{j})\cap im(f_{l})\ldots$ takes the form $[r-k_{i}-k_{j}-k_{l}\ldots,s-m_{i}-m_{j}-m_{l}\ldots]$. In general then, if $ker(f_i)=0$ for all $i$, we find that the dimension of the kernel of $F$ is:
\beq\label{many_to_one}
ker(F) = Max[0, \sum_i dim(V_i) -dim(V) , \sum_{a=2}^{N} \sum_{|M|=a} (-1)^a [r -\sum_{b \in M} k_b][s - \sum_{b \in M} m_b]]
\eeq
where $M$ is a subset of $\{1,2, \ldots N\}$ and $|M|$ denotes the length of the subset. 

To see that this line of reasoning produces the correct results, we shall demonstrate explicitly that $F$ is surjective according to \eref{image} if all of the component maps $f_l$ are. With $im(f_l)=dim(V)$ for all $l$, it follows that $im(f_l)\cap im(f_m)=dim(V)$ for all $l,m$. So clearly, the alternating sum of intersections in \eref{image} is
\beq
\ba{lcl}
im(F)& = &(N) dim(V) -(\text{no. of pairs})dim(V) +(\text{no. of triples})dim(V) - \dots \\
~& =& \{ N-\binom{N}{2}+\binom{N}{3}-\binom{N}{4} +\dots \}dim(V) \\
~& =& \{N -\displaystyle\sum_{j=2}^{N}(-1)^{j}\binom{N}{j} \}dim(V)
\ea
\eeq
but, the binomial series satisfies $\sum_{j=0}^{N}(-1)^{j}\binom{N}{j}=0$ (with $\binom{N}{0} \equiv 1$). Therefore,
\beq
im(F)=(n-(n-1))dim(V)=dim(V)
\eeq
as required.
 
\paragraph{ Case 3: ``One-to-Many"}
Next, we consider the case where $F:  V \rightarrow (V_1 \oplus V_2 ....\oplus V_N)$ and where the map $f$ is decomposable into component maps, $f_i : V \rightarrow V_i$. Let $V$ be composed of symmetric polynomials of bi-degree (l,p), then ($dim(V)=[l,p]$). Let the component maps have bi-degree $(k_i,m_i)$. Then, as in the above case, $ker(f_i)= Max[0, dim(V)-dim(V_i)]$. There are again two sub-cases to consider here.

First, we observe that in this case, the kernel of the map $F$ is simply the intersection of all the kernels of the component maps,
\beq 
\ker(F) = \bigcap_{i}ker(f_i)
\eeq
It follows from this that if $ker(f_i)=0$ for some $i$, $ker(F)=0$ (i.e. if any one map is injective, the total map is injective).

Thus, we have only to investigate the case for which $ker(f_i)>0$ for all $i$. In this case, the naive expectation would be that since the set of target spaces $V_i$ all consist of homogenous polynomials of (generically) different bi-degree, the images $im(f_i)$ do not intersect. That is,
\beq
im(F) =\sum_{i}^{N}im(f_i)
\eeq
But this fails to take into account ``relations" between these image spaces\footnote{This is related to syzygies at the level of modules.}. If  $f:  V \rightarrow V_1 \oplus V_2$ then we note that the images $im(f_i)$ are defined by
\bea
P_{f_{1}}P_{v}=P_{v_{1}} \\
P_{f_{2}}P_{v}=P_{v_{2}}
\eea
where $P_{v}$ is the same polynomial in each expression. But by inspection, we see that $P_{v_{1}}$ and $P_{v_{2}}$ must obey the relation,
\beq\label{rels}
P_{f_{1}} P_{v_{2}}=P_{f_{2}} P_{v_{1}}
\eeq
Hence the space $Im(F)$ has ${\sum_{i}}[l+k_i,p+m_i]$ degrees of freedom in  $P_{v_{1}}$ and $P_{v_{2}}$ subject to $[l+k_1+k_2,p+m_1+m_2]$ constraints in \eref{rels}. Therefore, in this case
\beq
im(F) ={\sum_{i}}dim(V_i) - [l+k_1+k_2,p+m_1+m_2]
\eeq
In full generality then,
\beq\label{one_to_many}
im(F) = {\sum_{i}}dim(V_i)-\sum_{a=2}^{N} \sum_{|M|=a} (-1)^a [r +\sum_b k_b][s +\sum_b m_b]
\eeq 
in the same notation as above. Written in terms of the kernel of $F$ this is
\beq
ker(f) = Max[0, (1-N)dimV + \sum_i ker(f_i) +{\sum_{a=2}}^{N} \sum_{|M|=a} (-1)^a [r +\sum_b k_b][s +\sum_b m_b]]
\eeq
It is worth noting at this stage, that the main information needed at each step of this analysis is the individual surjectivity/injectivity of the component maps. 

\paragraph{ Case 3: ``Many-to-Many"}
To compute line bundle cohomology using Leray tableau, we must in general be able to compute the kernel/image of a map $d_i$ \eref{leray_iterate} which maps $N$ ambient cohomology groups, or rather spaces of polynomials, to $M$ such spaces. That is, $F: (V_1 \oplus V_2 ....\oplus V_N) \rightarrow (U_1 \oplus U_2 ....\oplus U_M)$. We can represent such a map in matrix form as
\beq\label{matrix}
F_{ib}v_{b}=u_{i}
\eeq
where $v_{b} \in V_{b}$ ($b=1,\ldots N$) and $u_{i} \in U_{i}$ ($i=1,\ldots M$). Further, $F$ is a $M\times N$ matrix of polynomials, each component of which is defined as $f_{ib}: V_b \rightarrow U_i$ with bi-degree $(k_{ib},m_{ib})$ and $ker(f_{ib})=dim(V_b)-dim(V_i)$.

To understand the map $F$ we must combine the action of the two cases ``many-to-one" and ``one-to-many" described above. For example, we could define $V=V_1 \oplus V_2 ....\oplus V_N$ and consider the ``many-to-many" map as a ``one-to-many" map with $\bar{f}_{i}: V \rightarrow U_{i}$ via
$\bar{f}_{i}v=u_{i}$ and $v=v_{1}\oplus \ldots \oplus v_{N}$. Then according to \eref{one_to_many}, we can represent the total image, $im(F)$ as $\bigoplus_{i}im(\bar{f}_{i})$ minus relations of the form \eref{rels}.

 Equivalently, we could study the same map by defining $U = U_{1} \oplus \ldots \oplus U_{M}$  and beginning with the ``many-to-one" structure defined  in \eref{image}. In this way, the map $F$ may be thought of as a ``many-to-one" map of the form $\tilde{f}_{b}v_{b}=u$ with $u= u_1 + \ldots +u_{M}$. Then as in \eref{image} we have
\beq
im(F) = {\sum_i}^{N}im(\tilde{f}_i) - \sum_{i<j}im(\tilde{f}_i)\cap im(\tilde{f}_{j}) + \sum_{i<j<l}im(\tilde{f}_i)\cap im(\tilde{f}_j)\cap im(\tilde{f}_l) - \ldots 
\eeq
But what is the dimension of $im(\tilde{f}_i)\cap im(\tilde{f}_{j})$? The intersection of the images is determined by the expression
\beq
\tilde{f}_{i}v_{i}=\tilde{f}_{j}v_{j}
\eeq
To understand this however, requires knowledge of the particular structure of Leray maps. 

\paragraph{The structure of Leray maps} 
The maps $d_i$ in the Leray tableau \eref{leray_iterate} descend simply from the defining maps in the Calabi-Yau configuration matrix \eref{normal2}. When combined into a matrix as in \eref{matrix}, they no longer define a \emph{generic} matrix of polynomials, but rather a more specific one built from the configuration matrix

Consider a co-dimension one CICY with normal bundle $\cN=\cO(\bf{q})$ Then in this simple case, there is only one map $d_{1}^{1}$ in \eref{leray_iterate} which maps between $H^{j}(\cA, \cN^* \times \mathcal{L}) \stackrel{d_{1}^{1}}{\rightarrow} H^{j}(\cA,\mathcal{L})$. In this simple case, the map $d_{1}^{1}$ is given by the polynomial $P_{q}$ associated to a section of the normal bundle $\cN=\cO(\bf{q})$ (where the bi-degree of $P_{q}$ is given by the numbers $q^{r}$)
(see \cite{hubsch} for a derivation of this). For higher wedge powers $\wedge^k\cN^*$ the maps are likewise built from the defining polynomials of $X$. While it is possible to algorithmically generate such maps from a given normal bundle $\cN^*$ in arbitrary co-dimension, it is difficult to write down an analytic formula for the matrix of polynomials. To clarify this structure however, we will illustrate the produce with an example in the following section.

\section{An example}
As an example of the techniques above let us compute the cohomology of $\mathcal{L}=\cO(k,m)$ with $k,m >0$ on the CICY $X=\left[
\begin{array}
[c]{c}%
3\\
3
\end{array}
\left|
\begin{array}
[c]{ccc}%
0 & 0 & 1\\
2 & 2 & 2
\end{array}
\right.  \right]  $. We know the structure of the cohomology of $H^*(X,\mathcal{L})$ already on this space, by Kodaira vanishing and the index theorem \eref{ind}. However, it will be illustrative to derive the same result using the explicit map techniques described above. 

For this line bundle and manifold, following the construction of Leray tableau defined above \eref{leray} and \eref{leray_iterate} we have
\beq\label{sample_tableau}
E_{1}^{j,k}=
\left[
\begin{array}
[c]{ccccccc}%
E_1^{0,0} & \stackrel{d_{1}^{1}}{\leftarrow} & E_{1}^{0,1}& \stackrel{d_{1}^{2}}{\leftarrow} & E_{1}^{0,2}& \stackrel{d_{1}^{3}}{\leftarrow} & E_{1}^{0,3}\\
0 & & 0 & & 0 & & 0 \\
\vdots & & \vdots & & \vdots & & \vdots
\end{array}
\right] 
\eeq
Since there is only one non-zero row in \eref{sample_tableau}, the sequence converges at $E_{2}^{j,k}$ with
\beq\label{E2s}
\ba[c]{cc}
E_{2}^{0,0}=E_{1}^{0,0}-Im(d_{1}^{1}) &E_{2}^{0,1}=ker(d_{1}^{1})-Im(d_{1}^{2}) \\
E_{2}^{0,2}=ker(d_{1}^{2})-Im(d_{1}^{3}) & E_{2}^{0,3}=ker(d_{1}^{3})
\ea
\eeq
Written in terms of the Young tableau of the Bott-Borel-Weil Theorem (Thm \ref{BBW}) these cohomologies are given by
\beq
\ba[c]{ll}
E_1^{0,0}=H^0(\cA,\mathcal{L})=\binom{-k,0~~~~}{-m,0,0,0,0,0} & E_1^{0,1}=H^0(\cA,\cN^*\times \mathcal{L})=\binom{-k,0~~~~}{-m+2,0,0,0,0,0}^{\oplus 2}\oplus\binom{-k+2,0~~~~}{-m+2,0,0,0,0,0}  \\
E_1^{0,3}=H^0(\cA,\wedge^{3}\cN^*\times \mathcal{L})=\binom{-k+2,0~~~~}{-m+6,0,0,0,0,0} & E_1^{0,2}=H^0(\cA,\wedge^{2}\cN^*\times \mathcal{L})=\binom{-k,0~~~~}{-m+4,0,0,0,0,0}\oplus\binom{-k+2,0~~~~}{-m+4,0,0,0,0,0}^{\oplus 2} 
\ea
\eeq
Representing each tableau simply by it's dimension, i.e. $\binom{-k,0~~~~}{-m,0,0,0,0,0} \sim [k,m]$ as in \eref{dims}, the maps $d_{1}^{i}$ have the following structure:
\beq\label{a_random_label}
\ba[c]{l}
d_{1}^{1}: [k,m-2]^{\oplus 2} \oplus [k-2,m-2] \rightarrow [k,m] \quad \mbox{``many-to-one"} \\
d_{1}^{2}: [k,m-4]\oplus[k-2,m-4]^{\oplus 2}\rightarrow [k,m-2]^{\oplus 2} \oplus [k-2,m-2]  \quad \mbox{``many-to-many"} \\
d_{1}^{3}: [k-2,m-6] \rightarrow[k,m-4]\oplus[k-2,m-4]^{\oplus 2} \quad \mbox{``one-to-many"}
\ea
\eeq
where the descriptions classify the types of maps as discussed in Section \ref{big_hairy_maps}. In order to proceed further, we must understand the explicit form of the maps $d_{1}^{i}$ that arise in the the Leray and Koszul sequences. We will denote the component polynomials of the normal bundle $\cN=\cO(0,2)^{\oplus 2} \oplus \cO(2,2)$, as $P_{1}$ of bi-degree $(0,2)$, $P_{2}$ of bi-degree $(0,2)$ and $P_{3}$ of bi-degree $(2,2)$ respectively. We shall assume that each of $P_{i}$ is \emph{generic}. Then writing $d_{1}^{1}: V_{1}\oplus V_{2}\oplus V_{3} \rightarrow V$ with the $V_i$ as the polynomial spaces in \eref{a_random_label}, then by the definition of the Koszul and Leray sequences, the map $f_{i}: V_{i} \rightarrow V$ is $P_{i}$. Using the results for ``many-to-one" maps in \eref{many_to_one} the result for \eref{a_random_label} is
\beq
ker(d_{1}^{1})=[k,m-4]+[k-2,m-4]+[k,m-4]-[k-2,m-6]
\eeq
Similarly, we can write $d_{1}^{3}$ as a ``one-to-many" map: $d_{1}^{3}: U \rightarrow U_{1}\oplus U_{2}\oplus U_{3}$ with $U_{i}$ as in \eref{a_random_label} and $f_{i}: U \rightarrow U_{i}$ defined by $f_{i}=P_{i+2}$. But by inspection, in this case, the component maps $f_{i}$ are injective for all $i$. Therefore by \eref{one_to_many} it follows that
\beq
ker(d_{1}^{2})=0
\eeq
Finally, for the map $d_{1}^{2}$ we find that the map is defined in terms of the available polynomials as
\beq\label{matrix_map}
d_{1}^{2}=\left(
\begin{array}
[c]{ccc}%
P_{2} & P_{3} & 0 \\
P_{1} & 0 & P_{3} \\
0 & P_{1} & P_{2}
\end{array}
\right)
\eeq
The kernel of this map may be computed according to the discussion in the ``many-to-many" section above. It is given by
\beq
ker(d_{1}^{2})=[k-2,m-6]
\eeq
At last, combining the kernels/images of these maps with the $E_{2}^{j,k}$ in \eref{E2s} and computing the cohomology of $\mathcal{L}$ via \eref{hodge_converge} we find that
\beq
h^0(X,\mathcal{L})=E_{1}^{0,0}-E_{1}^{0,1}+E_{1}^{0,2}-E_{1}^{0,3}
\eeq
and $h^q(X,\mathcal{L})=0$ for all $q>0$ as expected by Kodaira vanishing.

\section{An outline of the algorithm}
Clearly, the necessary algorithm to compute line bundle cohomology is complex and best suited for machine implementation. We have written a Mathematica package to accomplish this and for clarity provide a schematic outline of the necessary algorithm here:

For a generic line bundle $\mathcal{L}=\cO_{X}(a,b)$ on $X$, the following steps must be implemented:
\begin{enumerate}
\item Define the Leray tableau for $\cO_{X}(a,b)$ as in \eref{leray} and \eref{leray_iterate}.
\item Using the Bott-Borel-Weil Theorem \ref{BBW}, re-write the entries in the Leray tableau, $E^{j,k}_{1}=H^{j}(\cA, \wedge^{k}\cN^* \times \mathcal{L})$, as Young-tableau. The result will be fully symmetric irreducible representations of  unitary groups.
\item Next, using the correspondence between totally symmetric tensors and polynomials, find a representation for each ambient cohomology group as a space of polynomials of a particular bi-degree according to \eref{toref1} and \eref{toref2}.
\item Formulate the explicit matrices of polynomials that correspond to the maps particular to the Koszul resolution. These in general will not be \emph{generic} matrices of polynomials, but defined by a specific set of maps (those in the configuration matrix \eref{normal2}) according to the Koszul sequence as in \eref{matrix_map}.
\item Use the results for Kernels/Images given above  to compute the ranks and kernels of the necessary maps \eref{many_to_one}, \eref{one_to_many}
\item Using the simple convergence of the Leray spectral sequence, $E^{j,k}_{\infty}=E^{j,k}_{2}$, for line bundles, compute the relevant cohomology groups on $X$ according to \eref{hodge_converge}
\end{enumerate}

With this important calculational tool in hand, we are now ready to address the question of stability of monad bundles. In the following Chapter we shall use the cohomology of line bundles to formulate a new stability criteria.

\chapter{Stability}\label{StabCh}

\section{Stable vector bundles in heterotic theories}
The problem of finding $N=1$ supersymmetric vacua in heterotic models is a long-standing one. The heterotic string provides one of the most promising avenues to realistic particle physics from string theory, but the inherent mathematical difficulty of the theory is a serious obstacle \footnote{See for example \cite{maria, Braun:2006ae,Gomez:2005ii}}. To build four-dimensional heterotic models it is necessary not only to specify a compact Calabi-Yau three-fold and two vector bundles, $V$ and $\tilde{V}$ over it, but also to to compute the topological data\footnote{i.e. bundle valued cohomologies, Chern classes, etc.} of $TX$, $V$ and $\tilde{V}$ in detail and show that these bundles produce a supersymmetric four-dimensional theory. 

With the powerful  Donaldson-Uhlenbeck-Yau theorem \cite{duy1, duy2} for vector bundles on Calabi-Yau $3$-folds, this is equivalent to proving that the vector bundles are \emph{stable}. This property regarding the sub-sheaves of a bundle $V$ is a notoriously difficult one to prove and as a result the stability criteria of this Chapter is among the more significant results of this work.

In the previous Chapter, we have generated thousands of  $SU(n)$ vector bundles $V$ on Calabi-Yau $3$-folds $X$ whose stability we wish to investigate. Since the slope of $V$ is $0$ for these bundles (since $c_1(V) = 0$ for $SU(n)$ bundles) in order for $V$ to be stable we must have that all proper subsheaves, $\mathcal{F}$, of $V$ have strictly negative slope.
Thus if $\mathcal{F} \subset V$ we require
\beq
\mu(\mathcal{F}) < 0
\eeq\label{neg_slope}
But what qualifies a sheaf $\mathcal{F}$ to be a subsheaf of $V$? This is simply the condition that it has smaller rank and the there exists an embedding $\mathcal{F} \hookrightarrow V$. The space 
of homomorphisms between $\mathcal{F}$ and $V$ is simply $\hom(\mathcal{F},V)$ which is, in turn, given by the space of global holomorphic sections $H^0(X, \mathcal{F}^* \otimes V)$. Hence, we 
have that
\beq\label{hom}
V ~{\rm stable } ~\Rightarrow \mu(\mathcal{F}) < 0 ~\forall \mathcal{F} ~{\rm s.t.}~
\rk(\mathcal{F}) < n ~{\rm and }~ H^0(X, \mathcal{F}^* \otimes V) \ne 0 \ .
\eeq
Recall that by definition \eref{slope} in Section \ref{stability}, the slope of a sheaf $\mathcal{F}$ is given by
\beq
\mu(\mathcal{F}) = \frac{1}{rk(\mathcal{F})}\int_{X} c_{1}(\mathcal{F}) \wedge J \wedge J
\eeq
where $J$ is the K\"ahler form on $X$. Expanding $J$ in a basis as $J=t^{s}J_{s}$ (with $t^{s} \geq 0$ forming the K\"ahler cone of a CICY $X$) this becomes
\bea
\mu(\mathcal{F}) =\frac{1}{rk(\mathcal{F})}\int_{X} c_{1}^{r}(\mathcal{F})J_{r} \wedge t^{s}J_{s} \wedge t^{u} J_{u} \\
=\frac{1}{rk(\mathcal{F})} c_{1}^{r}t^{s}t^{u}(\mathcal{F})\int_{X}J_{r} \wedge J_{s} \wedge J_{u} \\
=\frac{1}{rk(\mathcal{F})} d_{rsu}c_{1}^{r}(\mathcal{F})t^{s}t^{u}
\eea
where $d_{rsu}$ are the triple intersection numbers \eref{intersection} of the CICY. Relabeling this as
\beq\label{sdef}
s_{r} \equiv \frac{1}{rk(\mathcal{F})}d_{rsu}t^{s}t^{u}
\eeq
we can denote the slope as
\beq\label{ss-slope}
\mu(\mathcal{F})=s_{r}c_{1}^{r}(\mathcal{F})
\eeq
So, in this notation, if $V$ is stable for all $\mathcal{F} \subset V$, 
\beq\label{s-slope}
\mu(\mathcal{F}) = c_{1}^{r}(\mathcal{F})s_{r} <0
\eeq
This is a condition on the first Chern class of  $\mathcal{F}$. Since the K\"ahler cone for CICYs consists of all $t^{s} \geq 0$, and furthermore $d_{rsu} \geq 0$, it is clear that $s_{r} \geq 0$.

In the mathematics literature, when a bundle is called `stable' this is taken to mean stable with respect to \emph{all} possible $s_{r}$. A choice of such a vector is called a 'polarization'. That is, the bundle which is stable with respect to all polarizations is stable everywhere in the K\"ahler cone. However, viewed from the perspective of physics, this is actually a stronger condition than we require. In light of the ultimate goal of moduli stabilization in the four-dimensional effective theory, we expect eventually to restrict the theory to some point in K\"ahler moduli space. In order to satisfy the Hermitian Yang-Mills equations \eref{HYME} it is enough to show that the bundle is stable \emph{somewhere} in the K\"ahler cone (with the hope that we may eventually stabilize the moduli somewhere within this region). In this Chapter, we shall make use of this simplification to formulate a stability criteria for bundles defined over Calabi-Yau manifolds with $h^{1,1}(TX) >1$. This criteria will be a generalization of the necessary but not sufficient cohomology condition given in Hoppe's criteria \eref{hoppe}.

Because any discussion of stability is inherently somewhat convoluted, we shall attempt to outline the basic argument here. In Section \ref{line-bun} below, we demonstrate that rather than considering all proper sub-sheaves of a vector bundle $V$, it is possible to consider only sub-line bundles of appropriate antisymmetric tensor powers $\wedge^{k}V$. That is, if
\beq\label{intro1}
\mu(\mathcal{L}) < \mu(\wedge^{k}V)
\eeq
for all sub-line bundles $\mathcal{L} \subset \wedge^{k}V$ for $k=1, \ldots rkV-1$, then $V$ is stable.  In a straightforward generalization of Hoppe's criteria (whose proof we review in Section \ref{hoppe_proof}) we proceed to show that 
\beq\label{intro2}
H^0(X,\wedge^{k}V)=0
\eeq
for $k=1, \ldots rkV-1$. While this fact alone is not enough to prove that $V$ is stable, it will be make it possible to formulate additional cohomological criteria that are sufficient for stability.

As we stated above, if $\mathcal{L} \subset \wedge^{k}V$ then, we must have $hom(\mathcal{L}, \wedge^{k}V) \neq 0$. This fact together with \eref{intro2} implies that if we can explicitly demonstrate that for some open region of the K\"ahler cone, \eref{intro1} is satisfied at each $k$, for all line bundles for which
\bea\label{intro3}
hom(\mathcal{L},\wedge^{k}V) \neq 0 &\text{and}& h^0(X, \mathcal{L})=0
\eea
then we will have determined that $V$ is stable somewhere in the K\"ahler cone. In Section \ref{new_stab} we derive necessary cohomological bounds on the sub-line bundles such that \eref{intro1} and \eref{intro3} are satisfied for all 'potential' sub-line bundles $\mathcal{L}$. The stability criteria presented is valid for CICYs with $h^{1,1}(TX)=2$ which constitute 21 of the 36 manifolds with positive monads found in Chapter \ref{c:posCICY}. Including the 5 cyclic manifolds (for which we have Hoppe's criteria) this will provide a method of verifying stability for 26 manifolds. In Section \ref{more_kahler}, we shall discuss briefly how this may be generalized to the remaining 10 CICYs in our classification. It should be stressed that the new stability criteria presented here is a necessary but not sufficient condition. That is, any bundle that fails to satisfy the cohomology bounds presented in Section \ref{new_stab} below is not necessarily unstable. With these goals in mind, we turn now to the first of several simplifying observations.

\section{Sub-line bundles and Stability}\label{line-bun}
In general, to prove that a vector bundle $V$ is stable, it is necessary to demonstrate that
\beq
\mu(\mathcal{F}) < \mu(V)
\eeq\label{slope_condit}
for all proper sub-sheaves $\mathcal{F} \subset V$. For an arbitrary construction of vector bundles there are no techniques available for classifying all sub-sheaves (or computing their topological properties). As a result, stability is a difficult property to test for a bundle and presents a serious obstacle to building realistic models from heterotic compactifications. However, there are a few elementary properties that will prove to be surprisingly useful in analyzing stability. We 
turn here to the first of several important simplifications.

In this section, we will demonstrate that it is enough\footnote{See the Appendix of \cite{Donagi:2003tb} for a more detailed version of this argument.} to test the slope criteria for all \emph{sub line-bundles} $\mathcal{L}$ of certain anti-symmetric powers of the bundle $\wedge^{(k)}(V)$. We have already made use of this important result implicitly in Hoppe's Theorem \eref{hoppe} in Chapter \ref{MonCh1}. We shall make this explicit below and derive several useful results.

To begin, consider a rank-$n$ vector bundle $V$ over a projective variety $X$. If $\mathcal{F}$ is a sub-sheaf of $V$ then it injects into $V$ via the resolution
\beq\label{sub-sheaf}
0 \to \mathcal{F} \to V \to \mathcal{K} \to 0
\eeq
with $rk(\mathcal{F}) <  rk(V)$. We shall consider such sub-sheaves one rank at a time. First, we observe that since $V$ is a vector bundle, it is torsion-free and thus, has no rank-zero sub-sheaves. So, we begin with the case of a rank one sub-sheaf. Since $\mathcal{F}$ is torsion free there is an injection
\beq\label{double-dual}
\mathcal{F} \stackrel{i}{\longrightarrow} \mathcal{F}^{**}
\eeq
where $\mathcal{F}^{**}$ is the double-dual\footnote{For a locally free coherent sheaf, $\mathcal{L}$, the double dual is isomorphic, i.e. $\mathcal{L}^{**} \approx \mathcal{L}$.}. of $\mathcal{F}$. Since $\mathcal{F}^{**}$ is rank one and torsion free it can be shown that $\mathcal{F}$ is locally free, and hence a line bundle \cite{Friedmann}. Dualizing the sequence \eref{sub-sheaf} twice and using \eref{double-dual} we have
\beq
\mathcal{F} \subset \mathcal{F}^{**} \subset V^{**} \approx V
\eeq
It is straightforward to show that $\mu(\mathcal{F})=\mu(\mathcal{F}^{**})$. Thus, instead of checking the slope condition \eref{slope_condit} for all rank-one torsion-free subsheaves of $V$, it suffices to check it for all sub line-bundles. But what about sub-sheaves of higher rank?

Let $\mathcal{F}$ be a torsion free sub-sheaf of of rank $k$ ($<n$). Once again, we have an inclusion $0 \to \mathcal{F} \to V$ which in turn induces a mapping 
\beq
\wedge^{k}(\mathcal{F}) \to \wedge^{k}(V)
\eeq
which can be shown to be an injection as well \cite{Donagi:2003tb}. By definition of the anti-symmetric tensor power $\wedge^{k}$, $\wedge^{k}(\mathcal{F})$ is a rank one sheaf (with $\mathcal{F}$ rank $k$). Since $\mathcal{F}$ is torsion free, so are $\wedge^{k}(\mathcal{F})$ \cite{Kobayashi}. Next, by an argument similar to the one above, (\eref{double-dual}), we can argue that there is a line bundle associated to $\wedge^{k}\mathcal{F}$. Hence, for any torsion-free sub-sheaf of rank $k$, we have an associated line bundle. Note that in general for a rank $n$ bundle $V$,
\beq\label{wedgec1}
c_1(\wedge^{k}V)=\binom{n-1}{k-1}c_1(V)
\eeq
We observe here that for $SU(n)$ bundles with $c_1(V)=0$, $c_1(\wedge^{k}(V))=0$ as well. Likewise, we see that for a rank $k$ sub-line bundle, $\mathcal{F}$, \eref{wedgec1} gives us $\mu(\wedge^{k}\mathcal{F}) =\mu(\mathcal{F})$. Therefore, for each rank $k$ de-stabilizing sub-sheaf of $V$ we have a corresponding de-stabilizing sub-line bundle of $\wedge^{k}(V)$. Thus in proving stability of a rank $n$ $SU(n)$ vector bundle $V$ we need only to show that if $\mathcal{L} \subset \wedge^{k}(V)$, then
\beq
\mu(\mathcal{L}) < \mu(V)=0
\eeq
for all $k < n$. This is a dramatic simplification of the problem of stability and we shall make use of it to prove stability of the rank $3,4$ and $5$ monad bundles defined over certain CICYs.

\subsection{Hoppe's criteria}\label{hoppe_proof}
Before discussing stability on more general spaces, we remind the reader of the statement of Hoppe's Theorem \eref{hoppe} for cyclic projective varieties. Hoppe's condition simply states that for an $SU(n)$ bundle\footnote{for a generalization to $U(n)$ bundles, see \cite{huybrechts}.} $V$, if the bundle is stable then $h^{0}(X, \wedge^{k}V)=0$ for $k=1,\ldots rk(V)-1$. This  necessary (but not sufficient) condition for stability of vector bundles on `cyclic' projective varieties allowed us to prove stability of the $37$ positive monad bundles in Chapter \ref{MonCh1}. Before we attempt to formulate more general stability criteria, it will be illuminating to provide a sketch of the proof of Hoppe's theorem here.

Hoppe's criteria works as a `proof-by-contradiction': Let $X$ be a cyclic variety ($h^{1,1}(TX)=1$) and $V$ a rank $n$ (for our purposes $SU(n)$) vector over it. Suppose that there exists a rank $k$ de-stabilizing sub-sheaf $\mathcal{F} \subset V$. Then by the arguments of the previous section, we may define a line bundle via
\beq
\wedge^{k}\mathcal{F}^{**} \equiv \mathcal{O}(m)
\eeq
such that $\wedge^{k}\mathcal{F}^{**} \subset \wedge^{k}V$. Now since the slope of $V$ vanishes, in order for $\mathcal{F}$ to be de-stabilizing we must have that 
\beq
\mu(\wedge^{k}\mathcal{F}^{**}) = \mu(\cO(m)) \geq 0
\eeq
On a cyclic Calabi-Yau manifold, the single K\"ahler form $J$ allows us to classify all line bundles $\cO(m)$ by a single integer $m$ which defines their first Chern class $c_{1}(\cO(m))=mJ$ \cite{Distler:1987ee}. Clearly, from \eref{s-slope} if $\cO(m)$ is to be de-stabilizing, we must have $m>0$. Now, for $\cO(m)$ with $m>0$, the Kodaira vanishing theorem \eref{kodaira} and the index theorem tell us directly  that 
\beq
h^0(X, \cO(m))>0
\eeq
But by the sequence \eref{sub-sheaf},
\bea
0 \to \cO(m) \to \wedge^{k}(V) \to K \to 0 \\
0 \to H^0(X, \cO(m)) \to H^0(X,\wedge^{k}(V)) \to \ldots
\eea
We see that this implies that 
\beq
H^0(X, \wedge^{k}(V)) \neq 0
\eeq
Stated simply, if there exists a de-stabilizing sub-sheaf of $V$ on a cyclic Calabi-Yau manifold then it must be the case that $H^0(X \wedge^{k}(V)) \neq 0$. Therefore, we have arrived at Hoppe's criteria: If $H^0(X \wedge^{k}(V))=0$ for all $k=1, \ldots (n-1)$, then $V$ is stable. 

We observe here that the strength of this theorem followed from the fact that
\beq
\mu(\wedge^{k}\mathcal{F}^{**}) \geq 0 \Leftrightarrow h^{0}(X, \wedge^{k}\mathcal{F}^{**}) \neq 0
\eeq
This conclusion rests on the fact that $X$ is cyclic (and hence $c_1(\cO(m))$ is determined by a single integer $m$ for a line bundle $\cO(m)$). It is clear that if we are to examine stability on Calabi-Yau manifolds defined as complete intersection hypersurfaces in products of several projective spaces (those for which $h^{1,1}(TX)>1$), we will need something more.
\section{Cohomology of Sub-line Bundles}
Using the results of the previous section, we begin by considering a sub-line bundle $\mathcal{L}$ of $\wedge^{k}(V)$. Following from the observations of the previous section, we hope to be test stability of $V$ by investigating the sub-line bundles of $\wedge^{k}(V)$. We may ask, are there any simple characteristics that distinguish sub-line bundles for the monad bundles we consider? Using the results of the previous two sections, we see that are two simple tests of  `sub-line bundle-ness' for a line bundle $\mathcal{L} \subset V$ that will prove useful to us. 

First, as discussed in \eref{hom}, by definition, if $\mathcal{L} \subset \wedge^{k}(V)$ then 
\beq
Hom_{X}(\mathcal{L}, \wedge^{k}(V)) \neq 0
\eeq
Therefore, we have a non-trivial cohomology condition to check for any candidate sub-line bundle of $V$. Note that this is a necessary but not sufficient condition for a sub-line bundle. If $Hom_{X}(\mathcal{L}, \wedge^{k}(V))=0$ there are no maps between $\mathcal{L}$ and $V$. But if $Hom_{X}(\mathcal{L}, \wedge^{k}(V)) \neq 0$ this does not guarantee the existence of an injection, $f$, from $\mathcal{L}$ into $V$.

The second condition is a simple observation related to Hoppe's criteria \eref{hoppe}. For $\mathcal{L}$ a sub-line bundle of $V$, we have the short exact sequence
\beq\label{subseq}
0 \to \mathcal{L} \to V \to \mathcal{K} \to 0
\eeq
This in turn induces a long exact sequence in cohomology
\beq
0 \to H^0(X, \mathcal{L}) \to H^0(X, \wedge^{k}(V)) \to \ldots
\eeq
We observe now, that if a generalization of Hoppe's criteria were to told - i.e. If $H^0(X, \wedge^{k}(V))=0$ then 
\beq\label{coho}
H^0(X, \mathcal{L})=0
\eeq
for $\mathcal{L}$ satisfying \eref{subseq}. Beginning at the level of $k=1$, we see this simple observation is the reason why $H^0(X,V) = 0$ for stable $SU(n)$ bundles $V$ (as noted in Section \ref{s:het}). Noting that $H^0(X,V) = H^0(X, V \otimes \cO_X^*) \ne 0$, it is possible that $\cO_X$ could be a proper subsheaf of $V$. In fact, if $H^0(X,V) \neq 0$ then with a constant map taking sections to sections, it is clear that $\cO_X$ injects into $V$. Furthermore, $c_1(\cO_X) = 0$ and so $\mu(\cO_X) = 0$ and is not strictly negative, making $\cO_X$ a proper de-stabilizing subsheaf and $V$ unstable. Thus, we must have that $H^0(X,V)=0$ for stable $SU(n)$ bundles. In the following sections, it will be shown that if  $H^0(X, \wedge^{k}(V))=0$, then we can formulate a generalization of Hoppe's criteria. But first, we shall show below that this property does in fact hold for all positive monad defined bundles on CICYs.
\subsection{Proof that $H^0(X,\wedge^{(k)}V)=0$}\label{wedge-vanish}
While the Hoppe criteria \eref{hoppe} is insufficient to guarantee stability for vector bundles defined over complete intersection hypersurfaces in higher dimensional ambient spaces, the verification of a straightforward generalization of Hoppe's criteria is still useful for us. As we shall see in the following sections, if 
\bea\label{hoppe_gen}
H^{0}(X,\wedge^{(k)}V)=0 &  k=1, \ldots rk(V)-1
\eea
and $X$ is a complete intersection Calabi-Yau of co-dimension $K$, then we can generate a new stability criteria for manifolds with more than a single K\"ahler form.

We shall demonstrate that \eref{hoppe_gen} holds for the positive monad bundles of Chapter \ref{c:posCICY}.  To begin, recall that for $SU(n)$ bundles we have the useful isomorphisms
\beq\label{su(n)}
\wedge^{n-k}V \approx \wedge^{k}V^*
\eeq
Thus to demonstrate that $H^0(X,\wedge^{k}V)=0$, it may be easier to use  \eref{su(n)} and consider $H^0(X,\wedge^{n-k}V)=0$. We shall do this for each bundle rank separately.

\paragraph{Rank 3 Bundles} To begin, consider a rank $3$ $SU(n)$ positive monad bundle on $X$. By \eref{su(n)}, we have that $V \approx \wedge^{2}V^*$ and $\wedge^{2}V \approx V^*$.

Now, it was demonstrated in Section \ref{s:stable} that $H^{0}(X,V)=H^0(X,V^*)=0$ for all positive monad bundles, $V$. Thus, we have shown that the Hoppe condition is satisfied for all rank $3$ bundles.

\paragraph{Rank 4 Bundles} Once again, we have the results of Section \ref{s:stable} and \eref{su(n)}. That is, we have already demonstrated that $H^0(X, V)=0$ and $H^0(X, \wedge^{3}V)=H^0(X,V^*)=0$. Thus, for a rank $4$ bundle, it only remains to check that $H^{0}(X, \wedge^{2}V)=H^0(X,\wedge^{2}V^*)=0$. For this case we have the following long-exact sequence on $X$ obtained by taking the dual of the exterior power sequence of the monad \eref{monad}
\beq
0 \to S^{2}C^* \to C^* \otimes B^* \to \wedge^{2}B^* \to \wedge^{2}V^* \to 0
\eeq
By splitting this sequence into two and examining the associated long exact sequences in cohomology, a simple analysis shows us that if
\beq\label{criteria}
h^{p}(X,S^{p}C^* \otimes \wedge^{2-p}B^*)=0
\eeq
for $p=0,1,2$ then it is clear that $H^0(X,\wedge^{2}V)=0$. But since $S^{p}C^* \otimes \wedge^{2-p}B^*$ is a sum of strictly negative line bundle on $X$ (for a positive monad), by the Kodaira vanishing theorem \eref{kodaira}, the criteria, \eref{criteria} is clearly satisfied. Therefore, $H^0(X,\wedge^{2}V)=0$ and \eref{hoppe_gen} is satisfied.

\paragraph{Rank 5 Bundles} Finally we turn to rank $5$ bundles. Using the results of the previous sections, we have that $H^0(X,V)=H^{0}(X,\wedge^{4}V^*)=0$. Further, since $\wedge^{3}V \approx \wedge^{2}V^*$, by the argument of the preceding paragraph, $H^{0}(X, \wedge^{3}V)=0$. This leaves only one more cohomology $H^{0}(X,\wedge^{2}V)=H^0(X, \wedge^{3}V^*)$ to compute. To accomplish this, it will be useful to formulate the problem on the ambient space and then use Koszul sequence \eref{koszulA} to restrict to the Calabi-Yau.

Consider the exterior power sequence on the ambient space, $\cA$,
\beq
0 \to S^{3}\cC^* \to S^{2}\cC^* \otimes \cB^* \to \cC^* \otimes \wedge^{2}\cB^* \to \wedge^{3}\cB^* \to \wedge^{3}\cV^* \to 0
\eeq
We can now divide this into three short exact sequences,
\bea\label{triple}
 0 \to S^{3}\cC^* \to S^{2}\cC^{*}\otimes \cB^* \to K_{1} \to 0 \\
 0 \to K_{1} \to \cC^* \otimes \wedge^{2}\cB^* \to K_{2} \to 0 \\
 0 \to K_{2} \to \wedge^{3}\cB^* \to \wedge^{3}\cV^{*} \to 0
 \eea
We shall take $B$, $C$ to be positive line bundles, defined over the ambient space $\cA$ of a co-dimension $K$ CICY. In the next step - the Koszul sequences - we shall tensor each of the short exact sequences in \eref{triple} by $\wedge^{j}\cN^*$ for $j=0,1 \ldots K$ and compute the long exact sequences in cohomology. From the cohomology sequences we get
\beq
H^{l}(\cA, \wedge^{j}\cN^{*}\otimes\wedge^{3}\cV^{*})=H^{l+1}(\cA,\tilde{K}_{2})=H^{q+2}(\cA,\tilde{K}_{1}) 
\eeq
for $l=0,1,\ldots K$ and for $l=K+2, K+3$ we have the three exact sequences
\beq 
{\small
\ba{l}
 0 \to H^{3+K-1}(\cA,\wedge^{j}\cN^{*}\otimes\wedge^{3}\cV^{*}) \to H^{3+K}(\cA,\tilde{K}_{2}) \to H^{3+K}(\cA, \wedge^{j}\cN^{*}\otimes \wedge^{3}\cB^*) 
 \to H^{3+K}(\cA, \wedge^{j}\cN^{*}\otimes\wedge^{3}\cV^{*}) \to 0 \\
 0 \to  H^{3+K-1 }(\cA, \tilde{K}_{1}) \to H^{3+K}(\cA, \wedge^{j}\cN^{*}S^{3}\cC^*) \to H^{3+K}(\cA, \wedge^{j}\cN^{*}\otimes S^{2}\cC^{*}\otimes \cB^*) 
 \to H^{3+K}(\cA,\tilde{K}_{1}) \to 0 \\
 0 \to H^{3+K-1}(\cA,\tilde{K}_{2}) \to H^{3+K}(\cA,\tilde{K}_{1}) \to H^{3+K}(\cA, \wedge^{j}\cN^{*}\otimes \cC^* \otimes \wedge^{2}\cB^*)
 \to H^{3+K}(\cA, \tilde{K}_{2}) \to 0
 \ea
 }
 \eeq
Now to restrict to the Calabi-Yau $3$-fold $X$, we turn to the Koszul sequence for $\wedge^{3}V^*$. \beq\label{koszul_stability2}
0 \to \wedge^{3}\cV^{*} \otimes \wedge^K \cN^* \to \wedge^{3}\cV^{*} \otimes \wedge^{K-1} \cN^*
\to \ldots \to \wedge^{3}\cV^{*} \otimes \cN^* \to \wedge^{3}\cV^{*} \to \wedge^{3}\cV^{*}|_X \to 0 \ .
\eeq
We can decompose this long exact sequence into short exact sequences as:
\beq\label{three-seq}\ba{c}
\sseq{\wedge^{3}\cV^{*}\otimes\wedge^K\cN^*}{\wedge^{3}\cV^{*}\otimes\wedge^{K-1}\cN^*}{q_0};\\
\sseq{q_0}{\wedge^{3}\cV^{*}\otimes\wedge^{K-2}\cN^*}{q_1};\\
\vdots\\
\sseq{q_{K-3}}{\wedge^{3}\cV^{*}\otimes\cN^*}{q_{K-2}};\\
\sseq{q_{K-2}}{\wedge^{3}\cV^{*}}{\wedge^{3}\cV^{*}|_X} \ ,\\
\ea\eeq
where $q_j$ for $j=0,\ldots,K-2$ are appropriate (co)kernels. The last row and the object of our interest - reads
\beq
\sseq{q_{K-2}}{\wedge^{3}\cV^{*}}{\wedge^{3}\cV^{*}|_X} \ .
\eeq
This induces a long exact sequence in cohomology and to check the
vanishing of $H^0(X,\wedge^{3}V^*)$ it is clearly sufficient to show that
\beq
H^0(\cA, \wedge^{3}\cV^*) = H^1(\cA, q_{K-2}) = 0 \ .
\eeq
Arguing inductively, the vanishing of $H^1(\cA,
q_{K-2})$ could be guaranteed by the vanishing of $H^1(\cA, \wedge^{3}\cV^{*} \otimes \cN^*)$ and
$H^2(\cA, q_{K-3})$ and so on. In summary, we seek the vanishing of
\beq
H^j(\cA, \wedge^{3}\cV^* \otimes \wedge^j \cN^*) = 0 \ , \qquad
\forall~j=0, \ldots, K \ .
\eeq

Combining this result from our calculation of the ambient cohomology of $\wedge^{3}\cV^*$, we see that in order for $H^0(X,\wedge^{3}V^*)=0$, we require the vanishing of 
\beq
H^j(\cA, \wedge^{3}\cV^* \otimes \wedge^j \cN^*)=H^{j+2}(\cA, \tilde{K}_1)=0
\eeq
for $j=0,1 \ldots K$. But by inspection, $H^{j+2}(\cA, \tilde{K}_1)=0$ for $j=0,1,\ldots K-1$. So, it only remains to check that $H^{K+2}(\cA,\tilde{K}_1)=0$. From the sequence \eref{three-seq}, we find
\bea
h^{K+2}(\cA,\tilde{K}_1)=h^{3+K}(\cA, \wedge^{j}\cN^{*}\otimes S^{3}\cC^*) -Im(g) \\
g: H^{3+K}(\cA, \wedge^{j}\cN^{*}\otimes S^{3}\cC^*) \to H^{3+K}(\cA, \wedge^{j}\cN^{*}\otimes S^{2}\cC^{*}\otimes \cB^*) 
\eea
As a result, we can see that $H^0(X,\wedge^{3}V^*)=0$ whenever
\beq
h^{3+K}(\cA, \wedge^{j}\cN^{*}\otimes S^{2}\cC^{*}\otimes \cB^*) \geq H^{3+K}(\cA, \wedge^{j}\cN^{*}\otimes S^{3}\cC^*)
\eeq 
This simple cohomology check is satisfied for all the positive monad bundles defined in Chapter \ref{c:posCICY}. Thus, we have demonstrated that the generalized Hoppe criteria \eref{hoppe_gen} is satisfied for all the rank $5$ bundles and therefore for all positive monads in our list.
\section{A new stability criteria}\label{new_stab}
As we saw at the start of this Chapter in \eref{sdef} and \eref{ss-slope}, the slope of a sheaf $\mu(\mathcal{F})$ depends upon the K\"ahler parameters $t^{s}$. That is, stability of a sheaf/bundle is only a meaningful with respect to some 'polarization', $H=J\wedge J$.

In the mathematics literature, when a vector bundle is called ``stable" it is taken to mean stable with respect to all polarizations, that is, stable \emph{everywhere} in the K\"ahler cone. However, from a physics perspective, this is actually a stronger constraint than we require. In order to produce a $N=1$ supersymmetric theory in four-dimensions, it is sufficient for our purposes that the vector bundle $V$ be stable \emph{somewhere} in the K\"ahler cone (i.e. for some open set). We are therefore led to ask the question, can we formulate a generalization of Hoppe's criteria, that is, a necessary but not sufficient cohomological condition, which will guarantee that a monad-defined bundle $V$ is stable somewhere in the K\"ahler cone? With Hoppe's criteria available for spaces for which $h^{1,1}(TX)=1$, it is natural to ask this question for the next most complicated K\"ahler structure, that for which $h^{1,1}(TX)=2$. In the following sections we construct just such a method.
\subsection{Two K\"ahler Moduli}
We begin by considering favorable complete intersection Calabi-Yau manifolds which are defined as complete intersection hypersurfaces in a product of two projective spaces, $\mathbb{P}^{n_{1}}\times \mathbb{P}^{n_{2}}$. For these manifolds, $h^{1,1}(X) = 2$ and the favorable CICY has two  K\"ahler classes $J_{1,2}$ related directly to those on the ambient space. The K\"ahler cone is simply $\{ \IZ_{\ge 0} J_1 + 
\IZ_{\ge 0} J_2 \}$ and lives in a plane. A polarization $H$ is then 
described by \eref{ss-slope} as a vector $s_{r}=(a,b) \in \IZ^2_{\ge 0}$. 

Expanded in this basis of K\"ahler forms, the first chern class of an arbitrary sub-sheaf $\mathcal{F} \subset V$ is given by
\beq
c_{1}^{1}(\mathcal{F})J_{1} + c_{1}^{2}(\mathcal{F})J_{2}
\eeq
Thus, we can characterize the $c_1$ of a sub-sheaf by another integer pair, $(c_{1}^{1},c_{1}^{2}) \in \IZ^2$. In summary, stability of $V$ somewhere in the K\"ahler cone means that there exists an open region $\mathcal{R}$, in the positive quadrant, such that the dot product of $s_{r}$ and $c_{1}^{r}$ is strictly negative, i.e
\beq
s_{r}c_{1}^{r}(\mathcal{F}) <0
\eeq
for all sub-sheaves $\mathcal{F}$ and $s_{r} \in \mathcal{R}$. 

The central idea for a stability criteria is as follows: If it were possible to classify all potentially destabilizing sub-sheaves of a monad bundle $V$ and demonstrate explicitly that for all these, a region $R$ in the K\"ahler cone was left undisturbed, we would be able to prove that $V$ is stable in $R$. 

To this end, we utilize the observations of the preceding sections. We shall use our generalized notion of Hoppe's criteria and consider sub-line bundles of $\wedge^{k}(V)$. Since all line bundles on a Calabi-Yau $3$-fold are classified by their first Chern class, for each line bundle $\mathcal{L}$ we have a unique vector $(c_{1}^{1},c_{1}^{2})=(a,b)$ in the plane. By definition this is $\mathcal{L} \equiv \cO(a,b)$. For such a given line bundle we can determine the region of the first quadrant (i.e. those vectors $s_{r}$) which it destabilizes, via the expression
\beq\label{ninety}
\mu(\mathcal{L}) = s_{r}c_{1}^{r}(\mathcal{L}) = s_{1}a +s_{2}b >0
\eeq
Thus $\cO(a,b)$ destabilize all vectors $s_{r}=(s_{1},s_{2})$ in the first quadrant which have an angle less than $90^{\circ}$ with $(a,b)$. This is represented in Figure \ref{f:stab}. If the 'de-stabilized' regions defined by \eref{ninety} for some set of line bundles $\mathcal{L}_i$ span the set of all $s_{r}$ then the bundle is not stable. Note that the set of all $s_{r}$ is not the entire first quadrant of the plane shown in Figure \ref{f:stab}, but rather a conical subset of it, defined by the intersection numbers and K\"ahler moduli via \eref{sdef}. 

To check whether our monad bundles are stable, we shall proceed as follows: We shall classify all `potential' sub-line bundles $\cO(a,b)$ of a monad bundle $V$ and place bounds on the components $a,b$ of $c_{1}(\mathcal{L})$ via cohomological conditions. For $(a,b)$ in the allowed region, we shall demonstrate that the de-stabilized region is only a part of the K\"ahler cone (i.e. first quadrant). That is, that the monad bundle is stable \emph{somewhere} in the K\"ahler cone.

Consider a line bundle $\mathcal{L}=\cO(a,b)$ with $a,b$ arbitrary. When is $\mathcal{L}$ a sub-line of $\wedge^{k}V$ where $V$ is defined by \eref{defV}? As we demonstrated above,  in \eref{hom} if
\beq\label{test1}
hom(\mathcal{L}, \wedge^{k}V) =0
\eeq
Then clearly $\mathcal{L}$ is not a sub-line bundle of $\wedge^{k}V$. Likewise, we recall from section \ref{wedge-vanish} above that for the rank $3,4,5$ $SU(n)$ bundles defined in Chapter \ref{c:posCICY} over CICYs, $H^0(X,\wedge^{k}(V))=0$. Then, by \eref{sub-sheaf} if
\beq\label{test2}
H^0(X,\mathcal{L}) \neq 0
\eeq
it follows that $\mathcal{L}$ is not a sub-line bundle the bundle $\wedge^{k}V$. So we can immediately begin to eliminate regions of the $(a,b)$ plane as not corresponding to sub line bundles of $\wedge^{k}(V)$.
\begin{figure}[!h]
\centerline{\epsfxsize=3in\epsfbox{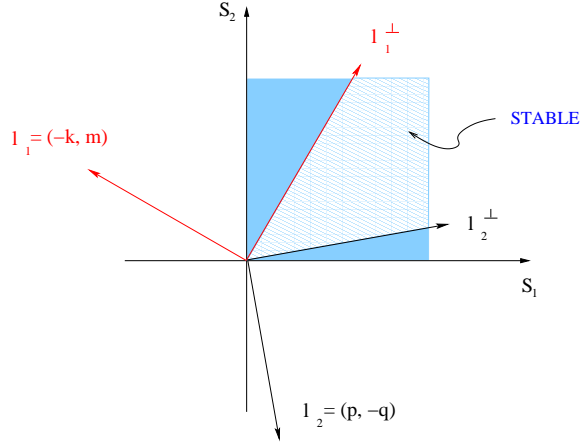}}
\caption{\emph{The K\"ahler cone and two potentially de-stabilizing line 
bundles $l_1$ and $l_2$.}}
\label{f:stab}
\end{figure}
\subsection{The cohomology conditions}
To begin, let $\wedge^{j}V$ be a bundle defined by a positive monad as in Chapter \ref{c:posCICY}. We shall now study the space of possible line bundles $\cO(a,b)$ and find constraints on $a,b$ such that $\cO(a,b)$ is potentially a sub-line bundle of $\wedge^{j}V$. First, consider $\mathcal{L}=\cO(a,b)$ with $a,b \geq 0$. Clearly, by the Kodaira vanishing theorem \eref{kodaira}, for these line bundles $H^{0}(X,\mathcal{L}) \neq 0$. Therefore by \eref{test1}, $\mathcal{L}$ is not a sub-line bundle of $\wedge^{j}V$. We can eliminate the entire first quadrant of the plane shown in Figure \ref{f:stab} as not sub-line bundles and hence of no interest in stability arguments.

Next, note that all $\mathcal{L}=\cO(a,b)$ with $a,b <0$ (i.e. those line bundles in the third quadrant) cannot de-stabilize $\wedge^{j}V$ anywhere in the K\"ahler cone. That is, \eref{intro1} is satisfied for all such line bundles. Therefore, the only potentially de-stabilizing line bundles that could concern us are those of the form
\beq\label{the_two}
\mathcal{L}_1 =\cO(-k,m) ~~\text{or} ~~\mathcal{L}_{2}=\cO(p,-q)
\eeq
with integers $k,m,p,q \geq 1$. For these, if 
\beq
hom(\wedge^{j}V,\mathcal{L})=H^0(X, \wedge^{j}V\otimes \mathcal{L}^{*})=0 ~~\text{or} ~~H^0(X, \mathcal{L}) \neq 0
\eeq
Then we eliminate $\mathcal{L}$ from consideration. For the first of these conditions, the exterior power sequences
\bea
0 \to \wedge^{j}B \to \wedge^{j}B \to \wedge^{j-1}B \otimes C \to \dots \to B\otimes S^{j-1}C \to S^{j}C \to 0 \\
0 \to S^{j}C^* \to S^{j-1}C^* \otimes B^* \to \dots \to C^* \otimes \wedge^{j-1}B^* \to \wedge^{j}V^* \to 0
\eea
provide us with cohomology conditions arising from the associated cohomology sequences. If
\beq\label{wedge_k}
h^0(X, \wedge^{j}B \otimes \mathcal{L}^*) = 0
\eeq
then $\mathcal{L}$ is not a sub-line bundle of $\wedge^{j}V$. Furthermore since \eref{su(n)} holds for $SU(n)$ bundles we also have that $\wedge^{j}V= \wedge^{(n-j)}V^*$ (with $rk(V)=n$) and thus if
\beq\label{wedge_k2}
H^{l}(X, \wedge^{(n-j-l)}B^* \otimes S^{l}C^* \otimes \mathcal{L}^*)=0 ~~\text{for} ~~l=0,1, \ldots (n-j)
\eeq
then once again, $\mathcal{L}$ is not a sub-line bundle of $\wedge^{j}V$. For the line bundles of the form \eref{the_two}, in general each of \eref{wedge_k} and \eref{wedge_k2} bound a single component, $k$ or $m$ (respectively $p,q$) of $\mathcal{L}_i$ ($i=1,2$). That is, despite the fact that both conditions test whether $\mathcal{L} \subset \wedge^{k}V$, they contain independent information. The procedure to check stability of $V$ is then as follows.

To test the stability of a rank $n$ monad bundle $V$ we must investigate sub-line bundles of $\wedge^{j}V$ for $j=1,\ldots n-1$ for each value of $j$. For a fixed value of $j$ we then seek line bundles of the form $\mathcal{L}_{1}=\cO(-k,m)$ or $\mathcal{L}_{2}=\cO(p,-q)$ which satisfy

\begin{enumerate}\label{list}
\item $h^0(X, \wedge^{j}B \otimes \mathcal{L}^*) \neq 0$
\item $H^{l}(X, \wedge^{(n-j-l)}B^* \otimes S^{l}C^* \otimes \mathcal{L}^*) \neq 0$ for at least one $l=0,1, \ldots (n-j)$
\item $H^0(X, \mathcal{L})=0$
\end{enumerate}
we call a line bundle satisfying the above conditions a `potential' sub-line bundle of $\wedge^{j}V$ (though it is not \emph{necessarily} a sub-line bundle of $\wedge^{j}V$) and the slope $\mu(\mathcal{L})$ must be checked in order to verify stability. 

For a sub-line bundle of the form $\mathcal{L}_1 =\cO(-k,m) \subset \wedge^{j}V$ satisfying these conditions, it will de-stabilize all vectors $s_{r}$ in the first quadrant according to \eref{ninety}. That is, those vectors $s_r$ with slope
\beq
\frac{s_2}{s_1} > \frac{k}{m}=\alpha(k)_{1} 
\eeq
Similarly, $\mathcal{L}_2 =\cO(p,-q)$ will destabilize those $s_{r}$ with slope
\beq
\frac{s_2}{s_1} < \frac{p}{q}=\alpha(k)_{2} 
\eeq
Since $k,m \geq 1$ (and $p,q>1$) it is clear that no single line bundle $\mathcal{L}_i$ can de-stabilize the \emph{entire} first quadrant (i.e. all $s_r$). Rather, in order for $\wedge^{j}V$ to be unstable everywhere in the K\"ahler cone there must exist a pair $(\mathcal{L}_{1},\mathcal{L}_{2})$ such that the angle between the two vectors $(-k,m)$ and $(p,-q)$ is less than or equal to $180^{\circ}$.

Considering all possible $\mathcal{L}_{i}$, if 
\beq\label{2d1}
\alpha(k)_{1}^{min} > \alpha(k)_{2}^{max}
\eeq
We shall say that the bundle $\wedge^{j}V$ is ``j"-stable for the open region of the K\"ahler cone defined by the vectors $s_{r}$ with slope $s_{2}/s_{1}$ in the set
\beq\label{line_slope}
[\alpha(k)_{1}^{min}, \alpha(k)_{2}^{max}]
\eeq  
A bundle can be shown to be stable if, for all $j$, there exists a non-zero region as in \eref{line_slope} and the intersection of all such sets is non-empty. Stated simply, if the set
\beq\label{interval}
[Min_{k}\{\alpha(k)_{1}^{min}\}, Max_{k}\{\alpha_{2}^{max} \}]
\eeq
is non-empty then the bundle is manifestly stable somewhere in an open subset of the K\"ahler cone. If $Min_{k}\{\alpha(k)_{1}^{min}\} < Max_{k}\{\alpha_{2}^{max} \}$ then the test fails and we cannot determine whether the bundle is stable or unstable. The main elements of this procedure are illustrated for a specific example in the following section.
\section{An Example}
As an example of the stability criteria described above, we can consider rank $3$ positive monad bundles on the manifold $\left[
\begin{array}
[c]{c}%
2\\
2
\end{array}
\left|
\begin{array}
[c]{ccc}%
3 \\
3
\end{array}
\right.  \right]$. For an $SU(3)$ bundle on this space, we have that
\beq
V \approx \wedge^{2}V^* ~~~ \wedge^{2}V \approx V^*
\eeq
Therefore, to demonstrate that $V$ is stable we must show that $V,\wedge^{2}V$ are both stable with respect to all sub-line bundles. That is, we must apply the procedure described in the previous section for $k=1,2$

To begin, we first consider line bundles in the 2nd and 4th quadrants (the only potentially de-stabilizing regions). That is, line bundles of the form $\mathcal{L}_{1}=\cO(-k,m)$ and $\mathcal{L}_{2}=\cO(p,-q)$. To determine the bounds on $k,m$ and $p,q$ we utilize the criteria defined in the previous section.

We begin by considering $j$-stability for $j=1$ and derive bounds on the potential sub-line bundles of $V \approx \wedge^{2}V^*$. First, we note that on this CICY, the line bundle $\mathcal{L}_{1}=\cO(-k,m)$ satisfies
\beq\label{cohob}
H^{0}(X, \mathcal{L}_{1}) =0 ~~\forall k,m \geq1
\eeq
(and by symmetry the same result holds for $\mathcal{L}_{2}=\cO(p,-q)$). 

As a result, this cohomology condition will not place a constraint on the line bundles. Next consider, $hom(\mathcal{L}_{1},V)$, this will vanish if
\beq
H^0(X, B \otimes \mathcal{L}_{1}^{*})=0
\eeq
Now by comparison with \eref{cohob}, we see that if $m > b_{2}^{max}$ where $b_{2}^{max}$ is  $Max_{i}\{\ b_{2}^{i}\}$ where $B= \bigoplus_{i}\cO(b_{r}^{i})$ and $i=1,\ldots rkB$. Therefore, we must have
\beq
m \leq b_{2}^{max}=\beta
\eeq
if $\mathcal{L}_{1}$ is to be a potential sub-line bundle of $V$. By symmetry, we see that this also implies that we must have $p \leq  \kappa = b_{1}^{max}$ in $\mathcal{L}_{2}=\cO(p,-q)$. But what about the other component, $k$? To bound this, we must use the second condition given in \ref{list} arising from $hom(\mathcal{L}_{1}, \wedge^{2}V^*)=0$. This is guaranteed if,
\bea
H^{0}(X, \wedge^{2}B^*\otimes \mathcal{L}_{1}^{*})=0 \\
H^{1}(X, B^*\otimes C^* \otimes \mathcal{L}_{1}^{*})=0 \\
H^{2}(X, S^{2}C^* \otimes \mathcal{L}_{1}^{*}) = 0
\eea
The above cohomologies will vanish if $k < \lambda=Min[\gamma_1, \gamma_{2}]$ where
\bea
\gamma_{1} = Min[(B \otimes C)_{\text{1st component}}] \\
\gamma_{2} =Min[(S^{2}C)_{\text{1st component}}]
\eea
and the minimum first component of $B \otimes C = \bigoplus_{j}\cO(e_{r}^{j})$ is given by $Min_{j}\{e_{1}^{j}\}$ with $j=1, \ldots rkB + rkC$. So, for a possible sub-line bundle we must have 
\beq
k \geq \lambda
\eeq
Once again, by symmetry of the CICY, it is clear that we must also have $q \geq \delta = Min[\tilde{\gamma}_1, \tilde{\gamma}_{2}]$ where $\tilde{\gamma}_{1} = Min[(B \otimes C)_{\text{2nd component}}]$  and $\tilde{\gamma}_{2} =Min[(S^{2}C)_{\text{2nd component}}]$.

So, for the $j=1$ step of the stability check, we have that for all line bundles $\mathcal{L}_{i}$ as above,
\beq
\frac{k}{m} \geq \frac{\lambda}{\beta}=\alpha(1)_{1}~~\text{and}~~ \frac{p}{q} \leq \frac{\kappa}{\delta}=\alpha(1)_{2}
\eeq
where $\lambda,\beta,\kappa,\delta$ and $\alpha(k)_{i}$ are defined as above. We note here that this bundle will be stable at the $k=1$ level (i.e. stable with respect to all rank $1$ sub-sheaves) if
\beq
\frac{k}{m} \geq \frac{\lambda}{\beta}> \frac{\kappa}{\delta} \geq \frac{p}{q}
\eeq
But by the definition of the monad sequence \eref{defV} and the fact that $b_{r}^{i} \leq c_{r}^{j}$ for all $r,i,j$ we have that
\beq
\delta\lambda = Min[\gamma_1, \gamma_{2}]Min[\tilde{\gamma}_1, \tilde{\gamma}_{2}] > b_{1}^{max}b_{1}^{max} =\kappa\beta
\eeq
for all rank $3$ monad bundles on the CICY. Therefore, these bundles are stable with respect to all potential sub-line bundles in the region of the K\"ahler cone defined by lines in the first quadrant with slope in $[\alpha(1)_{1},\alpha(1)_{2}]$.

Now, consider the next, $j=2$, level of the stability check for which we consider $\mathcal{L}_{i} \subset \wedge^{2}V \approx V^*$. Once again, we begin with $\mathcal{L}_{1}=\cO(-k,m)$ and observe that $hom(\mathcal{L}_{1},\wedge^{2}V^*)=0$ if
\beq
H^{0}(X, \wedge^{2}B \otimes \mathcal{L}_{1}^{*})=0
\eeq
and this is certainly true for $m>Max[(\wedge^{2}B)_{\text{2nd component}}]$. Therefore, a potential sub-line bundle $\mathcal{L}_1$ must satisfy
\beq\label{mbound}
m \leq Max[(\wedge^{2}B)_{\text{2nd component}}]=\sigma
\eeq
and by symmetry, $p \leq  Max[(\wedge^{2}B)_{\text{1st component}}] = \chi$. Finally, if we have
\bea
H^{0}(X, B^* \otimes  \mathcal{L}_{1}^{*})=0 \\
H^{1}(X, C^* \otimes \mathcal{L}_{1}^{*})=0 
\eea
then $ \mathcal{L}_{1}$ is not a sub-line bundle of $V^*$. The first part of this, $H^{0}(X, B^* \otimes  \mathcal{L}_{1}^{*})=0$ is trivially true for all line bundles $\mathcal{L}_{i}$ of the form described above. However, the second condition is more complicated. For a line bundle of the form $l = \cO(-a,b)$ on $X$, the Bott-Borel-Weil analysis of Chapter \ref{c:line_bundle_coho} shows that
$h^{1}(X, \mathcal{O}_{X}(-a,b))=0$ when (in the notation of Chapter  \ref{c:line_bundle_coho}), 
\beq\label{linef}
[a,b-3] >[a-3,b]~.
\eeq
Expanding out the binomial coefficients in \eref{linef} this is reduced to the simple inequality
\beq
(b-a)(-2+ab) >0
\eeq
Using this fact, we see that $h^1(X, C^* \otimes \mathcal{L}^*)=0$ when $k < c_{1}^{min}$ or when $k > c_{1}^{j}$ for some $j=1,\ldots rkC$ and
\beq\label{another_label}
(c_{2}^{j} +m -k+c_{1}^{j})((k-c_{1}^{j})(m+c_{2}^{j})-2)>0
\eeq
Now, by assumption for this case $k > c_{1}^{j}$ and $((k-c_{1}^{j})(m+c_{2}^{j})-2)$ is positive with $m+c_{2}^{j} \geq 2$. Therefore, the inequality \eref{another_label} will be satisfied so long as 
\beq
k < Min_{j}\{c_{1}^{j} +c_{2}^{j} +m \}
\eeq
Thus, in order for $\mathcal{L}_{1}$ to be a potential sub-line bundle, we must have
\beq
k \geq (c_{1}^{min}+c_{2}^{min}+\sigma )=\tau
\eeq
where $\sigma$ is defined by \eref{mbound}. In the analogous case, $ q \geq (c_{1}^{min}+c_{2}^{min} +\chi)= \rho$. Therefore,
for all potential sub-line bundles of $\wedge^{2}V \approx V^*$, these cohomology conditions lead to the following bounds on the components of the first chern class of $\mathcal{L}$
\beq
\frac{k}{m} \geq \frac{\tau}{\sigma}=\alpha(2)_{1}~~\text{and}~~ \frac{p}{q} \leq \frac{\chi}{\rho}=\alpha(2)_{2}
\eeq
and a simple calculation show that once again, $[\alpha(2)_{1}, \alpha(2)_{2}]$ defines a non-empty interval for all positive monad bundles on $X$ (by virtue of the fact that the entries in the defining line bundles of $B$ are less than or equal to those in $C$).

With these bounds in hand, we are now ready to check the complete stability criteria given in \eref{interval} in the previous section. To this end, we compute $Min_{k}[\alpha(k)_{1}]$ and $Min[\alpha(k)_{2}]$ and verify that the corresponding region in $s_{r}$ space has an open set of K\"ahler parameters, $t^{r}$ defining it. (i.e the stable region intersects a region of the space of vectors $s^r$ as defined in \eref{s-slope}). Direct computation yields that \eref{interval} is non-empty for the $63$ positive rank $3$ monad bundles on $X$. And thus, all the rank $3$ bundles will be stable in an open region of the K\"ahler cone. A similar analysis of the $611$ rank $3$ positive monads on the manifold  $\left[
\begin{array}
[c]{c}%
1\\
3
\end{array}
\left|
\begin{array}
[c]{ccc}%
2 \\
4
\end{array}
\right.  \right]$ demonstrates that these are all stable as well. However, for higher rank bundles or CICYs of higher co-dimension, similar calculations quickly becomes too lengthy to attempt by hand. To this end,   using the results of Chapter \ref{c:line_bundle_coho} to calculate the necessary cohomologies, we have written code to check the stability criteria for all the positive monad bundles on the $26$ CICYs with $h^{1,1}(TX)=2$. These scans are currently underway and the results will be presented in future work. 

It should be noted that the above analysis above does not imply that the bundles are unstable outside the open set defined by \eref{interval}. Indeed, it is possible that the bundles are stable everywhere in the K\"ahler cone as in the case of the cyclic manifolds considered in Chapter \ref{MonCh1}. However, from a physics perspective, we are content with the results above.  

\section{Conclusion and future work}\label{more_kahler}
In this chapter, we provided a new set of cohomology conditions that can be used to verify the stability of vector bundles over complete intersection Calabi-Yau manifolds. Generalizing the Hoppe criteria \eref{hoppe} of Chapter \ref{MonCh1}, we show that $H^0(X,\wedge^{k}V)=0$ for all positive monad bundles over CICYs. This insight allows us to simplify our analysis of the slopes \eref{ss-slope} of sub-sheaves and consider only sub-line bundles of $\wedge^{k}V$. Using the simple observations that sub-bundles of $\wedge^{k}V$ must have non-trivial maps into $\wedge^{k}V$ (i.e. $hom(\mathcal{L}, \wedge^{k}V) \neq 0$) and that such sub-line bundles must not have sections (i.e. $H^0(X, \mathcal{L})=0$), we demonstrate that the set of all potentially de-stabilizing line bundles is bounded and fails to de-stabilize the entire K\"ahler cone. We provide the necessary cohomology criteria for spaces with $h^{1,1}(TX)=2$.

The central principles of this analysis generalize straightforwardly to spaces with $h^{1,1}(TX)>2$. For instance, on the CICY defined by the normal bundle $\cN=\cO(2,2,2,2)$ in $\mathbb{P}^{1}\times \mathbb{P}^{1}\times \mathbb{P}^{1}\times\mathbb{P}^{1}$ it is clear that the potentially de-stabilizing line bundles are those of the form $\cO(m_{1},m_{2},m_{3},m_{4})$, where $m_{i}<0$ for at least one $i$. That is, rather than the two quadrants $\cO(-k,m)$ and $\cO(p,-q)$ considered in this Chapter, there are 14 regions of space that we must investigate (the remaining two - those for which $m_{i}$ is of definite sign for all $i$- are excluded by the same arguments as in the two-dimensional case). The cohomology conditions of \ref{list} will bound the components $m_i$. For instance, the condition that $hom(\mathcal{L},B) \neq 0$ bounds the components of potentially de-stabilizing mixed line bundles of the form $\cO(-m_{1},-m_{2},m_{3},m_{4})$, so that $m_{3} \leq b_{3}^{max}$ and $m_{4} \leq b_{4}^{max}$.
However, with potentially de-stabilizing line-bundles in $14$ regions, the analysis of of the geometry shown in Figure \ref{f:stab} is clearly considerably more complex for such spaces. It is not immediately apparent what the analog of conditions such as \eref{2d1} and \eref{line_slope} would be. The generalization of our results to spaces with $h^{1,1}(TX)>2$ is a topic of current investigation.

\chapter{Conclusion}

\section{Concluding remarks and future directions}

In this thesis, we have investigated two distinct approaches to string phenomenology through compactifications of M-theory and heterotic string theory. In the following paragraphs we will outline the key results of this work and discuss its potentially interesting extensions to future research.

In the first half of this thesis, we investigated the explicit construction of $M$-theory on spaces with co-dimension four orbifold singularities. In Chapter \ref{c:G2Ch1}, we developed for the first time the explicit coupling of $11$-dimensional supergravity to seven-dimensional super-Yang-Mills theory that describes M-theory near a co-dimension four ADE singularity. This construction provides the starting point for physically realistic compactifications of M-theory. In Chapter \ref{c:G2Ch2}, we performed one such compactification and explicitly derived the four dimensional theory from M-theory compactified on a $G_2$ space with co-dimension four $A$-type singularities.

However, as discussed in Chapter \ref{G2Intro}, to obtain realistic four-dimensional physics from M-theory compactified on a $G_2$ space, it is necessary to extend this work to include the effects of co-dimension seven singularities. Specifically, in order to obtain chiral matter from M-theory, the theory must be formulated near conical singularities. Using arguments from string duality, it is believed that at points of intersection of these four and seven singularities there should be chiral multiplets, charged under the appropriate non-abelian symmetries. 

In future work, we hope to consider the intersection of co-dimension four and seven singularities on $G_2$ spaces. There are several possible approaches to this goal. First, using inspiration from previous constructions of weak-$G_2$, and non-compact conical metrics, it may be possible to find the local modifications to the effective action. That is, a local (delta function) modification to the M-theory Lagrangian consistent with the enhanced symmetry and fields of the intersecting singularities. The explicit theory could then be embedded in an appropriate $G_2$ space and compactified to four-dimensions. To construct a truly physical compactification of M-theory would be a significant and exciting step towards a rigorous mathematical development of the theory, as well as an entirely new approach to the four-dimensional effective physics of M-theory. Furthermore, current work is underway \cite{joyce_talk} to produce \emph{compact} $G_2$ spaces with conical singularities. Compactification on such geometries will be a rich new area of investigation.

In this thesis we have investigated aspects of two out of the three main approaches to string phenomenology. While we have not discussed the final branch of string phenomenology-type II string compactifications, we shall mention a few possibilities for future work here. D-Brane models \cite{Blumenhagen:2005mu,Gmeiner:2005vz} in type II string theory have been studied in depth over the past decade and have provided numerous advances in string phenomenology, including the problem of compactifications with flux and moduli stabilization. Of particular interest to this work, intersecting D6-brane models also shed light on how chiral fermions arise in the $G_2$ compactifications described above. Such intersecting brane models can be uplifted to M-theory compactified on a $G_2$ space built by fibering a suitable Hyperk\"ahler four-manifold over a collection of supersymmetric $3$-cycles. Through string dualities, the uplift described above is known to exist, however no explicit examples have been traced through to both sides of the duality. That is, it is an open problem to develop a physically interesting $D6$-brane model in type IIA string theory that can be mapped to a $G_2$ compactification in a complete and lexicographic way. Using the results of this thesis, in future work we hope to investigate such $D6$-brane constructions and attempt to build such an explicit map to M-theory over singular $G_2$ spaces.

The second half of this thesis has investigated heterotic model building and the monad construction of vector bundles. Beginning in Chapter \ref{MonCh1}, we have laid out a program for algorithmically constructing large classes of heterotic models that can be scanned for physically desirable properties. Taking the viewpoint that any one heterotic model (i.e. Calabi-Yau $3$-fold $+$ vector bundle) is likely to fail when confronted with the detailed structure of the Standard Model, we have chosen to construct and explore large classes of models. To this end, we have developed new analytic and computational tools for computing features of the four-dimensional theory such as particle spectra and the presence of supersymmetric vacua. In Chapter \ref{MonCh1}, we introduced several of the main tools in this programme, including the monad construction of vector bundles, and developed the necessary mathematical technology on a simple class of `cyclic' Calabi-Yau manifolds. Continuing on to more general constructions, in Chapter \ref{c:posCICY} we extended the monad constructions to include bundles on the $4515$ favorable Complete Intersection Calabi-Yau manifolds. In addition to demonstrating that a finite class of suitable monad bundles exists on these spaces, we have explicitly constructed the $7118$ anomaly-free bundles which can produce three-generations of particles in the four-dimensional theory. In Chapter \ref{c:line_bundle_coho}, we develop new computational tools for determining the cohomology of line bundles over CICYs using a variant of the Bott-Borel-Weil theorem and the techniques of Leray and Spectral sequences. These results are essential to computing the particle spectra of our monad-defined bundles and to our overall program of model building. Finally in Chapter \ref{StabCh}, we provide a new criteria to test the stability of monad bundles over CICYs (and hence guarantee the existence of $N=1$ supersymmetric vacua in four-dimensions). As proving stability is a  notorious stumbling block for heterotic constructions, this criteria is a significant result .

The heterotic constructions described in this work are the first steps in a broader program of heterotic model building. With the tools of the previous Chapters in hand, we hope to eventually be able to scan through literally hundreds of billions of potential candidates in the vast landscape of string vacua. Current work is under way to extend our construction to include so called `semi-positive' monad bundles - i.e. those defined by sequences of the form \eref{defV} where the component line bundles of the monad may include bundles with zero as well as positive entries. While it is not clear that this class is finite, it is certainly the case that there are huge classes of anomaly free bundles with $3$-generations of matter. Initial scans for these bundles have produced millions of examples. Furthermore, for the semi-positive monad bundles in general $ind(V) \sim N_{gen}$ is smaller than for the positive monad bundles described in Chapter \ref{c:posCICY}. As a result, we expect that the introduction of discrete symmetries and Wilson lines in these models will produce many models with exactly three generations.

The next immediate phase of the constructions described in this thesis is to include Wilson lines to effect the Ôtwo-stageÕ symmetry breaking associated with heterotic model building. Once the study of Wilson lines in the monad construction is underway, it will be possible to use the extensive database and new techniques in computational algebraic geometry already developed to begin a scan for detailed phenomenology including exact particle spectra and symmetries. With the techniques now available to us, it may be possible to carry the computation of the four-dimensional effective field theory further than previous constructions. In particular, it may be possible to directly calculate quantities such as the Yukawa couplings, fermion mass terms, and the superpotential using the techniques of commutative algebra and graded modules. In addition, it would be interesting to use the monad construction to investigate several new features of heterotic compactifications including the inclusion of anti-five branes and flux-vacua. 

While the monad construction over the CICY manifolds produces a huge class of bundles, it still only makes use of a small subset of the known Calabi-Yau three-folds. The vast majority of Calabi-Yau manifolds are constructed as intersections in toric varieties and are known as Toric Calabi-Yau spaces \cite{MR1290195}. There are $473,800,776$ such manifolds known and to systematically explore their heterotic string vacua would substantially advance our understanding of particle phenomenology in string theory. Monad constructions of vector bundles on toric Calabi-Yau manifolds are currently under investigation.

Finally, as mentioned in Chapter \ref{HetIntro}, the $SU(n)$ vector bundles considered in this work are not the only choice of bundle available for heterotic model building. Another route to four-dimensional physics could be obtained by instead considering $U(n)$ bundles \cite{Blumenhagen:2006ux, Blumenhagen:2006wj}. Using these models it is possible to build vector bundles over simply-connected Calabi-Yau manifolds which can break $E_8$ directly to symmetries close to the Standard Model, thus by-passing the need for Wilson-lines. The monad construction described in this work could be readily applied to build $U(n)$ models. For example, any ``twisting" (i.e. tensoring by a line bundle) of one of the $SU(n)$ monad sequences described in this thesis would produce a $U(n)$ bundle. Furthermore, such a bundle would automatically be stable if the original $SU(n)$ bundle had been proven stable using the techniques of Chapter \ref{StabCh}. Investigations into such $U(n)$ monad bundles would be an interesting parallel development to the work already underway.

In the lines of investigation described above, we hope to develop further insight into the underlying geometry of string/M-theory and its physical predictions during this exciting era of fundamental physics. By exploring these new geometries it may be possible to find a view of string theory that looks profoundly similar to the world we observe.

\appendix
\chapter{Appendix1}

\section{Spinor Conventions} \label{spinors} 
In this section, we provide the conventions for gamma matrices and
spinors in eleven, seven and four dimensions and the relations between
them. This split of eleven dimensions into seven plus four arises
naturally from the orbifolds
$\mathbb{R}^{1,6}\times\mathbb{C}^{2}/\mathbb{Z}_{N}$ which we
consider in this paper. We need to work out the appropriate spinor
decomposition for this product space and, in particular, write
11-dimensional Majorana spinors as a product of seven-dimensional
symplectic Majorana spinors with an appropriate basis of
four-dimensional spinors. We denote 11-dimensional coordinates by
$(x^M)$, with indices $M,N,\ldots = 0,\ldots , 10$. They are split up
as $x^{M}=(x^{\mu},y^{A})$ with seven-dimensional coordinates $x^\mu$,
where $\mu ,\nu ,\ldots = 0,\ldots ,6$, on $\mathbb{R}^{1,6}$ and
four-dimensional coordinates $y^A$, where $A,B,\ldots = 7,\ldots ,10$, on
$\mathbb{C}^2/\mathbb{Z}_N$.\\

We begin with gamma matrices and spinors in 11-dimensions. The gamma-matrices,
$\Gamma^{M}$, satisfy the standard Clifford algebra
\begin{equation}
\{\Gamma^{M},\Gamma^{N}\}=2g^{MN},\label{clifford}%
\end{equation}
where $g^{MN}$ is the metric on the full space $\mathbb{R}^{1,6}%
\times\mathbb{C}^{2}/\mathbb{Z}_{N}$. We define the Dirac conjugate of an
11-dimensional spinor $\Psi$ to be
\begin{equation}
\bar{\Psi}=i\Psi^{\dagger}\Gamma^{0}.
\end{equation}
The 11-dimensional charge conjugate is given by
\begin{equation}
\Psi^{C}=B^{-1}\Psi^{\ast},
\end{equation}
where the charge conjugation matrix $B$ satisfies \cite{Tanii:1998px}
\begin{equation}
B\Gamma^{M}B^{-1}=\Gamma^{M\ast},\hspace{0.5cm}B^{\ast}B=\boldsymbol{1} _{32}.
\end{equation}
In this work, all spinor fields in 11-dimensions are taken to satisfy the
Majorana condition, $\Psi^{C}=\Psi$, thereby reducing $\Psi$ from 32 complex
to 32 real degrees of freedom.

Next, we define the necessary conventions for $SO(1,6)$ gamma matrices and
spinors in seven dimensions. The gamma matrices, denoted by $\Upsilon^\mu$,
satisfy the algebra
\begin{equation}
 \{\Upsilon^\mu ,\Upsilon^\nu\}=2g^{\mu\nu}\; ,
\end{equation}
where $g_{\mu\nu}$ is the metric on $\mathbb{R}^{1,6}$. The Dirac conjugate
of a general eight complex component spinor $\psi$ is defined by
\begin{equation}
\bar{\psi}=i\psi^{\dagger}\Upsilon^{0}\; .
\end{equation}
In seven dimensions, the charge conjugation matrix $B_8$ has the
following properties \cite{Tanii:1998px}
\begin{equation}
B_{8}\Upsilon^{\mu}B_{8}^{-1}=\Upsilon^{\mu\ast},\hspace{0.5cm}B_{8}^{\ast
}B_{8}=-\boldsymbol{1} _{8}\; .
\end{equation}
The second of these relations implies that charge conjugation, defined by
\begin{equation}
 \psi^c = B_8^{-1}\psi^\ast
\end{equation}
squares to minus one. Hence, one cannot define seven-dimensional
$SO(1,6)$ Majorana spinors.  However, the supersymmetry algebra in seven
dimensions contains an $SU(2)$ R-symmetry and spinors can be naturally
assembled into $SU(2)$ doublets $\psi^i$, where $i,j,\ldots =
1,2$. Indices $i,j,\ldots$ can be lowered and raised with the
two-dimensional Levi-Civita tensor $\epsilon_{ij}$ and $\epsilon^{ij}$,
normalized so that $\epsilon^{12}=\epsilon_{21}=1$. With these conventions
a symplectic Majorana condition
\begin{equation}
\psi_{i}=\epsilon_{ij}B_{8}^{-1}\psi^{\ast j},\label{symplectic majorana}%
\end{equation}
can be imposed on an $SU(2)$ doublet $\psi^i$ of spinors, where we have
defined $\psi^{\ast i}\equiv(\psi_{i})^{\ast}$. All seven-dimensional
spinors in this paper are taken to be such symplectic Majorana spinors.
Further, in computations with seven-dimensional spinors, the following
identities are frequently useful,
\begin{align}
\bar{\chi}^{i}\Upsilon^{\mu_{1}\ldots\mu_{n}}\psi^{j} &  =(-1)^{n+1}\bar{\psi
}^{j}\Upsilon^{\mu_{n}\ldots\mu_{1}}\chi^{i},\\
\bar{\chi}^{i}\Upsilon^{\mu_{1}\ldots\mu_{n}}\psi_{i} &  =(-1)^{n}\bar{\psi
}^{i}\Upsilon^{\mu_{n}\ldots\mu_{1}}\chi_{i}.
\end{align}

Finally, we need to fix conventions for four-dimensional Euclidean gamma
matrices and spinors. Four-dimensional gamma matrices, denoted by
$\gamma^A$, satisfy
\begin{equation}
 \{\gamma^A,\gamma^B\}=2g^{AB}\, ,
\end{equation}
with the metric $g_{AB}$ on $\mathbb{C}^2/\mathbb{Z}_N$. The chirality
operator, defined by 
\begin{equation}
\gamma=\gamma^{\underline{7}}\gamma^{\underline{8}}\gamma^{\underline{9}}
       \gamma^{\underline{10}}\, ,
\end{equation}
satisfies $\gamma^2 = {\bf 1}_4$. The four-dimensional charge conjugation
matrix $B_4$ satisfies the properties
\begin{equation}
B_4\gamma^AB_4^{-1}=\gamma^{A\ast}\; ,\qquad B_4^\ast B_4=-{\bf 1}_4\; .\label{B4}
\end{equation}
It will often be more convenient to work with complex coordinates
$(z^p,\bar{z}^{\bar{p}})$ on $\mathbb{C}^2/\mathbb{Z}_N$, where
$p,q,\ldots = 1,2$ and $\bar{p},\bar{q},\ldots = \bar{1},\bar{2}$.
In these coordinates, the Clifford algebra takes the well-known
``harmonic oscillator'' form
\begin{equation}
\left\{  \gamma^{p},\gamma^{q}\right\}  =0\;, \qquad
\left\{  \gamma^{\bar{p}},\gamma^{\bar{q}}\right\}  =0\; ,\qquad
\left\{  \gamma^{p},\gamma^{\bar{q}}\right\}     =2g^{p\bar{q}}\, ,
\end{equation}
with creation and annihilation ``operators'' $\gamma^p$ and $\gamma^{\bar{p}}$,
respectively. In this new basis, complex conjugation of gamma matrices~\eqref{B4}
is described by 
\begin{equation}
B_{4}\gamma^{\bar{p}}B_{4}^{-1}=\gamma^{p\ast}\; , \qquad B_{4}\gamma
^{p}B_{4}^{-1}=\gamma^{\bar{p}\ast}\; . \label{B4c}
\end{equation}
 A basis of spinors can be obtained by starting with the
``vacuum state'' $\Omega$, which is annihilated by $\gamma^{\bar{p}}$, that
is $\gamma^{\bar{p}}\Omega =0$, and applying creation operators to it.
This leads to the three further states 
\begin{equation}
\rho^{\underline{p}}=\frac{1}{\sqrt{2}}\gamma^{\underline{p}}\Omega\; ,\qquad
\bar{\Omega}=\frac{1}{2}\gamma^{\ul{1}}\gamma^{\ul{2}}\Omega\, .
\end{equation}
In terms of the gamma matrices in complex coordinates, the chirality operator
$\gamma$ can be expressed as
\begin{equation}
\gamma=-1+\gamma^{\bar{\ul 1}}\gamma^{\ul 1}+\gamma^{\bar{\ul 2}}\gamma^{\ul 2}-\gamma
^{\bar{\ul 1}}\gamma^{\ul 1}\gamma^{\bar{\ul 2}}\gamma^{\ul 2}\, .
\end{equation}
Hence, the basis $(\Omega ,\rho^{\ul p},\bar{\Omega})$ consists of chirality
eigenstates satisfying
\begin{equation}
\gamma\Omega=-\Omega\; , \qquad \gamma\bar{\Omega}=-\bar{\Omega}\; , \qquad
\gamma\rho^{\underline{p}} = \rho^{\underline{p}}\; . 
\end{equation}
For ease of notation, we will write the left-handed states as
$(\rho^i )=(\rho^{\ul{1}},\rho^{\ul{2}})$, where $i ,j ,\ldots =1,2$ 
and the right-handed states as $(\rho^{\bar{\imath}})=(\Omega ,\bar{\Omega})$
where $\bar{\imath},\bar{\jmath},\ldots =\bar{1},\bar{2}$. Note, it follows
from Eq.~\eqref{B4c} that
\begin{equation}
 B_4^{-1}\Omega^\ast = \bar{\Omega}\; ,\qquad
 B_4^{-1}\rho^{\ul{1}\ast} = \rho^{\ul{2}}\; .
\end{equation}
Hence $\rho^i$ and $\rho^{\bar{\imath}}$ each form a Majorana 
conjugate pair of spinors with definite chirality.\\

We should now discuss the four plus seven split of 11-dimensional gamma
matrices and spinors. It is easily verified that the matrices
\begin{equation}
\Gamma^{\mu}    =\Upsilon^{\mu}\otimes\gamma\; ,\qquad
\Gamma^{A}    =\boldsymbol{1}  _{8}\otimes\gamma^{A}, \label{gammas}
\end{equation}
satisfy the Clifford algebra~\eqref{clifford} and, hence, constitute a
valid set of 11-dimensional gamma-matrices. Further, it is clear that
an 11-dimensional charge conjugation matrix $B$ can be obtained from
its seven- and four-dimensional counterparts $B_8$ and $B_4$ by
\begin{equation}
 B=B_8\otimes B_4\; .
\end{equation}
A general 11-dimensional Dirac spinor $\Psi$ can now be expanded in terms
of the basis $(\rho^i,\rho^{\bar{\imath}})$ of four-dimensional spinors
as
\begin{equation}
\Psi =\psi_{i}(x,y)\otimes\rho^{i}+\psi_{\bar{\jmath}}(x,y)\otimes\rho
^{\bar{\jmath}}\; ,
\end{equation}
where $\psi_i$ and $\psi_{\bar{\jmath}}$ are four independent seven-dimensional
Dirac spinors. Given the properties of the four-dimensional spinor basis under
charge conjugation, a Majorana condition on the 11-dimensional spinor $\Psi$
simply translates into $\psi_i$ and $\psi_{\bar{\jmath}}$ each being
symplectic $SO(1,6)$ Majorana spinors.

\section{Some group-theoretical properties}
\label{Pauli} 
In this section we summarize some group-theoretical properties related
to the coset spaces $SO(3,n)/$ $SO(3)\times SO(n)$ of seven-dimensional
EYM supergravity. We focus on the parameterization of these coset
spaces in terms of 11-dimensional metric components, which is an
essential ingredient in re-writing 11-dimensional supergravity,
truncated on the orbifold, into standard seven-dimensional EYM
supergravity language.

We begin with the generic $\mathbb{C}^2/\mathbb{Z}_N$ orbifold, where
$N>2$ and $n=1$, so the relevant coset space is $SO(3,1)/SO(3)$. In
this case, it is convenient to use complex coordinates
$(z^p,\bar{z}^{\bar{p}})$, where $p,q,\ldots = 1,2$ and
$\bar{p},\bar{q},\ldots = \bar{1},\bar{2}$, on the orbifold.  After
truncating the 11-dimensional metric to be independent of the orbifold
coordinates, the surviving degrees of freedom of the orbifold part of
the metric can be described by the components ${e_p}^{\ul{p}}$ of the
vierbein, see Eqs.~\eqref{cond1}--\eqref{condn}. Extracting the
overall scale factor from this, we have a determinant one object
${v_p}^{\ul{p}}$, together with identifications by $SU(2)$ gauge
transformations acting on the tangent space index. Hence, ${v_p}^{\ul{p}}$ should
be thought of as parameterizing the coset $SL(2,\mathbb{C})/SU(2)$. This
space is indeed isomorphic to $SO(3,1)/SO(3)$. To work this out explicitly,
it is useful to introduce the map $f$ defined by
\begin{equation}
 f(u) = u_I\sigma^I
\end{equation}
which maps four-vectors $u_I$, where $I,J,\ldots =1,\ldots ,4$,
into hermitian matrices $f(u)$. Here the matrices $\sigma^I$ and their
conjugates $\bar{\sigma}^I$ are given by
\begin{equation}
(\sigma^{I})=(\sigma^{u},\boldsymbol{1}  _{2})\; ,\qquad
(\bar{\sigma}^{I})=(-\sigma^{u},\boldsymbol{1}  _{2})\; ,
\end{equation}
where the $\sigma^{u}$, $u=1,2,3$, are the standard Pauli matrices.
They satisfy the following useful identities
\begin{align}
\mathrm{tr}\left(  \sigma^{I}\bar{\sigma}^{J}\right)   &  =2\eta^{IJ},\\
\mathrm{tr}\left(  \bar{\sigma}^{I}\sigma^{(J}\bar{\sigma}^{\lvert K\rvert
}\sigma^{L)}\right)   &  =2\left(  \eta^{IJ}\eta^{KL}+\eta^{IL}\eta^{JK}%
-\eta^{IK}\eta^{JL}\right)\,  ,
\end{align}
where $I,J,\ldots$ indices are raised and lowered with the Minkowski metric
$(\eta_{IJ})=\rm{diag}(-1,-1,-1,+1)$. A key property of the map $f$ is that
\begin{equation}
 u_Iu^I={\rm det}(f(u))
\end{equation}
for four-vectors $u_I$. This property is crucial in demonstrating that
the map $F$ defined by
\begin{equation}
 F(v)u = f^{-1}\left( vf(u)v^\dagger\right)
\end{equation}
is a group homeomorphism $F:SL(2,\mathbb{C})\rightarrow SO(3,1)$.
Solving explicitly for the $SO(3,1)$ images ${\ell_I}^{\ul{J}}={(F(v))_I}^{\ul{J}}$
one finds
\begin{equation}
 {\ell_I}^{\ul{J}}=\frac{1}{2}{\rm tr}\left(\bar{\sigma}_Iv\sigma^Jv^\dagger\right)\; .
\end{equation}
This map induces the desired map $SL(2,\mathbb{C})/SU(2)\rightarrow SO(3,1)/SO(3)$
between the cosets.\\

The structure is analogous, although slightly more involved, for the orbifold
$\mathbb{C}^2/\mathbb{Z}_2$, where $n=3$ and the relevant coset space is
$SO(3,3)/SO(3)^2$. In this case, it is more appropriate to work with real
coordinates $y^A$ on the orbifold, where $A,B,\ldots =7,8,9,10$. The
orbifold part of the truncated 11-dimensional metric, rescaled to
determinant one, is then described by the vierbein ${v_A}^{\ul{A}}$ in real
coordinates, which parameterizes the coset $SL(4,\mathbb{R})/SO(4)$.
The map $f$ now identifies $SO(3,3)$ vectors $u$ with elements of the $SO(4)$ Lie algebra
according to
\begin{equation}
 f(u)=u_IT^I\; ,
\end{equation}
where $T^I$, with $I,J,\ldots = 1,\ldots ,6$ is a basis of
anti-symmetric $4\times 4$ matrices. We would like to choose these
matrices so that the first four, $T^1,\ldots ,T^4$ correspond to the
Pauli matrices $\sigma^1,\ldots , \sigma^4$ of the previous $N>2$
case, when written in real coordinates.  This ensures that our
result for $N=2$ indeed exactly reduces to the one for $N>2$ when
the additional degrees of freedom are ``switched off'' and,
hence, the action for both cases can be written in a uniform language.
It turns out that such a choice of matrices is given by
\begin{align}
T^{1}  &  =\left(
\begin{array}
[c]{cccc}%
0 & 0 & 0 & -1\\
0 & 0 & 1 & 0\\
0 & -1 & 0 & 0\\
1 & 0 & 0 & 0
\end{array}
\right)  ,\hspace{0.3cm}T^{2}=\left(
\begin{array}
[c]{cccc}%
0 & 0 & 1 & 0\\
0 & 0 & 0 & 1\\
-1 & 0 & 0 & 0\\
0 & -1 & 0 & 0
\end{array}
\right)  ,\hspace{0.3cm}\\
T^{3}  &  =\left(
\begin{array}
[c]{cccc}%
0 & -1 & 0 & 0\\
1 & 0 & 0 & 0\\
0 & 0 & 0 & 1\\
0 & 0 & -1 & 0
\end{array}
\right)  ,\hspace{0.3cm} T^{4}=\left(
\begin{array}
[c]{cccc}%
0 & -1 & 0 & 0\\
1 & 0 & 0 & 0\\
0 & 0 & 0 & -1\\
0 & 0 & 1 & 0
\end{array}
\right)\, .\hspace{0.3cm}%
\end{align}
The two remaining matrices can be taken as
\begin{equation}
T^{5}=\left(
\begin{array}
[c]{cccc}%
0 & 0 & 1 & 0\\
0 & 0 & 0 & -1\\
-1 & 0 & 0 & 0\\
0 & 1 & 0 & 0
\end{array}
\right)  ,\hspace{0.3cm}T^{6}=\left(
\begin{array}
[c]{cccc}%
0 & 0 & 0 & 1\\
0 & 0 & 1 & 0\\
0 & -1 & 0 & 0\\
-1 & 0 & 0 & 0
\end{array}
\right)  ,
\end{equation}
Note that $T^{1,2,3}$ and $T^{4,5,6}$ form the two sets of $SU(2)$ generators
within the $SO(4)$ Lie algebra. We may introduce a ``dual'' to the six $T^{I}$ matrices,
analogous to the definition of the $\bar{\sigma}^{I}$ matrices of the $N>2$ case, which will
prove useful in many calculations. We define
\begin{equation}
\left(  \overline{T}^{I}\right)  ^{AB}=-\frac{1}{2}\left(  T^{I}\right)
_{CD}\epsilon^{ABCD}%
\end{equation}
which has the simple form
\begin{equation}
(\overline{T}^{I})=(T^{u},-T^{\alpha}),
\end{equation}
where $u,v,\ldots =1,2,3$ and $\alpha ,\beta ,\ldots =4,5,6$. Indices $I,J,\ldots$ are raised
and lowered with the metric $(\eta_{IJ})={\rm diag}(-1,-1,-1,+1,+1,+1)$.
The matrices $T^{I}$ satisfy the following useful identities 
\begin{align}
\mathrm{tr}\left(  T^{I~}\overline{T}^{J}\right)   &  =4\eta^{IJ},\\
\left(  T^{I}\right)  _{AB}\left(  T^{J}\right)  _{CD}\eta_{IJ}  &
=2\epsilon_{ABCD},\\
\left(  T^{u}\right)  _{AB}\left(  T^{v}\right)  _{CD}\delta_{uv}  &
=\delta_{AC}\delta_{BD}-\delta_{AD}\delta_{BC}-\epsilon_{ABCD},\\
(T^{\alpha})_{AB}(T^{\beta})_{CD}\delta_{\alpha\beta}  &  =\delta_{AC}%
\delta_{BD}-\delta_{AD}\delta_{BC}+\epsilon_{ABCD}.
\end{align}
Key property of the map $f$ is
\begin{equation}
 (u_Iu^I)^2={\rm det}(f(u))
\end{equation}
for any $SO(3,3)$ vector $u_I$. This property can be used to show
that the map $F$ defined by
\begin{equation}
 F(v)u = f^{-1}\left( vf(u)v^T\right)
\end{equation}
is a group homeomorphism $F:SL(4,\mathbb{R})\rightarrow SO(3,3)$.
Solving for the $SO(3,3)$ images ${\ell_I}^{\ul{J}}={(F(v))_I}^{\ul{J}}$
one finds
\begin{equation}
 {\ell_I}^{\ul{J}}=\frac{1}{4}{\rm tr}\left( \bar{T}_IvT^Jv^T\right)\; .
\end{equation}
This induces the desired map between the cosets $SL(4,\mathbb{R})/SO(4)$
and $SO(3,3)/SO(3)^2$.


\section{Einstein-Yang-Mills supergravity in seven dimensions} \label{bigEYM}
In this section we will give a self-contained summary of minimal,
${\cal N}=1$ Einstein-Yang-Mills (EYM) supergravity in seven
dimensions. Although the theory may be formulated in two equivalent
ways, here we treat only the version in which the gravity multiplet
contains a three-form potential $C_{\mu\nu\rho}$ \cite{Park:1988id}, rather
than the dual formulation in terms of a two-index antisymmetric field
$B_{\mu\nu}$ which has been studied in
Refs.~\cite{Bergshoeff:1985mr,Han}. This three-form formulation is better
suited for our application to M-theory. The theory has an $SU(2)$
rigid R-symmmetry that may be gauged and the resulting massive
theories were first constructed in
Refs.~\cite{Townsend:1983kk,Mezincescu:1984ta,Giani:1984dw}. The seven-dimensional
supergravities we obtain by truncating M-theory are not massive and,
for this reason, we will not consider such theories with gauged
R-symmetry. The seven-dimensional pure supergravity theory can also be
coupled to $M$ vector multiplets~\cite{Park:1988id,Avramis:2004cn,Cremmer:1979up,Nicolai:1980td},
transforming under a Lie group $G=U(1)^n\times H$, where $H$ is
semi-simple, in which case the vector multiplet scalars parameterize
the coset space $SO(3,M)/SO(3)\times SO(M)$. In this Appendix, we will
first review seven-dimensional ${\cal N}=1$ EYM supergravity with such
a gauge group $G$. This theory is used in the main part of the paper
to construct the complete action for low-energy M-theory on the
orbifolds $\mathbb{R}^{1,6}\times\mathbb{C}^{2}/\mathbb{Z}_{N}$. The
truncation of M-theory on these orbifolds to seven dimensions leads to
a $d=7$ EYM supergravity with gauge group $U(1)^n\times SU(N)$, where
$n=1$ for $N>2$ and $n=3$ for $N=2$. Here, the $U(1)^n$ part of the
gauge group originates from truncated bulk states, while the $SU(N)$
non-Abelian part corresponds to the additional states which arise on
the orbifold fixed plane.  Since we are constructing the coupled
11-/7-dimensional theory as an expansion in $SU(N)$ fields, the
crucial building block is a version of $d=7$ EYM supergravity with
gauge group $U(1)^n\times SU(N)$, expanded around the supergravity and
$U(1)^n$ part. This expanded version of the theory is presented in the
final part of this Appendix.

\subsection{General action and supersymmetry transformations} \label{full}
The field content of gauged $d=7$, ${\cal N}=1$ EYM supergravity
consists of two types of multiplets. The first, the gravitational
multiplet, contains a graviton $g_{\mu\nu}$ with associated vielbein ${e_\mu}^{\ul{\nu}}$, a gravitino $\psi_\mu^i$,
a symplectic Majorana spinor $\chi^i$, an $SU(2)$ triplet of Abelian
vector fields ${{A_\mu}^i}_j$ with field strengths ${F^i}_j=\mathrm{d}{A^i}_j$, a three form field $C_{\mu\nu\rho}$ with field strength $G=\mathrm{d}C$,
and a real scalar $\sigma$. So, in summary we have
\begin{equation} \label{gravmult}
 \left( g_{\mu\nu},C_{\mu\nu\rho},{{A_\mu}^i}_j,\sigma ,\psi^i_\mu ,\chi^i\right)\; .
\end{equation}
Here, $i,j,\ldots = 1,2$ are $SU(2)$ R-symmetry indices. The second type is the vector multiplet, which contains 
gauge vectors $A_\mu^a$ with field strengths $F^a={\cal D}A^a$, gauginos $\lambda^{ai}$ and $SU(2)$ triplets of real scalars ${\phi^{ai}}_j$. In summary, we have
\begin{equation} \label{vecmult}
 \left( A_\mu^a, {\phi^{ai}}_j, \lambda^{ai}\right)\; ,
\end{equation}
where $a,b,\ldots = 4,\ldots ,M+3$ are Lie algebra indices of the gauge group $G$.

It is sometimes useful to combine all vector fields, the three Abelian ones in the gravity
multiplet as well as the ones in the vector multiplets, into a single $SO(3,M)$
vector
\begin{equation}
 (A_\mu^{\tilde{I}})=\left({{A_\mu}^i}_j,A_\mu^a \right)\; ,
\end{equation}
where $\tilde{I},\tilde{J},\ldots = 1,\ldots ,M+3$. Under this combination, the corresponding field strengths are given by
\begin{equation}
F_{\mu\nu}^{\tilde{I}}=2\partial_{\lbrack\mu}A_{\nu]}^{\tilde{I}}+{f_{\tilde
{J}\tilde{K}}}^{\ti{I}}A_{\mu}^{\tilde{J}}A_{\nu}^{\tilde{K}},
\end{equation}
where ${f_{bc}}^a$ are the structure constants for $G$ and all other components of ${f_{\tilde{J}\tilde{K}}}^{\ti{I}}$ vanish. 

The coset space $SO(3,M)/SO(3)\times SO(M)$ is described by a
$(3+M)\times (3+M)$ matrix
$L_{\tilde{I}}^{\phantom{I}\underline{\tilde{J}}}$, which depends on
the $3M$ vector multiplet scalars and satisfies the $SO(3,M)$
orthogonality condition
\begin{equation}
L_{\tilde{I}}^{\phantom{I}  \underline{\tilde{J}}}L_{\tilde{K}}^{\phantom{K}
\underline{\tilde{L}}}\eta_{\underline{\tilde{J}}\underline{\tilde{L}}}%
=\eta_{\tilde{I}\tilde{K}}
\end{equation}
with
$(\eta_{\tilde{I}\tilde{J}})=(\eta_{\underline{\tilde{I}}\underline
{\tilde{J}}})=\mathrm{diag}(-1,-1,-1,+1,\ldots,+1)$. Here, indices
$\ti{I},\ti{J},\ldots = 1,\ldots,(M+3)$ transform under
$SO(3,M)$. Their flat counterparts $\ti{\ul{I}},\ti{\ul{J}},\ldots$
decompose into a triplet of $SU(2)$, corresponding to the
gravitational directions and $M$ remaining directions corresponding to
the vector multiplets. Thus we can write $L_{\tilde{I}}^{\phantom{I}
\underline{\tilde{J}}}\to\left(
{L_{\tilde{I}}}^u,L_{\tilde{I}}{}^{a}\right)$, where $u=1,2,3$. The
adjoint $SU(2)$ index $u$ can be converted into a pair of fundamental
$SU(2)$ indices by multiplication with the Pauli matrices, that is,
\begin{equation}
L_{\tilde{I}}{}^{i}{}_{j}=\frac{1}{\sqrt{2}}L_{\tilde{I}}{}^{u}\left(
\sigma_{u}\right) ^{i}{}_{j}\;.
\end{equation}
There are obviously many ways in
which one can parameterize the coset space $SO(3,M)/SO(3)\times SO(M)$
in terms of the physical vector multiplet scalar degrees of freedom
${{\phi_a}^i}_j$. A simple parameterization of this coset in terms
of $\Phi\equiv({\phi_{a}}^u)$ is given by
\begin{equation}
L_{\tilde{I}}^{\phantom{I}  \underline{\tilde{J}}}=\left(  \exp\left[
\begin{array}[c]{cc}%
0 & \Phi^T\\
\Phi & 0
\end{array}
\right]  \right)  _{\tilde{I}}^{\phantom{I}  \underline{\tilde{J}}}\text{.}%
\end{equation}
In the final paragraph of this appendix, when we expand seven-dimensional
supergravity, we will use a different parameterization, which is better 
adapted to this task. The Maurer-Cartan form of the matrix $L$, defined by
$L^{-1}\mathcal{D}L$, is needed to write down the theory. The
components $P$ and $Q$ are given explicitly by
\begin{align}
P_{\mu a\phantom{i} j}^{\phantom{\mu a} i} &  =
L_{\phantom{I} a}^{\tilde{I}}\left( \delta_{\tilde{I}}^{\tilde{K}}
 \partial_{\mu}+f_{\tilde{I}\tilde{J}}{}^{\tilde{K}}A_{\mu}^{\ti{J}}\right)
 L_{\tilde{K}\phantom{i} j}^{\phantom{K} i},\\
Q_{\mu\phantom{i} j}^{\phantom{\mu} i} &  =
L_{\phantom{Ii} k}^{\tilde{I}i}\left( \delta_{\tilde{I}}^{\tilde{K}}
 \partial_{\mu}+f_{\tilde{I}\tilde{J}}{}^{\tilde{K}}A_{\mu}^{\ti{J}}\right)
 L_{\tilde{K}\phantom{k} j}^{\phantom{K} k}.
\end{align}
The final ingredients needed are the following projections of the structure constants
\begin{align}
D  &  =if_{ab}^{\phantom{ab}  c}L_{\phantom{ai}  k}^{ai}L_{\phantom{bj}
i}^{bj}L_{c\phantom{k}  j}^{\phantom{c}  k},\nonumber\\
D_{\phantom{ai}  j}^{ai}  &  =if_{bc}^{\phantom{bc}  d}L_{\phantom{bi}
k}^{bi}L_{\phantom{ck}  j}^{ck}L_{d}^{\phantom{d}  a},\nonumber\\
D_{ab\phantom{i}  j}^{\phantom{ab}  i}  &  =f_{cd}^{\phantom{cd}  e}%
L_{a}^{\phantom{a}  c}L_{b}^{\phantom{b}  d}L_{e\phantom{i}  j}^{\phantom{e}
i}.
\end{align}

It is worth mentioning that invariance of the theory under the gauge
group $G$ and the R-symmetry group $SU(2)$ requires that the
Maurer-Cartan forms $P$ and $Q$ transform covariantly. It can be shown
that this is the case, if and only if the ``extended'' set of
structure constants $f_{\tilde{I}\tilde{J}}{}^{\tilde{K}}$ satisfy the
condition
\begin{equation}
f_{\tilde{I}\tilde{J}}{}^{\tilde{L}}\eta_{\tilde{L}}{}_{\tilde{K}%
}=f_{[\tilde{I}\tilde{J}}{}^{\tilde{L}}\eta{}_{\tilde{K}]}{}_{\tilde{L}%
}.\label{gauge condition}
\end{equation}
For any direct product factor of the total gauge group, this condition
can be satisfied in two ways. Either, the structure constants are
trivial, or the metric $\eta_{\tilde{I}\tilde{J}}$ is the Cartan-Killing
metric of this factor. In our particular case, the condition~\eqref{gauge condition}
is satisfied by making use of both these possibilities. 
The structure constants vanish for the ``gravitational'' part of the
gauge group and the $U(1)^n$ part within $G$. For the semi-simple part $H$
of $G$, one can always choose a basis, so its Cartan-Killing metric is
simply the Kronecker delta.

With everything in place, we now write down the Lagrangian for the
theory. Setting coupling constants to one, and neglecting four-fermi
terms, it is given by \cite{Park:1988id}
\begin{align}
e^{-1}\mathcal{L}_{\mathrm{YM}}  &  =\frac{1}{2}R-\frac{1}{2}\bar{\psi}_{\mu
}^{i}\Upsilon^{\mu\nu\rho}\hat{\mathcal{D}}_{\nu}\psi_{\rho i}-\frac{1}%
{96}e^{4\sigma}G_{\mu\nu\rho\sigma}G^{\mu\nu\rho\sigma}-\frac{1}{2}\bar{\chi
}^{i}\Upsilon^{\mu}\hat{\mathcal{D}}_{\mu}\chi_{i}-\frac{5}{2}\partial_{\mu
}\sigma\partial^{\mu}\sigma\nonumber\label{EYM}\\
&  +\frac{\sqrt{5}}{2}\left(  \bar{\chi}^{i}\Upsilon^{\mu\nu}\psi_{\mu i}%
+\bar{\chi}^{i}\psi_{i}^{\nu}\right)  \partial_{\nu}\sigma+e^{2\sigma}%
G_{\mu\nu\rho\sigma}\left[  \frac{1}{192}\left(  \bar{\psi}_{\lambda}%
^{i}\Upsilon^{\lambda\mu\nu\rho\sigma\tau}\psi_{\tau i}+12\bar{\psi}^{\mu
i}\Upsilon^{\nu\rho}\psi_{i}^{\sigma}\right)  \right. \nonumber\\
&  \hspace{4.2cm}\left.  +\frac{1}{48\sqrt{5}}\left(  4\bar{\chi}^{i}%
\Upsilon^{\mu\nu\rho}\psi_{i}^{\sigma}-\bar{\chi}^{i}\Upsilon^{\mu\nu
\rho\sigma\tau}\psi_{\tau i}\right)  -\frac{1}{320}\bar{\chi}^{i}\Upsilon
^{\mu\nu\rho\sigma}\chi_{i}\right] \nonumber\\
&  -\frac{1}{4}e^{-2\sigma}\left(  L_{\tilde{I}\phantom{i}  j}^{\phantom{I}
i}L_{\tilde{J}\phantom{j}  i}^{\phantom{J}  j}+L_{\tilde{I}}^{a}L_{\tilde{J}%
a}\right)  F_{\mu\nu}^{\tilde{I}}F^{\tilde{J}\mu\nu}-\frac{1}{2}\bar{\lambda
}^{ai}\Upsilon^{\mu}\hat{\mathcal{D}}_{\mu}\lambda_{ai}-\frac{1}{2}%
P_{\mu\phantom{ia}  j}^{\phantom{\mu}  ai}P_{\phantom{\mu}  a\phantom{j}
i}^{\mu\phantom{a}  j}\nonumber\\
&  -\frac{1}{\sqrt{2}}\left(  \bar{\lambda}^{ai}\Upsilon^{\mu\nu}\psi_{\mu
j}+\bar{\lambda}^{ai}\psi_{j}^{\nu}\right)  P_{\nu a\phantom{j}
i}^{\phantom{\nu a}  j}+\frac{1}{192}e^{2\sigma}G_{\mu\nu\rho\sigma}%
\bar{\lambda}^{ai}\Upsilon^{\mu\nu\rho\sigma}\lambda_{ai}\nonumber\\
&  -ie^{-\sigma}F_{\mu\nu}^{\tilde{I}}L_{\tilde{I}\phantom{j}  i}%
^{\phantom{I}  j}\left[  \frac{1}{4\sqrt{2}}\left(  \bar{\psi}_{\rho}%
^{i}\Upsilon^{\mu\nu\rho\sigma}\psi_{\sigma j}+2\bar{\psi}^{\mu i}\psi
_{j}^{\nu}\right)  +\frac{3}{20\sqrt{2}}\bar{\chi}^{i}\Upsilon^{\mu\nu}%
\chi_{j}\right. \nonumber\\
&  \hspace{4.2cm}\left.  -\frac{1}{4\sqrt{2}}\bar{\lambda}^{ai}\Upsilon
^{\mu\nu}\lambda_{aj}+\frac{1}{2\sqrt{10}}\left(  \bar{\chi}^{i}\Upsilon
^{\mu\nu\rho}\psi_{\rho j}-2\bar{\chi}^{i}\Upsilon^{\mu}\psi_{j}^{\nu}\right)
\right] \nonumber\\
&  +e^{-\sigma}F_{\mu\nu}^{\tilde{I}}L_{\tilde{I}a}\left[  \frac{1}{4}\left(
2\bar{\lambda}^{ai}\Upsilon^{\mu}\psi_{i}^{\nu}-\bar{\lambda}^{ai}%
\Upsilon^{\mu\nu\rho}\psi_{\rho i}\right)  +\frac{1}{2\sqrt{5}}\bar{\lambda
}^{ai}\Upsilon^{\mu\nu}\chi_{i}\right] \nonumber\\
&  +\frac{5}{180}e^{2\sigma}\left(  D^{2}-9D_{\phantom{ai}  j}^{ai}%
D_{a\phantom{j}  i}^{\phantom{a}  j}\right)  -\frac{i}{\sqrt{2}}e^{\sigma
}D_{ab\phantom{i}  j}^{\phantom{ab}  i}\bar{\lambda}^{aj}\lambda_{i}%
^{b}+\frac{i}{2}e^{\sigma}D_{a\phantom{i}  j}^{\phantom{a}  i}\left(
\bar{\psi}_{\mu}^{j}\Upsilon^{\mu}\lambda_{i}^{a}+\frac{2}{\sqrt{5}}\bar{\chi
}^{j}\lambda_{i}^{\phantom{i}  a}\right) \nonumber\\
&  +\frac{1}{60\sqrt{2}}e^{\sigma}D\left(  5\bar{\psi}_{\mu}^{i}\Upsilon
^{\mu\nu}\psi_{\nu i}+2\sqrt{5}\bar{\psi}_{\mu}^{i}\Upsilon^{\mu}\chi
_{i}+3\bar{\chi}^{i}\chi_{i}-5\bar{\lambda}^{ai}\lambda_{ai}\right)
\nonumber\\
&  -\frac{1}{96}\epsilon^{\mu\nu\rho\sigma\kappa\lambda\tau}C_{\mu\nu\rho
}F_{\sigma\kappa}^{\tilde{I}}F_{\tilde{I}}{}_{\lambda\tau}.
\end{align}
The covariant derivatives that appear here are given by 
\bea
\hat{\mathcal{D}}_\mu\psi_{\nu i}&=&\partial_{\mu}\psi_{\nu i}+\frac{1}{2}Q_{\mu
i}{}^{j}\psi_{\nu j}-\Gamma^\rho_{\mu\nu}\psi_{\rho i} +\frac{1}{4}\omega_{\mu}^{\phantom{\mu}  \underline
{\mu}\underline{\nu}}\Upsilon_{\underline{\mu}\underline{\nu}}\psi_{\nu i}, \\
\hat{\mathcal{D}}_{\mu}\chi_{i}&=&\partial_{\mu}\chi_{i}+\frac{1}{2}Q_{\mu
i}{}^{j}\chi_{j}+\frac{1}{4}\omega_{\mu}^{\phantom{\mu}  \underline
{\mu}\underline{\nu}}\Upsilon_{\underline{\mu}\underline{\nu}}\chi_{i},\\
\hat{\mathcal{D}}_{\mu}\la_{ai}&=&\partial_{\mu}\la_{ai}+\frac{1}{2}Q_{\mu
i}{}^{j}\la_{aj}+\frac{1}{4}\omega_{\mu}^{\phantom{\mu}  \underline
{\mu}\underline{\nu}}\Upsilon_{\underline{\mu}\underline{\nu}}\la_{ai}+{f_{ab}}^cA_\mu^b\la_{ci}.
\eea
The associated supersymmetry transformations, parameterized by the spinor $\varepsilon_i$, are, up to cubic fermion terms, given by
\begin{align}
\delta\sigma &  =\frac{1}{\sqrt{5}}\bar{\chi}^{i}\varepsilon_{i}\text{ ,
}\nonumber\\
{\delta e_{\mu}}^{\underline{\nu}}  &  =\bar{\varepsilon}^{i}\Upsilon
^{\underline{\nu}}\psi_{\mu i}\text{ ,}\nonumber\\
\delta\psi_{\mu i}  &  =2\hat{\mathcal{D}}_{\mu}\varepsilon_{i}-\frac{1}%
{80}\left(  \Upsilon_{\mu}^{\phantom{\mu}  \nu\rho\sigma\eta}-\frac{8}%
{3}\delta_{\mu}^{\nu}\Upsilon^{\rho\sigma\eta}\right)  \varepsilon_{i}%
G_{\nu\rho\sigma\eta}e^{2\sigma}\text{ }\nonumber\\
&  +\frac{i}{5\sqrt{2}}\left(  \Upsilon_{\mu}^{\phantom{\mu}  \nu\rho}%
-8\delta_{\mu}^{\nu}\Upsilon^{\rho}\right)  \varepsilon_{j}F_{\nu\rho
}^{\tilde{I}}L_{\tilde{I}\phantom{i}  i}^{\phantom{I}  j}e^{-\sigma}%
-\frac{1}{15\sqrt{2}}e^{\sigma}\Upsilon_{\mu}\varepsilon_{i}D\text{
,}\nonumber\\
\delta\chi_{i}  &  =\sqrt{5}\Upsilon^{\mu}\varepsilon_{i}\partial_{\mu}%
\sigma-\frac{1}{24\sqrt{5}}\Upsilon^{\mu\upsilon\rho\sigma}\varepsilon
_{i}G_{\mu\nu\rho\sigma}e^{2\sigma}\text{ }-\frac{i}{\sqrt{10}}\Upsilon
^{\mu\nu}\varepsilon_{j}F_{\mu\nu}^{\tilde{I}}L_{\tilde{I}\phantom{i}
i}^{\phantom{I}  j}e^{-\sigma}+\frac{1}{3\sqrt{10}}e^{\sigma}\varepsilon
_{i}D\text{ ,}\nonumber\\
\delta C_{\mu\nu\rho}  &  =\left(  -3\bar{\psi}_{\left[  \mu\right.  }%
^{i}\Upsilon_{\left.  \nu\rho\right]  }\varepsilon_{i}-\frac{2}{\sqrt{5}}%
\bar{\chi}^{i}\Upsilon_{\mu\nu\rho}\varepsilon_{i}\right)  e^{-2\sigma}\text{
,} \label{7dsusy} \\
L_{\tilde{I}\phantom{i}  j}^{\phantom{I}  i}\delta A_{\mu}^{\tilde{I}}  &
=\left[  i\sqrt{2}\left(  \bar{\psi}_{\mu}^{i}\varepsilon_{j}-\frac{1}%
{2}\delta_{j}^{i}\bar{\psi}_{\mu}^{k}\varepsilon_{k}\right)  -\frac{2i}%
{\sqrt{10}}\left(  \bar{\chi}^{i}\Upsilon_{\mu}\varepsilon_{j}-\frac{1}{2}%
{}\delta_{j}^{i}\bar{\chi}^{k}\Upsilon_{\mu}\varepsilon_{k}\right)  \right]
e^{\sigma}\text{ ,}\nonumber\\
L_{\tilde{I}}^{\phantom{I}  a}\delta A_{\mu}^{\tilde{I}}  &  =\bar
{\varepsilon}^{i}\Upsilon_{\mu}\lambda_{i}^{a}e^{\sigma}\text{ ,}\nonumber\\
\delta L_{\tilde{I}\phantom{i}  j}^{\phantom{I}  i}  &  =-i\sqrt{2}%
\bar{\varepsilon}^{i}\lambda_{aj}L_{\tilde{I}}^{\phantom{I}  a}+\frac{i}%
{\sqrt{2}}\bar{\varepsilon}^{k}\lambda_{ak}L_{\tilde{I}}^{a}\delta_{j}%
^{i}\text{ ,}\nonumber\\
\delta L_{\tilde{I}}^{\phantom{I}  a}  &  =-i\sqrt{2}\bar{\varepsilon}%
^{i}\lambda_{j}^{a}L_{\tilde{I}\phantom{j}  i}^{\phantom{I}  j}\text{
,}\nonumber\\
\delta\lambda_{i}^{a}  &  =-\frac{1}{2}\Upsilon^{\mu\nu}\varepsilon_{i}%
F_{\mu\nu}^{\tilde{I}}L_{\tilde{I}}^{a}e^{-\sigma}+\sqrt{2}i\Upsilon^{\mu
}\varepsilon_{j}P_{\mu\phantom{aj}  i}^{\phantom{\mu }  aj}\text{ }-e^{\sigma
}\varepsilon_{j}D^{aj}{}_{i}\;. \nn 
\end{align}

\subsection{A perturbative expansion}\label{bigEYM2}
In this final section we expand the EYM supergravity of Section
\ref{full} around its supergravity and $U(1)^n$ part. The parameter
for the expansion is $h:=\kappa_7/g_\mathrm{YM}$, where $\kappa_7$ is
the coupling for gravity and $U(1)^n$ and $g_\mathrm{YM}$ is the
coupling for $H$, the non-Abelian part of the gauge group. To
determine the order in $h$ of each term in the Lagrangian, we need to
fix a convention for the energy dimensions of the fields. Within the
gravity and $U(1)$ vector multiplets, we assign energy dimension 0 to
bosonic fields and energy dimension 1/2 to fermionic fields.  For the
$H$ vector multiplet, we assign energy dimension 1 to the bosons and
3/2 to the fermions. With these conventions we can write
\begin{equation}
\mathcal{L}_{\mathrm{YM}}=\kappa_7^{-2}\left( \mathcal{L}_{(0)}+h^2\mathcal{L}_{(2)}+h^4\mathcal{L}_{(4)}+\ldots \right),
\end{equation}
where the $\mathcal{L}_{(m)}$, $m=0,2,4,\ldots$ are independent of
$h$. The first term in this series is the Lagrangian for EYM
supergravity with gauge group $U(1)^n$, whilst the second term
contains the leading order non-Abelian gauge multiplet terms. We will
write down these first two terms and provide truncated supersymmetry
transformation laws suitable for the theory at this order.

In order to carry out the expansion, it is necessary to cast the field
content in a form where the $H$ vector multiplet fields and the
gravity/$U(1)^n$ vector multiplet fields are disentangled. To this
end, we decompose the single Lie algebra indices
$a,b,\ldots=4,\ldots,M+3$ used in Section \ref{full} into indices
$\alpha,\beta,\ldots=4,\ldots,3+n$ that label the $U(1)$ directions
and redefined indices $a,b,\ldots=n+4,\ldots,M+3$ that are Lie algebra
indices of $H$. This makes the disentanglement straightforward for
most of the fields. For example, vector fields, which naturally
combine into the single entity $A_\mu^{\ti{I}}$, can simply be
decomposed as $A_\mu^{\ti{I}}=(A_\mu^I,A_\mu^a)$, where $A_\mu^I$,
$I=1,\ldots,n+3$, refers to the three vector fields in the gravity
multiplet and the $U(1)^n$ vector fields, and $A_\mu^a$ denotes the
$H$ vector fields. Similarly, the $U(1)$ gauginos are denoted by
$\la_{\alpha i}$, whilst the $H$ gauginos are denoted by
$\la_{ai}$. The situation is somewhat more complicated for the vector
multiplet scalar fields, which, as discussed, all together combine
into the single coset $SO(3,M)/SO(3)\times SO(M)$, parameterized by
the $SO(3,M)$ matrix $L$. It is necessary to find an explicit form for
$L$, which separates the $3n$ scalars in the $U(1)^n$ vector
multiplets from the $3(M-n)$ scalars in the $H$ vector multiplet. To
this end, we note that, in the absence of the $H$ states, the $U(1)^n$
states parameterize a $SO(3,n)/SO(3)\times SO(n)$ coset, described by
$(3+n)\times (3+n)$ matrices
${\ell_I}^{\underline{I}}=({\ell_I}^u,{\ell_I}^\alpha )$. Here,
$\ell\equiv (\ell_I^{\ph{I}u})$ are $(3+n)\times 3$ matrices where the
index $u=1,2,3$ corresponds to the three ``gravity'' directions and
$m\equiv ( \ell_I^{\ph{I}\alpha})$ are $(3+n)\times n$ matrices with
$\alpha =4,\ldots ,n+3$ labeling the $U(1)^n$ directions.  Let us
further denote the $SU(N)$ scalars by $\Phi\equiv
(\phi_a^{\ph{a}u})$. Then we can construct approximate representatives
$L$ of the large coset $SO(3,M)/SO(3)\times SO(M)$ by expanding, to
the appropriate order in $\Phi$, around the small coset
$SO(3,n)/SO(3)\times SO(n)$ represented by $\ell$ and $m$. Neglecting
terms of cubic and higher order in $\Phi$, this leads to
\begin{equation}\label{L}
L = \left( \begin{array}{ccc}
\ell+\frac{1}{2}h^2\ell\Phi^T\Phi & m & h\ell\Phi^T \\
h\Phi & 0 & \boldsymbol{1}_{M-n}+\frac{1}{2}h^2\Phi\Phi^T \\
\end{array} \right)\, .
\end{equation}
We note that the neglected $\Phi$ terms are of order $h^3$ and higher 
and, since we are aiming to construct the action only up to terms of order
$h^2$, are, therefore, not relevant in the present context. 

For the expansion of the action it is useful to re-write the coset parameterization~\eqref{L}
and the associated Maurer-Cartan forms $P$ and $Q$ in component form. We find
\begin{align}
{L_{I}{}^{i}}_{j}  &  ={\ell_{I}{}^{i}}_{j}+\frac{1}{2}h^2\ell_{I}{}^{k}{}_{l}\phi_{\phantom{bi}  k}^{al}\phi_{a}{}_{\phantom{bi}j}^{i}\; ,\\
{L_I}^\alpha &=h {\ell_I}^\alpha,  \\
L_{I}{}^{a}  &  =h\ell_{I}{}^{i}{}_{j}\phi_{\phantom{aj}i}^{aj},\\
L_{a}{}^{i}{}_{j}  &  =h\phi_{a}{}_{\phantom{b}j}^{i}\; ,\\
{L_a}^\alpha & = 0\; , \\
L_{a}{}^{b}  &  ={\delta_{a}}^{b}+\frac{1}{2}h^2\phi_{a}{}_{\phantom{i}j}^{i}
\phi^{bj}_{\phantom{bj}i}\; , \\
{P_{\mu\alpha}{}^{i}}_{j}  &  = {p_{\mu\alpha}{}^{i}}_{j}+\frac{1}{2}h^2{p_{\mu\alpha}}^{k}{}_{l}\phi^{a}{}^{l}{}_{k}{\phi_{a}{}^{i}}_{j}\; ,\\
{{P_{\mu a}}^i}_j  & = -h\hat{\mathcal{D}}_{\mu}{\phi_{a}{}^{i}}_{j}\; ,\\
Q_{\mu}{}^{i}{}_{j}  &  =q_{\mu}{}^{i}{}_{j}+\frac{1}{2}h^2\left( \phi^{a}{}^{i}{}%
_{k}\hat{\mathcal{D}}_{\mu}\phi_{a}{}^{k}{}_{j}-\phi_{a}{}^{k}{}_{j}\hat{\mathcal{D}}_{\mu}\phi^{a}{}^{i}{}_{k}\right)\; ,
\end{align}
where $p$ and $q$ are the Maurer-Cartan forms associated with the small coset matrix $\ell$. Thus
\begin{align}
p_{\mu\alpha\phantom{i}  j}^{\phantom{\mu\alpha }  i}  &  =\ell_{\phantom{I}
\alpha}^{I}\partial_{\mu}\ell_{I\phantom{i}  j}^{\phantom{\mu}  i}
,\\
q_{\mu\phantom{i}  j\phantom{k}  l}^{\phantom{\mu}  i\phantom{j}  k}  &
=\ell_{\phantom{Ii}  j}^{Ii}\partial_{\mu}\ell_{I\phantom{k}  l}%
^{\phantom{\mu}  k}\; ,\\
q_{\mu\phantom{i}  j}^{\phantom{\mu}  i}  &  =\ell_{\phantom{Ii}  k}%
^{Ii}\partial_{\mu}\ell_{I\phantom{k}  j}^{\phantom{\mu}  k}\; . %
\end{align}
The covariant derivative of the $H$ vector multiplet scalar ${{\phi_a}^i}_j$ is given by
\begin{equation}
\hat{\mathcal{D}}_{\mu}\phi_{a\phantom{i}j}^{\phantom{a}i}=\partial_{\mu
}\phi_{a\phantom{i}  j}^{\phantom{a}i}-q_{\mu\phantom{i}j\phantom{k}
l}^{\phantom{\mu}i\phantom{j}k}\phi_{a\phantom{l}  k}^{\phantom{a}
l}+f_{ab}^{\phantom{ab}c}A_{\mu}^{b}\phi_{c\phantom{i}j}^{\phantom{c}i}.
\end{equation}

Using the expressions above, it is straightforward to perform the expansion of $\mathcal{L}_{\mathrm{YM}}$ up to order $h^2\sim g_{\mathrm{YM}}^{-2}$. It is given by
\begin{align} \label{truncatedL}
\mathcal{L}_{\mathrm{YM}}\!  &  =\!\frac{1}{\kappa_{7}^{2}}\sqrt{-g}\left\{  \frac{1}{2}R-\frac{1}{2}\bar{\psi}_{\mu}^{i}\Upsilon^{\mu\nu\rho
}\hat{\mathcal{D}}_{\nu}\psi_{\rho i}-\frac{1}{4}e^{-2\sigma}\left(
\ell_{I\phantom{i}  j}^{\phantom{I}  i}\ell_{J\phantom{j}  i}^{\phantom{J}
j}+\ell_{I}^{\phantom{I}  \alpha}\ell_{J\alpha}\right)  F_{\mu\nu}^{I}%
F^{J\mu\nu}\right. \nonumber\\
&  \hspace{1.5cm}-\frac{1}{96}e^{4\sigma}G_{\mu\nu\rho\sigma}G^{\mu\nu\rho\sigma}-\frac{1}{2}\bar{\chi}^{i}\Upsilon^{\mu}\hat
{\mathcal{D}}_{\mu}\chi_{i}-\frac{5}{2}\partial_{\mu}\sigma\partial^{\mu
}\sigma+\frac{\sqrt{5}}{2}\left(  \bar{\chi}^{i}\Upsilon^{\mu\nu}\psi_{\mu
i}+\bar{\chi}^{i}\psi_{i}^{\nu}\right)  \partial_{\nu}\sigma\nonumber\\
&  \hspace{1.5cm}-\frac{1}{2}\bar{\lambda}^{\alpha i}\Upsilon^{\mu}%
\hat{\mathcal{D}}_{\mu}\lambda_{\alpha i}-\frac{1}{2}p_{\mu\alpha\phantom{i}
j}^{\phantom{\mu\alpha}  i}p_{\phantom{\mu\alpha j}  i}^{\mu\alpha j}%
-\frac{1}{\sqrt{2}}\left(  \bar{\lambda}^{\alpha i}\Upsilon^{\mu\nu}\psi_{\mu
j}+\bar{\lambda}^{\alpha i}\psi_{j}^{\nu}\right)  p_{\nu\alpha\phantom{j}
i}^{\phantom{\nu\alpha}  j}\nonumber\\
&  \hspace{1.5cm}+e^{2\sigma}G_{\mu\nu\rho\sigma}\left[  \frac{1}%
{192}\left(  12\bar{\psi}^{\mu i}\Upsilon^{\nu\rho}\psi_{i}^{\sigma}+\bar
{\psi}_{\lambda}^{i}\Upsilon^{\lambda\mu\nu\rho\sigma\tau}\psi_{\tau
i}\right)  +\frac{1}{48\sqrt{5}}\left(  4\bar{\chi}^{i}\Upsilon^{\mu\nu\rho
}\psi_{i}^{\sigma}\right.  \right. \nonumber\\
&  \hspace{3.5cm}\left.  \left.  -\bar{\chi}^{i}\Upsilon^{\mu\nu\rho\sigma
\tau}\psi_{\tau i}\right)  -\frac{1}{320}\bar{\chi}^{i}\Upsilon^{\mu\nu
\rho\sigma}\chi_{i}+\frac{1}{192}\bar{\lambda}^{\alpha i}\Upsilon^{\mu\nu
\rho\sigma}\lambda_{\alpha i}\right] \nonumber\\
&  \hspace{1.5cm}-ie^{-\sigma}F_{\mu\nu}^{I}\ell_{I\phantom{j}  i}%
^{\phantom{I}  j}\left[  \frac{1}{4\sqrt{2}}\left(  \bar{\psi}_{\rho}%
^{i}\Upsilon^{\mu\nu\rho\sigma}\psi_{\sigma j}+2\bar{\psi}^{\mu i}\psi
_{j}^{\nu}\right)  +\frac{1}{2\sqrt{10}}\left(  \bar{\chi}^{i}\Upsilon^{\mu
\nu\rho}\psi_{\rho j}-2\bar{\chi}^{i}\Upsilon^{\mu}\psi_{j}^{\nu}\right)
\right. \nonumber\\
&  \hspace{3.9cm}\left.  +\frac{3}{20\sqrt{2}}\bar{\chi}^{i}\Upsilon^{\mu\nu
}\chi_{j}-\frac{1}{4\sqrt{2}}\bar{\lambda}^{\alpha i}\Upsilon^{\mu\nu}%
\lambda_{\alpha j}\right] \nonumber\\
&  \hspace{1.5cm}+e^{-\sigma}F_{\mu\nu}^{I}\ell_{I\alpha}\left[  \frac{1}%
{4}\left(  2\bar{\lambda}^{\alpha i}\Upsilon^{\mu}\psi_{i}^{\nu}-\bar{\lambda
}^{\alpha i}\Upsilon^{\mu\nu\rho}\psi_{\rho i}\right)  +\frac{1}{2\sqrt{5}%
}\bar{\lambda}^{\alpha i}\Upsilon^{\mu\nu}\chi_{i}\right] \nonumber\\
&  \hspace{1.5cm}\left.  -\frac{1}{96}\epsilon^{\mu\nu\rho\sigma\kappa
\lambda\tau}C_{\mu\nu\rho}F_{\sigma\kappa}^{I}F_{I\lambda\tau}\right\} \nn \\
& +\frac{1}{g_{\mathrm{YM}}^{2}}\sqrt{-g}\left\{
-\frac{1}{4}e^{-2\sigma}F_{\mu\nu}^{a}F_{a}^{\mu\nu}-\frac{1}{2}%
\hat{\mathcal{D}}_{\mu}\phi_{a\phantom{i}  j}^{\phantom{a}  i}\hat
{\mathcal{D}}^{\mu}\phi_{\phantom{aj}  i}^{aj}-\frac{1}{2}\bar{\lambda}^{ai}\Upsilon^{\mu}\hat{\mathcal{D}}_{\mu}\lambda_{ai}\right. \nonumber\\
& \hspace{1.8cm} \left.  -e^{-2\sigma}\ell_{I\phantom{i}  j}^{\phantom{I}  i}%
\phi_{a\phantom{j}  i}^{\phantom{a}  j}F_{\mu\nu}^{I}F^{a\mu\nu}-\frac{1}%
{2}e^{-2\sigma}\ell_{I\phantom{i}  j}^{\phantom{I}  i}\phi_{a\phantom{j}
i}^{\phantom{a}  j}\ell_{J\phantom{k}  l}^{\phantom{J}  k}{\phi^{al}}_k
F_{\mu\nu}^{I}F_{\mu\nu}^{J}\right. \nonumber\\
& \hspace{1.8cm} -\frac{1}{2}p_{\mu\alpha\phantom{i}  j}^{\phantom{\mu\alpha}  i}%
\phi_{a\phantom{j}  i}^{\phantom{a}  j}p_{\phantom{\mu\alpha k}  l}^{\mu\alpha
k}\phi_{\phantom{al}  k}^{al}+\frac{1}{4}\phi_{a\phantom{i}  k}^{\phantom{a}
i}\hat{\mathcal{D}}_{\mu}{\phi^{ak}}_j\bar{\lambda
}^{\alpha j}\Upsilon^{\mu}\lambda_{\alpha i}\nonumber\\
& \hspace{1.8cm} -\frac{1}{\sqrt{2}}\left(  \bar{\lambda}^{\alpha i}\Upsilon^{\mu\nu}%
\psi_{\mu j}+\bar{\lambda}^{\alpha i}\psi_{j}^{\nu}\right)  \phi
_{a\phantom{j}  i}^{\phantom{a}  j}\phi_{\phantom{ak}  l}^{ak}p_{\nu
\alpha\phantom{l}  k}^{\phantom{\nu\alpha}  l}-\frac{1}{\sqrt{2}}\left(
\bar{\lambda}^{ai}\Upsilon^{\mu\nu}\psi_{\mu j}+\bar{\lambda}^{ai}\psi
_{j}^{\nu}\right)  \hat{\mathcal{D}}_{\nu}\phi_{a\phantom{j}  i}%
^{\phantom{ a}  j}\nonumber\\
& \hspace{1.8cm} +\frac{1}{192}e^{2\sigma}G_{\mu\nu\rho\sigma}\bar{\lambda}%
^{ai}\Upsilon^{\mu\nu\rho\sigma}\lambda_{ai}+\frac{i}{4\sqrt{2}}e^{-\sigma
}F_{\mu\nu}^{I}\ell_{I\phantom{j}  i}^{\phantom{I}  j}\bar{\lambda}%
^{ai}\Upsilon^{\mu\nu}\lambda_{aj}\nonumber\\
& \hspace{1.8cm} -\frac{i}{2}e^{-\sigma}\left(  F_{\mu\nu}^{I}\ell_{I\phantom{k}
l}^{\phantom{I}  k}\phi_{\phantom{al}  k}^{al}\phi_{a\phantom{i}
j}^{\phantom{a}  i}+2F_{\mu\nu}^{a}\phi_{a\phantom{j}  i}^{\phantom{a}
j}\right)  \left[  \frac{1}{4\sqrt{2}}\left(  \bar{\psi}_{\rho}^{i}%
\Upsilon^{\mu\nu\rho\sigma}\psi_{\sigma j}+2\bar{\psi}^{\mu i}\psi_{j}^{\nu
}\right)  \right. \nonumber\\
& \hspace{1.8cm} \left.  +\frac{3}{20\sqrt{2}}\bar{\chi}^{i}\Upsilon^{\mu\nu}\chi
_{j}-\frac{1}{4\sqrt{2}}\bar{\lambda}^{\alpha i}\Upsilon^{\mu\nu}%
\lambda_{\alpha j}+\frac{1}{2\sqrt{10}}\left(  \bar{\chi}^{i}\Upsilon^{\mu
\nu\rho}\psi_{\rho j}-2\bar{\chi}^{i}\Upsilon^{\mu}\psi_{j}^{\nu}\right)
\right] \nonumber\\
& \hspace{1.8cm} +e^{-\sigma}F_{\mu\nu}^{a}\left[  \frac{1}{4}\left(  2\bar{\lambda}%
^{ai}\Upsilon^{\mu}\psi_{i}^{\nu}-\bar{\lambda}^{ai}\Upsilon^{\mu\nu\rho}%
\psi_{\rho i}\right)  +\frac{1}{2\sqrt{5}}\bar{\lambda}^{ai}\Upsilon^{\mu\nu
}\chi_{i}\right] \nonumber\\
& \hspace{1.8cm} +\frac{1}{4}e^{2\sigma}f_{bc}^{\phantom{bc}  a}f_{dea}\phi_{\phantom{bi}
k}^{bi}\phi_{\phantom{ck}  j}^{ck}\phi_{\phantom{dj}  l}^{dj}\phi
_{\phantom{el}  i}^{el}-\frac{1}{2}e^{\sigma}f_{abc}\phi_{\phantom{bi}
k}^{bi}\phi_{\phantom{ck}  j}^{ck}\left(  \bar{\psi}_{\mu}^{j}\Upsilon^{\mu
}\lambda_{i}^{a}+\frac{2}{\sqrt{5}}\bar{\chi}^{j}\lambda_{i}^{\phantom{i}
a}\right) \nonumber\\
& \hspace{1.8cm} -\frac{i}{\sqrt{2}}e^{\sigma}f_{ab}^{\phantom{ab}  c}\phi_{c\phantom{i}
j}^{\phantom{c}  i}\bar{\lambda}^{aj}\lambda_{i}^{b}+\frac{i}{60\sqrt{2}%
}e^{\sigma}f_{ab}^{\phantom{ab}  c}\phi_{\phantom{al}  k}^{al}\phi
_{\phantom{bj}  l}^{bj}\phi_{c\phantom{k}  j}^{\phantom{c}  k}\left(
5\bar{\psi}_{\mu}^{i}\Upsilon^{\mu\nu}\psi_{\nu i}+2\sqrt{5}\bar{\psi}_{\mu
}^{i}\Upsilon^{\mu}\chi_{i}\right. \nonumber
\end{align}
\begin{align}
& \hspace{1.8cm} \left.  \left.  +3\bar{\chi}^{i}\chi_{i}-5\bar{\lambda}^{\alpha i}%
\lambda_{\alpha i}\right)  -\frac{1}{96}\epsilon^{\mu\nu\rho\sigma
\kappa\lambda\tau}C_{\mu\nu\rho}F_{\sigma\kappa}^{a}F_{a\lambda\tau}\right\}.
\end{align}
The associated supersymmetry transformations have an expansion similar
to that of the Lagrangian. Thus, the supersymmetry transformation of a
field $X$ takes the form
\begin{equation}
\delta X = \delta^{(0)}X+h^2\delta^{(2)}X+h^4\delta^{(4)}X+\ldots\;.
\end{equation}
We give the first two terms of this series for the gravity and $U(1)$
vector multiplet fields, and just the first term for the $H$ vector
multiplet fields. These terms are precisely those required to prove
that the Lagrangian given in Eq. \eqref{truncatedL} is supersymmetric
to order $h^2\sim g^{-2}_{\mathrm{YM}}$. They are
\begin{align}
\delta\sigma &  =\frac{1}{\sqrt{5}}\bar{\chi}^{i}\varepsilon_{i},\nonumber\\
\delta {e_{\mu}}^{\underline{\nu}}  &  =\bar{\varepsilon}^{i}%
\Upsilon^{\underline{\nu}}\psi_{\mu i},\nonumber\\
\delta\psi_{\mu i}  &  =2\hat{\mathcal{D}}_{\mu}\varepsilon_{i}-\frac{1}%
{80}\left(  \Upsilon_{\mu}^{\phantom{\mu}  \nu\rho\sigma\eta}-\frac{8}%
{3}\delta_{\mu}^{\nu}\Upsilon^{\rho\sigma\eta}\right)  \varepsilon_{i}%
G_{\nu\rho\sigma\eta}e^{2\sigma}+\frac{i}{5\sqrt{2}}\left(
\Upsilon_{\mu}^{\phantom{\mu}  \nu\rho}-8\delta_{\mu}^{\nu}\Upsilon^{\rho
}\right)  \varepsilon_{j}F_{\nu\rho}^{I}\ell_{I\phantom{i}  i}^{\phantom{I}
j}e^{-\sigma}\nonumber\\
&  +\frac{\kappa_{7}^{2}}{g_{\mathrm{YM}}^{2}}\left\{  \frac{1}{2}\left(
\phi_{ak}^{\phantom{ak}  j}\hat{\mathcal{D}}_{\mu}\phi_{\phantom{a}
i}^{a\phantom{i}  k}-\phi_{\phantom{a}  i}^{a\phantom{i}  k}\hat{\mathcal{D}%
}_{\mu}\phi_{ak}^{\phantom{ak}  j}\right)  \varepsilon_{j}-\frac{i}{15\sqrt
{2}}\Upsilon_{\mu}\varepsilon_{i}f_{ab}^{\phantom{ab}  c}\phi_{\phantom{al}
k}^{al}\phi_{\phantom{bj}  l}^{bj}\phi_{c\phantom{k}  j}^{\phantom{c}
k}e^{\sigma}\right. \nonumber\\
&  \left.  \hspace{1.1cm}+\frac{i}{10\sqrt{2}}\left(  \Upsilon_{\mu
}^{\phantom{\mu}  \nu\rho}-8\delta_{\mu}^{\nu}\Upsilon^{\rho}\right)
\varepsilon_{j}\left(  F_{\nu\rho}^{I}\ell_{I\phantom{k}  l}^{\phantom{I}
k}\phi_{\phantom{al}  k}^{al}\phi_{a\phantom{j}  i}^{\phantom{a}  j}%
+2F_{\nu\rho}^{a}\phi_{a\phantom{j}  i}^{\phantom{a}  j}\right)  e^{-\sigma
}\right\}  ,\nonumber\\
\delta\chi_{i}  &  =\sqrt{5}\Upsilon^{\mu}\varepsilon_{i}\partial_{\mu}%
\sigma-\frac{1}{24\sqrt{5}}\Upsilon^{\mu\upsilon\rho\sigma}\varepsilon
_{i}G_{\mu\nu\rho\sigma}e^{2\sigma}\text{ }-\frac{i}{\sqrt{10}%
}\Upsilon^{\mu\nu}\varepsilon_{j}F_{\mu\nu}^{I}\ell_{I\phantom{i}
i}^{\phantom{I}  j}e^{-\sigma}\nonumber\\
&  +\frac{\kappa_{7}^{2}}{g_{\mathrm{YM}}^{2}}\left\{  -\frac{i}{2\sqrt{10}%
}\Upsilon^{\mu\nu}\varepsilon_{j}\left(  F_{\mu\nu}^{I}\ell_{I\phantom{k}
l}^{\phantom{I}  k}\phi_{\phantom{al}  k}^{al}\phi_{a\phantom{j}
i}^{\phantom{a}  j}+2F_{\mu\nu}^{a}\phi_{a\phantom{j}  i}^{\phantom{a}
j}\right)  e^{-\sigma}\right. \nonumber\\
&  \left.  +\frac{i}{3\sqrt{10}}\varepsilon_{i}f_{ab}^{\phantom{ab}  c}%
\phi_{\phantom{al}  k}^{al}\phi_{\phantom{bj}  l}^{bj}\phi_{c\phantom{k}
j}^{\phantom{c}  k}e^{\sigma}\right\}  ,\nonumber\\
\delta C_{\mu\nu\rho}  &  =\left(  -3\bar{\psi}_{\left[  \mu\right.
}^{i}\Upsilon_{\left.  \nu\rho\right]  }\varepsilon_{i}-\frac{2}{\sqrt{5}}%
\bar{\chi}^{i}\Upsilon_{\mu\nu\rho}\varepsilon_{i}\right)  e^{-2\sigma
},\nonumber\\
\ell_{I\phantom{i}  j}^{\phantom{I}  i}\delta A_{\mu}^{I}  &  =\left[
i\sqrt{2}\left(  \bar{\psi}_{\mu}^{i}\varepsilon_{j}-\frac{1}{2}\delta_{j}%
^{i}\bar{\psi}_{\mu}^{k}\varepsilon_{k}\right)  -\frac{2i}{\sqrt{10}}\left(
\bar{\chi}^{i}\Upsilon_{\mu}\varepsilon_{j}-\frac{1}{2}{}\delta_{j}^{i}%
\bar{\chi}^{k}\Upsilon_{\mu}\varepsilon_{k}\right)  \right]  e^{\sigma
}\nonumber\\
&  +\frac{\kappa_{7}^{2}}{g_{\mathrm{YM}}^{2}}\left\{  \left(  \frac{i}%
{\sqrt{2}}\bar{\psi}_{\mu}^{k}\varepsilon_{l}-\frac{i}{\sqrt{10}}\bar{\chi
}^{k}\Upsilon_{\mu}\varepsilon_{l}\right)  \phi_{\phantom{al}  k}^{al}%
\phi_{a\phantom{i}  j}^{\phantom{a}  i}e^{\sigma}-\bar{\varepsilon}%
^{k}\Upsilon_{\mu}\lambda_{k}^{a}\phi_{a\phantom{i}  j}^{\phantom{a}
i}e^{\sigma}\right\}  ,\\
\ell_{I}^{\phantom{I}  \alpha}\delta A_{\mu}^{I}  &  =\bar{\varepsilon}%
^{i}\Upsilon_{\mu}\lambda_{i}^{\alpha}e^{\sigma},\nonumber\\
\delta\ell_{I\phantom{i}  j}^{\phantom{I}  i}  &  =-i\sqrt{2}\bar{\varepsilon
}^{i}\lambda_{\alpha j}\ell_{I}^{\phantom{I}  \alpha}+\frac{i}{\sqrt{2}}%
\bar{\varepsilon}^{k}\lambda_{\alpha k}\ell_{I}^{\phantom{I}  \alpha}%
\delta_{j}^{i}\nonumber\\
&  +\frac{\kappa_{7}^{2}}{g_{\mathrm{YM}}^{2}}\left\{  \frac{i}{\sqrt{2}%
}\left[  \bar{\varepsilon}^{k}\lambda_{\alpha l}\phi_{\phantom{al}  k}%
^{al}\phi_{a\phantom{i}  j}^{\phantom{a}  i}\ell_{I}^{\phantom{I}  \alpha
}+\bar{\varepsilon}^{l}\lambda_{ak}\phi_{\phantom{ai}  j}^{ai}\ell
_{I\phantom{k}  l}^{\phantom{I}  k}-\left(  \bar{\varepsilon}^{i}\lambda
_{aj}-\frac{1}{2}\delta_{j}^{i}\bar{\varepsilon}^{m}\lambda_{am}\right)
\phi_{\phantom{al}  k}^{al}\ell_{I\phantom{k}  l}^{\phantom{I}  k}\right]
\right\}  ,\nonumber\\
\delta\ell_{I}^{\phantom{I}  \alpha}  &  =-i\sqrt{2}\bar{\varepsilon}%
^{i}\lambda_{j}^{\alpha}\ell_{I\phantom{j}  i}^{\phantom{I}  j}+\frac{\kappa
_{7}^{2}}{g_{\mathrm{YM}}^{2}}\left\{  -\frac{i}{\sqrt{2}}\bar{\varepsilon
}^{i}\lambda_{j}^{\alpha}\phi_{\phantom{aj}  i}^{aj}\phi_{a\phantom{l}
k}^{\phantom{a}  l}\ell_{I\phantom{k}  l}^{\phantom{I}  k}\right\}
,\nonumber\\
\delta\lambda_{i}^{\alpha}  &  =-\frac{1}{2}\Upsilon^{\mu\nu}\varepsilon
_{i}F_{\mu\nu}^{I}\ell_{I}^{\phantom{I}  \alpha}e^{-\sigma}+\sqrt{2}%
i\Upsilon^{\mu}\varepsilon_{j}p_{\mu\phantom{\alpha j}  i}^{\phantom{\mu}
\alpha j}+\frac{\kappa_{7}^{2}}{g_{\mathrm{YM}}^{2}}\left\{  \frac{i}{\sqrt
{2}}\Upsilon^{\mu}\varepsilon_{j}\phi_{ai}^{\phantom{ai}  j}p_{\mu
\phantom{\alpha k}  l}^{\phantom{\mu}  \alpha k}\phi_{\phantom{al}  k}%
^{al}\right\}  ,\nonumber\\
\delta A_{\mu}^{a}  &  =\bar{\varepsilon}^{i}\Upsilon_{\mu}\lambda_{i}%
^{a}e^{\sigma}-\left(  i\sqrt{2}\psi_{\mu}^{i}\varepsilon_{j}-\frac{2i}%
{\sqrt{10}}\bar{\chi}^{i}\Upsilon_{\mu}\varepsilon_{j}\right)  \phi
_{\phantom{aj}  i}^{aj}e^{\sigma},\nonumber\\
\delta\phi_{a\phantom{i}  j}^{\phantom{a}  i}  &  =-i\sqrt{2}\left(
\bar{\varepsilon}^{i}\lambda_{aj}-\frac{1}{2}\delta_{j}^{i}\bar{\varepsilon
}^{k}\lambda_{ak}\right)  ,\nonumber\\
\delta\lambda_{i}^{a}  &  =\left(  -\frac{1}{2}\Upsilon^{\mu\nu}%
\varepsilon_{i}\left(  F_{\mu\nu}^{I}\ell_{I\phantom{j}  k}^{\phantom{I}
j}\phi_{\phantom{ak}  j}^{ak}+F_{\mu\nu}^{a}\right)  e^{-\sigma}-i\sqrt
{2}\Upsilon^{\mu}\varepsilon_{j}\hat{\mathcal{D}}_{\mu}\phi_{\phantom{a}
i}^{a\phantom{i}  j}-i\varepsilon_{j}f_{\phantom{a}  bc}^{a}\phi
_{\phantom{bj}  k}^{bj}\phi_{\phantom{ck}  i}^{ck}\right), \nn
\end{align}
\begin{align}
\delta A_{\mu}^{a}  &  =\bar{\varepsilon}^{i}\Upsilon_{\mu}\lambda_{i}%
^{a}e^{\sigma}-\left(  i\sqrt{2}\psi_{\mu}^{i}\varepsilon_{j}-\frac{2i}%
{\sqrt{10}}\bar{\chi}^{i}\Upsilon_{\mu}\varepsilon_{j}\right)  \phi
_{\phantom{aj}  i}^{aj}e^{\sigma},\nonumber\\
\delta\phi_{a\phantom{i}  j}^{\phantom{a}  i}  &  =-i\sqrt{2}\left(
\bar{\varepsilon}^{i}\lambda_{aj}-\frac{1}{2}\delta_{j}^{i}\bar{\varepsilon
}^{k}\lambda_{ak}\right)  ,\nonumber\\
\delta\lambda_{i}^{a}  &  =-\frac{1}{2}\Upsilon^{\mu\nu}\varepsilon_{i}\left(
F_{\mu\nu}^{I}\ell_{I\phantom{j}  k}^{\phantom{I}  j}\phi_{\phantom{ak}
j}^{ak}+F_{\mu\nu}^{a}\right)  e^{-\sigma}-i\sqrt{2}\Upsilon^{\mu}%
\varepsilon_{j}\hat{\mathcal{D}}_{\mu}\phi_{\phantom{a}  i}^{a\phantom{i}
j}-i\varepsilon_{j}f_{\phantom{a}  bc}^{a}\phi_{\phantom{bj}  k}^{bj}%
\phi_{\phantom{ck}  i}^{ck}.\nonumber
\end{align}
This completes our review of $\mathcal{N}=1$ EYM supergravity in seven dimensions.


\section{Review of orbifold based $G_2$ manifolds} \label{A}

\subsection{Construction and Classification of $G_2$ Orbifolds} \label{class2}
While the basic properties of $G_2$ manifolds was described in Chapter \ref{G2Intro}, in this section we describe the general idea~\cite{Joyce} of how to explicitly construct a class of compact
$G_2$ orbifolds and manifolds. Recall that a $G_2$ manifold is a
seven-dimensional Riemannian manifold admitting a globally defined
torsion free $G_2$ structure \cite{Joyce}. A $G_2$ structure
is given by a three-form $\varphi$ which can be written locally as \eref{flat_three_form}

Our starting point for constructing a compact
manifold of $G_2$ holonomy is an arbitrary seven-torus $\mathcal{T}^7$. We then
take the quotient with respect to a finite group $\Gamma$ contained in
$G_2$, such that the resulting orbifold has finite first fundamental
group. We shall refer to $\Gamma$ as the orbifold group. The result is
a $G_2$ manifold with singularities at fixed loci of elements of
$\Gamma$. Smooth $G_2$ manifolds can then be obtained by blowing up the
singularities. Loosely speaking, this involves removing a patch around
the singularity and replacing it with a smooth space of the same
symmetry. Note that, following this construction, the independent
moduli will come from torus radii and from the radii and orientation
of cycles associated with the blow-ups.

We now review, following Ref.~\cite{Barrett:2004tc}, a classification of
orbifold-based, compact $G_2$ manifolds in terms of the orbifold group
of the manifold. The classification deals with orbifold groups that
lead to a set of co-dimension four singularities, and that act in a
prescribed way on the underlying lattice that defines the
seven-torus. Essentially, each orbifold group element acts by rotating
two orthogonal two-dimensional sub-lattices of the seven-dimensional
lattice. Thus the matrix of an orbifold group element takes the form
\begin{equation}
\boldsymbol{1}_{3\times 3} \oplus \left( \begin{array}{cc}
\cos\theta_1 & -\sin\theta_1 \\
\sin\theta_1 & \cos\theta_1 \end{array} \right)
\oplus
\left( \begin{array}{cc}
\cos\theta_2 & -\sin\theta_2 \\
\sin\theta_2 & \cos\theta_2 \end{array} \right)
\end{equation}
in some coordinate frame. In addition to a rotation, the symmetries
can also contain a translation of the coordinates of the torus. For
such a symmetry to be compatible with a well-defined $G_2$ structure
on the orbifold, it must be possible to embed the
symmetry in $\mathrm{SU}(2)$, and thus we must have $\theta_2=\pm
\theta_1$. Such symmetries are only compatible with a seven-torus if
$\lvert\theta_1\rvert=2\pi/N$ for $N=2,3,4$ or 6.

Using the class of possible generators, one can obtain a class of
discrete symmetry groups from which compact manifolds of $G_2$
holonomy may be constructed. The conditions for some group $\Gamma$ to
be suitable are that there must exist both a seven-dimensional torus
$\mathcal{T}^7$ and $G_2$ structure $\varphi$ preserved by $\Gamma$,
and also that the first fundamental group of the orbifold
$\mathcal{T}^7/\Gamma$ is finite. There are no suitable orbifold
groups with fewer than three generators. The class found in
Ref.~\cite{Barrett:2004tc} lists all possibilities with precisely
three-generators. Let us present these groups in a form such that they
preserve the particular $G_2$ structure \eqref{flat_three_form}. Define
\bea
R_N&=&\boldsymbol{1}_{3\times 3} \oplus \left( \begin{array}{cc}
\cos(2\pi/N) & -\sin(2\pi/N) \\
\sin(2\pi/N) & \cos(2\pi/N) \end{array} \right)
\oplus
\left( \begin{array}{cc}
\cos(2\pi/N) & -\sin(2\pi/N) \\
\sin(2\pi/N) & \cos(2\pi/N) \end{array} \right) \, ,\\
P_N&=&(1)\oplus \left( \begin{array}{cc}
\cos(2\pi/N) & -\sin(2\pi/N) \\
\sin(2\pi/N) & \cos(2\pi/N) \end{array} \right)
\oplus
\left( \begin{array}{cc}
\cos(2\pi/N) & \sin(2\pi/N) \\
-\sin(2\pi/N) & \cos(2\pi/N) \end{array} \right)
\oplus \boldsymbol{1}_{2\times 2}\, , \\
Q_0 &=&  \mathrm{diag}(-1,1,-1,1,-1,1,-1)\,, \\
Q_1 &=&  \mathrm{diag}(-1,-1,1,1,-1,-1,1)\,, \\
Q_2& :& (x_1,x_2,x_3,x_4,x_5,x_6,x_7) \mapsto (-x_7,x_2,-x_5,x_4,x_3,x_6,x_1)\, , \\
Q_3& :& (x_1,x_2,x_3,x_4,x_5,x_6,x_7) \mapsto (x_3,x_2,-x_1,x_4,x_7,x_6,-x_5)\,, \\
Q_4& :& (x_1,x_2,x_3,x_4,x_5,x_6,x_7) \mapsto (-x_1,-x_2,x_3,-x_4,x_5,x_6,-x_7)\,,\\
Q_5& :& (x_1,x_2,x_3,x_4,x_5,x_6,x_7) \mapsto (x_1,-x_3,x_2,x_5,-x_4,x_6,x_7)\,,\\
Q_6 &:& (x_1,x_2,x_3,x_4,x_5,x_6,x_7) \mapsto (x_1,-x_5,x_4,-x_3,x_2,x_6,x_7)\, .
\eea
Then, one can take as generators $Q_0$ with one of the $P$s and one of the $R$s to obtain the following orbifold groups:
\begin{equation}
\begin{array}{l}
\mathbb{Z}_2 \times \mathbb{Z}_2 \times \mathbb{Z}_2, \\
\mathbb{Z}_2 \ltimes \left( \mathbb{Z}_2 \times \mathbb{Z}_3 \right), \\
\mathbb{Z}_2 \ltimes \left( \mathbb{Z}_2 \times \mathbb{Z}_4 \right), \\
\mathbb{Z}_2 \ltimes \left( \mathbb{Z}_2 \times \mathbb{Z}_6 \right) ,\\
\mathbb{Z}_2 \ltimes \left( \mathbb{Z}_3 \times \mathbb{Z}_3 \right) ,\\
\mathbb{Z}_2 \ltimes \left( \mathbb{Z}_3 \times \mathbb{Z}_6 \right) ,\\
\mathbb{Z}_2 \ltimes \left( \mathbb{Z}_4 \times \mathbb{Z}_4 \right) ,\\
\mathbb{Z}_2 \ltimes \left( \mathbb{Z}_6 \times \mathbb{Z}_6 \right) .\\
\end{array}
\end{equation}
One can take $Q_0$ with $Q_1$ and one of the $R$s to obtain
\begin{equation}
\mathbb{Z}_2^2 \ltimes \mathbb{Z}_N, \; \; N=3, \, 4 \,\, \mathrm{or} \,\, 6.
\end{equation}
Lastly, there are five other, more complicated groups, constructed as follows:
\bea
\mathbb{E}_1 &=:& \langle P_2,\, Q_2,\, R_4\, \lvert\, [P_2,Q_2]=1,\, [P_2,R_4]=1,\, Q_2^2R_4Q_2^2=R_4^{-1}  \rangle, \\
\mathbb{E}_2 &=:& \langle P_2,\, Q_3,\, R_4\, \lvert\,  [P_2,Q_3]=Q_3^2,\, [P_2,R_4]=1,\, Q_3^2R_4Q_3^2=R_4^{-1}  \rangle, \\
\mathbb{E}_3 &=:& \langle Q_4,\, Q_3,\, R_4\, \lvert\, [Q_4,Q_3]=Q_3^2,\, [Q_4,R_4]=R_4^2,\, Q_3^2R_4Q_3^2=R_4^{-1}  \rangle, \\
\mathbb{E}_4 &=:& \langle Q_5,\, Q_3,\, R_4\, \lvert\,  Q_5^2Q_3Q_5^2=Q_3^{-1},\, [Q_5,R_4]=1,\, Q_3^2R_4Q_3^2=R_4^{-1}  \rangle,\\
\mathbb{E}_5 &=:& \langle Q_6,\, Q_3,\, R_4\, \lvert\,  Q_6^2Q_3Q_6^2=Q_3^{-1},\, Q_6^2R_4Q_6^2=R_4^{-1},\, Q_3^2R_4Q_3^2=R_4^{-1}  \rangle.
\eea
 Note
that since we should really be thinking of orbifold group elements as
abstract group elements as opposed to matrices, the commutator is defined here by $[g,h]=g^{-1}h^{-1}gh$.
\\

\subsection{Properties of a class of $G_2$ manifolds} \label{manifolds}

In this sub-section we discuss properties of a general $G_2$
orbifold $\mathcal{Y}=\mathcal{T}^7/\Gamma$ with co-dimension four
fixed points, and a few details of its blown up analogue
$\mathcal{Y}^{\mathrm{S}}$. We assume that points on the torus that
are fixed by one generator of the orbifold group are not fixed by
other generators. Given an orbifold group, this can always be arranged
by incorporating appropriate translations into the generators, and
thus all of the examples of the previous sub-section are
relevant. Under this assumption we have a well-defined blow-up
procedure.

We begin then by discussing the orbifold
$\mathcal{Y}=\mathcal{T}^7/\Gamma$. Let us consider the
homology. There are no one-cycles on a $G_2$ manifold. If $\Gamma$ is
one of the orbifold groups listed in Section \ref{class2} then there
are no two-cycles consistent with its symmetries. We will allow for
more general orbifold groups $\Gamma$ in the following and indeed
the main part of the paper as long as they satisfy this condition.
It is then three-cycles that carry the important
information about the geometry of the space. Let us define
three-cycles by setting four of the coordinates $x^A$ to constants
(chosen so there is no intersection with any of the
singularities). The number of these that fall into distinct homology
classes is then given by the number of independent terms in the $G_2$
structure. Let us explain this statement. The $G_2$ structure can
always be chosen so as to contain the seven terms of the standard
$G_2$ structure \eqref{flat_three_form}, with positive coefficients multiplying
them. If we write $\mathcal{R}^A$ for the coefficient in front of the
$(A+1)^{\mathrm{th}}$ term in Eq.~\eqref{flat_three_form}, then by the number of
independent terms we mean the number of $\mathcal{R}^A$s that are not
constrained by the orbifolding. We then write $C^A$ for the cycle
obtained by setting the four coordinates on which the
$(A+1)^{\mathrm{th}}$ term in \eqref{flat_three_form} does not depend to
constants, for example,
\begin{equation} \label{bulkcycle}
C^0=\{x^4,x^5,x^6,x^7=\mathrm{const}\}.
\end{equation}
A pair of $C^A$s for which the corresponding $\mathcal{R}^A$s are
independent then belong to distinct homology classes. There is
therefore some subset $\mathcal{C}$ of $\{C^A\}$ that provides a basis for
$H_3(Y,\mathbb{Z})$. We can conclude that the third Betti number $b^3$ is dependent on the orbifold group $\Gamma$ and is in all
cases a positive integer less than or equal to seven. For the class of orbifold groups obtained in Section \ref{class2} it takes values as given in Table \ref{t:betti_third}. A description of
the derivation is given in the discussion below on constructing a $G_2$
structure on $\mathcal{Y}$. 

\begin{table}\label{t:betti_third}
\begin{center}
\begin{tabular}{c|c}
$\Gamma$ & $b^3(\Gamma)$ \\
\hline
$\mathbb{Z}_2\times\mathbb{Z}_2\times\mathbb{Z}_2$ & 7 \\
$\mathbb{Z}_2 \ltimes \left( \mathbb{Z}_2 \times \mathbb{Z}_3 \right)$ & 5 \\
$\mathbb{Z}_2 \ltimes \left( \mathbb{Z}_2 \times \mathbb{Z}_4 \right)$ & 5 \\
$\mathbb{Z}_2 \ltimes \left( \mathbb{Z}_2 \times \mathbb{Z}_6 \right)$ & 5\\
$\mathbb{Z}_2 \ltimes \left( \mathbb{Z}_3 \times \mathbb{Z}_3 \right)$ & 4\\
$\mathbb{Z}_2 \ltimes \left( \mathbb{Z}_3 \times \mathbb{Z}_6 \right)$ & 4\\
$\mathbb{Z}_2 \ltimes \left( \mathbb{Z}_4 \times \mathbb{Z}_4 \right)$ & 4\\
$\mathbb{Z}_2 \ltimes \left( \mathbb{Z}_6 \times \mathbb{Z}_6 \right)$ & 4\\
$\mathbb{Z}_2^2 \ltimes \mathbb{Z}_3$ & 5\\
$\mathbb{Z}_2^2 \ltimes \mathbb{Z}_4$ & 5\\
$\mathbb{Z}_2^2 \ltimes \mathbb{Z}_6$ & 5\\
$\mathbb{E}_1$ & 3\\
$\mathbb{E}_2$ & 3\\
$\mathbb{E}_3$ & 3\\
$\mathbb{E}_4$ & 2\\
$\mathbb{E}_5$ & 1\\
\end{tabular}
\end{center}
\caption{\emph{Third Betti numbers of $\mathcal{T}^7/\Gamma$ for different orbifold groups}}
\end{table}

We now present the most general Ricci flat metric and
$G_2$ structure on $\mathcal{Y}$. Given there are by assumption
no invariant two-forms, the symmetries restrict the metric to be
diagonal, and thus
\begin{equation} \label{bulkmetric}
\mathrm{d}s^2=\sum_{A=1}^7(R^A\mathrm{d}x^A)^2,
\end{equation}
where the $R^A$ are precisely the seven radii of the torus.

Under a suitable choice of coordinates the $G_2$ structure is obtained from
the flat $G_2$ structure \eref{flat_three_form} by rescaling $x^A\to
R^Ax^A$, leading to
\begin{eqnarray} \label{structure5}
\varphi & = & R^1R^2R^3\mathrm{d}x^1\wedge\mathrm{d}x^2\wedge\mathrm{d}x^3+R^1R^4R^5\mathrm{d}x^1\wedge\mathrm{d}x^4\wedge\mathrm{d}x^5-R^1R^6R^7\mathrm{d}x^1\wedge\mathrm{d}x^6\wedge\mathrm{d}x^7 \nonumber \\ & &+R^2R^4R^6\mathrm{d}x^2\wedge\mathrm{d}x^4\wedge\mathrm{d}x^6 +R^2R^5R^7\mathrm{d}x^2\wedge\mathrm{d}x^5\wedge\mathrm{d}x^7+R^3R^4R^7\mathrm{d}x^3\wedge\mathrm{d}x^4\wedge\mathrm{d}x^7 \nonumber \\ & & -R^3R^5R^6\mathrm{d}x^3\wedge\mathrm{d}x^5\wedge\mathrm{d}x^6.
\end{eqnarray}
For the orbifolding to preserve the metric some of the $R^A$ must
be set equal to one another. It is straightforward to check that if one of the orbifold symmetries
$\alpha$ involves a rotation in the $(A,B)$ plane by an angle not equal
to $\pi$, then we must set $R^A=R^B$. Following this prescription, it
is easy to find $b^3$ in terms of the orbifold group $\Gamma$.

To complete our description of $\mathcal{Y}$, we briefly mention its singularities. Recall that we are assuming that points on the torus fixed by one generator of the orbifold group are not fixed by other generators. Thus, if we use the index $\tau$ to label the generators of the orbifold group, which each have a certain number $M_\tau$ of fixed points associated with them, then the singularities of $\mathcal{Y}$ can be labelled by the pair $(\tau,s)$, where $s=1,\ldots,M_\tau$. Near a singular point we can then describe $\mathcal{Y}$ by saying it has the approximate form $\mathcal{T}^3_{(\tau,s)}\times\mathbb{C}^2/\mathbb{Z}_{N_{\tau}}$, where $\mathcal{T}^3_{(\tau,s)}$ is a three-torus.

We now move on to discuss the smooth $G_2$ manifold $\mathcal{Y}^{\mathrm{S}}$, constructed by blowing up the singularities of $\mathcal{Y}$. Blowing up a singularity heuristically involves the following. One firstly removes a four-dimensional ball centred around the singularity times the associated fixed three-torus
$\mathcal{T}^3_{(\tau,s)}$. Secondly one replaces the resulting hole by
$\mathcal{T}^3_{(\tau,s)}\times U_{(\tau,s)}$, where $U_{(\tau,s)}$ is the
blow-up of $\mathbb{C}^2/\mathbb{Z}_{N_{\tau}}$, as discussed in Ref.~\cite{Barrett:2004tc}. The blow-up $U_{(\tau,s)}$ has the same symmetry as $\mathbb{C}^2/\mathbb{Z}_{N_\tau}$ and is derived from a Gibbons-Hawking space (or gravitational multi-instanton) which approaches $\mathbb{C}^2/\mathbb{Z}_{N_\tau}$ asymptotically. Specifically, the central region of $U_{(\tau,s)}$ looks exactly like Gibbons-Hawking space, while the outer region looks exactly like $\mathbb{C}^2/\mathbb{Z}_{N_\tau}$.  Between the central and outer region $U_{(\tau,s)}$ can be thought of as interpolating between Gibbons-Hawking space and flat space. In this way, $\mathcal{Y}^{\mathrm{S}}$ remains smooth as one moves in or out of a blow-up region.

Gibbons-Hawking spaces provide a
generalization of the Eguchi-Hanson space and their different
topological types are labelled by an integer $N$ (where the case $N=2$
corresponds to the Eguchi-Hanson case). While the Eguchi-Hanson space
contains a single two-cycle, the $N^{\rm th}$ Gibbons-Hawking space
contains a sequence $\gamma_1,\ldots,\gamma_{N-1}$ of such cycles at
the ``centre'' of the space.  Only neighbouring cycles $\gamma_i$ and
$\gamma_{i+1}$ intersect and in a single point and, hence, the
intersection matrix $\gamma_i\cdot\gamma_j$ equals the Cartan matrix of
$A_{N-1}$. Asymptotically, the $N^{\rm th}$ Gibbons-Hawking space has
the structure $\mathbb{C}^2/\mathbb{Z}_{N}$.  Accordingly, we take
$N=N_\tau$ when blowing up $\mathbb{C}^2/\mathbb{Z}_{N_{\tau}}$.

The blown-up singularity $\mathcal{T}^3_{(\tau,s)}\times U_{(\tau,s)}$ contributes $N_\tau-1$ two-cycles and $3(N_\tau-1)$ three-cycles to the homology of $\mathcal{Y}^{\mathrm{S}}$. The two-cycles are simply the two-cycles on $U_{(\tau,s)}$, while the three-cycles are formed by taking the Cartesian product of one of these two-cycles with one of the three one-cycles on $\mathcal{T}^3_{(\tau,s)}$.
We can label these three-cycles by $C(\tau,s,i,m)$, where $i$ labels the two-cycles of
$U_{(\tau,s)}$ and $m$ labels
the direction on $\mathcal{T}^3_{(\tau,s)}$. We deduce the following formula for the third Betti number of $\mathcal{Y}^{\mathrm{S}}$:
\begin{equation}
b^3(\mathcal{Y}^{\mathrm{S}})=b^3(\Gamma)+\sum_\tau M_\tau \cdot 3(N_\tau-1).
\end{equation}

On one of the blow-ups $\mathcal{T}^3\times U$ (for convenience we suppress $\tau$ and $s$ indices) we use coordinates $\xi^m$ on $\mathcal{T}^3$ and complex coordinates $z^p$, $p=1,2$ on $U$. We write $R^m$ to
denote the three radii of $\mathcal{T}^3$, which will be the three $R^A$ in the
directions fixed by $\alpha$. The $G_2$ structure can be written as
\begin{equation} \label{structure4}
\varphi=\sum_m\omega^m\left(z^p,\boldsymbol{b}_1,\ldots,\boldsymbol{b}_{N_\tau}\right)\wedge R^m\mathrm{d}\xi^m-R^1R^2R^3\mathrm{d}\xi^1\wedge\mathrm{d}\xi^2\wedge\mathrm{d}\xi^3.
\end{equation}
Here the $\boldsymbol{b}_i$ are a
set of three-vectors, which parameterize the size of the two-cycles within the Gibbons-Hawking space, and also their orientation with respect to the bulk. The $\omega^m$ are a triplet
of two-forms that constitute a ``nearly'' hyperk\"ahler structure on
$U$. This $G_2$ structure makes a slight deviation from being torsion free in the region in which the space $U$ is interpolating between Gibbons-Hawking space and flat space. However, for sufficiently small blow-up moduli and a smooth and slowly varying interpolation this deviation is small \cite{Barrett:2004tc,Lukas:2003dn}. Consequently, this $G_2$ structure can reliably be used in M-theory calculations that work to leading non-trivial order in the blow-up moduli.

The metric corresponding to the $G_2$ structure \eqref{structure4} takes the form
\begin{equation}
\mathrm{d}s^2=\mathcal{G}_0\mathrm{d}\boldsymbol{\zeta}^2+\sum_{m=1}^{3}\mathcal{G}_m(\mathrm{d}\xi^m)^2,
\end{equation}
where $\mathrm{d}\boldsymbol{\zeta}$ is the line element on $U$, and the $\mathcal{G}$s
are conformal factors whose product is equal to 1.


\section{Reduction of M-theory on a $G_2$ manifold} \label{appred}
We have reviewed the reduction of M-theory on a smooth $G_2$ manifold in Section \ref{smooth_g2}. Here we shall apply these results to the manifolds discussed in Chapter \ref{c:G2Ch2} and give a formula for the K\"ahler potential for M-theory on the orbifold based $G_2$ manifolds we have used in this work.

We begin by presenting the K\"ahler potential for M-theory on the orbifold based $G_2$ manifold $\mathcal{Y}^{\mathrm{S}}$ described in Section \ref{manifolds}. First we note that the bulk metric moduli $a^A$ are given by
\begin{equation}
a^A=\int_{C^A}\varphi,
\end{equation}
where the $\{C^A\}$ are those cycles described in Section \ref{manifolds}; for example $C^0$ is given by \eqref{bulkcycle}. These simply evaluate to
\begin{equation}
\left. \begin{array}{cccc}
a^0=R^1R^2R^3, & a^1=R^1R^4R^5, & a^2=R^1R^6R^7, & a^3=R^2R^4R^6, \\
a^4=R^2R^5R^7, & a^5=R^3R^4R^7, & a^6=R^3R^5R^6. &  \, \\
\end{array} \right.
\end{equation}
Then the blow-up moduli, which are defined by
\begin{equation}
A(\tau,s,i,m)=\int_{C(\tau,s,i,m)}\varphi
\end{equation}
take the form
\begin{equation}
A(\tau,s,i,m)\sim R^m_{(\tau)}(b_{(\tau,s,i,m)}-b_{(\tau,s,i+1,m)})
\end{equation}
where $R^m_{(\tau)}$ denote the three radii of
$\mathcal{T}^3_{(\tau,s)}$, and $b_{(\tau,s,i,m)}$ are the parameters for the two-cycles within the blow-ups, consistent with the notation of equation
\eqref{structure4}. We denote the superfields associated with the bulk by $T^A$ and those associated with blow-ups by $U^{(\tau,s,i,m)}$, so
\begin{equation}
\mathrm{Re}(T^A)=a^A, \; \; \; \mathrm{Re}(U^{(\tau,s,i,m)})=A(\tau,s,i,m).
\end{equation}
The K\"ahler potential is given by the following formula, to leading
non-trivial order in the blow-up moduli, (taking a trivial reference
metric $g_0=1$):
\begin{equation} \label{K}
K =  -\frac{1}{\kappa_4^2}\sum_{A=0}^{6}\ln (T^A+\bar{T}^A) +\frac{2}{c_\Gamma\kappa_4^2}\sum_{s,\tau,m}\frac{1}{N_\tau}\frac{\sum_{i<j}\left(\sum_{k=i}^{j-1}(U^{(\tau,s,k,m)}+\bar{U}^{(\tau,s,k,m)})\right)^2}{(T^{A(\tau,m)}+\bar{T}^{A(\tau,m)})(T^{B(\tau,m)}+\bar{T}^{B(\tau,m)})} + \frac{7}{\kappa_4^2}\ln2.
\end{equation}
The index functions $A(\tau,m)$, $B(\tau,m)\in\{1,\ldots,7\}$ indicate
by which two of the seven bulk moduli $T^A$ the blow up moduli
$U^{(\tau,s,i,m)}$ are divided in the K\"ahler potential \eqref{K}. Their values depend only on the generator index $\tau$ and the
orientation index $m$. They may be calculated from the formula
\begin{equation}
a^{A(\tau,m)}a^{B(\tau,m)}=\frac{\left(R^m_{(\tau)}\right)^2\prod_AR^A}{\prod_bR^b_{(\tau)}}.
\end{equation}
The $\tau$ dependence is only through the fixed directions of the generator $\alpha_\tau$ and the possible values of the index functions are given in Table~4. We remind the reader that in many cases some of the $T^A$s are identical to each other and should be thought of as the same field. One follows the prescription given in Section \ref{manifolds} to determine which of these are identical. For the orbifold groups $\Gamma$ listed in Section \ref{class2}, the number $b^3(\Gamma)$ of distinct $T^A$ is given in Table \ref{t:betti_third}. Finally, $c_\Gamma$ is just a constant factor that depends on the orbifold group.
\\
\begin{table}[t]
\begin{center}
\begin{tabular}{|c|c|c|c|}
\hline
$\mathrm{Fixed \, directions\, of\, } \alpha_\tau$
&$(A(\tau,1),B(\tau,1))$ & $(A(\tau,2),B(\tau,2))$ &$ (A(\tau,3),B(\tau,3))$ \\
\hline
(1,2,3) & (1,2) & (3,4) & (5,6) \\
\hline
(1,4,5) & (0,2) & (3,5) & (4,6) \\
\hline
(1,6,7) & (0,1) & (3,6) & (4,5) \\
\hline
(2,4,6) & (0,4) & (1,5) & (2,6) \\
\hline
(2,5,7) & (0,3) & (1,6) & (2,5) \\
\hline
(3,4,7) & (0,6) & (1,3) & (2,4) \\
\hline
(3,5,6) & (0,5) & (1,4) & (2,3) \\
\hline
\end{tabular}
\caption{\emph{Values of the index functions $(A(\tau,m),B(\tau,m))$ that
appear in the K\"ahler potential.}}
\end{center}
\end{table}

\chapter{Appendix 2}\label{mon-appen}
%
%
\section{The Monad Construction of Vector Bundles}
We will briefly review here a powerful technique for constructing vector bundles, known as the \emph{Monad Construction}. This construction allows us to investigate holomorphic vector bundles with the methods of linear algebra. For a general complex manifold $X$, a \emph{monad} over $X$ is a complex
\beq\label{gen_mon}
0 \to A \stackrel{a}{\longrightarrow} B \stackrel{b}{\longrightarrow} C \to 0
\eeq
of holomorphic vector bundles over $X$ which is exact\footnote{A sequence $\ldots \stackrel{\alpha}{\rightarrow} V \stackrel{\beta}{\longrightarrow} \ldots$ is said to be exact at $V$ if $\text{Ker}(\beta)=\text{Im}(\alpha)$} at $A$ and at $C$ such that $Im(a)$ is a sub-bundle of $B$. The holomorphic vector bundle
\beq
E = \text{Ker}(b)/\text{Im}(a)
\eeq
is called the ``cohomology of the monad" and is the bundle defined by the monad sequence \eref{gen_mon}. Monads were first used by Horrocks to classify bundles over projective spaces \cite{Horrocks1,Horrocks2,barth}. Horrocks and Mumford famously showed that \emph{every bundle on} $\mathbb{P}^n$ could be defined via a monad sequence of the form \eref{gen_mon} where $A$ and $C$ are direct sums of line bundles and $B$ satisfies certain constraints on its cohomology. For instance, Horrocks demonstrated that every rank $2$ bundle on $\mathbb{P}^3$ can be obtained from a monad sequence of the form
\beq\label{horrocks}
0 \to \oplus_{i} \cO_{\mathbb{P}^3}(a_{i}) \to \oplus_{j} \cO_{\mathbb{P}^3}(b_{j}) \to  \oplus_{k} \cO_{\mathbb{P}^3}(c_{k}) \to 0
\eeq
Such constructions have been important to a number of physics applications including the ADHM construction of instantons \cite{ADHM}. The $1-1$ correspondence between vector bundles and monad sequences has been generalized \cite{malaspina-2006} and shown to hold on projective varieties, $X$ with $dim(X) \geq 3$.

A monad generates a so-called ``display'': a commutative diagram with exact rows and columns.
\beq\ba{ccccccccc}\label{display}
&&0&&0&&0&& \\
&&\downarrow&&\downarrow&&\downarrow&& \\
0&\to& A&\to& K &\to& E &\to&0  \\
&& || &&\downarrow&&\downarrow&&\\
0&\to& A &\to& B &\to& Q &\to&0 \\
&&\downarrow&&\downarrow&&\downarrow&& \\
0&\to& 0 &\to&  C &\to& C &\to&0 \\
&&\downarrow&&\downarrow&&\downarrow&& \\
&&0&&0&&0&& \\
\ea
\eeq
where $K=\text{Ker}(b)$ and $Q=\text{Coker}(a)$. 

In the following chapters, we will consider a more restricted class of monad, by taking $A$ in \eref{gen_mon} to be zero. With this constraint, the display \eref{display} simplifies to two copies of the short exact sequence
\beq\label{mon_app}
0 \to V \to B \stackrel{f}{\longrightarrow} C \to 0
\eeq
where $V=\text{Ker}(f)$. In this choice, given two bundles $B$ and $C$, and a map $f$ between them, we may define a new bundle $V$. Taking our building blocks to be the simplest vector bundles available, we shall take $B$ and $C$ to be direct sums of line bundles
\bea
B &:=& \bigoplus_{i=1}^{r_B} \cO_X(b_1^i, \ldots, b_k^i) \ , \quad
C := \bigoplus_{j=1}^{r_C} \cO_X(c_1^j, \ldots, c_k^j) \ .
\eea
We shall define monads of the form \eref{mon_app} over both an ambient space $\cA = \mathbb{P}^{n_1} \times \ldots \times \mathbb{P}^{n_m}$ and three dimensional projective varieties $X$ defined as complete intersection hypersurfaces (CICYs) in $\cA$.
\section{Monads, Sheaf Cohomology and Computational Algebraic Geometry} \label{appA}
In this Appendix, we briefly outline some basics of commutative algebra
as relevant for computing sheaf cohomology (see Refs.~\cite{schenck,AG1}). In
most computer algebra packages such as Macaulay2 \cite{m2}, 
of which we make extensive use in this paper, these
techniques are essential. Computational algebraic geometry has also been recently used in string phenomenology in \cite{Gray:2006jb, Gray:2006gn,Gray:2008zs,Gray:2007yq,Gray:2005sr} and the reader is referred to tutorials in these papers as well for a quick introduction.
\subsection{The Sheaf-Module Correspondence}
Since we are concerned with compact manifolds, we will focus on projective
varieties in $\IP^m$. A projective algebraic variety is the zero locus of a set of homogeneous
polynomials in $\IP^m$ with coordinates $[x_0:x_1:\ldots:x_m]$.
In the language of commutative algebra, 
projective varieties correspond to homogeneous ideals, $I$, in the polynomial ring
$R_{{\mathbb P}^n} = \IC[x_0, \ldots, x_m]$. An ideal $I\subset R_{{\mathbb P}^n}$,
associated to a variety,
is generated by the defining polynomials of the variety and consists of all
polynomials which vanish on this variety. The quotient ring $A=R_{{\mathbb P}^n}/I$ is called
the {\em coordinate ring} of the variety.

In general, a ring $R$ is called {\it graded} if
\[
R = \bigoplus\limits_{i \in \IZ} R_i, \qquad \mbox{such that }
r_i \in R_i, r_j \in R_j \Rightarrow r_ir_j \in R_{i+j} \; .
\]
For the polynomial ring $R_{{\mathbb P}^n}$ the $R_i$ consists of the homogeneous
polynomials of degree $i$.
In analogy to vector spaces over a field, one can introduce $R$-modules $M$ 
over the ring $R$. In practice, one can think of $M$ as consisting of vectors
with polynomial entries with $R$ acting by polynomial multiplication. 
A module is called {\em graded} if
\[
M =  \bigoplus\limits_{i \in \IZ} M_i, \qquad \mbox{such that }
r_i \in R_i, m_j \in M_j \Rightarrow r_i m_j \in M_{i+j} \ .
\] 
The graded ring $R$ is itself a graded $R$-module, $M(R)$.
Similarly, an ideal $I$ in a graded ring $R$ is a graded $R$-module
and a submodule of $M(R)$. Another important example of a 
graded $R$ module is $R(k)$ which denotes the ring $R$ with
degrees shifted by $-k$. For example, $x^2y \in R_{{\mathbb P}^n}$ is of degree 3, but 
seen as an element of the module $R_{{\mathbb P}^n}(-2)$, its degree is $3+2=5$.

Sheafs over a (projective) variety can also be described as a
module by virtue of the {\it sheaf-module correspondence}.
Given the graded ring $R$ and a finitely generated graded $R$-module
$M$, one defines an associated sheaf $\widetilde{M}$ as follows.
On an open set $U_g$, given by the complement of the zero locus of
$g \in R$, the sections over $U_g$ are
$\widetilde{M}(U_g) = \{ m/g^n |m\in M\, , \mbox{degree}(m) = \mbox{degree}(g^n)\}$.
On $\IP^m$, this looks concretely as follows. A sufficiently fine
open cover of $\IP^m$ is provided by $U_{x_i}$, the open sets where
$x_i\neq 0$. Let us first consider the module $M(R_{{\mathbb P}^n})$, that is, the ring
$R_{{\mathbb P}^n}$ seen as a module. Then $\widetilde{M(R_{{\mathbb P}^n})}(U_{x_i})=
\{f/x_i^m, f \mbox{ homogeneous of degree }n \}$ and, hence,
$\widetilde{M(R_{{\mathbb P}^n})}=\cO_{\IP^m}$, where $\cO_{\IP^m}$ is the trivial
sheaf on $\IP^n$. Similarly, for the modules $R_{{\mathbb P}^n}(k)$ one has
\[
\cO_{\IP^m}(k) \simeq \widetilde{R_{{\mathbb P}^n}(k)} \ .
\]
For projective varieties $X\subset \IP^m$ and associated ideal $I$,
the story is similar. Now, one needs to consider the graded modules over
the coordinate ring $A=R/I$. In particular, for line bundles $\cO_X(k)$
on $X$ one has
\[
 \cO_X(k)=\widetilde{A(k)}\; .
\]
\subsection{Constructing Monads using Computer Algebra}
Recall from \eref{monad2}, that we wish to construct bundles $V$
defined by
\beq\label{monad3}
0 \to V \stackrel{f}{\longrightarrow}
\bigoplus\limits_{i=1}^{r_B} \cO_X(b_i) 
\stackrel{g}{\longrightarrow}
\bigoplus\limits_{i=1}^{r_C} \cO_X(c_i) 
\longrightarrow 0 \ ,
\eeq
over the manifold $X$. In this subsection, we outline how one may
proceed with this construction using commutative algebra packages such
as \cite{m2} and applying the Sheaf-Module correspondence
discussed above. Let $A$ be the coordinate ring of $X$. For example,
for the quintic, $[4|5]$ we can write
\beq
A = \IC[x_0, \ldots, x_4] / \left( \sum_{i=0}^5 x_i^5 + \psi
x_0x_1x_2x_3x_4 \right) \ . \label{Aquintic}
\eeq
where the round brackets denote the ideal generated by the enclosed
polynomial. In practice, we will randomize $\psi$, the complex structure and in
fact work over the ground field $\IZ/p \IZ$ for some large prime $p$
instead of $\IC$ in order to speed up computation. The free modules
corresponding to the bundles $B$, $C$ are given by
$\oplus_{i=1}^{r_B}A(b_i)$, $\oplus_{i=1}^{r_B}A(c_i)$ with grading
$\{b_1, b_2, \ldots, b_{r_B} \}$, $\{c_1, c_2, \ldots, c_{r_C}\}$ and
ranks $r_B$, $r_C$~\footnote{
In most computer packages, the convention is to actually take the
grading to be negative, viz., $\{-b_1, -b_2, \ldots, -b_{r_B} \}$.}.
At the level of modules, the map $g$ can then be specified by an $r_C \times r_B$
matrix whose entries, $g_{ij}$ are homogeneous polynomials of degree $c_i - b_j$,
that is $g_{ij}\in\cO_X(c_i-b_j)$. Indeed, the degrees of the entries of $g$
are so as preserve the gradings of $B$ and $C$ and our choice $c_i \ge b_j$
ensures that such polynomials indeed exist.
Moreover, we choose these polynomials to be random; this corresponds to the
genericity assumption for $g$ used repeatedly in the main text.
%
\subsection{Algorithms for Sheaf Cohomology}
We shall not delve into the technicalities of this vast subject and
will only mention that for commutative algebra packages such as
\cite{m2}, there are built-in routines for computing
cohomology groups of sheafs (modules). The standard algorithm is based on
the so-called Bernstein-Gel'fand-Gel'fand correspondence and on Tate resolutions
of exterior algebras. The interested reader is referred to the books
\cite{schenck} and \cite{macaulay2} for details.

\subsection{A Tutorial}\label{ap:eg}
Let us explicitly present a Macaulay2 code \cite{m2}
for one of the examples from our classification. This will serve to illustrate the
power and relative ease with which computer algebra assists in the
proof of stability and the calculation of the particle spectrum.

Let us take the first rank 4 example for $X=[4|5]$ in Table 10,
which was further discussed around Eq.~\eref{5eg-1}. It is defined by
\beq\label{5eg}
B=\cO_X^{\oplus 2}(2) \oplus  \cO_X^{\oplus 4}(1)\; ,\qquad C= \cO_X^{\oplus 2}(4) \ .
\eeq
We work over the polynomial Ring $R_{\IP^4}$ with variables $x_0, \ldots, x_4$
and the ground field $\IZ/27449$. The (projective) coordinate ring
$A$ of a smooth quintic $X$ is then defined following Eq.~\eqref{Aquintic}.
In Macaulay this reads
{\sf \[\bt{rcl}
R &=& ZZ/27449[x\_\{0\}..x\_\{4\}]; \\
A &=& Proj( R/ideal(x\_\{0\}\verb|^|5 + x\_\{1\}\verb|^|5 +
x\_\{2\}\verb|^|5 + x\_\{3\}\verb|^|5 + x\_\{4\}\verb|^|5 + \\
&&2*x\_\{0\}*x\_\{1\}*x\_\{2\}*x\_\{3\}*x\_\{4\})); \\
\et\]}
Next, we define {\sf o}, the trivial sheaf (line-bundle) over $A$, and
the $A$-modules associated to the bundles {\sf B} and {\sf C}.
{\sf \[\bt{l}
o = OO\_(A);\\
B = module (o\verb|^|2 (2) ++ o\verb|^|4 (1)); \\
C = module (o\verb|^|2 (4));\\
\et\]}
Subsequently, a random, generic map, {\sf gmap}, can be constructed
between {\sf B} and {\sf C} (note that in Macaulay, maps are defined
backwards):
\[
\mbox{{\sf gmap = map(C, B, random(C, B));}}
\]
Finally, we can define $V^*$ as the co-kernel of the transpose of {\sf
  fmap}:
\[
\mbox{{\sf Vdual = sheaf coker transpose fmap;}}
\]
We can check that $V^*$ has the expected rank 4 using the command
\[
\mbox{{\sf print rank Vdual;}}
\]
The cohomologies of ${\sf Vdual}$ are easily obtained, for example,
{\sf \[
\mbox{print rank HH\^{~}\!\!2 Vdual;}
\]}
produces 90, precisely as expected. Likewise, one can verify that 
{\sf HH\verb|^|0 Vdual} gives $0$, as is required by stability.
To compute $n_{10}=h^1(X, \wedge^2 V^*)$, one only needs the following
command
\[
\mbox{{\sf print rank HH\^{~}\!\!1 exteriorPower(2, Vdual);}}
\]
which gives 0, as indicated in \tref{t:SO10}. 
For the non-generic map \eref{5eg-2}, one can define
{\sf \[\bt{rcl}
gmap &=& map (C, B, matrix\{\{4*x\_\{3\}\^{~}\!2, 9*x\_\{0\}\^{~}\!2 +
x\_\{2\}\^{~}\!2,  
8*x\_\{2\}\^{~}\!3,\\
&&2*x\_\{3\}\^{~}\!3,4*x\_\{1\}\^{~}\!3,9*x\_\{1\}\^{~}\!3\},
\{x\_\{0\}\^{~}\!2 + 10*x\_\{2\}\^{~}\!2, x\_\{1\}\^{~}\!2,\\
&&9*x\_\{2\}\^{~}\!3, 7*x\_\{3\}\^{~}\!3,9*x\_\{1\}\^{~}\!3 + x\_\{2\}\^{~}\!3,
x\_\{1\}\^{~}\!3 +7*x\_\{4\}\^{~}\!3 \} \} );
\et\]}
One can then check that the cohomologies of $V^*$
remain unchanged with respect to the generic case, that is,
$h^0(X,V^*)=h^1(X,V^*)=h^3(X,V^*)=0$
and $h^2(X,V^*)=90$ while {\sf rank HH\^{~}\!\!1 exteriorPower(2,
Vdual)} now results in $n_{10}=h^1(X,\Lambda^2V^*)=13$.

The singlets are also easy to compute. The group $H^1(X, V \otimes V^*)$ can be thought of as the global Ext-group $Ext^1(V,V) \simeq Ext^1(V^*, V^*)$; this is, again, implemented in \cite{m2}. The command 
``{\sf print rank Ext\^{~}\!\!1(Vdual, Vdual);}'' will give us 277.

\section{Some useful technical results}\label{proofs}
\subsection{Genericity of Maps}\label{ap:gen}
We first state a  helpful fact regarding the genericity of maps in the ambient space.
Consider a morphism $h:\cB \to \cC$ between two sums of line bundles
$\cB =\oplus_{i=1}^{r_B}\cO (b_i)$ and $\cC =\oplus_{i=1}^{r_C}\cO (c_i)$ on ${\cal A} ={\mathbb P}^m$.
The map $h$ can explicitly be specified by a $r_C\times r_B$ matrix $h_{ij}\in \cO (c_i-b_j)$
and it induces a map $\tilde{h}:H^0({\cal A},\cB )\rightarrow H^0({\cal A} ,\cC )$.
The induced map $\tilde{h}$ is also described by $h_{ij}$ acting on the sections
of $\cB$ and, hence, if the matrix $(h_{ij})$ has maximal rank (almost everywhere) then
$\tilde{h}$ has maximal rank.

\subsection{Proof that $n_1=h^1(X,\wedge^2 V^*)=0$ for the $SO(10)$
  Models}\label{SO(10)proof}
For simplicity, we provide here the argument for rank 4 bundles on the
quintic. As in the previous discussion, the proof is easily extended
to the other cases. We begin once again with the Koszul sequence in
the co-dimension 1 case, this time for $\wedge^2 V^*$:
\beq
0\rightarrow \cN^{* }\otimes \wedge ^{2}\cV^* \rightarrow \wedge
^{2}\cV^* \rightarrow  \wedge ^{2}V^* \rightarrow
0 \ .
\eeq
From this, we have the long exact sequence in cohomology,
\beq
...\rightarrow H^{1}(\mathcal{A},\cN^{* }\otimes \wedge ^{2}\cV^* )
\rightarrow H^{1}(\mathcal{A},\wedge ^{2}\cV^* )\rightarrow
H^{1}(X,\wedge ^{2}V^* )
\rightarrow H^{2}(\mathcal{A},\cN^{* }\otimes
\wedge ^{2}\cV^* )\rightarrow ...\  
\eeq
We will show that $h^{1}(X,\wedge ^{2}V^* )=0$ by proving that
$h^{1}(\mathcal{A},\wedge ^{2}\cV^* )$ and $h^{2}(\mathcal{A},\cN^{* }\otimes
\wedge ^{2}\cV^* )$ both vanish.

We begin with $h^{1}(\mathcal{A},\wedge ^{2}\cV^* )$. To proceed, we
have the exterior power sequences  
\bea
&& 0\to S^2\mathcal{C}^*\to\mathcal{C}^*\otimes
 \mathcal{B}^*\to\Lambda^2\mathcal{B}^*\to\Lambda^2
 \mathcal{V}^*\to 0 \ , \\
&&0\to \mathcal{N}^* \otimes S^2\mathcal{C}^*\to \mathcal{N}^*
\otimes \mathcal{C}^*\otimes \mathcal{B}^*\to \mathcal{N}^*
\otimes \Lambda^2 \mathcal{B}^*\to \mathcal{N}^* \otimes \Lambda^2
\mathcal{V}^*\to 0\; , 
\eea
which we can split into the short exact sequences:
\bea
0\to S^2\mathcal{C}^*\to \mathcal{C}^*\otimes
\mathcal{B}^*\to K_1\to 0 \ , \nn \\ 
0\to K_1\to \Lambda^2 \mathcal{B}^*\to\Lambda^2 \mathcal{V}^*\to
0 \ , 
\label{first}
\eea
and similarly,
\bea
0\to \mathcal{N}^* \otimes S^2\mathcal{C}^*\to \mathcal{N}^*
\otimes \mathcal{C}^*\otimes \mathcal{B}^*\to K_2\to 0 \ , \nn\\
0\to K_2\to \mathcal{N}^* \otimes \Lambda^2 \mathcal{B}^*\to {N}^*
\otimes \Lambda^2 \mathcal{V}^*\to 0\; .\ 
\label{second}
\eea
Each of these generates a long exact sequence in cohomology. Using the
familiar results for the cohomologies of positive and negative line
bundles on the ambient space, from \eqref{first} we immediately obtain
$h^1(\mathcal{A}, \Lambda^2 \mathcal{V}^*)=h^2(K_1)=0$. Likewise,
the cohomology sequence of \eqref{second} leads us to
$h^2(\mathcal{A}, \mathcal{N}^* \otimes\Lambda^2
\mathcal{V}^*)=h^3(K_2)=0$ and
\bea
0\to H^3(\mathcal{A},K_2)\to H^4(\mathcal{A},\mathcal{N}^* \otimes S^2
\mathcal{C}^* )\stackrel{f}{\longrightarrow}
H^4(\mathcal{A},\mathcal{N}^* \otimes\mathcal{C}^*\otimes
\mathcal{B}^* )\to H^4(\mathcal{A},K_2) \to 0 \ . \
\eea
Combining these results we find
\bea
h^2(\mathcal{A}, \mathcal{N}^* \otimes
\wedge^2\mathcal{V}^*)=h^4(\mathcal{A}, \mathcal{N}^* \otimes S^2
\mathcal{C}^*)-{\rm rk}(f) \ .
\eea
Now, as before we note that by maximal rank arguments of \ref{ap:gen}
and Serre duality, 
\beq
{\rm rk}(f)={\rm min}(h^4(\mathcal{A}, \mathcal{N}^* \otimes S^2
\mathcal{C}^*), h^4(\mathcal{A},\mathcal{N}^*
\otimes\mathcal{C}^*\otimes \mathcal{B}^* ))
\eeq 
By direct computation using \cite{m2} we find that $h^4(\mathcal{A},
\mathcal{N}^* \otimes S^2 \mathcal{C}^*)<
h^4(\mathcal{A},\mathcal{N}^* \otimes\mathcal{C}^*\otimes
\mathcal{B}^* )$ for all the bundles in our list. Thus,
$h^2(\mathcal{A}, \mathcal{N}^* \otimes \wedge^2\mathcal{V}^*)=0$ and
we may conclude that
\beq
h^1(\mathcal{A}, \wedge^2\mathcal{V}^*)=0 \ .
\eeq
The argument is the same in spirit for the other manifolds in our
list. The only key difference being the length of the starting Koszul
sequence (which will containing higher wedge powers of
$\mathcal{N}^*$). The resulting cohomology analysis follows
straightforwardly.


\section{Some Mathematical Preliminaries}
\paragraph{Serre Duality: }
For a vector bundle $V$ on a Calabi-Yau threefold $X$, the cohomology groups
of the bundle and its dual are related by Serre duality as: 
\beq\label{serre}
H^i(X,V) \simeq H^{3-i}(X, V^*) \qquad i=0,1,2,3 \ .
\eeq

\paragraph{Atiyah-Singer Index Theorem: }
For a unitary bundle $V$ on a Calabi-Yau threefold $X$, the index theorem relates the index, or the alternating sum of dimensions of the cohomology groups of $V$ with the characteristic classes of the bundle and the manifold:
\beq\label{AS}
\ind(V) = \sum\limits_{i=0}^3 (-1)^i h^i(X,V) = 
\int_X \ch(V) \wedge \td(X) = \frac12 \int_X  c_3(V)
\ ,
\eeq
where $\td(X)$ is the Todd class for the tangent bundle of $X$. Only in
the last equality have we used the fact the both $c_1(TX)$ and
$c_1(V)$ vanish.

\paragraph{Higher Exterior Powers: }
For $SU(n)$ bundles we have that \cite{AG1,AG2,FH}
\beq
\wedge^p V = \wedge^q V^* \qquad  p+q=n \ 
\eeq
and that (cf.Appendix B of \cite{Donagi:2004ia}),
\beq
c_3(\wedge^2 V) = (n-4) c_3(V) \ .
\eeq

\paragraph{The Bott Formula: }
The cohomology of line-bundles over projective space is given by the simple formula;
this is the Bott formula (cf.~e.g.,\cite{Distler:1987ee}), which dictates that
\beq\label{bott}
h^q(\IP^n, (\wedge^p T\IP^n) \otimes \cO(k)) =
\left\{\ba{lll}
{k+n+p+1 \choose p}{k+n \choose n-p} & q = 0 & k>-p-1,\\
1 & q=n-p & k=-n-1,\\
{-k-p-1 \choose -k-n-1}{-k-n-2 \choose p} & q = n & k<-n-p-1,\\
0 & \mbox{otherwise} & \ .
\ea\right.
\eeq

\paragraph{K\"unneth formula: }
The K\"unneth formula gives the cohomology of bundles over direct product of
spaces. For a product of projective spaces, it states that:
\beq\label{kunneth1}
H^n(\IP^{n_1} \times \ldots \times \IP^{n_m}, \cO(q_1, \ldots, q_m)) =
\bigoplus_{k_1+\ldots+k_m = n} H^{k_1}(\IP^{n_1},\cO(q_i)) \times
\ldots \times H^{k_m}(\IP^{n_m},\cO(q_m)) \ ,
\eeq

\paragraph{Kodaira Vanishing Theorem: }
On a K\"ahler manifold, $X$, and $P$ a positive line bundle
the Kodaira vanishing theorem \cite{AG1,AG2} states that 
\beq\label{kodaira}
H^{q}(X,P \otimes K_{X}) = 0 \quad \forall~q>0 \ ,
\eeq
where $K_{X}$ is the canonical bundle on $X$. For a Calabi-Yau manifold, $X$,
$K_{X}$ is trivial and therefore the only non-vanishing cohomology for
a positive line bundle, $P$, on $X$ is $H^{0}(X,P)$.

Parenthetically, let us see what the vanishing theorem looks like
explicitly on the ambient space; this will be useful later on. Now, on
$\cA = \IP^{n_1} \times \ldots \times \IP^{n_m}$, the canonical bundle
\cite{AG2} is $K_{\cA} = \cO_{\cA}(-n_1-1, \ldots, -n_m-1)$. Consider a line
bundle $\cL = \cO(k_1, \ldots, k_m)$ with $k_r \ge 0$ (i.e., not just with
entries strictly greater than 0 but may also admit 0 entries),
using the K\"unneth
formula in \eref{kunneth1}, we have 
\beq\label{kunneth2}\ba{rcl} 
&& H^q(\cA, \cO_{\cA}(k_1 - n_1 -1, \ldots, k_m - n_m - 1) \\ 
&=&  
\bigoplus\limits_{j_1+\ldots+j_m = q} H^{j_1}(\IP^{n_1},\cO(k_1 - n_1-1)) 
\times \ldots \times H^{j_m}(\IP^{n_m},\cO(k_m - n_m - 1)) \ .
\ea\eeq
However, \eref{kodaira} dictates that on each individual factor,
$H^q(\IP^{n_r}, \cO_{\IP^{n_r}}(k_r - n_r -1) = 0$, for all $q > 0$ if
$k_r > 0$. Therefore, since the only decomposition of $q=0$ is when all
$j_r = 0$, if there is a single $k_r$ which is strictly
positive, \eref{kunneth2} implies that $\cL$ has vanishing cohomology
for $q > 0$. In summary, on the ambient space
\beq\label{kodairaA}
h^q(\cA,\cO_{\cA}(k_1, \ldots, k_m)) = \left\{ \ba{ll}
\ind(\cO_{\cA}(k_1, \ldots, k_m)) & q = 0 \\
0 & q > 0 \ ,
\ea \right. \qquad k_r \ge 0 \mbox{ and at least one } k_{r'} > 0 \ .
\eeq

\section{More on Complete Intersection Calabi-Yau Threefolds}\label{a:cicy}
We have introduced the rudiments and set the notation for CICY's in the text,
in this appendix, we also present some detailed properties relevant to our
investigation. Much of the results can be found in \cite{hubsch} but some of the following are new.
\subsection{Chern Classes and Intersection Form}
The Chern classes are given as simple functions of the entries in the 
configuration matrix \cite{hubsch}. For the total Chern class
\beq
c = c_1^r J_r + c_2^{rs} J_r J_s + c_3^{rst} J_r J_s J_t \ ,
\eeq
where $J_r$ is the K\"ahler form in $\IP^{n_r}$, we have
\beq\label{chernX}
c_1^r = 0, \
c_2^{rs} = \frac12 \left[ -\delta^{rs}(n_r + 1) + 
  \sum_{j=1}^K q^r_j q^s_j \right], \
c_3^{rst} = \frac13 \left[\delta^{rst}(n_r + 1) - 
  \sum_{j=1}^K q^r_j q^s_j q^t_j \right] \ .
\eeq
The third and top chern class determines the Euler number of the threefold;
this is done with a notion of integration over $X$,
which can be defined with respect to a measure $\mu$ and
pulled back to a simpler integration over the ambient space $\cA$:
\beq\label{integration}
\int_X \cdot = \int_{\cA} \mu \wedge \cdot \ , \qquad
\mu := \wedge_{j=1}^K \left( \sum_{r=1}^m q^j_r J_r \right) \ .
\eeq
Now, the Euler number of $X$ is given by integrating the top Chern
class over $X$, which here implies that
\beq
\chi(X) = \mbox{Coefficient}(c_3^{rst}J_rJ_sJ_t \cdot \mu, \quad
\prod_{r=1}^m J_r^{n_r}) \ .
\eeq
We shall often place the Euler number as a subscript to the
configuration matrix. For example, the quintic is thus denoted as
$[5]_{-200}$. In summary, 
we shall often denote the topological data of a CICY explicitly as
\beq
X = [Q_{m \times K}]^{(h^{1,1},h^{2,1})}_{\chi} \ ,
\eeq
where $Q$ is the configuration matrix representing the embedding of
$K$ equations in a product of $m$ ambient projective space factors and
the Hodge numbers $(h^{1,1},h^{2,1})$ and the Euler number $\chi$ are
given respectively as super(sub)-scripts. We leave the method of computing
the hodge number to Appendix \ref{a:cicy}.

The final piece of topological data that we will need is the
triple intersection numbers $d_{rst}$ of $X$. 
These are defined by
\beq\label{intersection}
d_{rst}=\int_{X} J^{r}\wedge J^{s} \wedge J^{t}
\eeq
and can be computed using \eref{integration}. The complete list of complete
intersection configuration matrices, complete with their topological
data are available upon request and can also be found at \cite{cypage}.

\subsection{Redundancy in the CICY list}
It is worth observing that the list of CICYs found using the
techniques described in the previous sections is not a list of unique
Calabi-Yau manifolds. It was realised in \cite{Candelas:1990pi,Avram:1995pu,Candelas:1989ug} and more
recently in \cite{Candelas:2007ac} that CICYs, and perhaps Calabi-Yau
manifolds in general, are connected in moduli space. However, during
the investigations and resurrection of CICYs of late, it was further
realised that even the original list of 7890 may have redundancies
\cite{philip}. This is a relatively new observation and should be
pointed out. 

Now, Wall's theorem (cf.~\cite{He:1990pg}) states that for real
sixfolds, the intersection form and the second Potryagin class suffice
to distinguish non-isormophism. Though for complex threefolds, these
are not enough, the two quantities are good indicators (and will be enough to distinguish our heterotic models). Therefore, 
we propose a simple check for redundancy to be to check through the
basic topological invariants:
hodge numbers, euler
number, $c_2(TX)$, and intersection numbers, and look for any
CICYs with identical sets, up to permutation. 

Upon implementing such a
scan one finds, of the 7890 in the original list, that 
there are 378 sets of redundancies, consisting of equivalent pairs,
triples, or
even n-tuples for n as large as 6; these are expected to have
isomorphism. In all, 813 manifolds are involved; taking one
representative from each of the 378 sets, a total of 435 CICY seem
redundant. Throughout the rest of the paper, however, we will adhere
to the original identifier names of the manifolds to avoid confusion
and shall point out explicitly, where necessary, the equivalences. We turn now to determining the necessary topological data of the manifolds.

%
\subsection{Hodge Numbers}
We wish to know the full topological data of $X$ including the
Hodge numbers $h^{1,1}$ and $h^{2,1}$, whose difference, by the Index
Theorem \eref{AS}, is the Euler number $\chi(X)$:
\beq
h^{1,1} - h^{2,1} = \frac12 \chi(X) \ .
\eeq
Therefore, it suffices to compute either one of the two, which will
turn out to be a much more involved exercise and is the subject of
\cite{Green:1987cr}. Sadly, the actual data for these hodge numbers are lost,
but it is nevertheless expedient to review the methods of
\cite{Green:1987cr} and
reconstruct the hodge numbers because the
techniques can be readily applied to the monad bundles which we shall
study later.

Recalling that, by Hodge decomposition,
\beq
H^{p,q}(X) \simeq H^q(X, \wedge^p T^*X) \ ,
\eeq
where $T^*X$ is the cotangent bundle of $X$, we have that
\beq\label{h11h21}
H^{1,1}(X) = H^1(M, T^*X), \qquad
H^{2,1}(X) \simeq H^{1,2} = H^2(X, T^*X) \simeq H^1(X, TX) \ .
\eeq
In the second part of the above expression, we have used the nowhere
vanishing $(3,0)$-form to establish the isomorphism between
$H^{2,1}(X)$ and $H^{1,2}(X)$ as well as Serre duality, \eref{serre},
to establish the isomorphism between $H^2(X, T^*X)$ and $H^1(X, TX)$,
where $TX$, the dual to the cotangent bundle, is simply the tangent
bundle of $X$.

We can therefore concentrate on the computing $H^1(X, TX)$. We invoke
the Euler sequence which states that,
for an embedding of $X$ into ambient space $\cA$, there is a short exact
sequence
\beq
\sseq{TX}{T\cA|_X}{\cN_X} \ ,
\eeq
where $\cN_X$ is the normal bundle to $X$ embedded in $\cA$ and $T\cA|_X$ is
the restriction of the tangent bundle of $\cA$ to $X$. This induces
a long exact sequence in cohomology as
\beq
\ba{cccccccc}
0 
& \to & H^0(X, TX) & \to & H^0(X, TA|_X) & \to & H^0(X, N_X) & \to
\\
& \to & H^1(X, TX) & \stackrel{d}{\to} 
  & H^1(X, TA|_X) & \to & H^1(X, N_X) & \to \\
& \to & H^2(X, TX) & \to & \ldots &&&
\ea
\eeq
Using the fact $H^0(X, TX) = H^{1,3}(X) = 0$ because $X$ is Calabi-Yau, 
the relations \eref{h11h21}, and the fact that $\rk(d) = 0$ (cf.~eq (6.1)
of \cite{Green:1987cr}), we have the short exact sequence
\beq
\sseq{H^0(X, TA|_X)}{H^0(X, N_X)}{H^{2,1}(X)} \ .
\eeq
Whence,
\beq\label{h21}
h^{2,1}(X) = h^0(X, N_X) - h^0(X, TA|_X) \ .
\eeq

\subsection{Hodge Number Obstructions}
Making use of the essential techniques of Leray tableaux and Koszul resolutions described in Chapter \ref{c:line_bundle_coho}, it is in principle straightforward to compute the Hodge numbers of complete intersection 3-folds, the two terms in \eref{h21}. However, direct calculation shows that one quickly encounters certain obstructions to the computation which will naturally divide our set of 7890 configurations.
\paragraph{Trivial Direct Products}
First of all, we recognise that there are trivial cases in the list,
comprising of CICYs which are simply direct products of
lower-dimensional Calabi-Yau manifolds, viz., $K3 \times T^2$ and
$T^6$. These generically have reduced holonomy and we shall not
consider them.  Therefore, our list is
immediately reduced to be of length 7868.
\paragraph{Normal Bundle and Obstructions}
The $E_1^{j,k}(\cN_X)$ tableaux is readily established for the normal bundle
$\cN_X$ according to \eref{leray}, \eref{bott} and \eref{kunneth1}. It
turns out that if there exists $j \le j'$ in $[-K,0]$ such
that
\beq
E^{j,j} \ne  0 \mbox{ and } E^{j',j'-1} \ne 0 \ ,
\eeq
then, we would call such a case ``normal bundle obstructed,'' which
needs to be addressed separately \cite{Green:1987cr}.
For the remaining, the Leray
spectral sequence actually terminates at the $E_1$ stage and we can
read out the required cohomology as \cite{Green:1987cr}:
\beq\label{h0N_M}
h^0(X, N_X) = \sum_{j=0}^K e_1^{j,j}(N_X) + 
  \sum_{j=1}^K \sum_{l=0}^{j-1} (-1)^{j+l} e^{l,j}_1(N_X) \ .
\eeq
In the above, we have used, and shall henceforth adopt, the notation
that as $h^j$ denotes the dimension of the cohomology group $H^j$,
$e^{j,k}_r$ denotes the dimension of $E^{j,k}_r$.

Now, we find a total
of 12 of the normal bundle obstructed cases, 
with identifiers 1443, 1877, 2569, 2980, 3747, 4228,
4448, 4757, 6174, 6229, 7236 and 7243. For these, \cite{Green:1987cr}
gives a rule to replace the configuration matrix into a isomorphic one
and for which the obstruction is no longer encountered as one computes
$E^{j,k}_1$ and \eref{h0N_M} can be again directly applied.
\paragraph{Tangent Bundle and Obstructions}

Like the normal bundle spectral sequence, in general, the tangent bundle
spectral sequence can also be obstructed (i.e. one cannot compute $H^{q}(TX)$
without knowledge of specific maps). However, for the case of compete
intersection calabi-yau manifolds we are saved from a difficult analysis by
several useful results. 

The first such result is that for a particular class of configurations (those
without a decomposing $(n-1)$-leg, see \cite{Green:1987cr} for a description of the
dot/leg diagrams and notation), $E_{1}^{q+k,k}(V)$ vanishes for $q\geq n-1$
for any bundle $V$ on $X.$ It turns out that if a diagram representing a
Calabi-Yau $3$-fold has no decomposing $1$-legs, $H^{1}(X,T\mathcal{A})$ vanishes
and no decomposing $2$-legs implies that $H^{2}(X,N)=0$ so that the sequences%
\begin{align}
0  & \rightarrow H^{0}(X,T\mathcal{A})\rightarrow H^{0}(X,N_X)\rightarrow
H^{1}(X,TX)\rightarrow0\nonumber\\
0  & \rightarrow H^{1}(X,N_X)\rightarrow H^{2}(X,TX)\rightarrow H^{2}%
(X,T\mathcal{A})\rightarrow0
\end{align}
are exact \cite{Green:1987cr}.

For $3$-folds with decomposing $1$-legs the hodge numbers can be computed by
relying on the classification of complex surfaces (see eq.(2.4)in \cite{Green:1987cr}). Simple formulas for these
hodge numbers in terms of sub-diagrams were found in \cite{Green:1987cr},\cite{Green:1987zp}. For the bulk of cases, however, the diagrams have no decomposing $1$-legs.

Further, it can be shown that an $n$-fold configuration with the property of a
decomposing $(n-1)$-leg is equivalent to another one with no decomposing
$(n-1)$-leg \cite{Green:1987cr}. So in analysing configurations representing Calabi-Yau $3$-folds it is sufficient to look only at configurations with no decomposing 2-legs. This leads to the following structure%
\begin{align}
E_{1}^{0,0}(T\mathcal{A})  & =\bigoplus_{r=1}^{m}H^{0}(\mathbb{P}_{r}^{n_{r}%
},T(\mathbb{P}_{r}^{n_{r}}));\\
E_{1}^{K+2,K}(T\mathcal{A})  & \approx C^{m};\\
E_{1}^{q+k,k}(T\mathcal{A)}  & \backepsilon H^{0}(\mathbb{P}_{r}^{n_{r}%
},T(\mathbb{P}_{r}^{n_{r}})\otimes h_{r}^{-1})\approx C^{n_{r}+1},\\
\forall\{A,r  & :\sum_{a\in A}q_{a}^{r}=1,k=\left|  A\right|  +1\}\nonumber
\end{align}
where $\left|  \cA\right|$ denotes the cardinality of $\cA$, the set of indices
labeling a subset of constraints which act only in a $(q+k)$-dimensional
factor of the product $\cA$. With these results in hand, we can compute the
hodge numbers of $X$.

\section{Favourable CICYs and Positive Monads}\label{s:posCICY}
We found in Chapter \ref{c:posCICY} that positive monads are quite rare and
indeed, of the 4515 favourable CICYs, only 36 admit these. These
manifolds are presented in \tref{t:pos-cicy2}.
\begin{table}
$\ba{|l|l|l|l|l|} \hline
n \backslash K & 1 & 2 & 3 & 4 \\ \hline \hline
1 & [5] &  [3 \; 3 ], [4 \; 2 ] & [3 \; 2 \; 2 ] & 
  [2 \; 2 \; 2 \; 2] \\[0.2cm]  \hline
2 & \ba{l}
    \tconf{ 2 \cr 4 \cr  } , 
    \tconf{ 3 \cr 3 \cr  }
    \ea
    &
    \ba{l}
    \tconf{0 & 2 \cr 2 & 3 \cr  } ,
    \tconf{0 & 2 \cr 3 & 2 \cr  } , 
    \tconf{0 & 2 \cr 4 & 1 \cr  } , 
    \tconf{ 1 & 1 \cr 3 & 2 \cr  } , \\[0.2cm]
    \tconf{ 
    1 & 1 \cr 4 & 1 \cr  } , \tconf{ 2 & 1 \cr 1 & 3 \cr  } , \tconf{ 
    2 & 1 \cr 2 & 2 \cr  } , \tconf{ 2 & 1 \cr 3 & 1 \cr  }
    \ea
    &
    \ba{l}
    \tconf{ 
    0 & 0 & 2 \cr 2 & 2 & 2 \cr  } , 
    \tconf{ 
    0 & 0 & 2 \cr 3 & 2 & 1 \cr  } , 
    \tconf{ 
    0 & 1 & 1 \cr 2 & 2 & 2 \cr  } , \\[0.2cm]
    \tconf{ 
    0 & 1 & 1 \cr 2 & 3 & 1 \cr  } ,
    \tconf{ 
    0 & 1 & 1 \cr 3 & 2 & 1 \cr  } , 
    \tconf{ 
    0 & 1 & 1 \cr 4 & 1 & 1 \cr  } , \\[0.2cm]
    \tconf{ 
    0 & 2 & 1 \cr 2 & 2 & 1 \cr  } , 
    \tconf{ 
    1 & 1 & 1 \cr 3 & 1 & 1 \cr  } , 
    \tconf{ 
    2 & 1 & 1 \cr 2 & 1 & 1 \cr  } 
    \ea
    &
    \ba{l}
    \tconf{0 & 0 & 1 & 1 \cr 2 & 2 & 2 & 1 \cr  } \\[0.2cm] 
    \tconf{0 & 0 & 1 & 1 \cr 3 & 2 & 1 & 1 \cr  }
    \ea \\[0.2cm]  \hline
3 & \tconf{ 2 \cr 2 \cr 3 \cr  } &
    \ba{l} 
    \tconf{ 
      0 & 2 \cr 1 & 2 \cr 1 & 2 \cr  } , \tconf{ 
      1 & 1 \cr 0 & 2 \cr 1 & 3 \cr  } , \tconf{ 
      1 & 1 \cr 0 & 2 \cr 2 & 2 \cr  } , \tconf{ 
      1 & 1 \cr 1 & 1 \cr 1 & 3 \cr  } , \\[0.3cm] 
    \tconf{ 
      1 & 1 \cr 1 & 1 \cr 2 & 2 \cr  } , \tconf{ 
      1 & 1 \cr 1 & 2 \cr 0 & 3 \cr  } , \tconf{ 
      1 & 1 \cr 2 & 1 \cr 2 & 1 \cr  }  
    \ea
    &
    \tconf{1 & 1 & 0 \cr 1 & 0 & 1 \cr 3 & 1 & 1 \cr  } & - 
    \\[0.2cm]  \hline
4 & \tconf{2 \cr 2 \cr 2 \cr 2 \cr } &-&-&- \\ \hline
\ea$
\caption{\emph{The 36 favourable CICYs which admit positive monads. The rows,
indexed by $n$, signify the number of embedding product projective spaces 
and the columns, indexed by $K$, signify the co-dimension, i.e., the number
of defining equations.}}
\label{t:pos-cicy2}
\end{table}
We see that the 5 cyclic manifolds with $n=1$, as 
discussed in \cite{Anderson:2007nc},
comprise the first row of the list. The codimension $K=1$ CICYs are
transposes thereof; for the meaning of such a transposition, see
\cite{Candelas:2007ac}.

\addcontentsline{toc}{chapter}{Bibliography}
\bibliographystyle{hep}
\bibliography{thesis}

\end{document}